%% file: thesis_2side.tex
\documentclass[12pt,twoside]{report}
\pdfoutput=1
\usepackage{geometry}
\usepackage{setspace}
\usepackage{graphicx}
\usepackage{natbib}
\usepackage{amsmath}
\usepackage{amssymb}
\usepackage{lscape}
\usepackage{rotating}
\usepackage{longtable}
\begin{document}

\pagenumbering{roman}

\begin{titlepage}
\begin{center}
\Huge
\vspace*{1.0cm}
{\sc Ultracam} photometry of eclipsing \\ cataclysmic variable stars \\
\LARGE
\vspace{1.0cm}
William~James~Feline \\ \large \vspace{8.0cm} Department of Physics
and Astronomy \\ The University of Sheffield \\ 
\vspace{1.0cm} July, 2005 \\ 
\normalsize
\vspace{1.0cm}
{\it A thesis submitted to the University of Sheffield \\ in
  partial fulfilment of the requirements \\ of the degree of
  Doctor of Philosophy}
\end{center}
\end{titlepage}

\cleardoublepage

\bibliographystyle{apj}

\section*{Declaration}
\addcontentsline{toc}{section}{Declaration}
% * means this section not numbered
I declare that no part of this thesis has been accepted, or is
currently being submitted, for any degree or diploma or certificate or
any other qualification in this University or elsewhere.

This thesis is the result of my own work unless otherwise stated.
\cleardoublepage

\include{summary}
% * means this section not numbered
%insert summary here

\cleardoublepage
\markboth{CONTENTS}{ULTRACAM PHOTOMETRY OF ECLIPSING CVS}
\tableofcontents

\cleardoublepage
\markboth{LIST OF FIGURES}{ULTRACAM PHOTOMETRY OF ECLIPSING CVS}
\addcontentsline{toc}{section}{List of Figures}
\listoffigures

\cleardoublepage
\markboth{LIST OF TABLES}{ULTRACAM PHOTOMETRY OF ECLIPSING CVS}
\addcontentsline{toc}{section}{List of Tables}
\listoftables

\cleardoublepage
\markboth{ACKNOWLEDGMENTS}{ULTRACAM PHOTOMETRY OF ECLIPSING CVS}
\include{acknowledgments}

%insert thesis here(!)

%changes right-hand header to title of thesis
\cleardoublepage
\markboth{CHAPTER 1. INTRODUCTION}{ULTRACAM PHOTOMETRY OF ECLIPSING
  CVS}
\include{introduction}
%includes file called introduction.tex

\cleardoublepage
\markboth{CHAPTER 2. OBSERVATIONS \& DATA REDUCTION}{ULTRACAM
  PHOTOMETRY OF ECLIPSING CVS}
\include{observations}

\cleardoublepage
\markboth{CHAPTER 3. ANALYSIS TECHNIQUES}{ULTRACAM PHOTOMETRY OF
  ECLIPSING CVS}
\include{analysis}

\cleardoublepage
\markboth{CHAPTER 4. OU~VIR}{ULTRACAM PHOTOMETRY OF ECLIPSING CVS}
\include{ouvir_results}

\cleardoublepage
\markboth{CHAPTER 5. XZ~ERI \& DV~UMA}{ULTRACAM PHOTOMETRY OF
  ECLIPSING CVS}
\include{xzeridvuma}

\cleardoublepage
\markboth{CHAPTER 6. GY~CNC, IR~COM \& HT~CAS}{ULTRACAM PHOTOMETRY OF
  ECLIPSING CVS}
\include{gycncircomhtcas}

\cleardoublepage
\markboth{CHAPTER 7. CONCLUSIONS AND FUTURE WORK}{ULTRACAM PHOTOMETRY
  OF ECLIPSING CVS}
\include{conclusions}

\cleardoublepage
\markboth{BIBLIOGRAPHY}{ULTRACAM PHOTOMETRY OF ECLIPSING CVS}
\addcontentsline{toc}{chapter}{Bibliography}
\bibliography{/home/wf/paper/ouvir/abbrev,/home/wf/paper/ouvir/refs}

\end{document}

%% file: summary.tex
\section*{Summary}
\label{summary}
\addcontentsline{toc}{section}{Summary}

The accurate determination of the masses of cataclysmic variable stars
is critical to our understanding of their origin, evolution and
behaviour. Observations of cataclysmic variables also afford an
excellent opportunity to constrain theoretical physical models of the
accretion discs housed in these systems. In particular, the brightness
distributions of the accretion discs of eclipsing systems can be
mapped at a spatial resolution unachievable in any other astrophysical
situation. This thesis addresses both of these important topics via
the analysis of the light curves of six eclipsing dwarf nov\ae,
obtained using {\sc ultracam}, a novel high-speed imaging photometer.

The physical parameters of the eclipsing dwarf nov\ae\ OU~Vir, XZ~Eri
and DV~UMa are determined from timings of the white dwarf and bright
spot eclipses. For XZ~Eri and DV~UMa the physical characteristics are
also calculated using a parameterized model of the eclipse, and the
results from the two techniques critically compared. This work marks
the first accurate determination of the system parameters of both
OU~Vir and XZ~Eri. The mass of the secondary star in XZ~Eri is found
to be close to the upper limit on the mass of a brown dwarf.

The brightness distributions of the accretion discs of the six
eclipsing dwarf nov\ae\ OU~Vir, XZ~Eri, DV~UMa, GY~Cnc, IR~Com and
HT~Cas are determined using an eclipse mapping technique. The
accretion discs of the first five objects are undetected in the
observations, as expected for short-period quiescent dwarf nov\ae\
with low mass transfer rates. The
observations of HT~Cas, however, show significant changes in the
brightness distribution of the quiescent accretion disc between 2002
September and 2003 October, which are related to the overall system
brightness. These differences are caused by variations both in the
rate of mass transfer from the secondary star and through the
accretion disc. The disc colours indicate that it is optically thin in
both its inner and outer regions. I estimate the white dwarf
temperature of HT~Cas to be $15\,000\pm1000$~K in 2002 and
$14\,000\pm1000$~K in 2003.

%% file: acknowledgments.tex
\chapter*{{\em Acknowledgments}}
\label{acknowledgments}
\addcontentsline{toc}{chapter}{Acknowledgments}

First and foremost I must extend my sincere thanks to Vik Dhillon for
the excellent supervision he has given me both as an undergraduate and
a graduate student. His infectious enthusiasm, energy and knowledge
have been hugely appreciated throughout my studies. I would also like
to thank my `co-supervisor' Chris Watson for all his help. A worthy
opponent, too, at many sports (badminton excepted!). Another who has
more than earned his place in this section is Tom Marsh. My thanks to
him for
all of his software which he so expertly wrote and so generously
donated to the cause, and for his expert advice. This thesis would
have been much the worse (and much delayed) without the computer
genius of Paul `Pablo' Kerry and his expertise in keeping the systems
running smoothly. Thank you too, to everybody else in the Sheffield
astronomy group for their help throughout my Ph.D., particularly Tim
Thoroughgood and Stu Littlefair. I would also like to acknowledge the financial
support of PPARC and the contribution of everybody involved with the
{\sc ultracam} project, especially Mark `Stevo' Stevenson. Dr.~Mark
A.\ Garlick kindly permitted me to use his artwork `Magnetic
Accretion' (1998; figure~\ref{fig:polar}) and `Intermediate Polar'
(2001; figure~\ref{fig:intpolar}) in this thesis. More examples can be
found at {\tt www.space-art.co.uk}.

Of course, being a Ph.D.\ student is not just about writing the damn
thesis. On this count, I'd like to thank everyone else in the office
and department with whom I've spent my ill-gotten gains with down the
pub (and the Leadmill, and Balti King\ldots). A brief
mention of latex gloves, lemon jelly and mustard seeds is all that is
required, I think. I'd also like to thank my fellow members of
Sheffield University Bankers Hockey Club, although they'll never read
this, and to all my other friends, for being good mates throughout my
time in Sheffield and a welcome break from the office.

For nurturing my youthful curiosity, and for so many years of
support, advice and tolerance, I thank my parents, Timothy and
Jacqueline. Thanks also to my brother, David, and my sister, Eleanor.

My final thank-you goes, of course, to the lovely Laura, for all her
support, kind words and patience over the years.

\vspace{2.0cm}
\begin{center}
{\em For my parents.}
\end{center}

%% file: introduction.tex
\chapter{Introduction}
\label{ch:introduction}

\pagenumbering{arabic}

\section*{Context---}
\addcontentsline{toc}{section}{Context}

\begin{quotation}
  ``On the evening of December 15th, 1855, I remarked \ldots an object
  shining as a star of the ninth magnitude, with a very blue planetary
  light, which I have never seen before during the five years that my
  attention has been directed to this quarter of the heavens. On the
  next fine night, Dec.\ 18th, it was certainly fainter than on the
  15th by half a magnitude or more. Since that date I have not had an
  opportunity of examining it till last evening, January 10th, when
  its brightness was not greater than that of stars of the twelfth
  magnitude.''
\end{quotation}
This description, by J.~R.~\citet{hind1856}, marked the discovery of a
new class of variable star---the {\em dwarf nov\ae}. The star was soon
christened U~Geminorium, and in subsequent years became an exemplar of
its type. Significantly, Hind noted that the object appeared very
blue, implying high temperatures, which differentiated it from other
variables such as Algol (an eclipsing binary) and S~Cancri (a Mira
variable). Furthermore, U~Gem varied seemingly at random, a fact later
bemoaned by \citet{parkhurst1897}: ``Predictions with regard to it can
better be made after the fact.''

The U~Gem stars eventually became known as {\em dwarf nov\ae\/}
\citep{payne38}, by comparison to the even more spectacular {\em
nov\ae\/} which have been observed since antiquity, and are now called
classical, or old, nov\ae. These objects now form part of the group of
stars referred to as the {\em cataclysmic variables}.

The currently accepted model of cataclysmic variables was originally
developed by \citet{kraft59,kraft62}, who proposed that
\begin{quotation}
``\ldots all members of this group are spectroscopic binaries of short
period. \ldots the blue stars in these systems are probably white
dwarfs. The masses of the red components and their spectra \ldots seem
consistent with a
star of mass $\sim1\;{\rm M}_{\odot}$. \ldots the red stars overflow their
lobes of the inner Lagrangian surface; the ejected material forms, in
part, a ring, or disc, surrounding the blue star.''
\end{quotation}
This model has subsequently been expanded on \citep{warner71,smak71},
but remains essentially valid. In fact, any system which fits this
description can accurately be described as a cataclysmic variable.

\section{The canonical scheme}

\begin{figure}
\centerline{\includegraphics[width=12cm,angle=0]{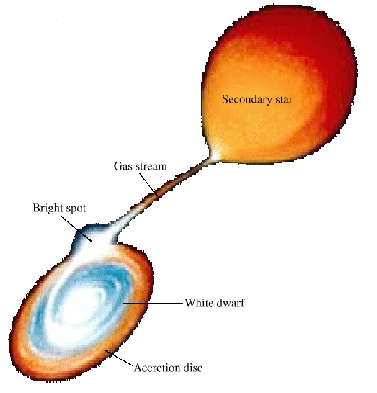}}
\caption[An artist's impression of a non-magnetic CV.]{An artist's
  impression of a non-magnetic CV, with the red dwarf secondary, gas
  stream, accretion disc and white dwarf primary marked.}
\label{fig:cv}
\end{figure}

Cataclysmic variable stars (CVs) are semi-detached binary systems,
with orbital periods of a few hours. The secondary star (of mass
$M_{2}$), usually a main-sequence star, transfers material to the
white dwarf primary (of mass $M_{1}$). In non-magnetic systems (in
which the magnetic field of the white dwarf is too weak to affect the
accretion flow), the
material is transferred via a gas stream, and then spirals round the
primary star, forming an accretion disc. The collision of the gas
stream with the accretion disc forms a so-called `bright spot,' a
shock-heated region of emission at the edge of the disc. At the inner
edge of the accretion disc, the disc material, orbiting in Keplerian
orbits, is decelerated to match the surface velocity of the white
dwarf in a {\em boundary layer}. If the white dwarf has a significant
magnetic field, however, the disc and boundary layer can be
partially (in the case of {\em intermediate polars\/}) or totally (in
the case of {\em polars\/}) disrupted and the accreting material
instead flows along magnetic field lines onto the surface of the
primary star (see \S~\ref{sec:magcvs}). Figure~\ref{fig:cv} shows an
artist's impression of a non-magnetic CV, with the main features
labelled.

The name cataclysmic comes from the violent but non-destructive
outbursts that first drew attention to these objects (see
\S~\ref{sec:novae}). These periodic 
outbursts mean that monitoring of CVs is a popular and fruitful task
among many amateur astronomers.

In systems which are inclined at large angles to our line of sight
($i\gtrsim70^{\circ}$), eclipses of the various components occur,
which can lead to fine structure in the eclipse morphology. Eclipses
of the white dwarf and bright spot are sharp (of the order of tens of
seconds), due to the compact nature of these regions, and are
superimposed on the more gradual eclipse of the extended accretion
disc. As the bright spot rotates into view, it can give rise to an
increase in the observed flux, resulting in an `orbital hump' in the
light curve.

\section{The Roche-lobe}
\label{sec:roche_lobe}

The orbital separation $a$ of the binary components is, from Newton's
form of Kepler's third law, a function of the mass of each component and
the orbital period, $P_{\rm{orb}}$:
\begin{equation}
\label{eq:kepler}
a^{3} = \frac{(M_{1}+M_{2})GP^{2}_{\rm{orb}}}{4\pi ^{2}},
\end{equation}
where $G$ is the gravitational constant. Given that the masses of the
stellar components of CVs are approximately solar, and that the
orbital periods are of the order a few hours, equation~\ref{eq:kepler}
implies binary separations of the order of one solar radius.

Such short orbital periods and close proximity mean that tidal forces
from the gravitational field of the primary and centrifugal forces
from the rotation cause the secondary star in CVs to be distorted into
a teardrop shape from the spherical shape that an isolated star would
assume. These tidal forces also ensure that the secondary is tidally
locked: it rotates at the same rate as it orbits. The time-scale for
synchronization is short, as material flowing into and out of the
tidal bulge will obviously expend a great deal of energy in doing
so. In contrast, the small radius of the primary means that it remains
effectively immune from such forces and its shape remains spherical.

\begin{figure}
\centerline{\includegraphics[width=10cm,angle=0]{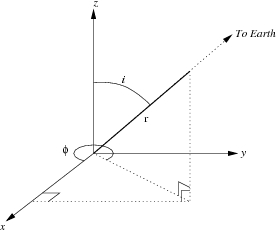}}
\caption[The Cartesian co-ordinate system used throughout this
thesis.]{The Cartesian co-ordinate system used throughout this
thesis. The frame is co-rotating with the binary system, with the
primary star at the origin. The {\em x\/}-axis is the line joining the
centres of the two stars, with {\em x\/} increasing towards the
secondary; the {\em y\/}-axis is in the binary plane, perpendicular to
the {\em x\/}-axis and in the direction of orbital motion; and the
{\em z\/}-axis is perpendicular to the binary plane. $i$ denotes the
orbital inclination, $\phi$ the orbital phase and $\rm{r}$ is the
length of the vector pointing towards Earth.}
\label{fig:axes}
\end{figure}

Before going into the details of the Roche geometry, I first define a
co-ordinate system to use. As is usual, I use a set of right-handed
Cartesian co-ordinates, with the {\em x\/}-axis being defined as the
line joining the centres of the two stars; the {\em y\/}-axis is in the
orbital plane, perpendicular to the {\em x\/}-axis and in the
direction of orbital motion; and the {\em z\/}-axis is perpendicular to
the binary plane. This co-ordinate system is illustrated in
figure~\ref{fig:axes}.

The total potential of the system is given by the sum of the
gravitational potentials of the two stars and the effective potential
of the centrifugal force. In the above co-ordinate system, the total
potential is therefore \citep*{kruszewski66,pringle85,frank85}
\begin{equation}
\Phi = -\frac{GM_{1}}{aR_{1}} - \frac{GM_{2}}{aR_{2}} - \frac{2 \pi
  ^{2}a^{2}}{P^{2}_{\rm orb}} \left [ \left (x-\frac{M_{2}}{M_{1}+M_{2}}
  \right )^{2} + y^{2} \right ],
\end{equation}
where $R_{1}$ and $R_{2}$ are the distances from the relevant star and
$y$ and $x$ are the distances along the relevant axes, all in units of
the orbital separation $a$.

\begin{figure}
\centerline{\includegraphics[width=15cm,angle=0]{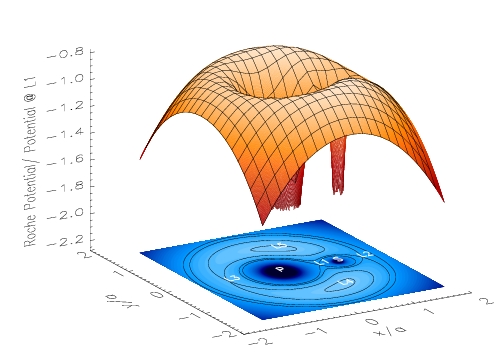}}
\caption[Roche potentials.]{Different representations of the Roche
  potential. The potential scale is normalised such that the potential
  at the ${\rm L}_{1}$ point is $-1.0$. The orange surface is a
  `rubber-sheet' representation of the Roche potential. The two deep
  depressions correspond to the two stars. This surface is
  colour--coded: darker colours indicate a more negative
  potential. The black grid superimposed on this surface shows the
  potential on lines of equal {\em x\/}- and {\em y\/}-values. The
  blue plane below this shows the same potential, but in two
  dimensions. The black curves superimposed on this plane are Roche
  equipotentials. The `critical potential' on which the ${\rm L}_{1}$
  point lies is one of them (the figure of eight). All the Lagrangian
  points are marked, as are the positions of the primary and secondary
  stars. The mass ratio $q$ is 0.175 and the orbital period is 1.74
  hours.}
\label{fig:3Dpotentials}
\end{figure}

Contours of equal potential, $\Phi=$~{\em const}, are known as {\em
Roche equipotentials}. The shapes of these equipotentials are
functions only of the mass ratio\footnote{Occasionally I will refer to
mass ratios $>1$. In these cases I remain consistent with the
definition of the secondary star being the mass donor.}
$q=M_{2}/M_{1}$, and their scale depends on the orbital
separation. The Lagrangian points 1--5 (${\rm L}_{1}\ldots {\rm
L}_{5}$), first discovered in 1772 by Lagrange, satisfy
\begin{equation}
\label{eq:lagrangianpts}
\frac{\partial \Phi}{\partial (x,y,z)} = 0,
\end{equation}
so a test particle at a Lagrangian point experiences no net force. As
figure~\ref{fig:3Dpotentials} illustrates, this is an unstable
equilibrium, as the Lagrangian points are potential maxima.

The largest closed equipotentials of each component meet at the inner
Lagrangian point ${\rm L}_{1}$ (see
figure~\ref{fig:3Dpotentials}). The surface defined for each component
by this equipotential is called the {\em Roche-lobe\/} of that star;
the potential defining the Roche-lobe is known as the {\em critical
potential}. Once the Roche-lobe is filled,
equation~\ref{eq:lagrangianpts} shows that the material at the ${\rm
L}_{1}$ point can easily transfer to the other star (see
\S~\ref{sec:gasstream}), with the initial impetus being given by the
gas pressure of the secondary star's atmosphere.

It is frequently useful to use the {\em volume radius\/} of the
Roche-lobe $R_{\rm{L}}$ as a measure of the size of the
secondary. This is defined as the radius of the sphere that would have
the same volume as the Roche-lobe \citep{eggleton83}:
\begin{eqnarray}
\label{eq:RL2}
R_{\rm{L}}=\frac{0.49aq^{\frac{2}{3}}}{0.6q^{\frac{2}{3}}+
  \ln\left(1+q^{\frac{1}{3}}\right)}\,,&&0<q<\infty
\end{eqnarray}
which is accurate to better than 1~per~cent.

\begin{figure}
\centerline{\includegraphics[width=3cm,angle=90]{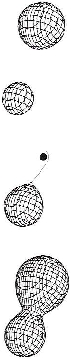}}
\caption[Types of close binary system.]{Different types of close
  binary systems. From left: detached binaries have both components
  within their Roche-lobes (e.g.\ NN~Ser); semi-detached binaries
  (including all CVs) have only one component within its Roche-lobe,
  the other fills its Roche-lobe and can transfer mass to the detached
  star; in contact binaries both stars overfill their Roche-lobes
  (e.g.\ W~UMa stars). From \citet{hellier01}.}
\label{fig:close_binaries}
\end{figure}

I have previously stated that CVs are `semi-detached binary systems.'
These are a sub-type of binary stars known collectively as `close
binary systems.' The defining characteristic of close binary stars is the
presence of an interaction between the two components other than that
of gravity. Alternatively and equivalently, close binaries can be
defined as systems in which the two stars affect each others'
evolution. This interaction can take the form of irradiation of one
star by the other, or as in the case of CVs, mass being transferred
from one star to the other. Close binaries come in three flavours,
illustrated in figure~\ref{fig:close_binaries}:
detached binaries have both stellar components contained within their
respective Roche-lobes; semi-detached binary stars (including all CVs)
have only one component within its Roche lobe, the other fills its
Roche-lobe and can transfer mass to the detached star; in the case of
contact binaries both stars overfill their Roche-lobes. 

\section{Classification of cataclysmic variables}
The classification of CVs is rooted in the historical observations of
these objects, which concentrated, for obvious reasons, on the
spectacular outbursts that characterise these stars and lend them
their name. The amplitude and duration of outbursts were obvious
parameters by which to classify CVs in the past, and by-and-large
remain so today.

\subsection{Classical and recurrent nov\ae}
\label{sec:novae}
The nov\ae\ that first drew the eye to CVs are dramatic increases in
brightness of these stars. The amplitude of these eruptions ranges
between 6 and $\sim19$ magnitudes, and last for a few days to
years. These eruptions are of such magnitude that to ancient
astronomers they appeared to be new stars. Ancient Chinese astronomers
dubbed them `guest stars,' whereas in the West they became known as
{\em nov\ae\ stella}.

{\em Classical nov\ae\/} have by definition only been {\em observed\/} to
go nova once. These are further subdivided on the basis of their
duration into fast nov\ae\ and slow nov\ae\ (which can last for
years). The nova duration is strongly correlated with the eruption
amplitude---the fastest nov\ae\ also have the greatest amplitudes.

{\em Recurrent nov\ae\/} are classical nov\ae\ that have been observed to
erupt more than once. The lack of definite nov\ae\ recurrences from
historical records \citep{duerbeck92} implies that, in general, the
recurrence time is $>1000$~years \citep[][page 258]{warner95}. The
ejection, at high velocity, of a substantial shell from recurrent
nov\ae\ permits them to be distinguished from dwarf nov\ae\ which do
not emit such a shell. (Dwarf nov\ae\ may, however, have an enhanced
stellar wind during outburst.)

The nova eruption is thought to be due to the accumulation of
hydrogen-rich material from the accretion disc on the surface of the
white dwarf. As material
is accumulated, the temperature and density of this layer
eventually become high enough for nuclear reactions to occur. Since
the accreted material is degenerate, the pressure is independent of
temperature\footnote{Degeneracy pressure
arises from the fact that when electrons are compressed into a very
small volume, Heisenberg's uncertainty principle means that since
their positions are well-known, their momenta must increase (since
the Pauli exclusion principle states that two electrons cannot occupy
exactly the same state, the momentum of one of the pair is forced to
increase). The increased momenta of the electrons results in a pressure,
supporting, in this case, the atmosphere of the white dwarf against
the pull of gravity. As degeneracy pressure arises from a quantum
mechanical effect, it is independent of temperature.} and a
thermonuclear runaway occurs in the accreted layer 
of hydrogen (the white dwarf itself is mainly composed of carbon and
oxygen). An exponential increase in energy generation, the nova
eruption, occurs until the Fermi temperature\footnote{The Fermi
temperature is the temperature
corresponding to the maximum energy a degenerate electron can
have. Above the Fermi temperature the momenta of the electrons due to
their thermal energy alone is sufficient to satisfy the Heisenberg
uncertainty principle.} is reached, and the degeneracy is lifted.

\subsection{Dwarf nov\ae}
\label{sec:dwarfnovae}
\begin{figure}
\centerline{\includegraphics[width=15cm,angle=0]{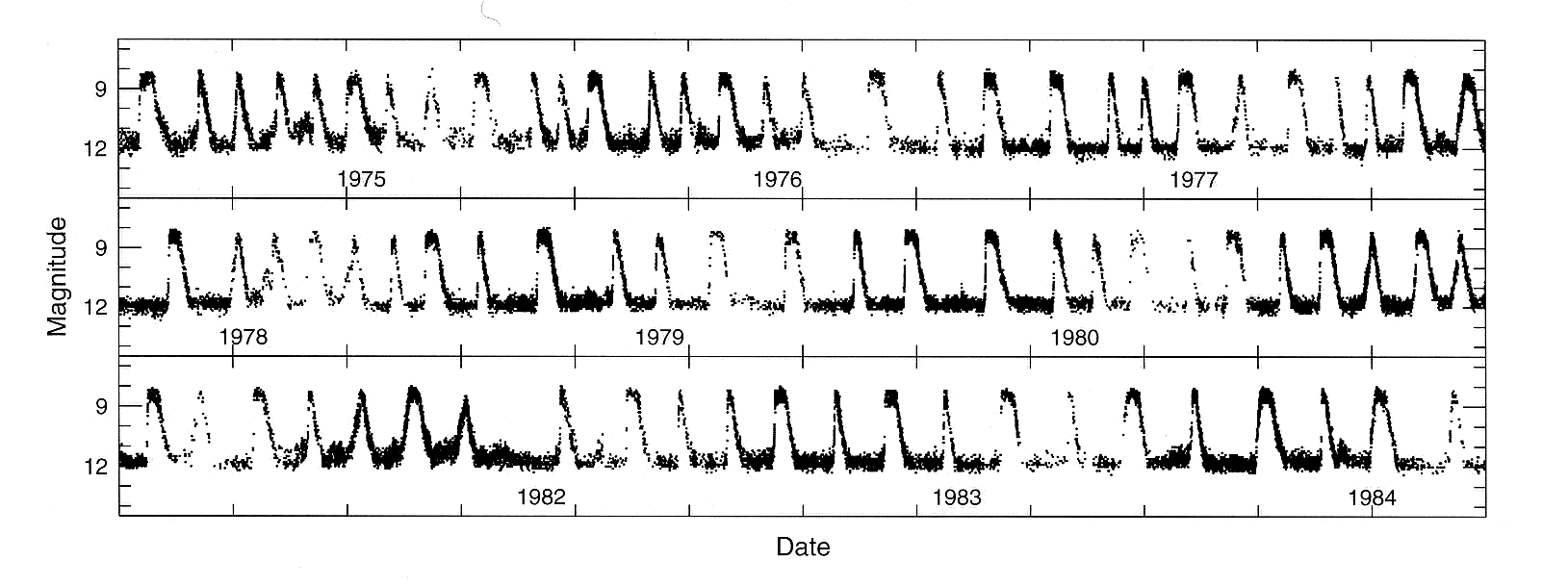}}
\caption[The ten-year light curve of SS~Cyg.]{The ten-year light curve
  of SS~Cyg, showing regular outbursts. Compiled by the American
  Association of Variable Star Observers (AAVSO), and reproduced from
  \citet{hellier01}. Tick-marks are at 100~day intervals.}
\label{fig:sscyg}
\end{figure}

The outbursts (discussed in more detail in \S~\ref{sec:outbursts})
that characterise dwarf nov\ae\ are rather less in amplitude
(typically between two and five magnitudes) than those of nov\ae,
hence the term {\em dwarf nov\ae}. Outbursts typically last for about
a week, with the interval between outbursts (which varies from ten days
to many years) being correlated with their duration. Both the
amplitude and duration of an outburst have well defined time-scales
for a particular object. The light curve of SS~Cyg, the brightest and
one of the best-studied dwarf nov\ae, is shown in
figure~\ref{fig:sscyg}, over a period of ten years.

There exist three distinct subtypes of dwarf nov\ae:
\begin{enumerate}
\item Z~Cam stars have light curves that show periods of rapid outburst
  activity interspersed with {\em standstills}, periods of constant
  brightness about 0.7 magnitudes below maximum light. These
  standstills last between tens of days and years. It is believed that
  Z~Cam stars have mass transfer rates close to that required to
  maintain the disc in permanent outburst, with occasional changes in
  the rate of mass transfer from the secondary star causing the onset
  of outbursts and subsequent return to standstill.
\item SU~UMa stars exhibit superoutbursts in addition to regular
  outbursts (see \S~\ref{sec:superoutbursts}). These superoutbursts
  are approximately 0.7 magnitudes brighter than normal outbursts, of
  longer duration and somewhat more regular. They often appear to be
  triggered by normal outbursts, as a pause before maximum
  superoutburst brightness is achieved reveals \citep[][page
  188]{warner95}.

  SU~UMa stars have another unique characteristic of their light
  curves---the superhump. These are periodic humps in the light curves
  of  dwarf nov\ae\ near the maximum of superoutburst. Superhumps have
  periods of a few percent longer than the orbital cycle, and their
  amplitude appears to be independent of the orbital
  inclination. SU~UMa stars are discussed in more detail in
  \S~\ref{sec:superoutbursts}.
\item U~Gem stars are the dwarf nov\ae\ that are neither Z~Cam nor
  SU~UMa stars.
\end{enumerate}

\begin{figure}
\centerline{\includegraphics[width=13cm,angle=0]{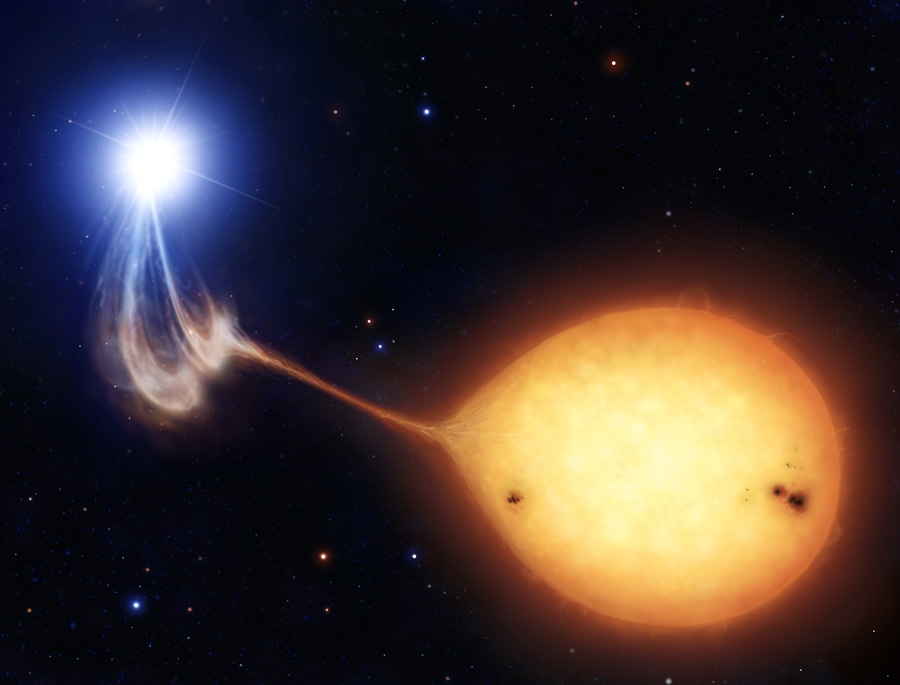}}
\caption[An artist's impression of a polar.]{An artist's impression of
  a polar. Image used by the kind permission of Mark A.\ Garlick.}
\label{fig:polar}
\end{figure}

\begin{figure}
\centerline{\includegraphics[width=13cm,angle=0]{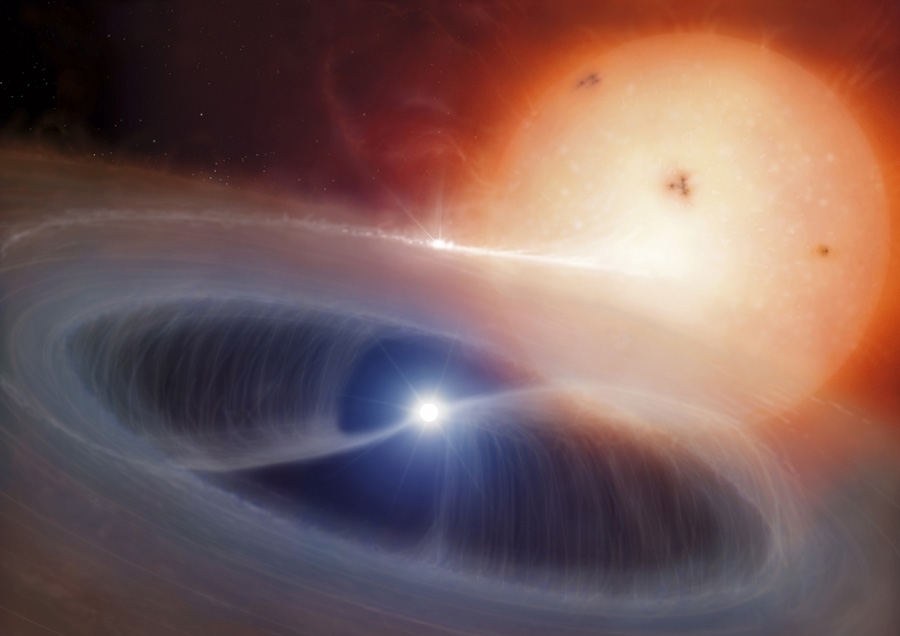}}
\caption[An artist's impression of an intermediate polar.]{An artist's
  impression of an intermediate polar. Image used by the kind
  permission of Mark A.\ Garlick.}
\label{fig:intpolar}
\end{figure}

\subsection{Novalikes}
Historically, nova-like variables were classified as such because they
were observed to be spectroscopically similar to the remnants of old
nov\ae, but had not been observed (yet) to undergo a nova
eruption. Old nov\ae\ are also classed as novalikes, and the category
contains all CVs with mass transfer rates sufficiently high to
maintain the disc in permanent outburst (see \S~\ref{sec:outbursts}).

\subsection{Magnetic CVs}
\label{sec:magcvs}
If the white dwarf has a strong magnetic field the accretion process
can be significantly affected. Depending on the strength of the field,
the gas stream and accretion disc can be partially or totally
disrupted.

{\em Polars}, otherwise known as AM~Her stars, are those CVs with the
strongest magnetic fields (typically a few tens of MGauss). The
magnetic field of the primary is so strong that the white dwarf's
rotation is tidally locked to the orbital period (i.e.\ the primary is
phase-locked or rotates synchronously) and the gas stream is
disrupted, splitting in two and flowing along the magnetic field lines
to the white dwarf (figure~\ref{fig:polar}). Synchronous rotation (at
least in the long-term; nova eruptions can temporarily knock the
system out of synchronization) is the defining characteristic of
polars.

{\em Intermediate polars\/} have significant magnetic fields that are
not strong enough to entirely disrupt the accretion process. If the
field is of such a strength that the gas stream becomes attached to
the field lines at a radius greater than the minimum circularisation
radius (\citealp{verbunt88}; see \S~\ref{sec:gasstream}) then a disc
cannot form. If the field is weaker and material begins to follow the
field lines within this radius, then a truncated disc structure is
formed instead (figure~\ref{fig:intpolar}).

Intermediate polars and
polars may be observationally distinguished due to asynchronous
rotation of the white dwarf manifesting itself in the form of
periodicities in the light curves of intermediate polars.

\section{Cataclysmic variable evolution}
\label{sec:cvevolmain}

Single star evolutionary theory predicts that white dwarfs form from
the cores of red giant stars. The radii of red giants are typically
between 50 to 500~R$_{\odot}$. How then do we reconcile the orbital
separations found in typical CVs of approximately 1~$R_{\odot}$ with the
giant progenitor stars of CVs? The answer is that the orbital
separations of CVs shrink over their lifetime as they evolve, from
initial separations greater than the radii of their progenitor stars
to the separations that we observe today. Proper consideration of the
effects of angular momentum is crucial to a complete understanding of
many stages of CV evolution, so I begin this section with a discussion
of the mechanisms of angular momentum loss in CVs.

\subsection{Angular momentum loss}
\label{sec:angularmomentumloss}

The total orbital angular momentum of the system $J$ is given by
\begin{equation}
\label{eq:totJ}
J= M_{1}a_{1}\frac{2\pi a_{1}}{P_{\rm{orb}}}+ M_{2}a_{2}\frac{2\pi
  a_{2}}{P_{\rm{orb}}},
\end{equation}
where $a_{1}$ and $a_{2}$ are the distances of the primary and
secondary stars, respectively, from the centre of mass of the
system. Since $a=a_{1}+a_{2}$, $M=M_{1}+M_{2}$ and
$a_{1}M_{1}=a_{2}M_{2}$, using Kepler's third law
(equation~\ref{eq:kepler}) leads to
\begin{equation}
\label{eq:angularmomentum}
J= M_{1}M_{2} \left ( \frac{Ga}{M} \right ) ^{\frac{1}{2}}.
\end{equation}
Differentiating equation~\ref{eq:angularmomentum} logarithmically with
respect to time and assuming that no mass is lost from the system as a
whole (i.e.\ $\dot{M}=0$) gives
\begin{equation}
\label{eq:adot}
\frac{\dot{a}}{a}= 2\frac{\dot{J}}{J}- 2\frac{\dot{M}_{2}}{M_{2}}
\left( 1- \frac{M_{2}}{M_{1}} \right),
\end{equation}
where the dot indicates the rate of change with respect to time (i.e.\
$\dot{M_{2}}$ is the secondary's rate of change of mass).

The above expression gives the response of the orbital separation to
mass transfer and angular momentum loss. One can use the approximate
relation of \citet{paczynski71} for the volume radius of the Roche
lobe
\begin{eqnarray}
\label{eq:rochelobe2}
R_{\rm{L}} = 0.462a \left( \frac{M_{2}}{M} \right)^{\frac{1}{3}}, && 0<q<0.3
\end{eqnarray}
which is accurate to 2~per~cent, with equation~\ref{eq:adot} to derive a
similar expression for the response of the Roche-lobe: logarithmically
differentiating equation~\ref{eq:rochelobe2} and combining with
equation~\ref{eq:adot} gives
\begin{equation}
\label{eq:roche_lobe_shrinkage}
\frac{\dot{R}_{\rm{L}}}{R_{\rm{L}}} = 2\frac{\dot{J}}{J}+\left(
2\frac{M_{2}}{M_{1}} - \frac{5}{3} \right) \frac{\dot{M}_{2}}{M_{2}}.
\end{equation}

The above is equally valid if the usual r\^{o}les of the primary and
secondary stars are reversed, that is, if the primary is the component
losing mass. The relevant subscripts in
equations~\ref{eq:totJ}--\ref{eq:roche_lobe_shrinkage} merely need to
be reversed (i.e.\ $M_{2}\rightarrow M_{1}$ etcetera).

Angular momentum loss in CVs is believed to occur via two main
mechanisms: magnetic braking and gravitational radiation.

\subsubsection{Magnetic braking}
The two essential components of magnetic braking are an ionized
stellar wind and a stellar magnetic field. We expect \citep{basri87}
both of these to occur for the secondary stars of CVs.

The ionized stellar wind from the secondary is forced to co-rotate with
the star due to coupling with the magnetic field lines. The stellar
wind thus exerts a braking torque on the rotation of the secondary
star. As tidal forces keep the secondary's rotation synchronous with
the orbital motion, the energy effectively comes from the orbital
motion. The orbital separation therefore shrinks due to the loss of
angular momentum to the stellar wind.

The standard picture of CV evolution \citep{rappaport83} has magnetic
braking as the main source of angular momentum loss until the
secondary star becomes fully convective, whereupon angular momentum
loss due to magnetic braking ceases (see also
\S~\ref{sec:disrupted}). However, \citet{andronov03} showed that if
the angular momentum loss properties of the secondary stars in CVs are
identical to those of single (or detached binary) stars
\citep{basri87}, then the time-scale for angular momentum loss due to
magnetic braking is two orders of magnitude greater than for the
`standard' model. This implies a much longer evolutionary time-scale
for CVs. The data used by \citet{basri87}, however, only includes
systems with orbital periods $\gtrsim17$~hr: it is not clear that the
results can be extrapolated to systems with shorter periods such as
most CVs. In a recent paper, \citet{andronov04} found that chemical
evolution of the secondary star affects its angular momentum loss
properties. The result is that the angular momentum loss rate in CV
secondaries may be greater than that of single stars, although it is
still predicted to be significantly less than in the standard
picture. The exact form of the angular momentum loss due to magnetic
braking remains uncertain.

\subsubsection{Gravitation radiation}
The small orbital separation of many CVs makes gravitational
radiation a significant source of angular momentum loss for these
systems. The rate of the angular momentum loss $\dot{J}$ is given by
\citep{landau58} 
\begin{equation}
\frac{\dot{J}}{J} = -\frac{32G^{3}}{5c^{5}}
\frac{M_{1}M_{2}(M_{1}+M_{2})}{a^4}.
\label{eq:gravitational_radiation}
\end{equation}
Gravitational radiation may be the dominant mechanism for
some short-period dwarf nov\ae\ and polars \citep[][page
447]{warner95}. Observations of orbital period decay in binary
pulsars \citep[e.g.\ PSR~1913+16;][]{taylor82} has provided
convincing observational evidence for the existence of gravitational
radiation. 

\subsection{Pre-common envelope evolution}
\label{pre-cv}
The progenitor stars of CVs stars start life as members of a wide
binary system. The more massive member of the binary naturally evolves
to the red giant phase more rapidly\footnote{Since luminosity $\propto
M^{3}$ and fuel reserves $\propto M$, the (main-sequence) lifetime
$\propto 1/M^{2}$}, expands to fill its Roche-lobe and mass transfer
(from the primary to the secondary star) begins. Mass is being
transferred farther from the centre of mass, and so in order to
conserve angular momentum (i.e. $\dot{J}=0$), the orbital separation
decreases (equation~\ref{eq:adot}) and the Roche-lobe shrinks in size
(equation \ref{eq:roche_lobe_shrinkage}).

The radius of a giant star is almost entirely governed by the mass of
its degenerate core; it does not depend on the mass of its outer
atmosphere. It follows that mass transfer from the primary to the
secondary star is unstable, as no stabilising reduction in radius of
the mass donor occurs. Mass transfer proceeds on the dynamical time
scale \citep[][page 450]{warner95}: $\dot{M} \sim M/\tau_{\rm{dyn}}$,
where
\begin{equation}
\tau_{\rm{dyn}} \sim \left( \frac{R^{3}}{GM} \right) ^{\frac{1}{2}}.
\label{eq:dynamical_timescale}
\end{equation}
However, the mass-receiving star can only adjust its structure on the
thermal, or Kelvin-Helmholtz time scale \cite[][page 450]{warner95}:
\begin{equation}
\tau_{\rm{KH}} = \frac{GM^{2}}{RL} \gg \tau_{\rm{dyn}},
\label{eq:kh_timescale}
\end{equation}
where $L$ is the stellar luminosity.
The end result is runaway mass transfer leading to the transferred
material forming an extended common envelope around both stars.

\subsection{Common envelope evolution}
\label{sec:commonenvelope}
The common envelope phase of CV evolution is when the vast majority of
the orbital separation shrinkage in CVs occurs. The pre-CV is
effectively orbiting within the atmosphere of a red giant. The drag
the stars encounter within the common envelope results in orbital
angular momentum being deposited within the envelope. The consequent
loss of energy from the orbit shrinks the binary separation to
$\sim1\;{\rm R}_{\odot}$ in approximately 1000 years. The energy
injected into the common envelope causes it to be ejected as a
planetary nebula, revealing a still-detached pre-cataclysmic star
consisting of a white dwarf primary and a red dwarf secondary
star\footnote{The endpoint of the common envelope phase can also be a
coalesced star, or a detached binary whose time-scale for evolution
into contact is so long that contact will never be achieved.}.

\subsection{Pre-cataclysmic variable evolution}
Evolution into contact and the formation of a CV occur due to angular
momentum loss from the system. Equation \ref{eq:roche_lobe_shrinkage}
illustrates that for zero angular momentum loss ($\dot{J}=0$) the
minimum radius of the secondary's Roche-lobe occurs for
$q=5/6$. Despite this, mass transfer clearly occurs in CVs, most of
which have $q<5/6$. To drive mass transfer hence requires the loss of
angular momentum from the system ($\dot{J}<0$).

For pre-CVs, magnetic braking is by far the most significant source of
angular momentum loss. For the more massive secondary stars with
$M_{2}\gtrsim1\;{\rm M}_{\odot}$, evolution of the secondary and
consequent expansion may cause it to come into contact with its
Roche-lobe before angular momentum loss
does. \citet{pylyser88a,pylyser88b} found that contact due to
evolution of the secondary star leads to mass transfer driven by
expansion of the secondary's envelope, an increasing orbital period
(and separation) and, ultimately, a detached system (occurring when
all the secondary's envelope has been lost). In this thesis, I
concentrate on the more common (as for most CV secondaries
$M_{2}\ll1\;{\rm M}_{\odot}$) scenario of angular momentum loss.

\subsection{Cataclysmic variable evolution}
\label{sec:cvevol}

The evolution of CVs relies on mechanisms of angular momentum loss in
order to drive mass transfer. In the case of CVs, mass transfer occurs
from the secondary star to the primary. Mass is therefore moving away
from the centre of mass and from equations~\ref{eq:adot} and
\ref{eq:roche_lobe_shrinkage} if angular momentum is conserved
($\dot{J}=0$), for $q<5/6$ (i.e.\ most CVs), this results in the
orbital separation and volume radius of the secondary's Roche-lobe
increasing, cutting off mass transfer. A mechanism of angular momentum
loss ($\dot{J}<0$) is therefore necessary for long-term mass transfer
to occur. Mass transfer through angular momentum loss must lead to the
volume radius of the secondary star's Roche-lobe and the orbital
separation shrinking, resulting (equation~\ref{eq:kepler}) in the
system evolving to shorter orbital periods.

For mass transfer to be stable, the secondary must be able to adjust
its radius quickly enough to remain within its Roche-lobe,
i.e. \citep[][page 458]{warner95}
\begin{equation}
\label{eq:xi}
\xi = \frac{\partial \ln R_{2}}{\partial \ln M_{2}} > \frac{d \ln
  R_{L}}{d \ln M_{2}}.
\end{equation}
Equations~\ref{eq:roche_lobe_shrinkage} and \ref{eq:xi}, for
conservative mass transfer ($\dot{J}=0$), require
\begin{equation}
\label{eq:qcrit}
q<\frac{1}{2}\xi + \frac{5}{6}
\end{equation}
for stable mass transfer to occur. For $M_{2}>M_{\odot}$, where
$\tau_{{\rm KH}}<\tau{_{\rm ML}}$ (where $\tau{_{\rm ML}}$ is the time
scale for mass loss) $\xi=0.87$ \citep[][page 458]{warner95}, this
leads to the
condition $q<1.26$ for mass transfer from a main-sequence donor star
to be stable. For $M_{2}>0.8\;M_{\odot}$, $\tau_{{\rm KH}}>\tau{_{\rm
ML}}$ and $\xi=-1/3$ \citep{paczynski65}, the corresponding condition
is $q<2/3$. If this condition is not satisfied (in either case) then
mass transfer is unstable and will proceed on the dynamical time-scale
(equation~\ref{eq:dynamical_timescale}), possibly leading to a second
common envelope phase.

It is easiest to think of the secular evolution of CVs as occurring in
a stepwise way (although the process is of course continuous in
practice). First, mass is transferred from the secondary star to the
primary. This causes, through equations~\ref{eq:adot} and
\ref{eq:roche_lobe_shrinkage}, the orbital separation and secondary
star's Roche-lobe to increase in size. The radius of the secondary
star becomes smaller in response to its reduced mass. Angular momentum
loss from the system then decreases both the orbital separation and
the size of the secondary star's Roche-lobe until mass transfer can
re-commence at a smaller orbital separation than previously.

\begin{figure}
\centerline{\includegraphics[width=10cm,angle=0]{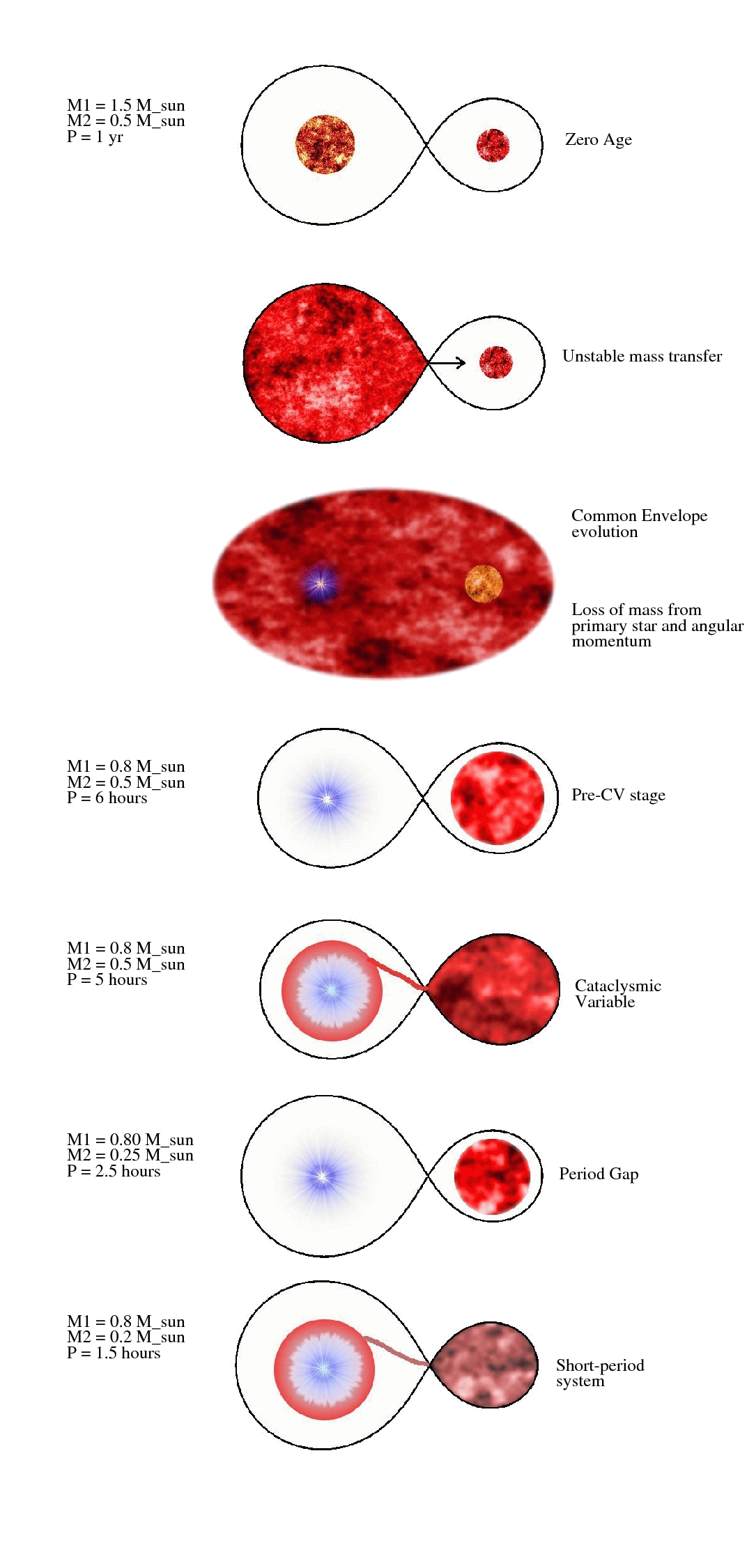}}
\caption[A schematic demonstrating the important stages in the
  evolution of a CV.]{A schematic demonstrating the important stages
  in the evolution of a CV. The primary star is on the left and the
  secondary on the right. From \citet{littlefair02}.}
\label{fig:evol}
\end{figure}

A schematic demonstrating the important stages
in the evolution of a CV is shown in figure~\ref{fig:evol}.

\subsection{The orbital period distribution}
\begin{figure}
\centerline{\includegraphics[width=8.0cm,angle=-90.]{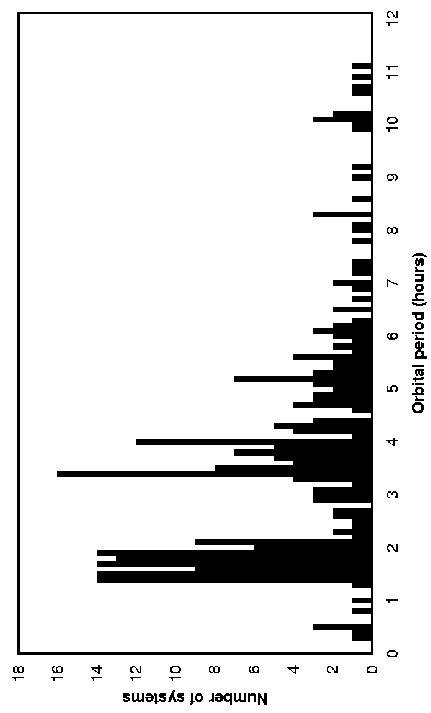}}
\caption[The orbital period distribution of CVs.]{The orbital period
  distribution of CVs for periods less than 12~hours, compiled from
  \citet{ritter98}.}
\label{fig:periods}
\end{figure}

As one of the most easily, and most accurately, determined physical
parameters, the distribution of the orbital periods of CVs potentially
provides a useful window into their evolution. The orbital period
distribution of CVs, shown in figure~\ref{fig:periods}, has three main
features: the minimum period, the long-period cut-off and the period
gap.

\subsubsection{The minimum period}
There is an observed minimum period for CVs of approximately
78~minutes\footnote{Those systems plotted on figure~\ref{fig:periods}
with an orbital period of less than this are AM~CVn systems, whose
secondary stars are composed chiefly of helium, leading to a more
compact secondary star.}. This is due to the secondary star becoming
degenerate for masses below $\sim0.08\;{\rm M}_{\odot}$. The
degenerate secondary effectively becomes a very low mass white dwarf,
and as such (if in thermal equilibrium) will obey the mass-radius
relationship for a white dwarf, for instance the analytical
approximation to the Hamada-Salpeter relation \citep{hamada61} of
\citet{nauenberg72}:
\begin{equation}
\label{eq:nauenberg}
R=7.795\times10^{6}\left[ \left(
  \frac{1.44\;{\rm M}_{\odot}}{M} \right) ^{\frac{2}{3}} - \left(
  \frac{M}{1.44\;{\rm M}_{\odot}} \right) ^{\frac{2}{3}} \right]
  ^{\frac{1}{2}}\rm{m}.
\end{equation}
Equation~\ref{eq:nauenberg} illustrates that decreasing the mass of a
degenerate secondary leads to expansion of its radius. For $q<2/3$
(see equation~\ref{eq:qcrit} and the following discussion),
this situation is, surprisingly enough, stable \citep[][pages 459 \&
462]{warner95}.

Once again, it is easiest to think of this continuous process as
occurring in two stages: as $q<5/6$, mass transfer leads to the
orbital separation and the size of the secondary's Roche-lobe
increasing (equations~\ref{eq:adot} and
\ref{eq:roche_lobe_shrinkage}). The continuing mechanisms of orbital
angular momentum loss then decrease the orbital separation and the
secondary's Roche-lobe until mass transfer can re-commence, but the
secondary has in the meantime expanded in response to its mass loss,
so mass transfer begins again at a slightly larger separation than
before.

\subsubsection{The long-period cut-off}
Above an orbital period of about six hours the number of CVs declines,
with very few observed with periods $>12$~hours. As seen in
\S~\ref{sec:cvevol}, stable mass transfer requires that
$q<1.26$. Since the mass-accreting star is a white dwarf, it has a
maximum mass of $1.44\;{\rm M}_{\odot}$ (the Chandrasekhar mass). Kepler's
third law (equation~\ref{eq:kepler}) requires that the orbital
separation increases as the orbital period increases. This results
(equation~\ref{eq:RL2}) in the size of the Roche-lobe, and therefore
the mass of the (dwarf) star required to fill it, increasing. The
constraint on the maximum mass of the white dwarf thus leads to an
upper limit on the orbital period of $\sim 12$~hours. The few systems
with $P_{\rm{orb}}\ge12$~hours have evolved secondaries. The fact that
few white dwarfs actually have the Chandrasekhar mass naturally
explains the gradual decline in the number of systems with
$P_{\rm{orb}}\gtrsim6$~hours.

\subsubsection{The period gap}
\label{sec:periodgap}
Between $2.2\leq P_{\rm{orb}}\leq 2.8$~hours there is a significant
deficiency of systems. (Polars do not show this gap, but intermediate
polars do.) A number of possibilities present themselves as
explanations for this period gap. As shown in \S~\ref{sec:cvevolmain},
CVs evolve to shorter orbital periods due to angular momentum
loss. This could imply that the CVs above and below the period gap are
actually two separate populations. The upper bound of the period gap
would then represent some minimum period for the long-period
population, and the lower bound a maximum period for the short-period
CVs. To produce a minimum orbital period of about three hours the
secondary stars would have to be degenerate and of very low
luminosity \citep{verbunt84}. This is not supported observationally,
however: such stars are observed to have normal main-sequence
luminosities \citep[][page 465]{warner95}.

The favoured scenario is that CVs evolve into the period gap from a
single population, but cease mass transfer whilst they are in it. This
obviously requires some sort of mechanism to halt mass transfer at the
upper bound of the period gap: the {\em disrupted magnetic braking
model}, the subject of the following section.

\subsection{The disrupted magnetic braking model}
\label{sec:disrupted}
The period gap is thought to be a consequence of a change in the
internal structure of the secondary star. As noted by
\citet{robinson81}, at an orbital period of around three hours, which
corresponds to a secondary mass of $\sim0.25\;{\rm M}_{\odot}$
\citep{smith98a}, the internal structure of the secondary changes from
a radiative core ($M_{2}>0.3\;{\rm M}_{\odot}$) to a convective core
($M_{2}<0.3\;{\rm M}_{\odot}$). It is suspected that this interferes
with the dynamo-generated magnetic field of the secondary
\citep{king88}, disrupting magnetic braking. Whatever the mechanism of
the reduction of magnetic braking, if it results in
$\tau_{\rm{KH}}<\tau_{\rm{MT}}$, where $\tau_{\rm{MT}}$ is the
time-scale for mass transfer, it gives the secondary time to readjust
its structure so that it can shrink back within its Roche-lobe,
causing mass transfer to cease and establishing the upper edge of the
period gap. (Remember that when mass transfer is occurring the
secondary is larger than its equilibrium main-sequence radius; it is
filling its Roche-lobe.)

Note that this explanation is rather speculative, since it assumes
that magnetic braking is the dominant cause of angular momentum loss
in systems above the period gap. \citet[see also
\S~\ref{sec:angularmomentumloss}]{andronov03} point out that it
is not clear that this is true and that there is no observational
evidence for a sudden cessation in magnetic braking at the point where
the secondary switches from a radiative to a convective core.

\section{The gas stream and accretion disc}

The presence of accretion discs in many CVs accounts for much of the
interest in these systems. Accretion discs are an incredibly
widespread astrophysical phenomenon, occurring in a variety of
locations and scales. They are present around young stars (T~Tauri
stars) where they assist the formation of the protostar by removing
angular momentum from material in the collapsing gas cloud. They are
also thought to be involved in the formation of planetary systems. At
the other extreme, accretion discs fuel the cores of active galaxies,
radiating the gravitational potential energy lost by material as it
falls towards a central massive black hole. Unfortunately, however,
accretion discs around young stars are shrouded by the dust and gas
from which these stars are forming, and those in the cores of active
galaxies are not only frequently obscured by the surrounding dust, gas
and stars, but are at great distances. Additionally, the accretion
discs present in CVs evolve more quickly than those in either T~Tauri
stars or the cores of active galaxies. One final advantage that
observations of the discs in CVs have over those of T~Tauri discs is
that CV discs are hotter, and therefore more luminous at optical
wavelengths. It is in CVs, therefore, specifically in eclipsing
systems, that accretion discs are most profitably observed.

\subsection{Gas stream dynamics}
\label{sec:gasstream}

\begin{figure}
\centerline{\includegraphics[width=12cm,angle=-90]{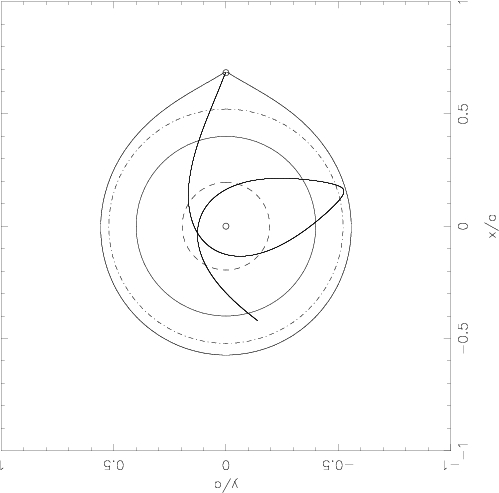}}
\caption[The trajectory of the gas stream for $q=0.15$.]{The
  trajectory of the gas stream for $q=0.15$. The stream originates at
  the inner Lagrangian point with a small initial velocity with
  respect to the binary frame. The minimum circularisation radius
  \citep[][their equation 13]{verbunt88} is shown as the dashed
  circle, a disc radius of
  $0.4a$ is shown as a solid line, the dot-dash line is the tidal
  radius \citep{paczynski77} and the Roche-lobe of the primary is the
  tear-drop shaped solid line. The position of the white dwarf at the
  origin is marked with an open circle, as is the inner Lagrangian
  point. The secondary star is to the right of the frame. Note that
  the stream does not re-cross the Roche-lobe.}
\label{fig:gasstream}
\end{figure}

From the inner Lagrangian point to the point where it impacts the
disc, the gas stream follows a ballistic trajectory, shown in
figure~\ref{fig:gasstream}. The equations of motion for a point mass
in a Cartesian co-ordinate system co-rotating with the binary as
described in \S~\ref{sec:roche_lobe} are
\citep[e.g.][]{flannery75,dhillon90}:
\begin{equation}
\label{eq:xacceleration}
\ddot{x} = \frac{GM_{1}}{r_{1}^{2}} \frac{(x_{1}-x)}{r_{1}} +
\frac{GM_{2}}{r_{2}^{2}} \frac{(x_{2}-x)}{r_{2}} + 2\omega \dot{y} +
  \omega^{2}r_{\rm{cm}} \frac{(x-x_{\rm{cm}})}{r_{\rm{cm}}}
\end{equation}
and
\begin{equation}
\label{eq:yacceleration}
\ddot{y} = \frac{GM_{1}}{r_{1}^{2}} \frac{(y_{1}-y)}{r_{1}} +
\frac{GM_{2}}{r_{2}^{2}} \frac{(y_{2}-y)}{r_{2}} + 2\omega \dot{x} +
  \omega^{2}r_{\rm{cm}} \frac{(y-y_{\rm{cm}})}{r_{\rm{cm}}},
\end{equation}
where the subscript `cm' denotes the distance to the centre of mass of
the system and $\omega=2\pi/P_{\rm{orb}}$ is the angular
frequency. The first two terms in equations \ref{eq:xacceleration} and
\ref{eq:yacceleration} are from the gravitational influence of the
primary and secondary stars, respectively; the third is from the
Coriolis force and the fourth from the fictitious centrifugal
force.

The position of such a test particle at a given time is an example of
the well-known three-body problem. As there is no (known) explicit
solution, the problem must be solved by numerical integration.  The
position, velocity and acceleration of a test particle are calculated
at time intervals $\Delta t$. As this interval becomes smaller, the
calculation becomes more accurate, but more CPU-intensive. High
accuracy is necessary near the primary star, where the potential
gradient is steep (see figure~\ref{fig:3Dpotentials}), but wasteful
where the potential is relatively flat. A good compromise can be
achieved by adjusting the time interval according to the relation
\begin{equation}
\Delta t = \Delta t_{\rm{in}} \left( \frac{R}{R_{{\rm L}1}} \right)^{2},
\end{equation}
where $\Delta t_{\rm{in}}$ is the initial time interval and $R$ and
$R_{\rm L1}$ are the distances of the test particle and ${\rm L}_{1}$
point from the primary star, respectively. This decreases the time
interval as a function of the square of the distance from the primary
star, which is appropriate since the force exerted on the particle by
the gravitational attraction of the primary also varies with the
square of the distance from the star.

An additional constraint is that the energy of the particle is
conserved along its path, so that the quantity
\begin{equation}
\label{eq:jacobi}
E_{J} = \dot{x}^{2} + \dot{y}^{2} -2\Omega(x,y),
\end{equation}
where
\begin{equation}
\Omega(x,y) = \frac{1}{2} \left[ \frac{M_{2}R^{2}_{2}}{M_{1}+M_{2}} +
  \frac{M_{1}R^{2}_{1}}{M_{1}+M_{2}} \right] +
  \frac{M_{2}}{(M_{1}+M_{2})R_{2}} + \frac{M_{1}}{(M_{1}+M_{2})R_{1}},
\end{equation}
called the {\em Jacobi energy}, remains constant \citep{warner72}. In
practice, the Jacobi energy is subject to the constraint
\begin{equation}
\frac{\Delta E_{J}}{E_{J}} < {\rm tol},
\end{equation}
where tol is the fractional accuracy required (typically $10^{-4}$).

From the constraint on the Jacobi energy, it follows that the stream
cannot re-cross the critical potential, and always approaches it with
a low velocity. If the Jacobi energy is not conserved for a given
time-step calculation, then the time interval $\Delta t$ is reduced by
a factor of two and the step re-calculated until
equation~\ref{eq:jacobi} is satisfied.

The accuracy of the calculation can be further improved by use of a
second-order Runge-Kutta technique. This involves calculating the
acceleration of the particle at the start and end of the time
interval, and then applying the mean of these to the particle over
the time-step.

Due to its angular momentum, the gas stream will pass by the white
dwarf and eventually loop back around and collide with itself. This
impact will give rise to turbulent shocks which dissipate much of the
kinetic energy of the stream. Angular momentum, however, is not so
easily lost and the material will therefore settle into the lowest
energy orbit for a given angular momentum: a circular one.

The minimum outer radius of the accretion disc can be derived by
calculating the radius around the white dwarf at which orbiting
material has the same angular momentum as material at the inner
Lagrangian point. This {\em circularisation radius\/} $R_{\rm{min}}$ is
given by \citep[their equation 13]{verbunt88}:
\begin{eqnarray}
R_{\rm{min}} \approx  0.0883 + 0.04858 \log q^{-1} + 0.11489 \log^{2}
q^{-1} \nonumber \\ - 0.020475 \log^{3} q^{-1}\,. && 10^{-3}<q<1
\end{eqnarray}

The maximum possible radius of the disc can be determined from
consideration of simple periodic particle orbits. Particle
trajectories for radii approaching the radius of the primary's Roche
lobe become significantly non-circular due to the gravitational
influence of the secondary star. Assuming that the largest orbit that
does not intersect with any others is the maximum radius of the
accretion disc (this is sensible because larger orbits that intersect
will dissipate energy and prevent growth of the disc) gives the
so-called {\em tidal radius}, $R_{\rm{tidal}}$, an approximate relation
for which is (\citealp{paczynski77}; \citealp[][page 57]{warner95})
\begin{eqnarray}
\label{eq:tidal}
\frac{R_{\rm{tidal}}}{a} = \frac{0.6}{1+q}\,. && 0.03<q<1
\end{eqnarray}

\subsection{The radial temperature profile}
\label{sec:radialtemp}
The velocities of the material in the accretion disc can usually be
assumed to be negligibly different from Keplerian velocities, as at
these distances from the primary the gravitational influence of the
secondary star is slight. Viscosity in the disc occurs due to
interaction between particles in slightly different orbits. Particles
in smaller orbits orbit faster than material further out, and
via viscous interaction speed up this outer material, transferring
angular momentum to it, and {\em vice versa}. Angular momentum is
therefore transferred outwards in the disc, resulting in a net flow of
material inwards \citep{lyndenbell74}, driving accretion  onto the
white dwarf.

If we assume that all the potential energy $E_{p}$ lost by some mass
$m$ as it spirals towards a mass $M$ from an initial radius $R$ to a
radius $R-dR$ is radiated away as blackbody radiation, then the rate
of energy release is
\begin{eqnarray}
\label{eq:energy}
\Delta E_{p} = -GM\dot{m} \left ( \frac{1}{R}-\frac{1}{R-dR} \right).
\end{eqnarray}
If we assume that the annulus defined by the radii $R$ and $R-dR$
emits as a blackbody, then we have
\begin{equation}
\label{eq:blackbody}
\Delta E_{p} = 4\pi R \, dR \, \sigma T^{4}.
\end{equation}
Combining equations \ref{eq:energy} and \ref{eq:blackbody} and setting
$R-dR \approx R$ gives
\begin{equation}
\label{eq:radialtemp1}
T^{4} = \frac{GM\dot{m}}{4\pi R^{3}\sigma},
\end{equation}
i.e.
\begin{equation}
T \propto R^{-3/4}.
\end{equation}
This derivation neglects two effects. First, as material spirals to
smaller radii its Keplerian velocity increases, so some of the
gravitational energy released goes into the kinetic energy of the
particles. Second, as material accretes onto the white dwarf, it
decelerates to the rotation speed of the white dwarf in a boundary
layer between the white dwarf and inner edge of the accretion
disc. This deceleration results in kinetic energy being converted into
thermal energy. With these factors taken into account, equation
\ref{eq:radialtemp1} becomes \citep{bath81,horne85a}
\begin{equation}
\label{eq:radialtemp2}
T^{4} = \frac{3GM\dot{m}}{8\pi R^{3}\sigma} \left(
\sqrt{1-R_{1}/R} \right ),
\end{equation}
where $R_{1}$ is the radius of the white dwarf.

In dwarf nov\ae\ in outburst and long-period novalikes, this simple
$R^{-3/4}$ radial temperature profile is indeed observed by eclipse
mapping experiments \citep{horne85a,horne85b,rutten92b}. In quiescent
dwarf nov\ae\ a much flatter profile is observed
\citep[e.g.][]{wood89a}. This is thought to be because the disc does
not achieve a steady state  in quiescence (in a steady state the disc surface
density does not evolve with time; $\partial\Sigma/\partial t=0$; see
the following section).

\subsection{Outbursts and the disc instability model}
\label{sec:outbursts}

Dwarf nova outbursts have been observed and studied for well over a
century. Outbursts of U~Gem were first discovered in 1855
\citep{hind1856}. Unsurprisingly, early attempts to explain dwarf nova
outbursts tried to associate them with nov\ae\ and recurrent
nov\ae. For several years the idea that dwarf nov\ae\ were just that,
miniature nov\ae\ eruptions, was popular. In this scenario the
outbursts are caused by a thermonuclear runaway of the hydrogen in the
white dwarf envelope. \citet[][page 167 on]{warner95} gives an
excellent summary of early models of dwarf nov\ae\ outbursts.

The disc instability model was proposed by \citet{osaki74}. It
attributes dwarf nov\ae\ outbursts to ``sudden gravitational energy
release due to {\em intermittent\/} accretion of material onto the white
dwarf component \ldots\ from the surrounding disc.'' This intermittent
accretion is triggered by an instability in the accretion
disc. \citeauthor{osaki74} suggested that the secondary star transfers
material at a constant rate, which is greater than the mass transfer rate
through the disc. This would result in material accumulating in the
accretion disc until some critical density were reached, whereupon the
viscosity in the disc would increase greatly, enhancing the accretion
rate onto the white dwarf (see \S~\ref{sec:radialtemp}). The viscous
heating of the disc results in an increase in the luminosity of the
system. This follows from equation \ref{eq:radialtemp2}
and the Stefan-Boltzmann equation:
\begin{equation}
L=4\pi R^{2}\sigma T^{4},
\end{equation}
which gives the bolometric luminosity of a blackbody of radius $R$ and
temperature $T$, where $\sigma=5.67\times10^{-8}$~Wm$^{-2}$K$^{-4}$ is
the Stefan-Boltzmann constant.

\begin{figure}
\centerline{\includegraphics[width=9cm,angle=-90]{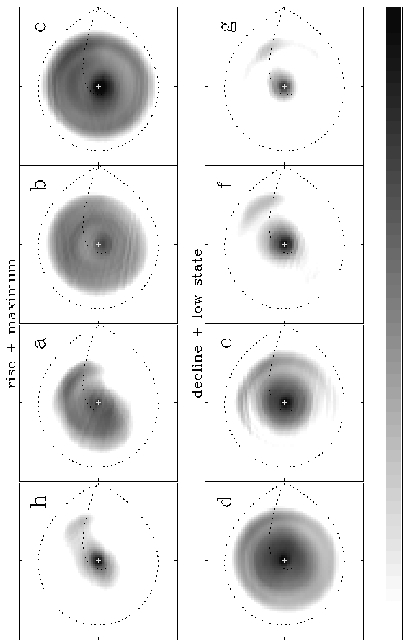}}
\caption[A sequence of eclipse maps of the dwarf nova EX~Dra through
  the outburst cycle.]{A sequence of eclipse maps of the dwarf nova
  EX~Dra showing the rise (a--b) from quiescence (h) to maximum light
  (c), through the decline (d--f) to a low brightness state (g) that
  the system enters before recovering its quiescent brightness. The
  Roche-lobe and gas stream are shown as dotted lines, and the white
  dwarf at the centre is marked with a cross. The scale is logarithmic,
  with brighter regions indicated in black. From \citet{baptista01}.}
\label{fig:outburstmap}
\end{figure}

Another consequence of the increase in the rate of angular momentum
transportation is the expansion of the accretion disc, due to the
conservation of angular momentum. Eclipse mapping of dwarf nov\ae\
during outburst has demonstrated this increase in the size of the
accretion disc (figure~\ref{fig:outburstmap}).

\subsubsection{The viscosity of accretion discs}
The exact mechanism of viscosity in accretion discs is
non-obvious. Simple molecular viscosity is too weak to explain the
rates of mass transfer through the disc necessary to reproduce the
observed behaviour of accretion discs. Turbulence in the accretion
disc could increase the viscosity, by causing globules of material to
move to different orbits, transporting angular momentum between
orbits at different radii.

\citet{shakura73} characterised the turbulence by the {\em alpha
viscosity\/} $\alpha$. The alpha viscosity parameterizes the efficiency
of the mechanism of angular momentum transport, and has a
maximum\footnote{\citet{shakura73} point out that for $\alpha>1$ the
turbulence is supersonic, leading to rapid heating of the disc
material and the subsequent reduction of the alpha viscosity to
$\alpha\ll1$.} of $\alpha=1$.  The maximum size of the turbulent
eddies is of the order of the disc thickness, $z_{0}$.  The alpha
viscosity prescription allows the viscosity $\nu$ to be quantified as
\begin{equation}
\label{eq:alphaviscosity}
\nu = \alpha c_{\rm{s}} z_{0},
\end{equation}
where $c_{\rm s}$ is the sound velocity.

This parameterization of the viscosity allows theoretical models of
accretions discs, known as {\em alpha discs}, to be constructed by
combining the alpha viscosity with the equations of gas dynamics. This
leads to (\citealp{shakura73}; \citealp[][page 47]{warner95})
\begin{equation}
\label{eq:discheight}
z_{0}\propto r^{9/8},
\end{equation}
assuming that $\alpha$ is independent of
radius. Equation~\ref{eq:discheight} shows that alpha discs are
concave, flaring out at their outer edges. Alpha discs are also `thin
discs,' i.e.\ their heights are small compared to their
radii. Comparison of alpha disc models to observations shows that
during outburst, values for $\alpha$ range from approximately 0.1 to
0.5, and during quiescence, from 0.01 to 0.05 (\citealp{mineshige89};
\citealp[][page 179]{warner95};
\citealp*{hellier01,lasota01,schreiber03}).

\subsubsection{The disc instability model}
The alpha viscosity itself gives no clue as to the cause of the
turbulence. The key to this was developed in the 1990s, and is based
on magnetic instabilities resulting in turbulence in the disc
\citep{balbus91,hawley91}. Ionized material in the disc couples to the
magnetic field in the disc, which forces material rotating more slowly
(at larger radii) outwards and material rotating more quickly (at
smaller radii) inwards. This stretches the field lines, amplifying
them, and eventually leads to magnetic turbulence in the disc. The
Balbus-Hawley or magneto-rotational instability has recently
been demonstrated in the laboratory \citep{sisan04}.

The Balbus--Hawley instability provides a theoretical explanation of
the trigger of dwarf nov\ae\ outbursts, in that it only operates when
material in the disc is ionized. The disc of a dwarf nova in outburst
is hot and highly viscous, whereas during quiescence it is cold and
less viscous. All that is now required is some mechanism to cause
heating of the disc in order to trigger the Balbus--Hawley instability
and subsequently an outburst.

\begin{figure}
\centerline{\includegraphics[width=13cm,angle=-90]{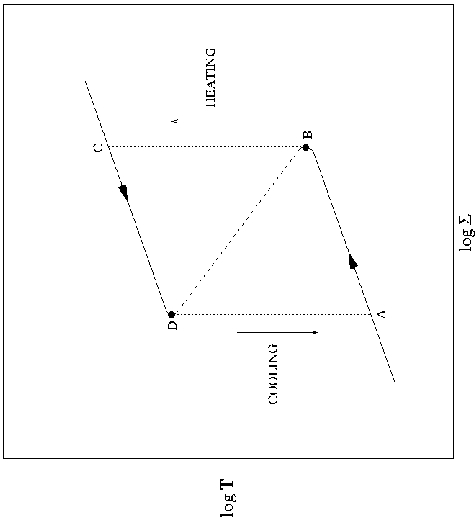}}
\caption[The $\Sigma-T$ relation and the resulting thermal limit
  cycle for dwarf nov\ae.]{The $\Sigma-T$ relation and the
  resulting thermal limit cycle for dwarf nov\ae. At point A, the
  dwarf nova is at quiescence, with a cool, unionized disc. Material
  builds up in the disc, increasing the disc temperature, and the
  dwarf nova moves to point B. Here, the hydrogen in the disc becomes
  partially ionized, leading to a large opacity increase and
  subsequent runaway temperature increase to point C---the dwarf nova
  enters outburst. At point C, the hydrogen has become fully ionized,
  so the disc opacity becomes insensitive to temperature, halting the
  runaway temperature increase. Enhanced mass transfer through the
  disc onto the white dwarf due to the Balbus-Hawley instability leads
  to the surface density of the disc
  decreasing, and with it the temperature, until the hydrogen begins
  to recombine at point D. At this point the thermal instability
  begins to operate again, and the disc rapidly cools to point A. The
  S-curve comprises the solid and dashed lines. From \citet{watson02a}.}
\label{fig:scurve}
\end{figure}

This mechanism is the {\em thermal instability}. The line of thermal
equilibrium in the $\Sigma-T$ plane, where $\Sigma$ is the
surface density of the material, is known as the `S-curve' (shown in
figure~\ref{fig:scurve}). On the S-curve, heating from viscous forces
balances the radiation from the surface. A system located off this
line of thermal equilibrium will heat or cool, as appropriate, until
equilibrium is established on the S-curve. Not all equilibrium states
are stable, however, only those that satisfy $dT/d\Sigma>0$. Those
states with negative gradients ($dT/d\Sigma<0$) are unstable: a small
positive perturbation of $\Sigma$ leads to an increase in $T$, which
takes the state away from the S-curve of thermal equilibrium. To
regain stability the system would have to migrate to lower $T$.

Consider first of all an unionized, cold annulus within an accretion
disc. If the rate of mass transfer into the unionized annulus is
greater than the rate at which material flows through it, it will
inevitably begin to fill up, increasing the surface density. The
greater surface density of the disc results in an increase in the
viscosity, which in turn increases the temperature via viscous
heating. The increase in viscosity also causes the mass transfer rate
through the annulus to increase. As unionized hydrogen has a low
opacity $\kappa$ to radiation, and this is not strongly dependent on
temperature, the energy released by the viscosity will escape and the
system will tend to stabilise itself on the lower branch of the
S-curve. If, however, the annulus becomes hot enough ($\sim7000$~K)
for the hydrogen to become partially ionized, the situation
changes. Unlike unionized hydrogen, the opacity of partially ionized hydrogen
has a strong dependence on temperature: $\kappa \propto T^{10}$. The
temperature rise therefore causes a massive increase in the opacity of
the annulus, trapping the energy released by the viscosity, and
further increasing the temperature. Although the viscosity increase
means that the surface density of the region is being reduced, the
effect this has on the temperature increase is vastly outweighed by
the increase in the opacity-trapped energy.

Once the annulus is completely ionized, the opacity is no longer
highly sensitive to temperature, and the annulus settles into
equilibrium at a much higher temperature on the upper branch of the
S-curve. The Balbus-Hawley instability can now kick in as the magnetic
field is able to couple to the ionized material in the annulus, and
the viscosity will increase. This means, however, that the rate of
mass transfer through the annulus is greater than the rate of mass
transfer into it, so the surface density of the annulus gradually
decreases, and with it the temperature, until the hydrogen in the
annulus becomes partially ionized again. The temperature-opacity
dependence returns, and the opacity rapidly drops as the temperature
falls, until the hydrogen becomes unionized and the viscosity returns
to normal. This cycle is known as the {\em thermal limit cycle}, and
is illustrated in figure~\ref{fig:scurve}; see \citet[][page 173
on]{warner95} for a fuller discussion.

The thermal instability described above begins in a certain annulus
and then proceeds to the rest of the disc by distributing hot material
to adjacent annuli. This {\em heating wave\/} can start either in the
inner disc, in which case the resulting outburst is known as {\em
inside-out}, or in the outer disc, leading to an {\em outside-in\/}
outburst. Which type occurs depends on the mass transfer rate from the
secondary star. At lower mass transfer rates, the material has time to
filter through the disc, and accumulates at smaller radii. If the mass
transfer rate from the secondary star is large, however, the material
tends to build up nearer the outer edge of the disc. In the former
case, an inside-out outburst results; in the latter, an
outside-in. Inside-out outbursts tend to have a slower rise to
outburst maximum than outside-in outbursts. This is due to three main
factors:
\begin{enumerate}
\item Viscosity causes more material to flow inwards than outwards;
\item Inner radii have smaller surface densities ($\Sigma \propto
  R^{1.05}$, \citealp{cannizzo88}), so there is less material to
  spread outwards;
\item Outer radii are larger, so the surface density of material
  moving to larger radii is reduced, hampering the progress of the
  heating wave. Material travelling inwards has its surface density
  increased, thus aiding the progress of the heating wave.
\end{enumerate}

\subsection{Superoutbursts and superhumps: SU UMa stars}
\label{sec:superoutbursts}

The definition of an SU~UMa star (see \S~\ref{sec:dwarfnovae}) is a dwarf
nova that also exhibits superhumps. To date, no star has yet been
found that exhibits superhumps but not superoutbursts, or {\em vice
versa}, so I shall presume that the presence of one of these
phenomena implies the other. The orbital period distribution of SU~UMa
stars is pronounced: they all (with the exception of TU~Men) lie
below the period gap. It is suspected that all dwarf nov\ae\ below the
period gap are SU~UMa stars \citep[][page 127]{warner95}.

The cause of superhumps is the precession of an elliptical disc
\citep{vogt82}. The precession period $P_{\rm{prec}}$ of such a disc
will create a beat, or superhump, period $P_{\rm{sh}}$ with the
orbital period $P_{\rm{orb}}$:
\begin{equation}
\label{eq:superhump}
\frac{1}{P_{\rm{sh}}} = \frac{1}{P_{\rm{orb}}} -
\frac{1}{P_{\rm{prec}}}.
\end{equation}

\begin{figure}
\centerline{\includegraphics[width=14cm,angle=0]{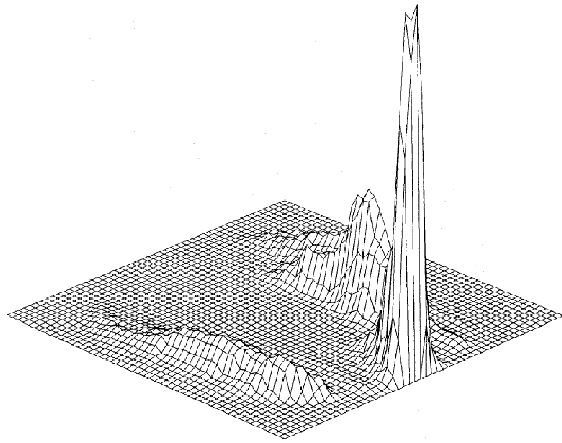}}
\caption[The superhump light distribution across the disc of Z~Cha
  during superoutburst.]{The superhump light distribution across the
  disc of the SU~UMa star Z~Cha during superoutburst. The primary is
  at the centre of the grid (which has sides equal to the orbital
  separation) and the secondary off the grid at bottom right. The
  superhump light originates from three regions at the rim of the disc
  where particle orbits intersect. The plot has been produced by
  maximum-entropy eclipse mapping, and is taken from
  \citet{odonoghue90}.}
\label{fig:superhumpmap}
\end{figure}

The origin of the elliptical disc is tidal resonances of particles in
the outer disc with the secondary star
\citep{whitehurst88,whitehurst91}. Particles with orbital periods in
resonance with the orbital period of the secondary are forced, due to
the gravitational interaction with the secondary, to follow
non-circular orbits (see also \S~\ref{sec:gasstream}). The particles
cannot, however, follow these orbits exactly, because they intersect
both with neighbouring circular orbits and themselves. The
non-circular orbit cannot be uniformly populated due to these
self-interactions, so a precessing arc of material is
formed. Interactions between this arc and the disc itself are thought
to produce the superhump light (see figure~\ref{fig:superhumpmap}).

\subsection{Spiral shocks}

{\em Spiral shocks\/} are another manifestation of the tidal influence
of the secondary star on the accretion disc. They are density waves
formed in the disc when particles in intersecting, non-circular orbits
interact \citep{sawada86a,sawada86b}. The shock waves form a two-armed
spiral pattern in the disc and are capable of transporting
angular momentum through the disc without the need for viscosity (see
\S~\ref{sec:outbursts}). Spiral shocks were first detected in the
outburst accretion disc of the dwarf nova IP~Peg
\citep{steeghs97,steeghs98} and have since been observed in many other
CVs, including V347~Pup \citep{still98}, EX~Dra \citep{joergens00},
U~Gem \citep{groot01} and WZ~Sge \citep{baba02}. Until recently,
spiral shocks had only been observed in either outburst or a high
state, but there is some recent evidence for spiral shocks in the
quiescent discs of IP~Peg \citep{neustroev02a} and U~Gem
\citep{moralesrueda04,undasanzana05}.

\section{Methods of mass determination}
\label{sec:massdetermination}

The mass ratio and the component masses are, apart perhaps from the
orbital period, the most fundamental physical parameters of any binary
star system. A knowledge of the component masses is central to our
understanding the origin, evolution and behaviour of CVs. Population
synthesis models \citep[e.g.][]{dekool92} and the disrupted magnetic
braking model (\S~\ref{sec:disrupted}) of CV evolution are just two
crucial aspects that require reliable masses in order to be
observationally tested.  Unfortunately, at present reliable CV mass
estimates are limited to approximately 20 systems, partially due to
the intrinsic difficulties in obtaining such masses (see
\citet{smith98a} for a review).

\subsection{Mass-orbital period relations}
Reasonable estimates of the masses of the secondary stars in CVs can
be obtained from the mass-orbital period relationship
(\citealp{robinson73,robinson76}; \citealp[][page
106]{warner95}). Note that equation~\ref{eq:kepler} can be rewritten
as
\begin{equation}
\frac{M_{2}}{{\rm M}_{\odot}} = 0.358 \left( \frac{1+q}{q}
\right)^{\frac{1}{2}} \left(\frac {R_{\rm{L}}}{a}
\right)^{\frac{3}{2}} \left( \frac{M_{1}}{{\rm M}_{\odot}} 
\frac{{\rm R}_{\odot}}{R_{\rm{L}}} \right)^{\frac{3}{2}}
P_{\rm{orb}}(\rm{h}).
\end{equation}
The first two terms in brackets are virtually independent of the mass
ratio $q$. The last can be determined from any mass-radius
relationship. Following \citet[][page 106]{warner95}, I adopt the
empirical result for low-mass ($<0.5\;{\rm M}_{\odot}$) main-sequence
stars of \citet{caillaut90}, as appropriate for the secondary stars in
CVs:
\begin{equation}
\frac{R}{{\rm R}_{\odot}} = 0.918 \left( \frac{M}{{\rm M}_{\odot}}
\right) ^{0.796}.
\end{equation}
Combining these last two equations with equation~\ref{eq:rochelobe2}
gives an approximate mass-period relationship for the secondary stars
in CVs:
\begin{equation}
\label{eq:massradius}
\frac{M_{2}}{{\rm M}_{\odot}} \approx 0.091 P^{1.44}_{\rm{orb}} (\rm{h}).
\end{equation}
\citet{smith98a} obtained an empirical mass-period relation for the
secondary star from CVs with well-determined component masses:
\begin{equation}
\label{eq:massradius2}
\frac{M_{2}}{{\rm M}_{\odot}}=(0.126\pm0.011)P_{\rm{orb}}({\rm h}) -
(0.11\pm0.04).
\end{equation}

Although such a method of mass determination is obviously not entirely
satisfactory, mass-period relations such as
equations~\ref{eq:massradius} and \ref{eq:massradius2} provide a means
of estimating the secondary mass in a CV from what is often the only
available observational constraint: the orbital period. Unfortunately,
they provide no clue as to the mass of the primary star.

\subsection{Radial velocity mass determination}
\label{sec:radmass}
The masses of the primary and secondary star can be directly
determined if the radial velocities of the stellar components $K_{1}$
and $K_{2}$ and the orbital period and inclination are accurately known.

\begin{figure}
\centerline{\includegraphics[width=12cm,angle=-90]{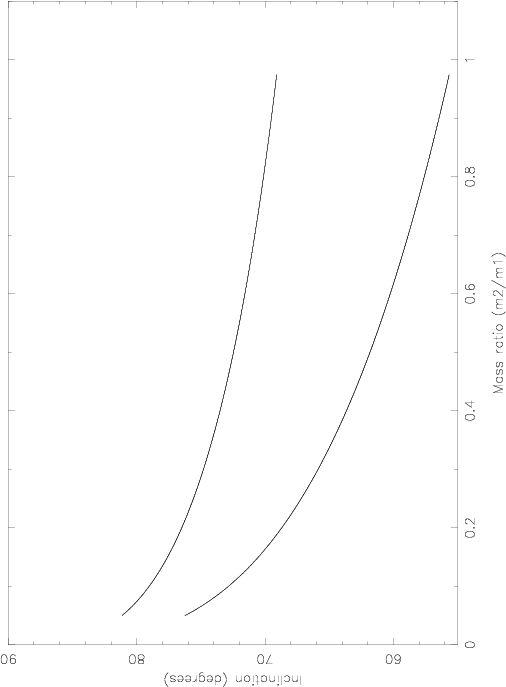}}
\caption[The relationship between the mass ratio $q$ and orbital
  inclination $i$ for grazing eclipses of the bright spot and white
  dwarf.]{The relationship between the mass ratio $q$ and orbital
  inclination $i$ for grazing eclipses ($\Delta\phi=0$) of (bottom)
  the bright spot and (top) the white dwarf. The disc radius is
  $0.35a$.}
\label{fig:qi}
\end{figure}

Observed at an inclination $i$, the radial velocity amplitude of the
primary is
\begin{equation}
K_{1}=\frac{2\pi a_{1}}{P_{\rm{orb}}} \sin i
\end{equation}
and that of the secondary is
\begin{equation}
K_{2}=\frac{2\pi a_{2}}{P_{\rm{orb}}} \sin i.
\end{equation}
From
\begin{equation}
a=a_{1} \left( \frac{M_{1}+M_{2}}{M_{2}} \right)
\end{equation}
and Kepler's third law (equation~\ref{eq:kepler}), we then obtain the
standard relationships for the mass functions $f$ of the primary
\begin{subequations}
\label{eq:fm1}
\begin{eqnarray}
f(M_{1}) & = & \frac{P_{\rm{orb}}K^{3}_{2}}{2\pi G}\\
 & = & \frac{(M_{2} \sin i)^{3}}{(M_{1}+M_{2})^2}\\
 & = & M_{1} \left( \frac{q}{1+q} \right) ^{2} \sin ^{3} i
\end{eqnarray}
\end{subequations}
and of the secondary
\begin{subequations}
\begin{eqnarray}
\label{eq:fm2}
f(M_{2}) & = & \frac{P_{\rm{orb}}K^{3}_{1}}{2\pi G}\\
 & = & \frac{(M_{2} \sin i)^{3}}{(M_{1}+M_{2})^2}\\
 & = & M_{2} \left( \frac{1}{1+q} \right) ^{2} \sin ^{3}i.
\end{eqnarray}
\end{subequations}
Note that the mass functions are functions of the observable
quantities $K$ and $P_{\rm{orb}}$ only. As the mass ratio $q$ is given
by
\begin{equation}
q=K_{1}/K_{2}=M_{2}/M_{1},
\end{equation}
the orbital inclination is all that is needed to gain accurate
individual masses if the mass functions are known. However, the
inclination is generally only reliably determined in eclipsing
systems. The fact that a system exhibits eclipses at all constrains
the inclination to $60^{\circ}\lesssim i <90^{\circ}$
(figure~\ref{fig:qi}). An eclipse of the white dwarf further limits
the inclination to $75^{\circ}\lesssim i <90^{\circ}$. The presence of
distinct eclipses of the white dwarf and bright spot allows, as
discussed in the following section, accurate determination of both the
mass ratio and orbital inclination.

The orbital inclination can also be estimated in non-eclipsing systems
from modelling of ellipsoidal variations \citep{russell45} from the
secondary star \citep[e.g.][]{sherrington82,berriman83,mcclintock83},
but this requires accurate modelling of reflection effects and
gravity- and limb-darkening
\citep[e.g.][]{lucy67,pantazis98,claret03}. \citet{smith98a} describe
various other methods by which estimates of stellar masses in CVs may
be made, but in general eclipsing systems are the only ones for which
we know reliable inclinations \citep[][page 103]{warner95}.

Measuring the radial velocity amplitude of the white dwarf is
problematic. The white dwarf (absorption) lines usually only dominate
in the ultra-violet, and are hence generally unobservable from
below-atmosphere sites. \citet{sion98} obtained a direct measurement
of the white dwarf radial velocity amplitude of U~Gem from Hubble
Space Telescope (HST) observations using the Si {\sc iii} absorption
line. In general, however, $K_{1}$ has to be measured from optical
emission lines of the accretion disc. For an azimuthally symmetric
disc, these lines trace the orbital motion of the primary. Problems
arise when the disc departs significantly from this symmetry, due
primarily to interactions between the disc and gas stream in the
region of the bright spot. This gives rise to phase shifts and
apparent orbital eccentricities in the radial velocity
curves. Techniques that determine the radial velocity amplitude from
the emission line wings (e.g.\ the double-Gaussian method of
\citealp{schneider80}) presume that most of the asymmetric emission
originates in the outer disc \citep{horne86b} where the disc-stream
interaction should be maximised. Frequently, however, this does not
solve the problem \citep[e.g.][]{thoroughgood04}, and the emission
lines are asymmetric at all radii. The radial velocity distortions can
be either due to asymmetric brightness distributions
\citep[e.g.][]{stover81}, perhaps due to the gas stream overflowing
the edge of the disc and impacting the disc at smaller radii
\citep{lubow75}, and/or to non-Keplerian velocity distributions
\citep[e.g.][]{schoembs83,marsh87}. Equation~\ref{eq:fm2} shows that
the mass function is highly dependent on an accurate determination of
$K_{1}$; for the most part making the use of uncertain values of
$K_{1}$ in mass determinations unwise.

For the above reasons, radial velocity studies generally concentrate
on $K_{2}$. Although absorption lines from the secondary can be
observed at the red end of the spectrum, the continuum emission is
often dominated by the disc. Cross-correlation of the CV spectrum with
a template of a cool dwarf spectrum \citep{stover80} is often used to
obtain the radial velocity amplitude of the secondary. Summing
phase-binned spectra followed by such cross-correlation can reveal
many weak absorption lines from the secondary \citep{horne86b}. A
technique known as skew-mapping, which also makes use of
cross-correlation of CV and template star spectra, is useful in cases
where the signal-to-noise ratio is poor. \citet{vandeputte03} describe
the method in detail; it entails finding the peak of the line integral
of the cross-correlation function in the $(K_{1},K_{2})$ plane.

Rotational broadening of the secondary's absorption lines can also be
used to determine the value of $v \sin i$ for the secondary, where $v$
is the rotational velocity of the secondary. As the secondary star is
tidally locked, its rotation period is identical to the orbital
period, yielding (\citealp{friend90a}; \citealp*{horne93})
\begin{equation}
\frac{R_{2}}{a} (1+q) = \frac{v\sin i}{K_{2}}.
\end{equation}
Using equation~\ref{eq:RL2} for the volume radius of the Roche-lobe
together with the above expression allows the mass ratio to be
determined \citep[e.g][]{horne93,thoroughgood01}. The inclination can
then be determined from the unique relation between the mass ratio and
orbital inclination for a given eclipse width of the primary
$\Delta\phi$ (\citealp{bailey79}; see figure~\ref{fig:qi}), which can
be understood as follows:
\begin {enumerate}
\item At smaller orbital inclinations a larger secondary radius
$R_{2}$ is required in order to produce a given eclipse width.
\item The secondary radius is defined by the mass ratio because the
secondary fills its Roche lobe.
\item Therefore for a specific white dwarf eclipse width
$\Delta\phi$ the inclination is known as a function of the mass ratio.
\end{enumerate}
Remember that the shape of the system does not depend on the orbital
separation $a$: this just determines the scale.

The relation between the eclipse width, orbital inclination and
secondary radius is often approximated by the eclipse of a point
source by a spherical body, which gives the analytical expression
\citep{dhillon91}
\begin{equation}
\left( \frac{R_{2}}{a} \right) ^{2} = \sin ^{2}(\pi\Delta\phi)+ \cos
^{2}(\pi\Delta\phi) \cos ^{2} i.
\end{equation}
For an axi-symmetric disc the white dwarf eclipse phase width
$\Delta\phi$ is approximately equal to the full phase width at half
maximum of the disc eclipse $\Delta\phi_{1/2}$
\citep[e.g.][]{dhillon90}. If the disc eclipse is symmetrical about
phase zero then this is a good approximation. The radius of the
secondary can then be determined from the mass ratio. Kepler's third
law (equation~\ref{eq:kepler}) then allows the rest of the system
parameters, including the stellar masses, to be determined.

Such mass estimates from the radial velocity of the secondary,
however, can run into problems if the secondary flux is not uniform
across the surface of the star. Effects such as gravity- and
limb-darkening need to be accounted for, but the major problem is
irradiation from the primary. This results in absorption features
being eliminated from (or reduced in strength on) the illuminated face
of the secondary, a problem made more complex by the fact that regions
of the secondary star will be shadowed by the accretion disc, so will
still show absorption lines. This scenario results in an asymmetric line
flux distribution across the secondary star, which can adversely
affect the measurement of $K_{2}$. For example, if the absorption
lines are reduced in strength on the side of the secondary facing the
primary, then the flux from the absorption lines will be centred
towards the far side of the secondary. The measured value of $K_{2}$
will therefore be larger than the true, dynamical, value. As an
example of how this can be corrected for, \citet{thoroughgood04} used
model CV spectra with varying  numbers of vertical slices across the
inner hemisphere of the secondary star's Roche-lobe omitted in order to
model the irradiation of the inner face of the star, leading to a
corrected value of $K_{2}$.

\subsection{The photometric method of mass determination}
\label{sec:phomass}
In a number of eclipsing objects, the component masses may be
determined from photometry combined with a mass-radius relation for the
primary \citep[e.g.][]{wood86b}. This results in a purely photometric
model of the system, untroubled by concerns about the contamination of
$K_{1}$ or $K_{2}$. This technique is thus both a valuable method of
determining the masses in itself, and a useful check of spectroscopic
results. Table~\ref{tab:q_comparison} compares the results achieved by
reliable spectroscopic techniques to those found via photometry
alone. The agreement between the values quoted is good, however, many
spectroscopic determinations of the mass ratio for other dwarf nov\ae\
have been excluded due to unreliable techniques being employed. The
conclusion to be drawn is that both methods can produce reliable and
accurate values, but that great care must be taken when using
spectroscopic data (see \citealp{smith98a} for a detailed discussion of
the techniques of mass determination).

\begin{table}
\begin{center}
\caption[A comparison of purely spectroscopic and purely photometric
  determinations of the mass ratios of selected dwarf nov\ae.]{A
  comparison of purely spectroscopic and purely photometric
  determinations of the mass ratios of selected dwarf nov\ae. Only
  those dwarf nov\ae\ with mass ratios accurately determined by both
  spectroscopy alone and photometry alone are shown, as the use of
  inappropriate data and/or techniques can result in wildly inaccurate
  results. See
  \citet{smith98a} for a critical discussion of the (then) available
  mass estimates of CVs. The photometric determinations listed below
  all use the method described in \S~\ref{sec:phomass}. The
  spectroscopic determinations used the following
  techniques. V2051~Oph: Ref.~1 used $K_{1}$ from H$\beta$ and
  H$\gamma$, corrected for disc asymmetry, the radial velocities of
  the disc emission lines and the eclipse width from the continuum
  light curve. IP~Peg: Ref.~3 used $K_{2}$ from TiO absorption
  features combined with the projected rotational velocity of the
  secondary star as determined by Catal\'{a}n (private communication);
  ref.~4 used Roche tomography. EX~Dra: Ref.~6 used $K_{1}$ measured
  from H$\beta$, H$\gamma$ and H$\delta$ using the double-Gaussian
  method of \citet{schneider80} and $K_{2}$ measured from calcium
  absorption lines.}
\vspace{0.3cm}
\small
\begin{tabular}{lclcl}
\hline
Object & \multicolumn{2}{c}{Spectroscopic} &
\multicolumn{2}{c}{Photometric} \\
 & $q$ & Ref.\ & $q$ & Ref.\ \\
\hline
V2051~Oph & $0.26\pm0.04$ & 1 & $0.19\pm0.03$ & 2 \\
IP~Peg & $0.322^{+0.075}_{-0.037}$, $0.43$ & 3, 4 & $0.35<q<0.49$ & 5 \\
EX~Dra & $0.75\pm0.01$ & 6 & $0.72\pm0.06$ & 7 \\
\hline
\end{tabular}
\vspace{0.3cm}
\begin{tabular}{ll}
References: & 1.\ \citet{watts86}, 2.\ \citet{baptista98}, \\
 & 3.\ \citet{beekman00}, 4.\ \citet{watson03}, \\
 & 5.\ \citet{wood86a}, 6.\ \citet{fiedler97}, \\
 & 7.\ \citet{baptista00}.
\end{tabular}
\normalsize
\label{tab:q_comparison}
\end{center}
\end{table}

\begin{figure}
\centerline{\includegraphics[width=12cm,angle=0]{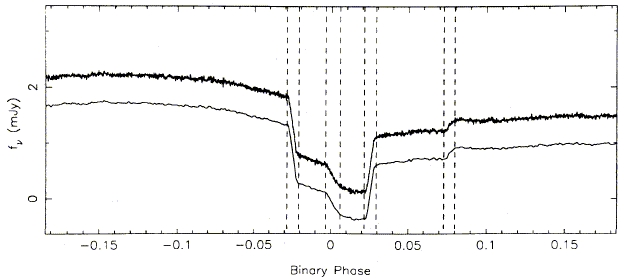}}
\caption[An eclipse of OY~Car, illustrating the distinct eclipses of
  the white dwarf and bright spot.]{An eclipse of OY~Car, illustrating
  the distinct eclipses of the white dwarf and bright spot. From left,
  marked by pairs of vertical dashed lines (i.e.\ the start and end):
  the ingress of the white dwarf; the ingress of the bright spot; the
  egress of the white dwarf and the egress of the bright spot. The
  upper curve is the raw data and the lower curve the data after
  smoothing. Adapted from \citet{wood89a}.}
\label{fig:contactphases}
\end{figure}

The most fundamental requirement of the photometric technique is the
presence of clear and distinct eclipses of the white dwarf and bright
spot. An example of such a system is OY~Car, illustrated in
figure~\ref{fig:contactphases}. The method then proceeds by utilising
the unique relationship between $q$ and $i$ for a given $\Delta\phi$
(see the previous section and figure~\ref{fig:qi}).

If it is assumed that the ballistic gas stream, the trajectory of
which can be determined as described in \S~\ref{sec:gasstream}, passes
through the position of the bright spot, by calculating and comparing
the ingress and egress phases of each point along the gas stream for
different values
of $q$ to those of the bright spot eclipse timings the mass ratio and
orbital inclination can be uniquely determined. The additional
assumption that the bright spot lies on the outer edge of the
accretion disc gives the radius of the disc. To obtain the component
masses, a mass-radius relation for the primary is required, for
instance the Nauenberg approximation
(equation~\ref{eq:nauenberg}). This method of determining the system
parameters is discussed in more detail in \S~\ref{sec:derivative}.

\section{This thesis}
Chapter~\ref{ch:observations} describes the observations and the
reduction procedure employed on the data obtained. I give a detailed
review of the {\sc ultracam} instrument, give full details of the
observations discussed in the latter portions of this thesis and
elucidate the procedures used to reduce the data. In
chapter~\ref{ch:analysis}, I describe the analysis techniques I
applied to this data and a detailed comparison of two distinct methods
of photometric parameter determination. I also describe the
eclipse mapping method. The structuring of the three results chapters
mirrors that of the three papers that this thesis is based upon, and
is roughly chronological. The results for the system parameters for
OU~Vir are given in chapter~\ref{ch:ouvir} and those for XZ~Eri and
DV~UMa in chapter~\ref{ch:xzeridvuma}. Observations of GY~Cnc and
IR~Com are described and discussed in
chapter~\ref{ch:gycncircomhtcas}. The main subject of this latter
chapter is, however, a discussion of the results of eclipse mapping of
the quiescent disc of HT~Cas in two distinct states in 2002 and
2003. The results of eclipse mapping experiments for the other objects
are given at the end of the relevant chapter. I conclude in
chapter~\ref{ch:conclusions} with a discussion of the main results of
this thesis, and suggest some appropriate future work.

%% file: observations.tex
\chapter{Observations and data reduction}
\label{ch:observations}

All the data presented in this thesis come from observations made with
{\sc ultracam} on the 4.2-m William Herschel Telescope (WHT) at the Isaac
Newton Group of Telescopes, La Palma. The {\sc ultracam} instrument is
described in detail below; see also \citet{dhillon01b}, \citet{beard02},
\citet{dhillon05}, \citet{stevenson05}.

\section{{\sc Ultracam}}
\label{sec:ultracam}

\begin{figure}
\centerline{\includegraphics[width=14cm,angle=0]{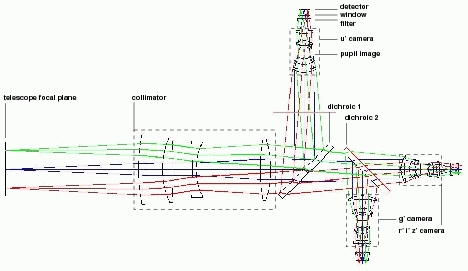}}
\caption[A ray-trace through {\sc ultracam}.]{A ray-trace through {\sc
  ultracam}, showing the major optical components: the collimator,
  dichroics, cameras, filters and detector windows. From
  \citet{dhillon05}.}
\label{fig:raytrace}
\end{figure}

\begin{figure}
\centerline{\includegraphics[width=12cm,angle=0]{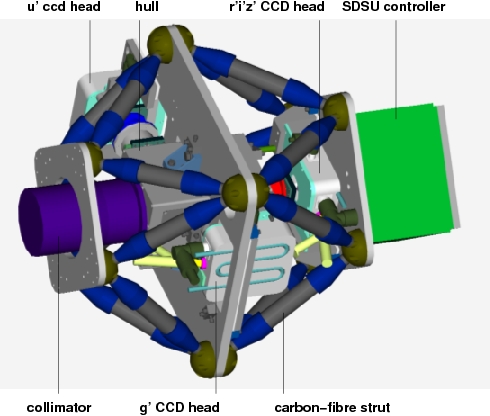}}
\caption[A CAD image of the opto-mechanical design of {\sc
  ultracam}.]{A CAD image of the opto-mechanical design of {\sc
  ultracam}, highlighting some of the components discussed in the text
  and shown in the ray-trace of figure~\ref{fig:raytrace}. From
  \citet{dhillon05}.}
\label{fig:cad}
\end{figure}

\begin{figure}
\centerline{\includegraphics[width=12cm,angle=0]{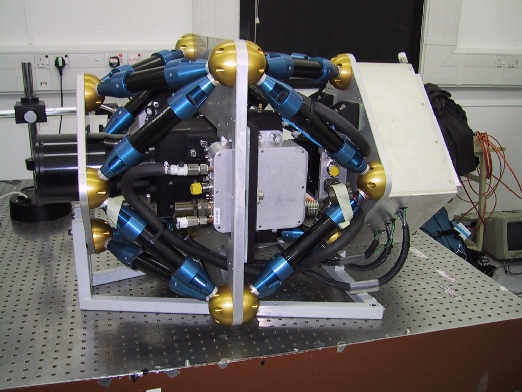}}
\caption[A photograph of {\sc ultracam} in the test focal station of the WHT
  just prior to commissioning on the telescope.]{A photograph of
  {\sc ultracam} in the test focal station of the WHT just prior to
  commissioning on the telescope (courtesy of Sue Worswick).}
\label{fig:ucam_photo}
\end{figure}

{\sc Ultracam} is an ultra-fast, triple-beam CCD camera. A ray-trace of
{\sc ultracam} is shown in figure~\ref{fig:raytrace}. A CAD image of
the opto-mechanical design of {\sc ultracam} is shown in
figure~\ref{fig:cad} and a photograph of the instrument itself is
shown in figure~\ref{fig:ucam_photo}. Light from the telescope first
passes through a collimator (which is interchangeable, allowing {\sc
ultracam} to be mounted on a variety of telescopes). It then
encounters the first of two dichroic beam-splitters, which reflects
light short-wards of $\sim390$~nm through $90^{\circ}$, allowing
longer wavelengths through. The longer-wavelength light then strikes
the second dichroic. The cut-point for this beam-splitter is
$\sim550$~nm, and light of a wavelength shorter than this is reflected
through $90^{\circ}$ (in the opposite direction to the blue light),
while the rest is transmitted. The light is now split into three
wavelength-dependent components. Each passes through re-imaging optics
and the relevant filter (discussed below) before falling onto one of
the three CCD detectors.

\begin{figure}
\centerline{\includegraphics[width=12cm,angle=0]{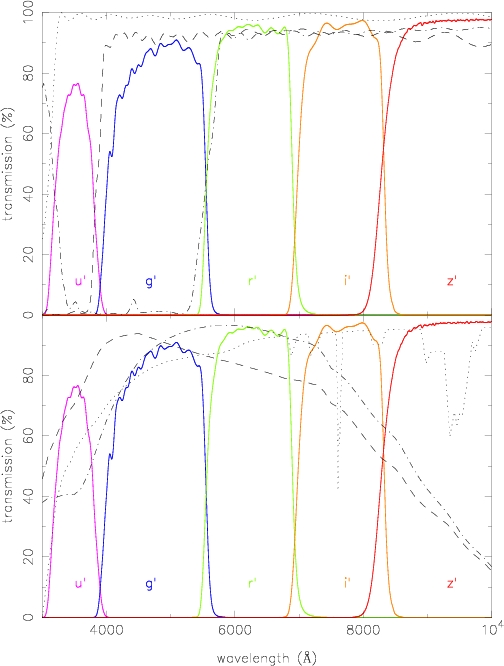}}
\caption[Transmission profiles of the {\sc ultracam} optical
  components.]{Top: Transmission profiles of the {\sc ultracam} SDSS
  filter-set (solid lines), the anti-reflection coating used on the
  {\sc ultracam} lenses (dotted line), and the two dichroics (dashed line
  and dashed-dotted line). Bottom: Transmission profiles of the
  {\sc ultracam} SDSS filter-set (solid lines) and the atmosphere for unit
  airmass (dotted line). Also shown are the quantum efficiency curves
  of the {\em u}$^\prime$ and {\em g}$^\prime$ CCDs (dashed line) and
  the {\em r\/}$^\prime${\em i\/}$^\prime${\em z\/}$^\prime$ CCD
  (dashed-dotted line). From \citet{dhillon05}.}
\label{fig:transmission}
\end{figure}

{\sc Ultracam} uses the {\em u\/}$^{\prime}${\em g\/}$^{\prime}${\em
r\/}$^{\prime}${\em i\/}$^{\prime}${\em z\/}$^{\prime}$ filter system
defined by the Sloan Digital Sky Survey (SDSS; see
\citealp{fukugita96,smith02}). The filter transmission functions are
shown in figure~\ref{fig:transmission}. The importance of this choice of
filter system is three-fold:
\begin{enumerate}
\item The {\em u\/}$^{\prime}${\em g\/}$^{\prime}${\em
  r\/}$^{\prime}${\em i\/}$^{\prime}${\em z\/}$^{\prime}$ filter
  system is likely to become the dominant filter system in the future,
  as the SDSS will survey the sky in unprecedented depth and detail.
\item Overlaps between the filters are minimised compared to the {\em
  UBVRI} system. This is of vital importance considering the dichroic
  beam-splitters used in {\sc ultracam}.
\item The SDSS {\em r\/}$^{\prime}$ filter has a curtailed red wing
  compared to the Cousins {\em R\/} filter. This eliminates fringing
  with thinned chips in the {\em r\/}$^{\prime}$ filter.
\end{enumerate}
The SDSS filter system is also useful in that it is a very broad band
system---the bandpasses are significantly wider than other filter
systems \citep{fukugita96}. This ensures high efficiency, which is
useful when observing faint targets.

The ability of {\sc ultracam} to observe {\em simultaneously\/} in three
colours is a crucial aspect of its design. Three colours enables a
stellar spectrum to be distinguished from a blackbody. Simultaneous
observations also eliminate the problem of the source varying between
filter changes, crucial for observations of rapidly varying targets
such as close binary stars.

The CCDs used in {\sc ultracam} are key to its high
time-resolution. The chips used are three Peltier-cooled (see
\S~\ref{sec:dark} for a discussion of why this is necessary),
back-illuminated, anti-reflection coated (see
figure~\ref{fig:transmission}), thinned EEV 47-20 frame-transfer CCDs
with an imaging area of $1024\times1024$ pixels ($13.3\times13.3$~mm),
giving a plate scale of $0.3^{\prime\prime}/{\rm pixel}$ on the
WHT. This gives a field-of-view of $5^{\prime}$ on the WHT. At a
Galactic latitude of $30^{\circ}$ (the all-sky average), this means
that the probability of finding a comparison star brighter than
13~magnitudes is 0.96 (calculated using the on-line {\sc ultracam}
comparison star probability calculator, written by Vik Dhillon). 

Frame-transfer CCDs have a masked-off storage area. Charge from the
exposed portion of the chip is shifted (or vertically {\em clocked\/})
into this region before being horizontally clocked and digitized. This
has the advantage that horizontal clocking and digitization can occur
whilst the next exposure is taking place. This can vastly reduce the
dead-time between exposures, since the digitization time is typically
much greater than the clocking times (the digitization time is
$\sim6$~sec for the  full-frame for {\sc ultracam}). The vertical
clocking speed for the chips used in {\sc ultracam} is $24\;\mu {\rm
s}\,/\,$row, which comes to $\sim24$~ms for the total 1024~rows of pixels
of the {\sc ultracam} CCDs. As long as the exposure time is longer
than the sum of the horizontal clocking and digitization times, the
dead-time is reduced to the vertical clocking time.

Another useful, indeed crucial, aspect of the design of {\sc ultracam} is
the ability to only read out selected parts of the CCDs, called {\em
windows}. This reduces the digitization time, and therefore the
readout time, enabling higher frame rates to be achieved.

{\sc Ultracam} also has a mode known as drift mode. In this mode a
pair of windows are vertically clocked until they are just within the
masked-off region of the chip, whereupon another pair of windows are
exposed. In this way a vertical stack of windows is produced, and the
dead time between exposures is much reduced. The stack of windows are
continually shifted down the exposed and masked areas of the chip
before being read out at the bottom. New windows are continually added
at the top of the exposed area of the chip. Drift mode allows frame
rates of up to 500~Hz to be realised. \citet{stevenson05} discusses
this in detail.

\section{Journal of observations}
\label{sec:journal}
Table~\ref{tab:journal} presents a full journal of observations. All
the objects are dwarf nov\ae\ observed in quiescence, although OU~Vir
shows some evidence (see chapter~\ref{ch:ouvir}) for being on the
descent from superoutburst in both 2002 May and 2003 May. In the case
of the 2003 May observations, this is supported by the detection of a
superoutburst on May 2 by \citet{kato03}.

The observers in 2002 May (the commissioning run) were Vik Dhillon,
Tom Marsh, Mark Stevenson, Paul Kerry, Carolyn Brinkworth, David
Atkinson and Andy Vick. In 2003 May, the observers were Vik Dhillon,
Tom Marsh, myself, Carolyn Brinkworth and Paul Kerry. In
2003 November, the observers were Tom Marsh and myself.

\begin{sidewaystable}
\begin{center}
\caption[Journal of observations.]{Journal of observations. All dates
  are the start of the nights' observing. The {\em r\/}$^{\prime}$
  extinction is taken from the ING website, www.ing.iac.es, and is
  measured nightly by the Carlsberg Meridian Telescope. Values marked
  by a colon (:) are uncertain, probably due to the night being
  non-photometric.}
\vspace{0.3cm}
\small
\begin{tabular}{ccccccccccc}
Target & Date & Filters & UT start & UT end & Exposure &
Seeing & Data & Cycle & Eclipses & $r^{\prime}$
extinction \\
 & (yyyy mm dd) & & (hh:mm) & (hh:mm) & time (sec) & (arcsec) & points
 & & & (mag/airmass) \\
\hline
OU Vir & 2002 05 16 & {\em u\/}$^{\prime}${\em g\/}$^{\prime}${\em
 r\/}$^{\prime}$ & 23:38 & 02:18 & 0.5 & 1.2 & 1685 & 9442--9443
 & 2 & 0.124 \\

 & 2002 05 18 & {\em u\/}$^{\prime}${\em g\/}$^{\prime}${\em
 r\/}$^{\prime}$ & 00:20 & 00:25 & 1.7 & 2.1 & 59 & 1538 & 0 &
 0.107 \\

 & 2002 05 18 & {\em u\/}$^{\prime}${\em g\/}$^{\prime}${\em
 r\/}$^{\prime}$ & 00:26 & 02:10 & 4.0 & 2.1 & 1538 & 9470 & 1 &
 0.107 \\

 & 2003 05 19 & {\em u\/}$^{\prime}${\em g\/}$^{\prime}${\em
 z\/}$^{\prime}$ & 01:30 & 02:35 & 9.2 & 4.5 & 430 & 14504 & 0 &
 0.188: \\

 & 2003 05 20 & {\em u\/}$^{\prime}${\em g\/}$^{\prime}${\em
 i\/}$^{\prime}$ & 01:14 & 02:36 & 5.2 & 1.2 & 933 & 14518 & 1 &
 0.092: \\

 & 2003 05 22 & {\em u\/}$^{\prime}${\em g\/}$^{\prime}${\em
 i\/}$^{\prime}$ & 22:58 & 23:25
 & 4.2 & 0.8 & 373 & 14544 & 1 & 0.234 \\

 & 2003 05 25 & {\em u\/}$^{\prime}${\em g\/}$^{\prime}${\em
 i\/}$^{\prime}$ & 00:05 & 00:41 & 4.2 & 1.5 & 548 & 14586 & 1 &
 0.115 \\\\

DV UMa & 2003 05 20 & {\em u\/}$^{\prime}${\em g\/}$^{\prime}${\em
 i\/}$^{\prime}$ & 23:06 & 23:40 & 5.9 & 1.3--2.0 & 339 & 69023 &
 1 & 0.092: \\

 & 2003 05 22 & {\em u\/}$^{\prime}${\em g\/}$^{\prime}${\em
 i\/}$^{\prime}$ & 22:26 & 22:54 & 4.9 & 1.2 & 345 & 69046 & 1 &
 0.234 \\

 & 2003 05 23 & {\em u\/}$^{\prime}${\em g\/}$^{\prime}${\em
 i\/}$^{\prime}$ & 23:03 & 23:08 & 3.9 & 1.0 & 60 & 69058 & 0 &
 0.383 \\

 & 2003 05 23 & {\em u\/}$^{\prime}${\em g\/}$^{\prime}${\em
 i\/}$^{\prime}$ & 23:08 & 23:44 & 3.9 & 1.0 & 540 & 69058 & 1 &
 0.383 \\\\

XZ Eri & 2003 11 13 & {\em u\/}$^{\prime}${\em g\/}$^{\prime}${\em
 i\/}$^{\prime}$ & 23:25 & 01:48 & 7.0 & 1.0--2.0 & 1225 &
 4733--4734 & 2 & 0.073 \\\\

GY Cnc & 2003 05 19 & {\em u\/}$^{\prime}${\em g\/}$^{\prime}${\em
 z\/}$^{\prime}$ & 21:00 & 22:20 & 2.1 & $\gtrsim3$ & 2256 & 6826 & 1 &
 0.188: \\

 & 2003 05 23  & {\em u\/}$^{\prime}${\em g\/}$^{\prime}${\em
 i\/}$^{\prime}$ & 22:02 & 23:01 & 1.6 & 1 & 2150 & 6849 & 1 & 0.383

\end{tabular}
\normalsize
\label{tab:journal}
\end{center}
\end{sidewaystable}

\setcounter{table}{0}
\begin{sidewaystable}
\begin{center}
\caption[{\em Continued}. Journal of observations.]{{\em
  Continued}. Journal of observations. The {\em i\/}$^{\prime}$
  sensitivity was lost during the 2002 September 14 eclipse of HT~Cas
  due to a technical problem with this band. The GPS signal, used for
  time-stamping each exposure, was lost for the HT~Cas data of 2003
  October 29. This means that the absolute time of each exposure was
  incorrectly recorded, although the relative timing within the run
  remains accurate. The cycle number for the data of 2003 October 29
  is therefore estimated from times in the observing log. Due to poor
  weather, the extinction could not be measured for this night, and is
  therefore assumed to be 0.1~mag/airmass in the {\em r\/}$^{\prime}$
  band, the mean of the previous and subsequent nights'.}
\vspace{0.3cm}
\small
\begin{tabular}{ccccccccccc}
Target & Date & Filters & UT start & UT end & Exposure &
Seeing & Data & Cycle & Eclipses & $r^{\prime}$
extinction \\
 & (yyyy mm dd) & & (hh:mm) & (hh:mm) & time (sec) & (arcsec) & points
 & & & (mag/airmass) \\
\hline
IR Com & 2003 05 21 & {\em u\/}$^{\prime}${\em g\/}$^{\prime}${\em
 i\/}$^{\prime}$ & 23:49 & 00:26 & 3.2 & 1 & 676 & 37857 & 1 & 0.197 \\
 & 2003 05 23 & {\em u\/}$^{\prime}${\em g\/}$^{\prime}${\em
 i\/}$^{\prime}$ & 23:50 & 00:29 & 3.2 & 1 & 720 & 37880 & 1 & 0.383 \\
 & 2003 05 25 & {\em u\/}$^{\prime}${\em g\/}$^{\prime}${\em
 i\/}$^{\prime}$ & 21:39 & 22:28 & 3.2 & 1.5 & 901 & 37902 & 1 & 0.115 \\\\

HT Cas & 2002 09 13 & {\em u\/}$^{\prime}${\em g\/}$^{\prime}${\em
 i\/}$^{\prime}$ & 23:13 & 01:00 & 1.1 & 1.2 & 5651 & 119537 & 1 &
 0.089 \\ & 2002 09 14 & {\em u\/}$^{\prime}${\em g\/}$^{\prime}${\em
 i\/}$^{\prime}$ & 22:43 & 00:23 & 0.97--1.1 & 1.3--2.3 & 5470 &
 119550 & 1 & 0.071\\
 & 2003 10 29 & {\em u\/}$^{\prime}${\em
 g\/}$^{\prime}${\em i\/}$^{\prime}$ & -- & -- & 1.3 & 1.4 & 4659 &
 125116 & 1 & 0.1: \\ & 2003 10 30 & {\em u\/}$^{\prime}${\em
 g\/}$^{\prime}${\em i\/}$^{\prime}$ & 19:25 & 22:01 & 1.3 & 1.0--1.5
 & 6930 & 125129--125130 & 2 & 0.084\\

\end{tabular}
\normalsize
\end{center}
\end{sidewaystable}

\section{Data reduction}
\label{sec:reduction}
In this section I describe the data reduction procedure. Excepting
GY~Cnc, all the data were reduced by myself using the optimal
extraction algorithm \citep{naylor98} incorporated in Tom Marsh's {\sc
ultracam} pipeline data reduction software, which resulted in a
significant improvement in the signal-to-noise ratio over `normal'
extraction at low count rates (the {\em u\/}$^{\prime}$ data in
particular; see \S~\ref{sec:extraction}). The data for GY~Cnc were
reduced using the pipeline data reduction software with normal
aperture photometry, due to a problem with the optimal extraction at
high count rates\footnote{Note first that this problem does not affect
the rest of the data presented in this thesis, and second that this
problem is (believed to be) unrelated to the fact that so-called
`optimal' photometry is only optimised for low count rates where the
noise is sky-limited (see \S~\ref{sec:optimal}).}.

Subsequent analysis was conducted by myself. Transparency variations
were removed by dividing the target counts by that of a comparison
star. Times were converted from modified Julian dates on the UTC
time-scale (MJD) to heliocentric Julian dates (HJD) using the {\sc
fruit} Fortran subroutine, written by Peter Young. The comparison star
counts were converted to SDSS magnitudes using observations of
standard stars \citep{smith02}. All the data were corrected to zero
airmass using the procedure discussed in detail in
\S~\ref{sec:fluxcal}.

\subsection{Bias frames}
The presence of readout noise, which occurs when digitizing charge in
each pixel of the CCD, necessitates the addition of a (near)
constant number of counts to each pixel: the {\em bias}. If this bias
were not added, a systematic error would result when reading out low
counts, since negative counts are not recorded by the
analogue-to-digital converter. A bias frame can be obtained by taking
a zero second exposure. In practice, with {\sc ultracam}, an exposure
time of 1~ms (the minimum exposure time permitted by the camera
control software), in dark conditions, was used to take bias frames.

The high time-resolution of {\sc ultracam} means that many bias frames
were taken. These were combined using a clipped mean (at the $3\sigma$
level) to create a high signal-to-noise master bias frame free of
cosmic rays. The master bias was subtracted from each data frame
automatically by the pipeline software.

\subsection{Flat fielding}
The sensitivity of the CCD may vary across its surface. This variation
occurs on both small scales (pixel-to-pixel variations in area) and
large scales (e.g.\ vignetting, dust on the chip). These problems can
be corrected by using a {\em flat field}.

A flat is created by taking an exposure of a uniform field
of light. The best such field is usually an empty area of the twilight
sky. Due to the inevitable presence of some (faint) field stars, the
telescope was stepped in a spiral pattern during the flat field
exposures. The resulting flats were then combined into a master flat 
field for use by the pipeline software by the following procedure.

First, the master bias was subtracted from each individual flat
frame. The individual frames were then combined into a master flat
using the `makeflat' procedure in the pipeline software. This ignores
all frames with counts above or below a certain level ($30\,000$ and
$7\,000$, respectively). High counts may mean that the exposure is
saturated, distorting the shape of the flat. Also, above counts of
$\sim30\,000$ a `peppering' effect occurs on the {\sc ultracam}
CCDs. This effect is when adjacent pixels have high and low count
levels, resulting in a chequered appearance across the chip. The cause
of this effect is currently unknown \citep{stevenson05}. At the other
extreme, low counts will introduce excessive noise into the master
flat. The makeflat procedure first determines the mean levels of all
input frames and then averages them, again using a clipped mean at the
$3\sigma$ level, in groups of similar mean level. The averages are
then co-added. The makeflat procedure correctly weights low and high
signal flats. Each CCD is then normalised by its mean to produce the
master flat.

The data frames are then corrected for the sensitivity variations and
vignetting by dividing through by the resulting master flat after bias
subtraction.

\subsection{Dark frames}
\label{sec:dark}
Even in the complete absence of photons, the CCD would register a
signal greater than that of the bias. This is due to thermal
excitation of electrons in the semiconductor, known as {\em dark
current}. This can be corrected for by exposing the CCD in the absence
of light for a significant length of time. The resulting frames can be
combined (correcting for the exposure time) and subtracted from the
data and the flats.

Dark current with {\sc ultracam} is less than 0.1 electrons/pixel/sec, a
factor of approximately 50 less than the photon rate from the {\em
u\/}$^{\prime}$-band sky on the WHT. This is achieved due to the cooling
of the CCDs to $-40^{\circ}$C. Dark frames are difficult to
obtain in practice, since eliminating all sources of light in the WHT
dome proved problematical, so they were not used in the reduction
of these data. Given the very low level of the dark current with {\sc
ultracam}, the use of dark frames would have made a negligible
difference to the final data.

\subsection{Extraction}
\label{sec:extraction}
The end-product of an {\sc ultracam} observation is a spatially
resolved image of the star field, containing many stellar profiles,
some of which, especially in crowded fields, may overlap. There are
two principal techniques used to measure the brightness of a star in
such a field: aperture photometry, or `normal' extraction, and profile
fitting, or `optimal' extraction.

\subsubsection{Normal extraction}
Aperture photometry quite simply adds up all the counts within a
defined software aperture to obtain the counts from the star. Three
circles, of differing radii, are usually defined. The inner aperture
contains the star. The outer two are used to define an annulus in
which to measure the sky counts. The average sky brightness is
estimated (using a clipped mean) from the counts within the sky
annulus, and this value is subtracted from each of the pixels lying
within the star aperture. The standard deviation $\sigma$ of the sky
brightness $\eta$ can either be estimated from the photon noise
($\sigma^{2}=\eta$ for Poissonian statistics) or from the variance
($\sigma^{2}=\frac{(\eta-\bar{\eta})^{2}}{N-1}$, where $N$ is the number
of pixels). That is, the error in the former case comes from Poissonian
statistics and in the latter from the measured variation in the
sky photons within the sky annulus. For the data in this thesis, the sky
error was estimated using the variance.

The choice of aperture size is crucial to obtain the best
signal-to-noise. If the aperture size is too small, then too few
counts from the star will be included. For differential photometry
this is not a problem with regard to estimating the flux from the
target because the same fraction of counts will be lost from the
comparison star as from the target star. The signal-to-noise will not
be maximised, however, as the total counts will be lower. This of
course assumes that the point-spread function (PSF) is constant across
the field and that the apertures are both accurately centred on the
stars. Other noise sources such as read-out noise from the detector
will be minimised. Use of too small an aperture can, however, give
rise to so-called {\em aperture noise}. This arises because the flux
depends on which pixels are summed, and therefore on the size and
position of the seeing disc \citep{shahbaz94}. If an aperture is
chosen that is too large, then excess noise from the sky and read-out
noise will be included. In practice, several different sizes of
aperture are reduced and the one with the best signal-to-noise ratio
used. Variable seeing means that the size of this aperture may change
during the course of one set of observations (although this can be
allowed for by fitting a PSF to a bright comparison star and scaling the
target aperture by a set amount of the resulting full-width at half
maximum (FWHM); see below).

\subsubsection{Optimal extraction}
\label{sec:optimal}
Two major problems exist with simple aperture photometry. The first,
and most obvious, is encountered with crowded fields. This can lead to
the flux from a neighbouring star contaminating the target
aperture. Disentangling 
the flux from the two (or more) stars requires {\em profile
fitting}. This models the sky-subtracted flux as a series of point
sources convolved with the PSF. The PSF is either fitted by an
analytical function \citep[e.g.][]{penny86} or by fitting the observed
profile with an empirical profile \citep[e.g.][]{stetson87}. The
second problem was alluded to above: it is one of optimisation.

The `optimal' extraction algorithm for imaging photometry was
developed by \citet{naylor98}. It aims to extract the best possible
signal-to-noise from the data. For faint sources, the noise is
sky-limited, that is, the noise from the sky is much greater than
photon noise from the source. (It is in fact sky and/or readout noise
limited, but readout noise is also dependent on the number of pixels
in the aperture, so I shall for simplicity's sake lump the (constant)
readout noise and the (variable) sky noise together and refer only to sky
noise when meaning both.) In this case, the best aperture is small, so
that as little sky is included as possible. But how small can this
aperture be before too few counts from the source are included?

The solution is to sum the counts from the star using a weight
function for each pixel determined from the PSF measured using a
bright comparison star. The procedure recommended by \citet{naylor98}
is as follows:
\begin{enumerate}
\item Use a bright, non-saturated single comparison star in the image
  to fit the PSF. The integral of this PSF is normalised to
  one. The PSF is then integrated over individual pixels in order to
  resample it to the pixel grid $(i,j)$ of the detector, forming the
  estimated PSF, $P^{\rm{E}}$.
\item The stars of interest (the target and any comparison stars) are
  then fitted using $P^{E}$, with the parameters fixed, in order to
  accurately determine their positions.
\item The normalisation from the PSF fit to the faintest star defines
  $F^{\prime}$, the flux from the faintest star, used to define the
  variances $\sigma^{2}_{i,j}$ of each pixel:
\begin{equation}
\label{eq:pixelvariance}
\sigma^{2}_{i,j}=\sigma^{2}_{\rm{s}} +
\frac{F^{\prime}P^{\rm{E}}_{i,j}}{\sqrt{g}},
\end{equation}
where $\sigma_{\rm{s}}$ is the variance for a single pixel
containing only sky and $g$ is the gain.
\item The result from equation~\ref{eq:pixelvariance} is then used to
  define a weight function $W_{i,j}$ for each star:
\begin{equation}
\label{eq:weights}
W_{i,j}=\frac{P^{\rm{E}}_{i,j}}{\sigma^{2}_{i,j}
  \sum_{k,l}[(P^{\rm{E}}_{k,l})^{2}/\sigma^{2}_{k,l}]}.
\end{equation}
This weights the flux contribution from each pixel according to the
variance ($\sigma^{2}_{i,j}$) of that pixel.  The weight function
defined by equation~\ref{eq:weights} can then be used to estimate the
flux from each star
\begin{equation}
F=\sum_{i,j} W_{i,j}(D_{i,j}-S_{i,j}),
\end{equation}
where $D_{i,j}$ and $S_{i,j}$ are the total counts and estimated sky
counts for each pixel. The standard deviation of the measured flux is
then
\begin{equation}
\label{eq:optimalsigma}
\sigma=\sqrt{\sum_{i,j}W^{2}_{i,j}\sigma^{2}_{i,j}}.
\end{equation}
\end{enumerate}
Since the weight mask, mathematically, extends out to infinity, a
choice must be made about where to terminate it. This should obviously
be before any neighbouring stars make any significant contribution to
the flux or the edge of the detector area is reached. A sensible
termination point is $2\times$FWHM, by which time the weights are so small
that very little difference is made to the signal-to-noise ratio
(remember that the seeing profile is almost always a good approximation
to a Gaussian).

The PSF is fitted using a function which describes the seeing disc. A
Gaussian function is frequently used, but in practice a Moffat
profile \citep{moffat69} often produces a better fit, and in any case
approximates a Gaussian well in poor seeing. The intensity $I(r)$ at a
distance $r$ from the location of the peak intensity of the profile
$I_{0}$ for a Moffat profile is given by
\begin{equation}
\label{eq:moffat}
\frac{I(r)}{I_{0}}=\frac{1}{[1+(r/R)^{2}]^{\beta}},
\end{equation}
where $R$ is known as the {\em width parameter\/} and $\beta$ is a
dimensionless parameter which dictates the shape of the profile, and
is typically 3--5. Both of these parameters depend on the seeing.

Optimal photometry can also be used to disentangle the fluxes of two
stars whose profiles overlap by using the measured
PSF. \citet{naylor98} describes the procedure, but it will not be
discussed here, as it was not necessary for the work contained in this
thesis.

Optimal photometry is designed for cases where the variances are sky
or readout-noise limited, so for high signal-to-noise data it is not
necessarily the best method. For the data presented in this thesis,
optimal photometry was found in all cases (except for GY~Cnc; see
\S~\ref{sec:reduction}) to significantly improve the
{\em u\/}$^{\prime}$ signal-to-noise, frequently improving the {\em
r\/}$^{\prime}$, {\em i\/}$^{\prime}$ or {\em z\/}$^{\prime}$ band also, and
made negligible difference in the {\em g\/}$^{\prime}$. It was therefore
used to reduce all the data (except for GY~Cnc). As previously
mentioned, optimal 
extraction only gives the correct {\em ratio\/} of target counts to
comparison counts. To correct this ratio to actual fluxes and
magnitudes, the comparison flux must be measured using normal
extraction during photometric conditions.

For the work contained in this thesis, the principal advantages of
optimal photometry are as follows.
\begin{enumerate}
\item Ease of use. Optimal photometry automatically extracts the best
  (or near-best) signal-to-noise ratio for each frame. It is not necessary to
  manually compare the results from different aperture sizes as with
  normal extraction.
\item The most obvious advantage over normal aperture photometry is
  the increase in signal-to-noise for low count rates. This is at
  least 10~per~cent, and can be much greater \citep{naylor98}.
\end{enumerate}

\subsection{Flux calibration}
\label{sec:fluxcal}

There are three main steps in the reduction of raw (sky- and
bias-subtracted and properly flat-fielded) counts to correctly
calibrated fluxes: correction of extinction due to scattering and
absorption in the Earth's atmosphere, an instrumental correction due
to the deviation of the detecting apparatus from the standard and
conversion of the calibrated magnitudes into flux.

\subsubsection{Atmospheric extinction}
Absorption and scattering processes in the Earth's atmosphere
significantly reduce the flux from a star. There are three main
sources of this extinction: Rayleigh scattering by molecules in the
atmosphere, molecular absorption (by, for example, ozone, water vapour
and carbon dioxide) and aerosol scattering (by dust for instance). The
exact extinction correction depends on the current atmospheric
conditions, the airmass, $X$, of the target and the wavelength or
filter. The airmass is the length of the column of air passed through
by the light 
relative to that at the zenith. It increases as the angular distance
from the zenith (where $X=1$) increases. This relationship can be well
approximated by assuming the atmosphere to be plane-parallel, yielding
\begin{equation}
X=\sec Z,
\end{equation}
where $Z=90^{\circ}-altitude$ is the zenith distance. This
approximation breaks down for $Z\gtrsim60^{\circ}$, where a more
complex formula must be used \citep[e.g.][]{kristensen98}.

The flux received at the telescope is usually corrected to above the
Earth's atmosphere, i.e.\ at airmass zero. The instrumental magnitude
$m_{\lambda}$ can be corrected to the instrumental magnitude at
airmass zero $m_{\lambda0}$ by the following equation
\citep[e.g.][]{fukugita96}:
\begin{subequations}
\begin{eqnarray}
\label{eq:airmass}
m_{\lambda0} & = & m_{\lambda}-(\kappa^{\prime}_{\lambda}+
\kappa^{\prime\prime}_{\lambda} C)X\\
\label{eq:airmass2}
 & = & -2.5\log_{10}\left(\frac{c}{t}\right)-(\kappa^{\prime}_{\lambda}+
\kappa^{\prime\prime}_{\lambda} C)X,
\end{eqnarray}
\end{subequations}
where $\kappa^{\prime}_{\lambda}$ is the primary extinction
coefficient, $\kappa^{\prime\prime}_{\lambda}$ is the secondary
extinction coefficient, $C$ is the colour index (e.g.\ {\em
u}$^{\prime}-${\em g\/}$^{\prime}$) of the target, $c$ is the number of
received counts and $t$ is the exposure time.

The data in this thesis have been corrected for first-order extinction
effects only. That is, the $\kappa^{\prime\prime}_{\lambda}C$ term has
been neglected in equations~\ref{eq:airmass} and \ref{eq:airmass2}, as
$\kappa^{\prime\prime}_{\lambda}$ is significantly 
smaller than $\kappa^{\prime}_{\lambda}$ and $C$ is usually
$\lesssim1$. For example, \citet{smith02}
quote $\kappa^{\prime\prime}_{g^{\prime}}=-0.016\pm0.003$, whereas
$\kappa^{\prime}_{g^{\prime}}$ is typically an order of magnitude
greater than the modulus of this.  This simplifies matters: the
airmass correction now depends only on the primary extinction
coefficient (usually measured in mag/airmass) and the airmass. All
data were corrected to airmass zero (i.e.\ above the atmosphere) using
the nightly extinction coefficients measured by the Carlsberg Meridian
Telescope on La Palma in the {\em r\/}$^{\prime}$ filter, and
converted to other colour bands using the procedure described by
\citet{king85} and the effective wavelengths of the filters given by
\citet{fukugita96}.

The $\kappa^{\prime\prime}_{\lambda}$ term in
equations~\ref{eq:airmass} and \ref{eq:airmass2} is necessary for
extremely accurate
photometry because of the wavelength-dependent nature of the
scattering processes. This means that within a filter's bandpass, some
wavelengths will suffer more extinction than others. In general,
shorter wavelengths are scattered and absorbed more than longer
ones. It also corrects for any differences between the spectral energy
distributions of the standard and target stars, as well as any
instrumental differences between the system used to define the
magnitude scale and the one used for the observations (e.g.\ the
filter transmission or the CCD quantum efficiency) that may lead to
colour-dependent offsets. Neglecting this term will therefore give
rise to a (small; of the order of $\kappa^{\prime\prime}_{\lambda}$)
systematic error in the fluxes \citep[for a detailed discussion,
see][]{smith02}.

\subsubsection{Instrumental correction}
No two instrumental set-ups are ever exactly the same,
unfortunately. This means that in order to be able to directly compare
flux or magnitude measurements made with different detector and filter
systems, each must be corrected to an agreed standard. For the SDSS
photometric system used by {\sc ultracam}, this standard is defined by the
standard star network as measured by the 60-cm SDSS Monitor Telescope at
Apache Point Observatory using the {\em u\/}$^{\prime}${\em
g\/}$^{\prime}${\em r\/}$^{\prime}${\em i\/}$^{\prime}${\em z\/}$^{\prime}$
filter system with a thinned, back-illuminated,
UV-anti-reflection-coated CCD \citep{fukugita96,smith02}. To transform
an extinction-corrected magnitude to a standard magnitude,
$m_{\lambda0s}$, a {\em zero-point\/} $\rho$ for each filter must be
determined:
\begin{eqnarray}
\label{eq:zeropoint}
m_{\lambda0s} & = & m_{\lambda0} + \rho.
\end{eqnarray}

Combining equations~\ref{eq:airmass2} and \ref{eq:zeropoint}, the
following expression for the zero-point is obtained:
\begin{equation}
\label{eq:zeropoint2}
\rho = m_{\lambda0s} + 2.5\log_{10}\left(\frac{c}{t}\right) + 
\kappa^{\prime}_{\lambda} X.
\end{equation}

The zero-points for each filter are obtained from observations of SDSS
standard stars. These were reduced using the `normal' photometry option in
the pipeline software (optimal photometry only yields a valid flux
ratio; the individual flux estimates are unreliable) during
photometric conditions. The zero-point for each filter was found by
taking the weighted mean of all the zero-points for individual
exposures. A summary of the standard stars observed for this purpose
and the zero-points thus determined is given in
table~\ref{tab:standards}. 

\begin{sidewaystable}
\begin{center}
\caption[Determination of zero-points from {\sc ultracam} SDSS
  standard stars.]{Determination of mean zero-points from {\sc
  ultracam} SDSS standard stars. Standard star magnitudes are given in
  \citet{smith02}, and the zero-point $\rho$ is as defined in
  equation~\ref{eq:zeropoint2}.}
\vspace{0.3cm}
\small
\begin{tabular}{lllccccc}
Target & Date & Filters & \multicolumn{5}{c}{Zero-point $\rho$} \\
 & (yyyy mm dd) & & {\em u\/}$^{\prime}$ & {\em g\/}$^{\prime}$ & {\em
 r\/}$^{\prime}$ & {\em i\/}$^{\prime}$ & {\em z\/}$^{\prime}$ \\
\hline

BD +82015 & 2003 11 12 & {\em u\/}$^{\prime}${\em g\/}$^{\prime}${\em
 z\/}$^{\prime}$ & $25.00\pm0.39$ & $26.91\pm0.34$ & -- & -- &
 $25.28\pm0.33$ \\

G 93--48 & 2003 11 3--4 & {\em u\/}$^{\prime}${\em g\/}$^{\prime}${\em
 r\/}$^{\prime}$ & $24.70\pm0.41$ & $26.62\pm0.39$ &
 $26.18\pm0.42$ & -- & -- \\

Feige 22 & 2002 9 19 & {\em u\/}$^{\prime}${\em g\/}$^{\prime}${\em
  r\/}$^{\prime}$ & $24.860\pm0.061$ & $26.736\pm0.058$ &
  $26.242\pm0.060$ & -- & -- \\

PG 1047 +003A & 2003 5 21--25 & {\em u\/}$^{\prime}${\em g\/}$^{\prime}${\em
  i\/}$^{\prime}$ & $25.133\pm0.055$ & $26.983\pm0.049$ & -- &
  $26.093\pm0.047$ & -- \\

RU 149B & 2003 11 3 & {\em u\/}$^{\prime}${\em g\/}$^{\prime}${\em
  r\/}$^{\prime}$ & $24.96\pm0.33$ & $26.77\pm0.29$ & $26.37\pm0.29$ &
  -- & -- \\

RU 152 & 2003 11 1 & {\em u\/}$^{\prime}${\em g\/}$^{\prime}${\em
  r\/}$^{\prime}$ & $24.92\pm0.29$ & $26.79\pm0.29$ & $26.37\pm0.30$ &
  -- & -- \\

SA 113 339 & 2003 5 21--24 & {\em u\/}$^{\prime}${\em g\/}$^{\prime}${\em
  i\/}$^{\prime}$ & $25.02\pm0.17$ & $26.89\pm0.17$ & -- &
  $25.98\pm0.17$ & -- \\

SA 95 190 & 2002 9 19 & {\em u\/}$^{\prime}${\em g\/}$^{\prime}${\em
  r\/}$^{\prime}$ & $24.952\pm0.11$ & $26.789\pm0.096$ &
  $26.295\pm0.096$ & -- & -- \\

SA 115 516 & 2003 10 30 & {\em u\/}$^{\prime}${\em g\/}$^{\prime}${\em
  i\/}$^{\prime}$ & $25.09\pm0.70$ & $26.87\pm0.58$ & -- &
  $26.11\pm0.52$ & -- \\

BD 353 659 & 2002 9 9, 12 & {\em u\/}$^{\prime}${\em g\/}$^{\prime}${\em
  i\/}$^{\prime}$ & $25.1113\pm0.0017$ & $26.9164\pm0.0015$ & -- &
  $26.1226\pm0.0014$ & -- \\

Hilt 190 & 2002 9 9 & {\em u\/}$^{\prime}${\em g\/}$^{\prime}${\em
  i\/}$^{\prime}$ & $24.88\pm0.38$ & $26.79\pm0.34$ & -- &
  $26.03\pm0.31$ & -- \\

GJ 745A & 2002 9 9 & {\em u\/}$^{\prime}${\em g\/}$^{\prime}${\em
  i\/}$^{\prime}$ & $26.11\pm0.22$ & $26.93\pm0.28$ & -- & $26.11\pm0.22$
  & -- \\

\hline

Weighted mean & -- & -- & $25.1111\pm0.0017$ & $26.9163\pm0.0015$ &
   $26.26\pm0.05$ & $26.1226\pm0.0014$ & $25.28\pm0.33$

\end{tabular}
\normalsize
\label{tab:standards}
\end{center}
\end{sidewaystable}

As mentioned in \S~\ref{sec:optimal}, optimal extraction only yields a
valid ratio of counts. In order to determine the magnitude of the
target star $m_{\lambda0s}^{{\rm tar}}$, the magnitude of the
comparison star $m_{\lambda0s}^{{\rm comp}}$ was therefore determined
using `normal' photometry using the above techniques, in a similar way
as for the standard stars, using the zero-points obtained
therefrom. The magnitude of the target star is then simply given by
\begin{equation}
m_{\lambda0s}^{{\rm tar}} = -2.5\log_{10}\left(\frac{c^{{\rm
      tar}}}{c^{{\rm comp}}}\right) +
m_{\lambda0s}^{{\rm comp}}.
\end{equation}

\subsubsection{Absolute calibration}

The above corrections place the observations on the {\em AB\/}
magnitude system \citep{oke83}. One of the major advantages of this
system is that the magnitude is directly related to the flux per unit
frequency $f_{\nu}$ \citep{fukugita96}:
\begin{equation}
f_{\nu}({\rm Jy}) = 3631\times10^{(-0.4m_{\lambda0s})}.
\end{equation}

%% file: analysis.tex
\chapter{Analysis techniques}
\label{ch:analysis}

\section{Phasing the data}

Before any of the following analyses can proceed, the data must first
be phased, that is, the {\em  x\/}-axis must be converted from units
of time to units that express at what point the star is in its
orbit. The orbital phase usually runs from 0 to 1 over an orbital
cycle (numbers greater than 1 indicate subsequent cycles). Phase zero
is usually defined, and I follow this convention, as the mid-point of
the white dwarf eclipse. For a symmetrical white dwarf eclipse,
implying a symmetrical light distribution over the white dwarf, this
is equivalent to the superior conjunction of the white dwarf. Some
authors define phase zero as the point of minimum light, which in
systems with asymmetric disc emission (e.g.\ from the bright spot) can
lead to a systematic disagreement of a few seconds with the definition
I have adopted.

The times of white dwarf mid-ingress $T_{\rm{wi}}$ and mid-egress
$T_{\rm{we}}$ were determined by locating the times when the minimum
and maximum values, respectively, of the light curve derivative
occurred (see \S~\ref{sec:derivative}), using the techniques
described by \citet{wood85,wood86b,wood89a}. A
median filter was used to smooth the data, the derivative of which was
then calculated numerically (a median filter preserves the shape of
the original light curve better than a box-car, or running mean,
filter). A box-car filter (which reduces the noise more than a median
filter would) was applied to this derivative, and simple searches were
made to locate the minimum and maximum values of the derivative
corresponding to the midpoints of ingress $T_{\rm{wi}}$ and egress
$T_{\rm{we}}$. (In fact this method locates the steepest part of the
ingress and egress, but one would expect these to correspond to the
midpoints unless the light distribution of the white dwarf is
asymmetrical.) If a bright spot eclipse is also present, the ingress
and egress times of the white dwarf must be visually inspected to
ensure that they are not confused with those of the bright spot. The
times of mid-eclipse were then determined by assuming the white dwarf
eclipse to be symmetric around phase zero and taking
$T_{\rm{mid}}=(T_{\rm{we}}+T_{\rm{wi}})/2$. This technique locates the
time of mid-eclipse to an accuracy comparable to the time-resolution
of the data.

The orbital ephemeris was then determined from all available
mid-eclipse times and cycle numbers for each target by a linear
least-squares fit. The errors adopted for the times of mid-eclipse
taken from the literature, where not explicitly stated, were estimated
from the number of significant figures quoted or the time-resolution
of the data, whichever gave the larger error. The resulting orbital
ephemerides are of the form
\begin{equation}
HJD = HJD_{0} + P_{\rm{orb}} E,
\end{equation}
where $E$ is the cycle number and $HJD$ and $HJD_{0}$ are the
heliocentric Julian date of the midpoint of the eclipse of cycles $E$
and 0. All HJD times quoted, throughout this thesis, are co-ordinated
universal time, UTC, corrected to the heliocentre (i.e.\ not
barycentric dynamical time, TDB).

\section{Derivative method}
\label{sec:derivative}

This method of determining the system parameters of an eclipsing dwarf
nova was originally developed by \citet{wood86b}. It relies upon the
fact that there is a unique relationship between the mass ratio
$q=M_{2}/M_{1}$ and orbital inclination $i$ for a
given eclipse phase width $\Delta\phi$ \citep{bailey79}, as discussed
in \S~\ref{sec:radmass} and \S~\ref{sec:phomass}.

The first requirement of this technique is accurate timings of the
white dwarf and bright spot eclipse contact phases. Here and hereafter
the midpoints of ingress and egress are denoted by $\phi_{\rm{i}}$
and $\phi_{\rm{e}}$, respectively, the eclipse contact phases
corresponding to the start and end of the ingress by
$\phi_{1}$ and $\phi_{2}$ and the start and end of the egress by
$\phi_{3}$ and $\phi_{4}$. In the discussion that follows I use the
suffixes `$\rm{w}$' and `$\rm{b}$' to denote the white dwarf and
bright spot contact phases, respectively (e.g.\ $\phi_{\rm{wi}}$
means the midpoint of the white dwarf ingress). The midpoints of the
eclipse ingresses and egresses were determined as described in the
previous section, with the sole change being that the data is now
phased. The eclipse contact phases $\phi_{1}$\ldots$\phi_{4}$ were
determined by locating the points
where the derivative differs significantly from a spline fit to the
more slowly varying component (for instance, a disc eclipse or the
orbital hump). Throughout this thesis, the eclipse phase width quoted
is the full-width at half-maximum, given by
\begin{equation}
\label{eq:phasewidth}
\Delta \phi = \phi_{\rm{we}}-\phi_{\rm{wi}}.
\end{equation}

The trajectory of the gas stream originating from the inner Lagrangian
point ${\rm L}_{1}$ is calculated by solving the equations of motion
(equations~\ref{eq:xacceleration} and \ref{eq:yacceleration}) using a
second-order Runge--Kutta technique and conserving the Jacobi Energy
(equation~\ref{eq:jacobi}) to 1 part in $10^{4}$. This assumes that
the gas stream follows a ballistic path. As $q$ decreases, the path of
the stream moves away from the white dwarf. For a given mass ratio $q$
each point on the stream has a unique phase of ingress and egress. The
eclipse (or not) of a point in the system by the secondary star was
determined using {\sc blink}, a Fortran subroutine written by Keith
Horne and Tom Marsh. {\sc Blink} tests for occultation by the
secondary star of a given location in the co-rotating binary system at
a given phase using the procedure described by \citet{horne85}.

\begin{figure}
\centerline{\includegraphics[width=12cm,angle=0]{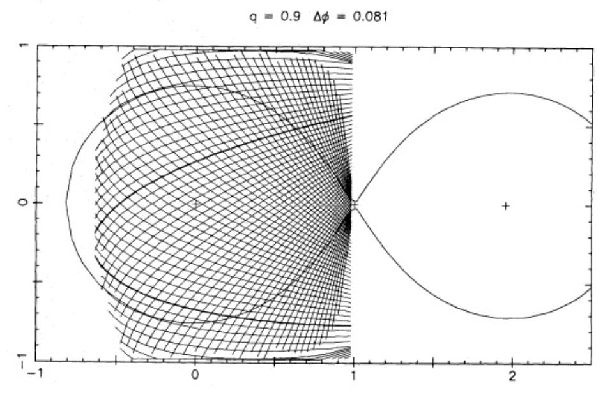}}
\caption[Phase arcs at different orbital phases.]{At each phase, when
projected along the line of sight onto
a given plane (here, and usually, the orbital plane) the limb of the
secondary star forms an arc. Each point along such a `phase arc' is
eclipsed at the same moment. The figure shows various phase arcs at
different orbital phases for $q=0.9$ and $\Delta\phi=0.081$. The axes
are in units of the ${\rm L}_{1}$ distance. with the white dwarf at
the origin. Figure from \citet{horne85}.}
\label{fig:phasearc}
\end{figure}

For each phase, the limb of the secondary forms an arc when projected
along the line of sight onto a given plane (hereafter referred to as a
phase arc): each point on an individual phase arc is eclipsed at the
same time (see figure~\ref{fig:phasearc}). The intersection of the
phase arcs corresponding to the
respective eclipse contact phases can be used to constrain the size of
the white dwarf and the structure of the bright spot. The light
centres of the white dwarf and bright spot must lie at the
intersection of the phase arcs corresponding to the relevant phases of
mid-ingress and mid-egress, $\phi_{\rm{i}}$ and $\phi_{\rm{e}}$. The
phase arcs were calculated using full Roche lobe geometry rather than
an approximate calculation, using the {\sc blink} subroutine.

As previously discussed in \S~\ref{sec:phomass}, the mass ratio---and
hence the inclination---may be determined by comparing the bright
spot light centres corresponding to the measured eclipse contact
phases $\phi_{\rm{wi}}$ and $\phi_{\rm{we}}$ with the theoretical
stream trajectories for different mass ratios $q$. This requires the
assumption that the gas stream passes directly through the light
centre of the bright spot. I constrain the light centre of the bright
spot to be the point where the gas stream and outer edge of the disc
intersect, so that the distance from the primary at which the gas
stream passes through the light centre of the bright spot gives the
outer disc radius $R_{\rm{d}}/a$. For data covering multiple eclipses,
the uncertainties on these parameters may be determined from the
root mean square (rms) variations of the measured contact phases.

\begin{figure}
\centerline{\includegraphics[width=12cm,angle=0]{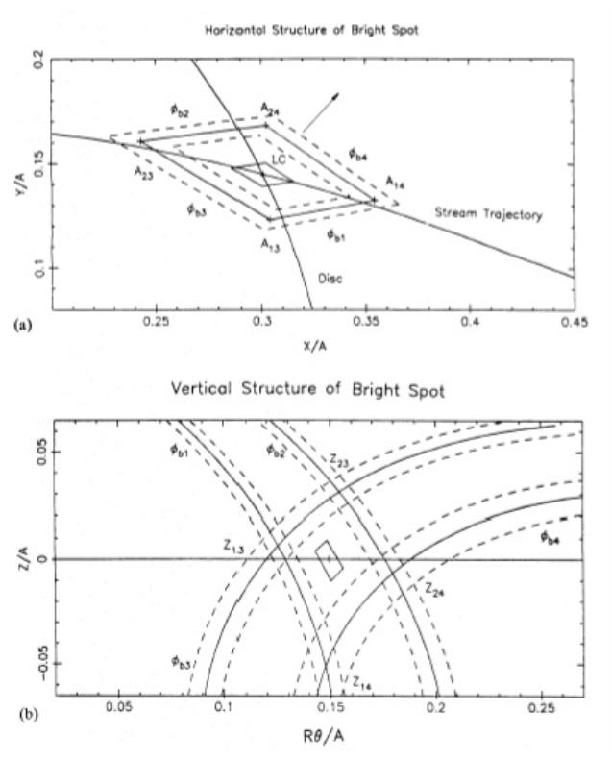}}
\caption[The horizontal and vertical structure of the bright spot
  of Z~Cha and the definition of $A_{jk}$ and $Z_{jk}$.]{The
  horizontal and vertical structure of the bright spot
  of Z~Cha for $q=0.1495$. The top panel (a) shows the region on the
  orbital $(x,y)$ plane on which the bright spot lies. The light
  centre of the bright spot {\em LC\/} is shown surrounded by a solid
  box corresponding to the rms variations in the phases of mid-ingress
  $\phi_{\rm bi}$ and mid-egress $\phi_{\rm be}$. Surrounding this is
  the solid box corresponding to the eclipse contact phases
  $\phi_{{\rm b}j}\ldots\phi_{{\rm b}k}$ (the vertices of which define
  $A_{jk}$) and their rms variations (dotted boxes). The accretion
  disc of radius $R_{\rm d}/a=0.334$ is also plotted, as is the stream
  trajectory. Panel (b) is
  similar, differing in that it shows the projection of the phase arcs
  onto the vertical cylinder of radius equal to that of the disc,
  $0.334a$, i.e.\ in the plane $(R\theta,z)$. $\theta$ increases in the
  direction of orbital motion and is zero at the line joining the
  centres of the two stars. Intersections of the phase arcs
  corresponding to the contact phases of the bright spot are labelled
  $Z_{jk}$. Figure from \citet{wood86b}.}
\label{fig:wood86bs}
\end{figure}

The eclipse constraints on the structure of the bright spot can be
used to determine upper limits on the angular size and the radial and
vertical extent of the bright spot. Defining $A_{jk}$ and $Z_{jk}$
graphically in figure~\ref{fig:wood86bs} as the positions of the
intersections of the phase arcs $\phi_{{\rm b}j}$ and $\phi_{{\rm
b}k}$ in the $(x,y)$ and $(R\theta,z)$ planes respectively, one can
define
\begin{subequations}
\begin{eqnarray}
\label{eq:theta}
\Delta \theta & = &
(\theta_{23}+\theta_{24}-\theta_{13}-\theta_{14})/2\\
\label{eq:r}
\Delta R_{\rm{d}} & = & (R_{24}+R_{14}-R_{23}-R_{13})/2\\
\label{eq:z}
\Delta Z & = & (H_{23}-H_{14})/2\\
\label{eq:z2}
\Delta Z_{2} & = & H_{23},
\end{eqnarray}
\end{subequations}
where $R_{jk}$ and $\theta_{jk}$ are the radius and azimuth of
$A_{jk}$ and $H_{jk}$ the height of $Z_{jk}$ above the orbital
plane. Equations~\ref{eq:theta} and \ref{eq:r} are defined as by
\citet{wood86b}. Note that the definition of $\Delta Z$ in equation
\ref{eq:z} differs slightly to that defined in \citet{wood86b}: this
is in order to be more consistent with the definitions of $\Delta
\theta$ and $\Delta R_{\rm d}$ in equations \ref{eq:theta} and
\ref{eq:r}. $\Delta Z_{2}$, defined in equation~\ref{eq:z2}, is
identical to $\Delta Z$ as defined by \citet{wood86b}, and is included
here for ease of comparison.

The eclipse constraints on the radius of the white dwarf can be used,
together with the mass ratio and orbital inclination to determine the
radius of the white dwarf. An alternative possibility is that the
sharp eclipse is caused by a bright inner disc region or boundary
layer surrounding the white dwarf like a belt. Another possibility is
that the lower hemisphere of the white dwarf is obscured by an
optically thick accretion disc, which would result in a larger white
dwarf radius than that measured (see, for example,
figure~\ref{fig:ouvir_whitedwarf}). The latter can be checked, as if
the contact phases $\phi_{\rm wi}$ and $\phi_{\rm we}$ lie half-way
through the white dwarf ingress and egress, the light distribution
must be symmetrical.

The following analysis assumes that the eclipse is solely of a white
dwarf. If the eclipse is actually of a belt and the white dwarf itself
is not visible, then the white dwarf radius must be somewhat smaller
than the radius of the belt. If the white dwarf does contribute
significantly to the eclipsed light, then the white dwarf radius
derived is actually an upper limit, so that the white dwarf mass
determined from it is actually a lower limit
(equation~\ref{eq:nauenberg}). The only way to verify the assumption
that the central light source is the white dwarf alone is to measure
the semi-amplitude of the radial velocity curve of the secondary star,
$K_{2}$, and compare the resulting mass to that predicted by the
photometric model. One could also check if this assumption is true
using a longer baseline of quiescent observations, as one might expect
eclipse timings of an accretion belt to be much more variable than
those of a white dwarf. I note, however, that the white dwarf masses
of OU~Vir and XZ~Eri, given in chapters~\ref{ch:ouvir} and
\ref{ch:xzeridvuma} respectively, are consistent with the mean white
dwarf mass of $0.69\pm0.13\;{\rm M}_{\odot}$ for CVs below the period gap
\citep{smith98a}. Although the white dwarf in DV~UMa is unusually
massive, the assumption that we are observing the white dwarf and
not the boundary layer around the primary cannot cause this, as the
white dwarf mass derived would be in this case a lower limit. Also,
\citet{baptista00} point out that in short-period dwarf nov\ae\
(specifically OY~Car, Z~Cha and HT~Cas; \citealp{wood90}) like OU~Vir,
XZ~Eri and DV~UMa the boundary layer is faint (or absent),
whereas longer-period dwarf nov\ae\ such as IP Peg \citep{wood86a} and
EX Dra \citep{baptista00} usually have detectable boundary layers. As an
illustrative example, in the case of EX~Dra, \citet{baptista00}
conclude that the white dwarf is surrounded by an extended boundary
layer on the basis of the implausibly low white dwarf masses implied
by assuming otherwise and the observed variability of both the flux
and duration of the primary eclipse. Throughout this thesis I
assume that the central eclipsed object is indeed a white dwarf.

\subsubsection{Light curve deconvolution}
\label{sec:deconvolution}
\begin{figure}
\centerline{\includegraphics[width=12cm,angle=0]{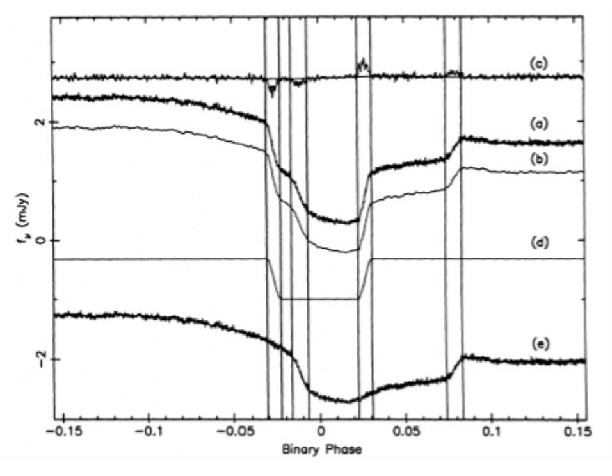}}
\caption[Deconvolution of the white dwarf light curve of Z~Cha from
  the mean light curve.]{Deconvolution of the white dwarf light curve
  of Z~Cha from
  the mean light curve. (a) shows the original mean light curve; (b)
  the smoothed light curve, offset downwards by 0.5~mJy; (c) the
  derivative of the smoothed light curve, with a spline fit to phases
  outside ingress and egress superimposed, both offset upwards by
  2.75~mJy and multiplied by a factor of 10; (d) the reconstructed
  white dwrf eclipse, offset downwards by 1~mJy; and (e) the original
  mean light curve after subtraction of the white dwarf light curve,
  offset downwards by 3~mJy. The vertical lines mark the contact
  phases of the white dwarf and bright spot. Figure from
  \citet{wood86b}.}
\label{fig:wood86decon}
\end{figure}

Once the white dwarf eclipse contact phases have been found, the white
dwarf light curve can be reconstructed and subtracted from the overall
light curve, as illustrated in figure~\ref{fig:wood86decon}. The
procedure is as follows. The white dwarf flux is
assumed to be zero between the contact phases $\phi_{2}$ and
$\phi_{3}$, as here the white dwarf is totally eclipsed. The
derivative between the contact phases is then numerically
integrated. The white dwarf flux is assumed to be constant outside
eclipse, and is determined from the mean of the integrated flux at
contact phases $\phi_{1}$ and $\phi_{4}$. The result is symmetrized
about phase zero, and smoothed to obtain a noise-free estimate of the
white dwarf light curve. This can then be subtracted from the overall
light curve to give the light curve of the bright spot and disc.  The
white dwarf flux thus determined can be used to determine its
temperature and distance (see \S~\ref{sec:fitting}).

\section{{\sc Lfit} method}
\label{sec:lfit}
Another way of determining the system parameters is to use a physical
model of the binary system to calculate eclipse light curves for each
of the various components. I used the technique developed by
\citet{horne94} and described in detail therein. This model assumes
that the eclipse is caused by the secondary star, which completely
fills its Roche lobe. A few changes were necessary in order to make
the model of \citet{horne94} suitable for my data. The most important
of these was the fitting of the secondary flux, prompted by the
detection of a significant amount of flux from the secondary in the
{\em i\/}$^{\prime}$ band of DV~UMa. The secondary flux is very small in
all the other bands.

The 10 parameters that control the shape of the light curve are as
follows:
\begin{enumerate}
\item The mass ratio $q$.
\item The eclipse phase full-width at half-depth $\Delta\phi$.
\item The outer disc radius $R_{\rm{d}}/a$.
\item The white dwarf limb darkening coefficient $U_{1}$.
\item The white dwarf radius $R_{1}/a$.
\item The bright spot scale $SB/a$. The bright spot is modelled as a
linear strip passing through the intersection of the gas stream and
disc. The relative intensity distribution along this strip is given by
$(X/SB)^{2}e^{-X/SB}$, where $X$ is the distance along the strip, the
maximum being at the intersection of the bright spot and disc, at
$X=2SB$. The relative intensity $I$ of the bright spot is then
modulated according to the orbital phase by a sine function. For
DV~UMa this function was
\begin{subequations}
\begin{equation}
I = \left\{ \begin{array}{ll}
    \sin (\theta_{\rm{B}} + \phi) & \mbox{if $\sin (\theta_{\rm{B}} +
    \phi)>0$} \\
    0 & \mbox{otherwise}
    \end{array}
  \right. ,
\end{equation}
where $\theta_{\rm{B}}$ is as defined below. For XZ~Eri a better fit
was achieved using
\begin{equation}
I = \left\{ \begin{array}{ll}
    \sin ^2(\theta_{\rm{B}} + \phi) & \mbox{if $\sin (\theta_{\rm{B}} +
    \phi)>0$} \\
    0 & \mbox{otherwise}
    \end{array}
  \right. .
\end{equation}
\end{subequations}
\item The line along which, on the edge of the disc, the bright spot
lies is tilted by an angle $\theta_{\rm{B}}$,
measured relative to the line joining the white dwarf and the
secondary star. The bright spot does not, therefore, necessarily emit
its anisotropic light (see below) in a direction normal to the edge of the
accretion disc. This allows adjustment of the phase of the orbital hump.
\item The fraction of bright spot light which is isotropic
$f_{\rm{iso}}$. The bright spot emits a fraction of its light
isotropically, in all directions, and the remainder anisotropically,
in a direction which determines the phase of the orbital hump
maximum.
\item The disc exponent $b$, describing the power law of the radial
intensity distribution of the disc. The relative intensity $I$ at a
radius $R$ of the disc is $I \propto R^{b+1}$.
\item A phase offset $\phi_{0}$.
\end{enumerate}

The light curve $D(\phi)$ was modelled as a sum of multiple components (the
white dwarf, bright spot, accretion disc and red dwarf), the
contribution of the first three of which can vary with the orbital
phase $\phi$:
\begin{equation}
\label{eq:modelsum}
M(\phi) = \sum_{i=1}^{n} L_{i}(\phi),
\end{equation}
where $M(\phi)$ is the model flux at phase $\phi$ and $L_{i}(\phi)$ is
the flux of component $i$ at phase $\phi$. Fitting of ellipsoidal
variations made no
significant improvement to the overall fit, so I have assumed the
flux from the secondary star to be constant. The {\sc amoeba} algorithm
(downhill simplex; \citealp{press86}) was used to adjust selected
parameters to find the best fit. The algorithm attempts to minimise
$\chi^{2}$, the usual goodness-of-fit statistic:
\begin{equation}
\label{eq:chisquared}
\chi^{2} = \sum_{j=1}^{n} \left(
\frac{D_{j}-M_{j}}{\sigma_{j}} \right)^{2}.
\end{equation}
It is often useful to define a {\em reduced} $\chi^{2}$,
\begin{equation}
\chi^{2}_{\rm R}=\frac{\chi^{2}}{n-\nu},
\label{eq:reducedchisquared}
\end{equation}
where $\nu$ is the number of degrees of freedom.
At each evaluation of the function $M(\phi)$ the light curves of the
individual components were scaled using a linear regression, the shape
of each light curve being set by the values of the parameters at that
time. A positivity constraint was imposed: whenever a negative flux
was found for a component, the flux of that component was set to zero
and the fit was repeated to determine the flux for the other
components. The procedure did not iterate to $\chi^{2}_{\rm R}=1$ due
to the presence of flickering and
other variability in the light curve not allowed for in the
model. Consequently, the algorithm was run until the parameters output
no longer changed significantly between iterations (i.e.\ the
parameter change was less than the typical uncertainty on each
parameter; see below). Typically, a minimum
of $10\,000$ iterations were performed to produce the parameters
presented in this thesis.

The $1\sigma$ error on an individual parameter of a {\em
M}-dimensional model fit is given \citep{lampton76} by the
perturbation of that parameter necessary to increase the $\chi^{2}$ of
the fit by 1, i.e.
\begin{equation}
\chi^{2}-\chi_{\rm{min}}^{2}=\Delta\chi^{2}=1.
\end{equation}
This is, of course, equivalent to finding the root of
\begin{equation}
\label{eq:chisqerr}
f(\chi^{2})=\chi^{2}-\chi_{\rm{min}}^{2}-1.
\end{equation}
The procedure I employed to determine the errors on each parameter was
as follows. I perturbed the parameter of interest from its best fit
value by an arbitrary amount (initially 5~per~cent) and re-optimised
the rest of them (holding the parameter of interest, and any others
originally kept constant, fixed). The {\sc amoeba} algorithm was allowed to
iterate for $2100$ iterations, then the new value of $\chi^{2}$ was
computed. If the root was not bracketed by the two values of
$\chi^{2}$, i.e. $\chi^{2}-\chi_{\rm{min}}^{2}<1$, then the
perturbation was increased by a factor of $4$ until it was. A
bisection method \citep{press86} was then used to find the value of
the parameter in question which gave the root of
equation~\ref{eq:chisqerr}, with the value of $\chi^{2}$ at each step
being computed as described previously in this paragraph. The
difference between the final, perturbed, value of the parameter and
its best fit value gave the $1\sigma$ error on that parameter.

\section{Comparing the derivative and {\sc lfit} methods}
The methods discussed in the previous two sections, the derivative and
{\sc lfit} techniques, were compared with each other using fake
light curves.

Fake, noise-free light curves were kindly produced by Dr.~Chris
Watson using his {\sc rochey} code to specifications set out by
myself. All other work in this section was conducted by myself under
the supervision of Dr.~Vik Dhillon. Gaussian noise was then added to
the resultant light curve. The errors were of the form
\begin{equation}
\sigma_{{\rm i}} = E \sqrt{f_{i} f_{2}} + \frac{E f_{2}}{10},
\end{equation}
where $E$ is an arbitrary number controlling the fractional error,
$f_{i}$ the flux on data\-point $i$ and $f_{2}$ the flux on the
second datapoint (chosen arbitrarily because it is out of eclipse and
not contaminated by the orbital hump). The first term scales the error
with the square root of the flux, as expected for shot noise, with the
$\sqrt{f_{2}}$ term ensuring that the signal-to-noise ratio of the
second datapoint is $\sim\frac{1}{E}$. The final term prevents the
error on a zero flux point being zero, which is both unphysical and
causes coding problems (mainly with dividing by zero). The noise added
to the data was
\begin{equation}
f_{i} = f_{i}+R\sigma_{i},
\end{equation}
where $R$ is a normally-distributed random number with zero mean and
unit variance. Flickering was modelled by adding additional (Gaussian)
noise to the light curve but not increasing the errors. For all the
fake light curves used in this section, $E=0.04$ for the addition of
both noise and flickering. The amplitude of the flickering scaled with
the flux level so that flickering during mid-eclipse was much less
than that out of eclipse. This is designed to reproduce the real
behaviour of flickering, which is observed to be greatly reduced
during eclipse \citep[e.g.][]{patterson81,bruch00b,baptista04}.

The parameters used to produce the fake light curves used for the
comparison of the derivative and {\sc lfit} methods are given in
table~\ref{tab:fake_in_param}. The properties of the various fake
light curves I used are given below:
\begin{enumerate}
\item A `normal' light curve, with the ingress and egress of the white
  dwarf and bright spot clear and distinct.
\item As 1, but with the ingress of the white dwarf and bright spot
  merged together, as seen for IP~Peg (whose parameters have been used
  to produce this light curve).
\item The aim of these light curves is to investigate the case of the
  compact objects (the white dwarf and bright spot) being relatively
  faint. I would expect that if the level of noise and/or flickering
  becomes greater than the amplitude of the eclipse, then
  determination of the system parameters would become
  difficult/impossible.
  \begin{enumerate}
  \item Faint white dwarf.
  \item Faint bright spot.
  \end{enumerate}
\item The aim of these light curves is to investigate what happens to
  the estimated parameters when the assumption that the bright spot
  lies at the intersection of the accretion disc and gas stream is
  broken. To this end:
  \begin{enumerate}
  \item The bright spot is ahead of the impact region.
  \item The bright spot is behind the impact region.
  \end{enumerate}
\item These light curves have different accretion disc radii to check
  that both techniques are effective over a range of parameters.
  \begin{enumerate}
  \item A smaller disc radius, of $0.25a$.
  \item A larger disc radius, of $0.36a$.
  \end{enumerate}
\end{enumerate}

\begin{figure}
\centerline{\includegraphics[width=12cm,angle=-90]{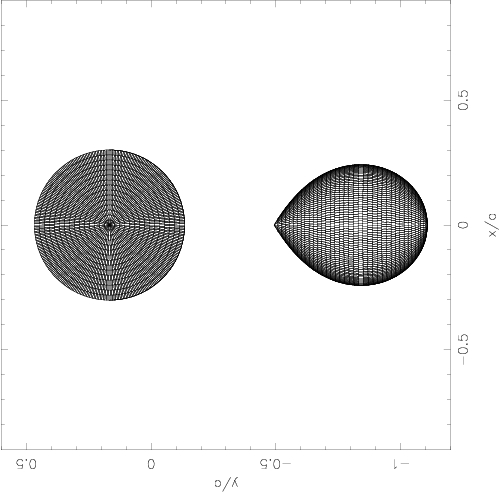}}
\caption[The model grid used to produce the fake data for comparing
  the results of the derivative and {\sc lfit} techniques.]{The model
  grid used to produce the fake data for comparing the results of the
  derivative and {\sc lfit} techniques. The origin is at the centre of
  mass; otherwise the co-ordinate system is as described in
  chapter~\ref{ch:introduction}. Figure by Dr.~Chris Watson.}
\label{fig:grid}
\end{figure}

The model used to produce the fake light curves was a grid of the
accretion disc and secondary star, with each element in the grid array
having an adjustable intensity. The secondary star was assumed to fill its
Roche lobe, so the surface of the secondary star is defined by the
critical potential. The visibility of each grid element, or tile, was
tested at a given phase for a given mass ratio and orbital
inclination, and a light curve built up in this manner. The grid used
for model 1 (described in table~\ref{tab:fake_in_param}) is
illustrated in figure~\ref{fig:grid}. The disc has a height of
$0.0002a$ above the orbital plane, with the rim of the disc being
subdivided into tiles which can be used to model the orbital hump. The
white dwarf was modelled by increasing the intensity of the disc tiles
at the centre of the disc. The bright spot was modelled by a disc rim
only. Modelling the bright spot as a combination of the disc rim and
tiles in the outer annulus of the disc led to incorrect determinations
of the disc radius and mass ratio. This is because the disc rim is at
the outer radius of the accretion disc, whereas, since a tile is only
flagged as eclipsed when its centre is eclipsed, the radius of the
outer annulus is slightly (the difference being half the width of the
outer annulus) smaller than the radius
of the disc rim. The intensity of the secondary star was set to zero
in all models, as it is usually very faint in faint short-period dwarf
nov\ae.

\begin{figure}
\centerline{\includegraphics[width=12cm,angle=-90]{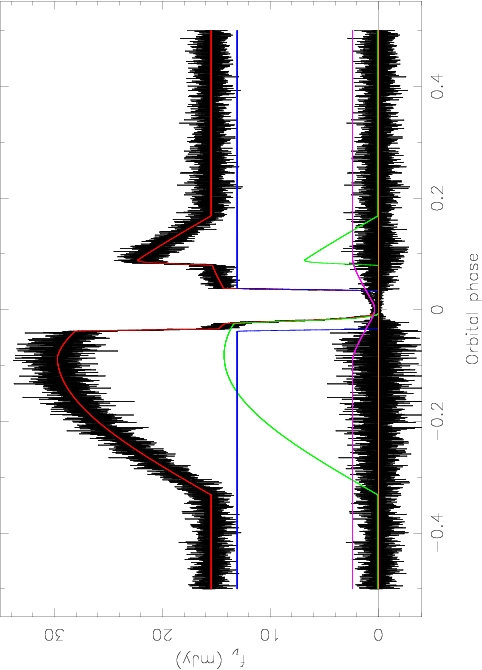}}
\caption[Fake light curve produced and fitted as described in the text using the
  parameters given in table~\ref{tab:fake_in_param} for model 1.]{Fake
  light curve produced as described in the text using the parameters
  given in table~\ref{tab:fake_in_param} for model 1.}
\label{fig:fake_1}
\end{figure}

\begin{figure}
\centerline{\includegraphics[width=12cm,angle=-90]{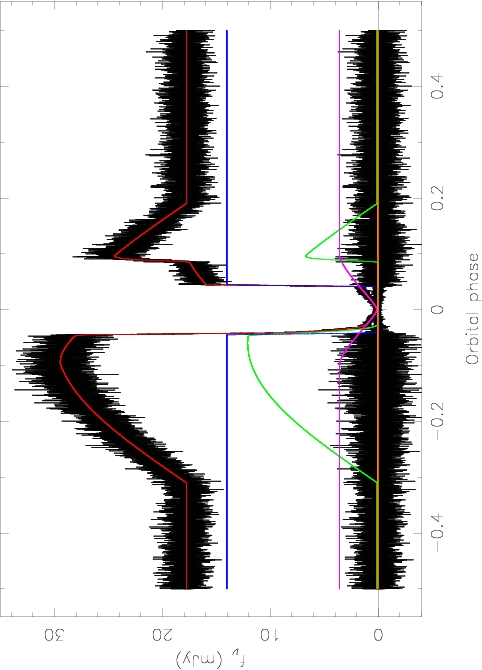}}
\caption[Fake light curve produced and fitted as described in the text using the
  parameters given in table~\ref{tab:fake_in_param} for model 2.]{Fake
  light curve produced as described in the text using the parameters
  given in table~\ref{tab:fake_in_param} for model 2.}
\label{fig:fake_2}
\end{figure}

\begin{figure}
\centerline{\includegraphics[width=12cm,angle=-90]{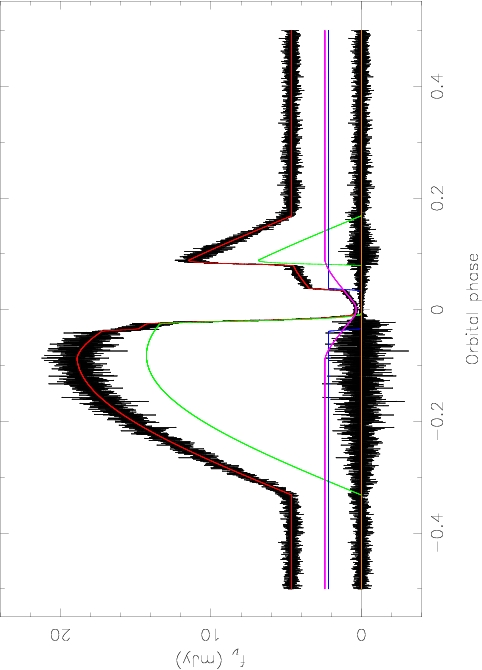}}
\caption[Fake light curve produced and fitted as described in the text using the
  parameters given in table~\ref{tab:fake_in_param} for model 3a.]{Fake
  light curve produced as described in the text using the parameters
  given in table~\ref{tab:fake_in_param} for model 3a.}
\label{fig:fake_3a}
\end{figure}

\begin{figure}
\centerline{\includegraphics[width=12cm,angle=-90]{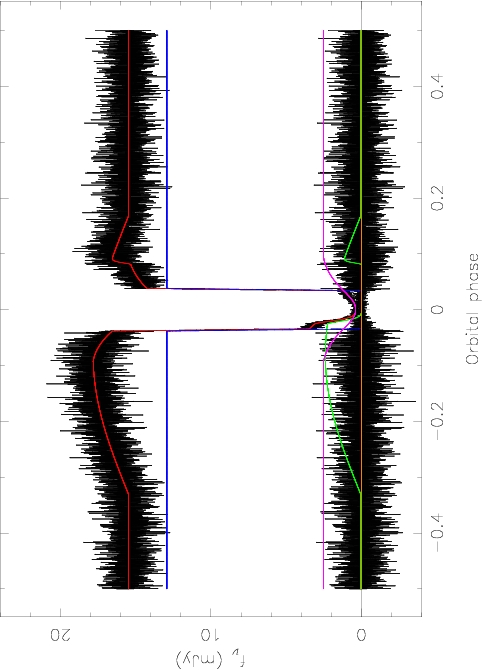}}
\caption[Fake light curve produced and fitted as described in the text using the
  parameters given in table~\ref{tab:fake_in_param} for model 3b.]{Fake
  light curve produced as described in the text using the parameters
  given in table~\ref{tab:fake_in_param} for model 3b.}
\label{fig:fake_3b}
\end{figure}

\begin{figure}
\centerline{\includegraphics[width=12cm,angle=-90]{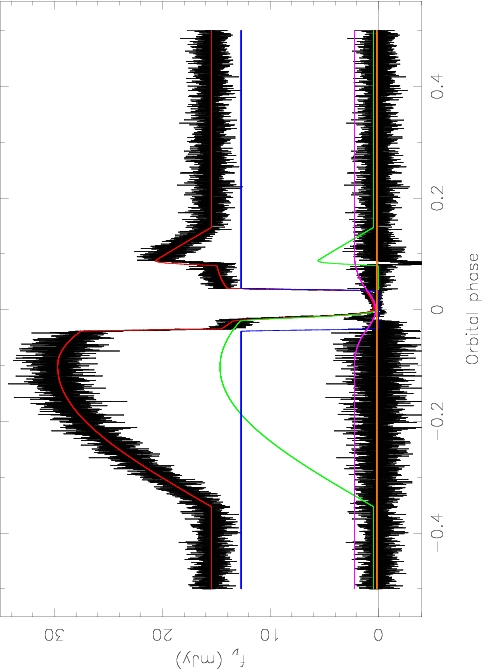}}
\caption[Fake light curve produced and fitted as described in the text using the
  parameters given in table~\ref{tab:fake_in_param} for model 4a.]{Fake
  light curve produced as described in the text using the parameters
  given in table~\ref{tab:fake_in_param} for model 4a.}
\label{fig:fake_4a}
\end{figure}

\begin{figure}
\centerline{\includegraphics[width=12cm,angle=-90]{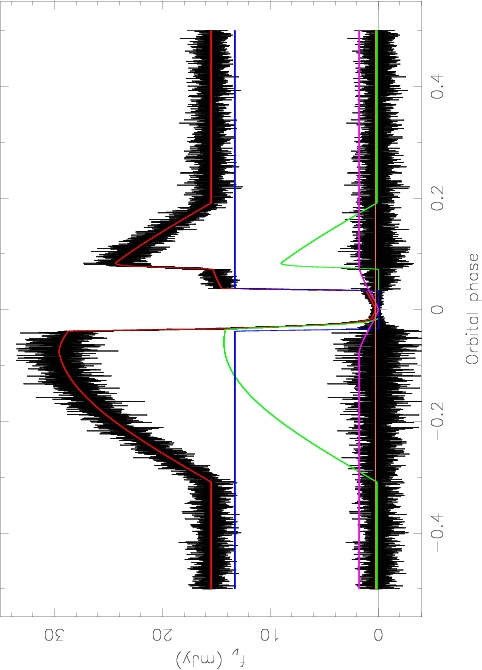}}
\caption[Fake light curve produced and fitted as described in the text using the
  parameters given in table~\ref{tab:fake_in_param} for model 4b.]{Fake
  light curve produced as described in the text using the parameters
  given in table~\ref{tab:fake_in_param} for model 4b.}
\label{fig:fake_4b}
\end{figure}

\begin{figure}
\centerline{\includegraphics[width=12cm,angle=-90]{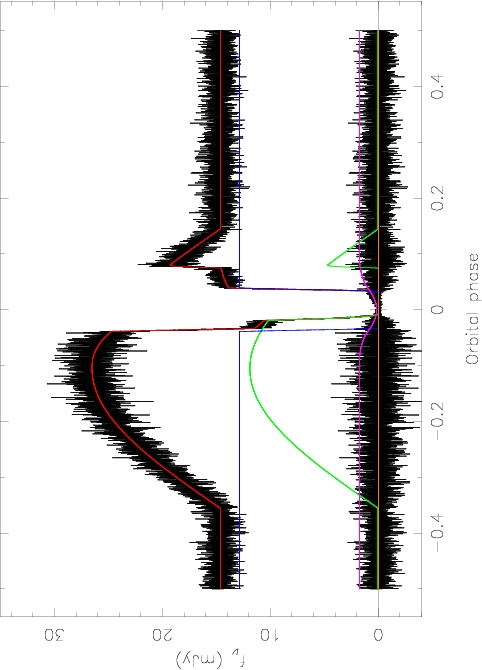}}
\caption[Fake light curve produced and fitted as described in the text using the
  parameters given in table~\ref{tab:fake_in_param} for model 5a.]{Fake
  light curve produced as described in the text using the parameters
  given in table~\ref{tab:fake_in_param} for model 5a.}
\label{fig:fake_5a}
\end{figure}

\begin{figure}
\centerline{\includegraphics[width=12cm,angle=-90]{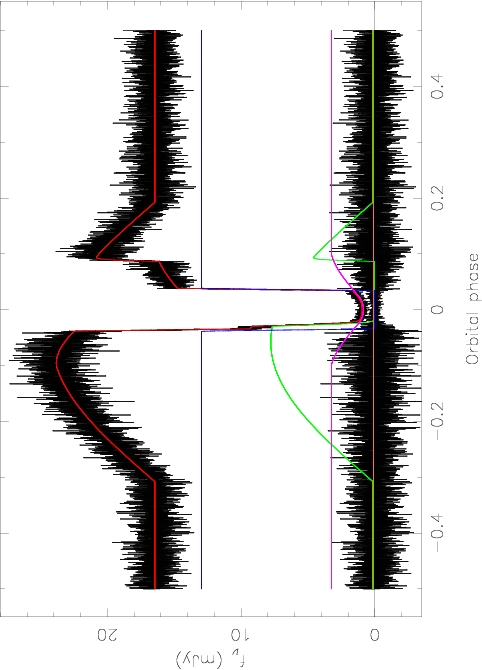}}
\caption[Fake light curve produced and fitted as described in the text
  using the
  parameters given in table~\ref{tab:fake_in_param} for model 5b.]{Fake
  light curve produced as described in the text using the parameters
  given in table~\ref{tab:fake_in_param} for model 5b.}
\label{fig:fake_5b}
\end{figure}

The parameters recovered from the various models are given in
tables~\ref{tab:fake_out_param_lfit} and \ref{tab:fake_out_param_der}
for the {\sc lfit} and derivative techniques, respectively. The
light curve fits from the {\sc lfit} program are shown in
figures~\ref{fig:fake_1}--\ref{fig:fake_5b}. 

\begin{sidewaystable}
\begin{center}
\caption[The input parameters of the fake light curves used
  for comparison of the derivative and {\sc lfit} methods.]
  {The input parameters of the fake light curves used
  for comparison of the derivative and {\sc lfit} methods. The
  white dwarf has in all cases a radius $R_{1}=0.02a$ and the
  bright spot $R_{\rm{bs}}=0.04a$. The disc radius is
  $R_{\rm{d}}=0.3a$ for all the light curves except models 5a and
  5b. The relative contributions (in arbitrary units) of the white
  dwarf (WD), bright spot (BS), accretion disc (AD) and red dwarf (RD)
  are also given.}
\vspace{0.3cm}
\small
\begin{tabular}{lcccccccccc}
\hline
 & & & & & \multicolumn{2}{c}{Bright spot position} &
 \multicolumn{4}{c}{Flux (arbitrary units)} \\
Model & $q$ & $i (\deg)$ & $\Delta\phi$ & $P$~(sec) & $x/a$ & $y/a$ & WD
 & BS & AD & RD \\ 
\hline
1 & 0.20 & $85.0^{\circ}$ & 0.0725 & 7200 & 0.2663 & 0.1381 & 13.00678
 & 14.29689 & 2.45069 & 0 \\
2 & 0.43 & $82.2^{\circ}$ & 0.0863 & 13669 & 0.2826 & 0.1006 & 13.9795
 & 11.6358 & 3.79294 & 0 \\
3a & 0.20 & $85.0^{\circ}$ & 0.0725 & 7200 & 0.2663 & 0.1381 & 2.1678
 & 14.29689 & 2.45069 & 0 \\
3b & 0.20 & $85.0^{\circ}$ & 0.0725 & 7200 & 0.2663 & 0.1381 &
 13.00678 & 2.3828 & 2.45069 & 0 \\
4a & 0.20 & $85.0^{\circ}$ & 0.0725 & 7200 & 0.2472 & 0.1700 &
 13.00677 & 14.30843 & 2.45069 & 0 \\
4b & 0.20 & $85.0^{\circ}$ & 0.0725 & 7200 & 0.2828 & 0.1000 &
 13.00677 & 14.08273 & 2.45069 & 0 \\
5a & 0.20 & $85.0^{\circ}$ & 0.0725 & 7200 & 0.2006 & 0.1491 &
 12.91617 & 11.92371 & 1.6998 & 0 \\
5b & 0.20 & $85.0^{\circ}$ & 0.0725 & 7200 & 0.3396 & 0.1195 &
 12.92509 & 7.42059 & 3.51082 & 0 \\
\hline
\end{tabular}
\normalsize
\label{tab:fake_in_param}
\end{center}
\end{sidewaystable}

\begin{table}
\begin{center}
\caption[Reconstructed parameters from {\sc lfit}.]{Reconstructed
  parameters from {\sc lfit}.}
\vspace{0.3cm}
\small
\begin{tabular}{l|c|c|cc}
\hline
 & \multicolumn{4}{c}{Model number} \\ \cline{2-5}
Parameter & 1 & 2 & \multicolumn{2}{|c}{3} \\
 & & & a & b \\
\hline
Error (per cent) & 4 & 4 & 4 & 4 \\
Flickering & & & & \\
\hspace{0.1cm} (per cent) & 4 & 4 & 4 & 4 \\
Inclination $i$ & $84.8^{\circ}\pm0.1^{\circ}$ & $81.6^{\circ}\pm0.1$
& $84.77^{\circ}\pm0.03^{\circ}$ & $86.2^{\circ}\pm0.4^{\circ}$ \\
Mass ratio $q$ & $0.2035$ & $0.455$ & $0.20579$ & $0.181$ \\
& $\pm0.0010$ & $\pm0.004$ & $\pm0.00014$ & $\pm0.006$ \\
Eclipse phase & $0.07255$ & $0.08629$ & $0.072643$ & $0.07252$ \\
\hspace{0.1cm} width $\Delta\phi$ & $\pm0.00005$ & $\pm0.00004$ &
$\pm0.00015$ & $\pm0.00005$ \\
Outer disc & $0.2983$ & $0.2912$ & $0.29558$ & $0.3215$ \\
\hspace{0.1cm} radius $R_{\rm{d}}/a$ & $\pm0.0013$ & $\pm0.0023$ &
$\pm0.00026$ & $\pm0.0004$ \\ 
White dwarf & & & & \\
\hspace{0.1cm} limb & $0.5$ & $0.5$ & $0.5$ & $0.5$ \\
\hspace{0.1cm} darkening $U_{\rm{1}}$ & & & & \\
White dwarf & $0.0132$ & $0.0106$ & $0.0135$ & $0.01376$ \\ 
\hspace{0.1cm} radius $R_{1}/a$ & $\pm0.0003$ & $\pm0.0003$ &
$\pm0.0006$ & $\pm0.00029$ \\
Bright spot & $0.01179$ & $0.01135$ & $0.01169$ & $0.0123$ \\ 
\hspace{0.1cm} scale $SB/a$ & $\pm0.00018$ & $\pm0.00026$ &
$\pm0.00006$ & $\pm0.0006$ \\
Bright spot & $119.42^{\circ}$ & $111.51^{\circ}$ & $119.41^{\circ}$ &
$119.3^{\circ}$ \\
\hspace{0.1cm} orientation $\theta_{\rm{B}}$ &
$\pm0.08^{\circ}$ & $\pm0.16^{\circ}$ &
$\pm0.012^{\circ}$ & $\pm0.5^{\circ}$ \\
Isotropic flux & $0.0017$ & $0.009$ & $0.0006$ & $0.013$ \\ 
\hspace{0.1cm} fraction $f_{\rm iso}$ & $\pm0.0028$ & $\pm0.008$ &
$\pm0.0013$ & $\pm0.015$ \\
Disc exponent $b$ & $0.16\pm0.12$ & $0.10\pm0.3$ & $0.10\pm0.18$ &
$-0.49\pm0.12$ \\
Phase offset $\phi_{0}$ & $0$ & $0$ & $0$ & $0$ \\
$\chi^{2}$ of fit & $14498$ & $25940$ & $18688$ & $13127$ \\  
Number of & & & & \\
\hspace{0.1cm} datapoints $\nu$ & $7199$ & $13668$ & $7199$ & $7199$ \\
\hline
Flux & & & & \\
\hspace{0.1cm} (arbitrary units) & & & & \\
\hspace{0.1cm} White dwarf & $13.08\pm0.04$ & $14.039\pm0.038$ &
$2.199\pm0.022$ & $12.94\pm0.03$ \\
\hspace{0.1cm} Accretion disc & $2.37\pm0.04$ & $3.59\pm0.04$ &
$2.422\pm0.028$ & $2.50\pm0.03$ \\
\hspace{0.1cm} Secondary & $0$ & $0.042\pm0.010$ & $0.006\pm0.008$ &
$0$ \\
\hspace{0.1cm} Bright spot & $14.288\pm0.026$ & $12.094\pm0.024$ &
$14.289\pm0.011$ & $2.395\pm0.020$ \\
\hline
\end{tabular}
\normalsize
\label{tab:fake_out_param_lfit}
\end{center}
\end{table}

\setcounter{table}{1}
\begin{table}
\begin{center}
\caption[{\em Continued}. Reconstructed parameters from {\sc
    lfit}.]{{\em Continued}. Reconstructed parameters from {\sc
    lfit}. No error estimates could be determined for some parameters
    of model 4a, since the fit became unphysical if left to iterate to
    convergence.}
\vspace{0.3cm}
\small
\begin{tabular}{l|cc|cc}
\hline
 & \multicolumn{4}{c}{Model number} \\ \cline{2-5}
Parameter & \multicolumn{2}{|c}{4} & \multicolumn{2}{|c}{5} \\
 & a & b & a & b \\
\hline
Error (per cent) & 4 & 4 & 4 & 4 \\
Flickering & & & & \\
\hspace{0.1cm} (per cent) & 4 & 4 & 4 & 4 \\
Inclination $i$ & $89.7^{\circ}$ & $77.3^{\circ}\pm0.1^{\circ}$ &
$84.4^{\circ}\pm0.1^{\circ}$ & $85.6^{\circ}\pm0.2^{\circ}$ \\
Mass ratio $q$ & $0.157$ & $0.5254$ & $0.2128$ & $0.1902$ \\
& & $\pm0.0024$ & $\pm0.0013$ & $\pm0.0020$ \\ 
Eclipse phase & $0.0725$ & $0.07258$ &
$0.07259$ & $0.07255$ \\
\hspace{0.1cm} width $\Delta\phi$ & &
$\pm0.00005$ & $\pm0.00006$ & $\pm0.00006$ \\ 
Outer disc & $0.2908$ & $0.2635$ & $0.2439$ &
$0.3629$ \\
\hspace{0.1cm} radius $R_{\rm{d}}/a$ & $\pm0.0011$ & $\pm0.0007$ &
$\pm0.0007$ & $\pm0.0008$ \\
White dwarf & & & & \\
\hspace{0.1cm} limb & $0.5$ & $0.5$ & $0.5$ & $0.5$ \\
\hspace{0.1cm} darkening $U_{\rm{1}}$ & & & & \\
White dwarf & $0.01420$ & $0.01137$ & $0.0174$ &
$0.0137$ \\ 
\hspace{0.1cm} radius $R_{1}/a$ & $\pm0.00026$ & $\pm0.00021$ &
$\pm0.0003$ & $0.0004$ \\
Bright spot & $0.0160$ & $0.01448$ & $0.00928$ & $0.00612$ \\ 
\hspace{0.1cm} scale $SB/a$ & $\pm0.0003$
& $\pm0.00016$ & $\pm0.00016$ & $0.00020$ \\ 
Bright spot & $126.95^{\circ}$ & $111.08^{\circ}$ & $127.94^{\circ}$ &
$110.76^{\circ}$ \\
\hspace{0.1cm} orientation $\theta_{\rm{B}}$ &
$\pm0.09^{\circ}$ & $\pm0.08^{\circ}$ &
$\pm0.11^{\circ}$ & $\pm0.16$ \\
Isotropic flux & $0.030$ & $0.0091$ & $0.0013$ &
$0.018$ \\ 
\hspace{0.1cm} fraction $f_{\rm iso}$ & $\pm0.003$ & $\pm0.0026$ &
$\pm0.0027$ & $\pm0.006$ \\
Disc exponent $b$ & $-0.70\pm0.23$ & $0.2\pm0.4$ & $-0.93\pm0.13$ &
$-0.08\pm0.15$ \\
Phase offset $\phi_{0}$ & $0$ & $0$ & $0$ & $0$ \\
$\chi^{2}$ of fit & $16399$ & $14523$ & $14385$ & $13538$ \\  
Number of & & & & \\
\hspace{0.1cm} datapoints $\nu$ & $7199$ & $7199$ & $7199$ & $7199$ \\
\hline
Flux & & & & \\
\hspace{0.1cm} (arbitrary units) & & & & \\
\hspace{0.1cm} White dwarf & $12.71\pm0.06$ & $13.28\pm0.05$ &
$12.86\pm0.04$ & $12.97\pm0.06$ \\
\hspace{0.1cm} Accretion disc & $2.14\pm0.08$ & $1.79\pm0.06$ &
$1.76\pm0.04$ & $3.25\pm0.09$ \\
\hspace{0.1cm} Secondary & $0.175\pm0.017$ & $0.273\pm0.013$ & $0$ &
$0.09\pm0.03$ \\
\hspace{0.1cm} Bright spot & $14.658\pm0.025$ & $14.303\pm0.028$ &
$11.898\pm0.023$ & $7.78\pm0.03$ \\
\hline
\end{tabular}
\normalsize
\end{center}
\end{table}

\begin{table}
\begin{center}
\caption[Reconstructed parameters from the derivative
  method.]{Reconstructed parameters from the derivative method.}
\vspace{0.3cm}
\small
\begin{tabular}{l|c|c|cc}
\hline
 & \multicolumn{4}{c}{Model number}\\
\cline{2-5}
Parameter & 1 & 2 & \multicolumn{2}{|c}{3} \\
 & & & a & b \\
\hline
Error (per cent) & 4 & 4 & 4 & 4 \\
Flickering (per cent) & 4 & 4 & 4 & 4 \\
Inclination $i$ & $84.8^{\circ}$ & $81.9^{\circ}$ & $85.4^{\circ}$ &
 $88.0^{\circ}$ \\
Mass ratio $q$ & 0.206 & 0.470 & 0.200 & 0.169 \\
$\Delta\phi$ & 0.072649 & 0.088308 & 0.073344 & 0.073205 \\
$\Delta R_{{\rm d}}/a$ & 0.0326 & 0.0692 & 0.0314 & 0.1589 \\
$\Delta \theta$ & $9.4^{\circ}$ & $11.8^{\circ}$ & $16.1^{\circ}$ &
 $12.5^{\circ}$ \\
$\Delta Z/a$ & -- & -- & -- & -- \\
$\Delta Z_{2}/a$ & 0.0347 & 0.0586 & 0.0539 & 0.0987 \\
$R_{{\rm d}}/a$ & 0.3043 & 0.2944 & 0.3076 & 0.3394 \\
$\theta$ & $26.3^{\circ}$ & $19.4^{\circ}$ & $26.3^{\circ}$ &
 $23.5^{\circ}$ \\
\hline
\end{tabular}
\normalsize
\label{tab:fake_out_param_der}
\end{center}
\end{table}

\setcounter{table}{2}
\begin{table}
\begin{center}
\caption[{\em Continued}. Reconstructed parameters from the derivative
  method.]{{\em Continued}. Reconstructed parameters from the
  derivative method.}
\vspace{0.3cm}
\small
\begin{tabular}{l|cc|cc}
\hline
 & \multicolumn{4}{c}{Model number}\\
\cline{2-5}
Parameter & \multicolumn{2}{|c}{4} & \multicolumn{2}{|c}{5} \\
 & a & b & a & b \\
\hline
Error (per cent) & 4 & 4 & 4 & 4 \\
Flickering (per cent) & 4 & 4 & 4 & 4 \\
Inclination $i$ & -- & $77.3^{\circ}$ & $86.8^{\circ}$ & $85.9^{\circ}$ \\
Mass ratio $q$ & $<0.164$ & 0.53 & 0.195 & $0.185$ \\
$\Delta\phi$ & 0.073344 & 0.072927 & 0.075149 & 0.072510 \\
$\Delta R_{{\rm d}}/a$ & -- & 0.0332 & 0.0293 & 0.0406 \\
$\Delta \theta$ & -- & $21.2^{\circ}$ & $21.5^{\circ}$ & $5.1^{\circ}$ \\ 
$\Delta Z/a$ & -- & 0.0346 & -- & 0.0515 \\
$\Delta Z_{2}/a$ & -- & 0.0296 & 0.0421 & 0.0549 \\ 
$R_{{\rm d}}/a$ & -- & 0.2369 & 0.2460 & 0.3633 \\ 
$\theta$ & -- & $26.7^{\circ}$ & $37.9^{\circ}$ & $19.6^{\circ}$ \\
\hline
\end{tabular}
\normalsize
\end{center}
\end{table}

The errors on the parameters as determined by {\sc lfit} are clearly
too small when compared to the input parameters given in
table~\ref{tab:fake_in_param}. As such, they likely represent the {\em
reproducibility\/} of the fit to the data, rather than the standard
deviation from the true values. It is probable that there is a
systematic error present between the model used to generate the
light curves and the model used to reconstruct them. This highlights
the fact that any such model-fitting approach to the determination of
the system parameters will suffer from the applicability of the model:
that is, one does not know how far from reality the model deviates. One
such likely discrepancy is in the form of the bright spot intensity
distribution. {\sc Lfit} models this in the eminently reasonable form
given in \S~\ref{sec:lfit}. If the intensity distribution of the
bright spot differs from this, however, then a systematic error will
be introduced into the estimate of the mass ratio due to the position
of the centre-of-light of the bright spot differing from that
expected. Another possible explanation of the small errors is that
they may be due to the program having difficulty in iterating to
$\chi^{2}_{{\rm min}}$ during the bisection used to determine the
errors.

I did not estimate uncertainties for the parameters produced by the
derivative technique, as first, the dominant source of error is
likely to be systematic differences between the models used to produce
and analyse the light curves and second, due to the prohibitive amount of
time it would have taken to produce and analyse the multiple light
curves necessary to determine the rms variations in the contact phases.

Models 1, 2, 3a, 4b, 5a, 5b and arguably 3b were satisfactorily
reconstructed by both the {\sc lfit} and derivative methods. Model 4a
was found to produce an unphysical light curve with the derivative
technique, since the constraints placed upon the eclipse phase width
$\Delta\phi$ (by the white dwarf eclipse width) and the mass ratio $q$
(by the positions of the bright spot ingress and egress) were not
consistent with an orbital inclination $i\leq90^{\circ}$. The {\sc
lfit} method produced a very small mass ratio (as might be expected,
since a reduction in $q$ moves the path of the gas stream further from
the white dwarf), but also a very high orbital inclination, which
meant that the solution found by the iterative procedure became
unphysical if left to iterate towards an optimum result. Model 4b was
reconstructed with a lower orbital inclination, which was caused by
the model fitting a larger mass ratio in order to place the bright
spot at the correct location. Models 5a and 5b were adequately
reconstructed by each technique, confirming the reliability of both
over a range of different accretion disc radii.

Comparison of the recovered parameters for models 3a and 3b leads to
the conclusion that the visibility of the bright spot is more
important than the visibility of the white dwarf. This is perhaps due
to the fact that the white dwarf eclipse is symmetrical, so the twin
constraints of ingress and egress lead it to being much less
susceptible to noise in the data than the a symmetric light curve of
the bright spot. The times of ingress and egress of the bright spot
are dependent on more parameters than those of the white dwarf: the
bright spot times depend on the mass ratio, the disc radius, the phase
offset and the orbital inclination, as opposed to the eclipse phase
width, the phase offset and the orbital inclination for the white dwarf.

The fluxes recovered for each component by the {\sc lfit} method were
generally in good agreement with the input fluxes, as can be seen by
comparison of the relevant figures in tables~\ref{tab:fake_in_param} and
\ref{tab:fake_out_param_lfit}. The largest discrepancies occur for
models 4a and 4b, as might be expected, since one of the basic
assumptions of the method is deliberately broken for both these
models. Again, it appears that systematic errors frequently dominate,
a fact that should be borne in mind when fluxes derived in this way
are used.

In conclusion, both techniques reproduced the system parameters well
in all of the realistic light curves (models 1, 2, 3a, 3b, 5a
and 5b). The parameters for model 3b were rather less accurate, as the
bright spot was to some extent lost amongst the noise. The errors
determined by the {\sc lfit} method (by increasing $\chi^{2}$ by 1)
were too small, suggesting that systematic errors introduced by
differences between the assumed model and the real case are the
dominant source of error. Agreement between the two methods was good
for all the models investigated. Breaking the assumption that the
bright spot was at the intersection of the gas stream and accretion
disc's outer edge had predictable results: placing it ahead of
the intersection led to too small a mass ratio; behind too large a
mass ratio.

These two techniques were also compared using real data, as discussed
in \S~\ref{sec:xzeridvumacomp}.

\newpage

\clearpage

\clearpage

\clearpage

\clearpage

\section{Mass determination}
\label{sec:massdet}
The derivative and {\sc lfit} techniques yield the system parameters
relative to the orbital separation (or the L$_{1}$ distance).
To determine the absolute system parameters I have used the
Nauenberg mass-radius relation for a cold, non-rotating white dwarf
(equation~\ref{eq:nauenberg})\footnote{From here on, I refer to
parameters measured in units of the orbital separation or L$_{1}$
distance as {\em relative} and those measured in solar units as {\em
absolute}.}. If one sets $R_{1}/a=y$, Kepler's third law
(equation~\ref{eq:kepler}) can be rewritten in terms of the parameters
$R_{1}$ and $y$, giving another restriction on the white dwarf radius:
\begin{equation}
\label{eq:kepler2}
R_{1}=y \left( \frac{GM_{1}(1+q)P^{2}_{\rm{orb}}}{4\pi^{2}} \right)
^{\frac{1}{3}}.
\end{equation}
Equations \ref{eq:nauenberg} and \ref{eq:kepler2} can be easily solved
to give the system parameters. The secondary radius $R_{2}$ can be
calculated by approximating it to the volume radius of the Roche
lobe (equation~\ref{eq:RL2}).

As the \citet{nauenberg72} mass-radius relation assumes a cold white
dwarf, I have attempted to correct this relation to the approximate
temperature given by a fit to the deconvolved white dwarf fluxes (see
below). \citet{wood89a} and \citet{koester86} note that the radius of
a white dwarf at $10^{4}$~K is about 5 per cent larger than a cold
white dwarf. To correct to the appropriate temperature from $10^{4}$~K
the white dwarf cooling curves of \citet{wood95} have been used.

\section{White dwarf model atmospheres}
\label{sec:fitting}

The temperature and distance of the white dwarf component can be
determined by fitting the fluxes from three filters to the spectrum of
a blackbody or model atmosphere.

The expected flux $f$ from a blackbody $B_{\nu}(\lambda , T)$ in a
passband with transmission function $P(\lambda)$ is
\citep[e.g.][]{wood89a}
\begin{equation}
\label{eq:flux}
f= \frac{\int P(\lambda)B_{\nu}(\lambda , T)d\lambda/\lambda}{\int
  P(\lambda)d\lambda/\lambda}\cdot \frac{\pi R^{2}_{1}}{D^{2}},
\end{equation}
where $D$ is the distance to the star. By fitting a blackbody function
to the white dwarf flux in each passband the white dwarf temperature
$T_{1}$ and distance can be determined. As a white dwarf spectrum is
one of the closest astronomical approximations to a blackbody, this
procedure is reasonable.

The white dwarf fluxes were also fitted to the hydrogen-rich, $\log
g=8$ white dwarf model atmospheres of \citet{bergeron95} by
$\chi^{2}$ minimisation. The colour indices quoted therein were
converted to the SDSS system using the observed transformations of
\citet{smith02}. This procedure determined the temperature of the
white dwarf but not the distance, since the absolute magnitude (and
hence the distance) is much more heavily dependent on the exact value
of $\log g$ than the colours (which give the temperature). This latter
method, however, will determine the white dwarf temperature more
accurately than a blackbody fit, as it will allow, for example, for
the Balmer jump in the white dwarf spectrum.

\section{Eclipse mapping}
\label{sec:em}
In \S~\ref{sec:gasstream} I discussed various theoretical models and
predicted properties of accretion discs, and alluded to a technique
whereby the intensity distribution across the disc could be uncovered.
This technique, called {\em eclipse mapping}, is the subject of this
section.

The eclipse mapping method was developed by \citet{horne85}. It
enables the light distribution across the disc (including the
contributions of the white dwarf and bright spot) of eclipsing systems
to be mapped. The shape of the eclipse produced by the occultation of
the accretion disc by the red dwarf depends on the light distribution
across the disc. Unfortunately, as the eclipse light curve is
one-dimensional, and the light distribution across the disc is
two-dimensional (in the approximation of a flat disc), the solution is
not unique. One possible solution to this is to use a model-fitting
approach, but this has the obvious disadvantage that it is
model-dependent. This problem becomes more acute when it is considered
that our present knowledge of accretion disc physics is
incomplete. The departure of accretion discs in CVs from axi-symmetry,
due to the impact of the gas stream with the edge of the disc, provide
another problem for model-fitting procedures.

The above concerns naturally lead to the eclipse-mapping approach. In
this method, the intensity at each point of the accretion disc is an
independent parameter. In the simplest implementation of the method,
the disc is modelled as a simple Cartesian grid, co-rotating with the
binary system. In this form, the eclipse mapping method makes three
basic assumptions:
\begin{enumerate}
\item The secondary star fills its Roche-lobe. There is ample evidence
  (e.g.\ ellipsoidal variations from the distorted secondary star;
  \citealp{allan96}) for mass transfer via Roche-lobe overflow
  (the very presence of the accretion disc and bright spot imply it),
  so this assumption seems reasonably valid.
\item The intensity distribution is two-dimensional; it is constrained
  to the orbital plane. \citet{rutten98} found that this is a good
  assumption provided that the inner disc regions are not obscured by
  the disc rim. The opening angles of accretion discs in CVs are
  typically $\sim 5^{\circ}$ (for example, the disc in the SW~Sex star
  DW~UMa has an opening angle of $\geq8^{\circ}$; \citealp{knigge00}), so
  this assumption is (reasonably) valid for orbital inclinations
  $i\lesssim85^{\circ}$.
\item The emission is phase-independent (apart from the eclipse by the
  secondary star). This last assumption is the most problematic. Many
  CVs have orbital humps due to anisotropic radiation from the bright
  spot, which violates this assumption. In the following section I
  discuss how this may be accounted for.
\end{enumerate}

The eclipse-mapping technique has since been further developed to
allow spectral eclipse mapping \citep{rutten93,rutten94a}, modelling
of a three-dimensional accretion disc \citep{rutten98}, fitting of
multi-colour light curves by physical properties of a model disc
(so-called {\em physical parameter\/} eclipse mapping; see
\citealp{vrielmann02}) and mapping the spatial location of the
flickering source in the discs of CVs
\citep{welsh95,bruch00b,baptista04}.

\subsection{Theory}
This section follows the derivations given by \citet{skilling84},
\citet{gull89,gull91} and \citet{watson02a}.

Eclipse mapping requires that we use the observed data $D$ to make
inferences about the various possible intensity distributions across
the disc $A, B, C\ldots$ etc. Letting $h$ represent any of the
hypotheses $A, B, C\ldots$, we wish to calculate
\begin{equation}
P(h|D),
\end{equation}
which is the probability of $h$ occurring, given $D$. The data,
however, give
\begin{equation}
P(D|h),
\end{equation}
the likelihood of $D$ occurring, given $h$. In order to reverse
$P(D|h)$ to obtain $P(h|D)$, note that the probability of {\em both\/}
$h$ and $D$ occurring is
\begin{subequations}
\begin{eqnarray}
P(h,D) & = & P(h)P(D|h) \\
 & = & P(D)P(h|D),
\end{eqnarray}
\end{subequations}
where $P(h)$ and $P(D)$ are the probabilities of the prior $h$ and the
evidence $D$. The prior probability term includes our prior
expectations about possible intensity distributions across the disc
$h$ {\em before\/} acquiring the data $D$. Rearranging the above
equations, we obtain Bayes' theorem
\begin{equation}
\label{eq:bayes}
P(h|D) = \frac{P(h)P(D|h)}{P(D)}.
\end{equation}
The selection of $h$ proceeds by choosing that intensity distribution
which maximises the entropy (that is, is maximally
non-committal). 
The entropy $S(h)$ is defined in this thesis as
\begin{equation}
S(h) = \sum_{j=1}^{n} \left[ h_{j} - d_{j} - h_{j}\ln
  \left(\frac{h_{j}}{d_{j}}\right) \right],
\end{equation}
where $h_{j}$ is the intensity of element $j$ and $d_{j}$ is a {\em
  default\/} image to which the reconstruction
will default in the absence of data. The default image may
be used to include prior knowledge of the likely distribution of the
light across the disc. This is discussed in more detail in the next
section. The entropy measures the deviation of the reconstructed
intensity map $h_{j}$ with respect to the default map $d_{j}$. The
global maximum of the entropy is therefore at $h=d$, where $S=0$.

It can be shown \citep{gull89,gull91} that
\begin{equation}
\label{eq:probh}
P(h)\propto \exp (\alpha S),
\end{equation}
where $\alpha$ is some constant. In order to be able to solve
equation~\ref{eq:bayes} we now require only $P(D|h)$.

If the only variations in the observed data (the light curve) are due
to the eclipse of the disc by the secondary star, then the data
$D(\phi)$ must be related to the intensity distribution across the
disc $h(j)\equiv h_{j}$ by
\begin{equation}
\label{eq:datasum}
D(\phi) = \sum_{j=1}^{n} V(j,\phi)h(j) \pm \sigma(\phi),
\end{equation}
where $V(j,\phi)$ is the fractional visibility of element $j$ at phase
$\phi$ and $\sigma(\phi)$ is
the error on $D(\phi)$. The above can be simplified to
\begin{equation}
o_{j} = p_{j} + \sigma_{j},
\end{equation}
where $o_{j}$ and $p_{j}$ are the observed and predicted intensities
for element $j$ and $\sigma_{j}$ is the error on $o_{j}$. If we
assume that the errors are normally distributed, then
\begin{equation}
\label{eq:probdh}
P(D|h) = \prod_{j=1}^{n} (2\pi\sigma^{2}_{j})^{-\frac{1}{2}}
\exp\left(\frac{-(p_{j}-o_{j})^{2}}{2\sigma^{2}_{j}}\right),
\end{equation}
and therefore
\begin{equation}
\label{eq:probdhchi}
P(D|h) \propto \exp(-\chi^{2}_{\rm R}),
\end{equation}
where $\chi^{2}_{\rm R}$ is as defined in
equation~\ref{eq:reducedchisquared}.

Equation~\ref{eq:bayes} demonstrates that to find the most probable
image $h$, the probability $P(h|D)$ must be
maximised. Equations~\ref{eq:probh} and \ref{eq:probdhchi} show that
this is equivalent to minimising the quantity
\begin{equation}
\label{eq:maxent}
\chi^{2}_{\rm R}-\alpha S.
\end{equation}

By minimising equation~\ref{eq:maxent}, the selected solution $h$ is
the one with the maximum entropy consistent within the constraints
imposed by the data. No additional assumptions or biases are
introduced by this method; this is the Principle of Maximum Entropy.

\subsection{Practice}
\label{sec:em_practice}
The development of the eclipse mapping code used in this thesis has
been my own work, except for the use of the {\sc memsys} package,
which was written by \citet{skilling84}. I chose to write my own
eclipse mapping code in order to better understand the processes
involved. I gratefully acknowledge the assistance of Dr.~Chris Watson
during the early stages of development.

In this thesis, I use a simple Cartesian grid centred on the white
dwarf, with the co-ordinate system as defined in
\S~\ref{sec:roche_lobe}. The length of the grid side $R_{\rm{g}}$ is
chosen to be of the order of or greater than the tidal radius
$R_{\rm{tidal}}$ (\S~\ref{sec:gasstream},
equation~\ref{eq:tidal}). The centre of the $i^{\rm{th}}$ tile is then
given by
\begin{subequations}
\begin{equation}
x(i) = \left(i-0.5-n\;{\rm aint} \left[ \frac{i-1}{n} \right] \right)
  \left(\frac{2R_{\rm{g}}}{n} \right) - R_{\rm{g}}
\end{equation}
\begin{equation}
y(i) = -\left(0.5+{\rm aint} \left[ \frac{i-1}{n} \right] \right)
  \left(\frac{2R_{\rm{g}}}{n} \right) + R_{\rm{g}},
\end{equation}
\end{subequations}
where $n$ is the number of tiles per side and `aint' denotes that its
argument is truncated ({\em not\/} rounded) to an integer.

The visibility function $V(i,\phi)$ in equation~\ref{eq:datasum} is
determined by use of the {\sc blink} subroutine discussed in
\S~\ref{sec:derivative}. This requires computation of the Earth vector
$\hat{E}$, which is the vector pointing towards Earth from the grid
element in question. In the current co-ordinate system, the Earth
vector is given by
\begin{subequations}
\begin{equation}
\hat{E}_x = \cos (2\pi\phi) \sin i
\end{equation}
\begin{equation}
\hat{E}_y = -\sin (2\pi\phi) \sin i
\end{equation}
\begin{equation}
\hat{E}_z = \cos i .
\end{equation}
\end{subequations}

Using the position of the point at the centre of each tile to assess
its visibility results in
\begin{equation}
V(i,\phi) = \left\{ \begin{array}{ll}
  0 & \mbox{if eclipsed} \\
  1 & \mbox{otherwise}
  \end{array}
\right. .
\end{equation}
The accuracy of the visibility function $V(i,\phi)$ can be improved by
either subdividing the tiles into sub-tiles or increasing the number of
tiles in the grid. In the former case, the visibility function of a
given tile at a given phase, $V(i,\phi)$, is  given by the fraction of
its sub-tiles that are not eclipsed at their centre at that phase. Each
{\sc memsys} iteration requires computation of the intensity of each
tile, so this subdivision has a computational advantage over merely
increasing the number of tiles in the grid, as the visibility function
$V(i,\phi)$ only has to be calculated once, at the start of the
eclipse mapping procedure. As discussed below, subdivision of the
tiles can also eliminate memory problems encountered when dealing with
large grid sizes.

The fineness of the grid has an optimum value, equal to the distance
that the projected shadow of the secondary star moves across the
centre of the disc in one phase step of the light curve
\citep{baptista93}. If the grid is too coarse, then we recover less
information than is possible; if it is too fine, then we are
attempting to fit the data in more detail than is warranted by the
data, leaving room for noise to be propagated into the reconstructed
maps (and it is a waste of time and effort). The optimum number of
tiles per side of the grid $N$ is given by \citep{baptista93}
\begin{equation}
N = \frac{Ra(0.5-0.227\log q)\sin i}{R_{\rm{L1}}\tan \Delta\phi},
\end{equation}
where $R$ is the length of the side of the grid, $R_{\rm{L1}}$ is the
distance to the inner Lagrangian point from the white dwarf and
$\Delta\phi$ is the phase resolution of the data. Due to constraints
on the maximum size of the arrays used in my eclipse mapping code, the
maximum grid size was $77\times77$. If the optimum number of tiles was
greater than this number, the tiles were subdivided, as discussed
above, and the number of tiles reduced accordingly.

\subsubsection{The default map}
\begin{figure}
\centerline{\includegraphics[width=12cm,angle=0]{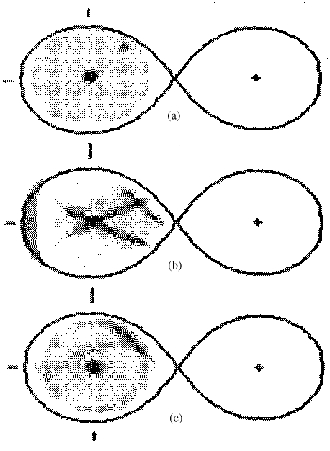}}
\caption[Suppression of arcs in the reconstructed image.]{Suppression of
  arcs in the reconstructed image. (a) The original accretion disc
  image, with two Gaussian spots superimposed on a uniform
  background. (b) The image obtained using the uniform default
  map. (c) The image obtained using the default of full azimuthal
  smearing. Adapted from \citet{horne85}.}
\label{fig:arcs}
\end{figure}

The choice of default map has a critical impact on the quality of the
reconstructions obtained \citep[see, for
example][]{horne85,baptista01a}. At first glance, the most obvious
choice for the default map is a uniform one. This is not, however, the
best choice, as it leads to a reconstructed image that is heavily
distorted by criss-crossed arcs. These result from the twin
constraints on the reconstruction: the entropy and the eclipse of the
disc. The maximisation of entropy means that extreme intensities are
suppressed; the flux from a compact source such as the bright spot or
the white dwarf is distributed over a larger area. The constraints
provided by the eclipse result in the flux being spread along ingress
and egress arcs that pass through the true location of the compact
source, as illustrated in figure~\ref{fig:arcs}.

Choosing a non-uniform default map allows prior expectation of the
likely disc intensity map to be included in the
procedure. \citet{baptista01a} discusses some useful prescriptions for
the default, listed in table~\ref{tab:defaults}.

The default $D_{j}$ usually takes the form of a weighted average of
the intensities $I_{k}$ of the elements in the grid:
\begin{equation}
\label{eq:default}
D_{j} = \frac{\sum_{k} \omega_{jk}I_{k}}{\sum_{k} \omega_{jk}},
\end{equation}
where $\omega_{jk}$ is the user-defined {\em weight function}. It is
ultimately via the weight function that {\em a priori\/} information
about the disc is included in the reconstruction. The weight function
is usually defined as a Gaussian point-spread function of width
$\Delta$.

\begin{table}
\begin{center}
\caption[Prescriptions for weight functions
  $\omega_{jk}$.]{Prescriptions for weight functions
  $\omega_{jk}$. $d_{jk}$ is the distance between pixels $j$ and $k$;
  $R_{j}$ and $R_{k}$ are the distances of pixels $j$ and $k$ from the
  origin, respectively; $\theta_{jk}$ is the azimuthal angle between
  pixels $j$ and $k$; and $s_{jk}=|R_{j}\theta_{j} - R_{k}\theta_{k}|$
  is the arc length between pixels $j$ and $k$. 1--5 from
  \citet{baptista01a}.}
\vspace{0.3cm}
\small
\begin{tabular}{ccc}
1) & Most uniform map & $\omega_{jk}=1$ \\\\
2) & Smoothest map &
$\omega_{jk}=\exp\left(-\frac{d^{2}_{jk}}{2\Delta^{2}}\right)$ \\\\
3) & Most axi-symmetric map &
$\omega_{jk}=\exp\left[-\frac{(R_{j}-R_{k})^{2}} {2\Delta^{2}_{R}}
  \right]$ \\
 & (full azimuthal smearing) \\
4) & Limited azimuthal smearing &
$\omega_{jk}=\exp \left[-\frac{1}{2} \left\{ \left(\frac{R_{j}-R_{k}}
  {\Delta_{R}} \right)^{2} + \left( \frac{\theta_{jk}}{\Delta_{\theta}}
  \right)^{2} \right\} \right]$ \\
 & (constant angle $\theta$) \\
5) & Limited azimuthal smearing &
$\omega_{jk}=\exp\left[-\frac{1}{2} \left\{ \left(\frac{R_{j}-R_{k}}
  {\Delta_{R}} \right)^{2} + \left( \frac{s_{jk}}{\Delta_{s}}
  \right)^{2} \right\} \right]$ \\
 & (constant arc length $s$) \\
6) & Limited azimuthal smearing & $\omega_{jk}=\exp \left[-\frac{1}{2} \left(
  \frac{\theta_{jk}}{\Delta_{\theta}} \right)^{2} \right]$ \\
 & (for disc rim) 
\end{tabular}
\normalsize
\label{tab:defaults}
\end{center}
\end{table}

One would expect the discs in CVs to be roughly axi-symmetric since
the material in the disc is, to a first approximation at least, in
Keplerian orbits. By using the prescriptions for the default map given
in table~\ref{tab:defaults} this expectation can be included in the
default map. The effect of this is that the entropy becomes
insensitive to structure on scales larger than $\Delta$. Structure on
small scales  will be suppressed by the entropy, whereas large-scale
structure will be freely determined by the data. This can be seen from
study of the weight functions given in table~\ref{tab:defaults}. If
the numerator is much smaller than $\Delta$ then the weight will be
large; if the numerator is larger than the denominator $\Delta$ then
the weight will be small. Figure~\ref{fig:arcs} shows the effect of
the default of {\em full azimuthal smearing\/} (see
table~\ref{tab:defaults}). Some remnants of the spurious arcs remain,
but they are greatly reduced in amplitude.

\begin{figure}
\centerline{\includegraphics[width=10cm,angle=0]{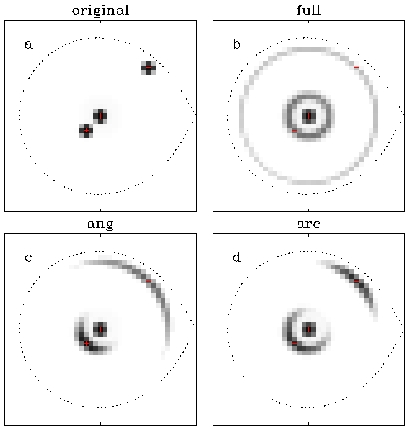}}
\caption[The effect of different weight functions on the default
  map.]{The effect of different weight functions on the default
  map. (a) shows the original map, with three Gaussian spots. The
  remaining panels show the default maps (note: {\em not\/} the
  reconstructed maps) produced from the map illustrated in panel (a)
  by applying the weight functions (see table~\ref{tab:defaults}) of
  (b) full azimuthal smearing, (c) constant angle $\theta$ and (d)
  constant arc length $s$. From \citet{baptista96}.}
\label{fig:weights}
\end{figure}

The default of full azimuthal smearing can result in distorted
reconstructions of discrete structures in the disc. For instance, the
bright spot is often smeared out into a ring of the same radial
distance from the white dwarf as the bright spot. In order to limit
this distortion, a default map of {\em limited azimuthal smearing\/}
can be used. Two methods of achieving this have been proposed: a
weight function of constant angles \citep{rutten93} and a weight
function of constant arc length \citep{baptista96}. The default of
constant angles gives better resolution in the inner parts of the
disc, whereas the default of constant arc length gives better
resolution in the outer parts of the disc (see \citet{baptista01a} and
figure~\ref{fig:weights}). For the data presented in this thesis, it
was found that the difference between the maps reconstructed using the
defaults of constant angles and constant arc length was negligible,
and therefore the default used throughout this thesis was that of
limited azimuthal smearing (constant angles).

A different default map was used for the disc rim to reflect its
constant radial distance from the centre of the grid and to allow the
rim intensity to be independent of the intensity distribution of the
$(x,y)$ grid. In order to fulfil this latter criterion, the weight of
each pixel in the disc rim with respect to each pixel in the $(x,y)$
grid was zero, and {\em vice versa}. The default chosen for the disc
rim was number 6 in table~\ref{tab:defaults}, with
$\Delta_{\theta}({\rm rim})=\Delta_{\theta}({\rm disc})$.

\subsubsection{The orbital hump}

The first method I attempted to use to correct for the orbital hump
followed \citet{baptista00}, who fitted a spline function to phases
outside eclipse, divided the light curve by the fitted spline and
scaled the result to the value of the spline function at phase zero, in
order to scale the light curve back to the original flux level
\citep[see also][]{horne85}. This technique worked well on the light
curve of IP~Peg in outburst to which \citet{baptista00} applied it,
but it is not suitable for the light curves of quiescent dwarf nov\ae\
which are the subject of this thesis. This is because by fitting and
dividing by a spline function we assume that all parts of the
accretion disc contribute equally to the anisotropic flux (the orbital
hump). This is not the case: the bright spot produces anisotropic
radiation, whereas the white dwarf and accretion disc do not
\citep{horne85,bobinger97}.

The method I eventually employed in order to model the orbital hump
was to introduce a disc rim \citep[e.g.][]{bobinger97}. This rim is
assumed to be of negligible height (so that I still assume a flat
disc) and is divided into $m$ segments. The visibility function
$V(i,\phi)$ for the rim is given by
\begin{equation}
V(i,\phi) = \left\{ \begin{array}{ll}
  0 & \mbox{if eclipsed} \\
  0 & \mbox{if $\sin (\theta + \phi)<0$} \\
  \sin^{2} (\theta + \phi)  & \mbox{otherwise}
  \end{array}
\right. ,
\end{equation}
where $\theta$ is the angle between the position on the disc rim and
the positive {\em x\/}-axis, measured clockwise. The intensity of each
element of the disc rim was fitted along with the intensity of each of
the Cartesian grid elements, and was found to model the orbital hump
very effectively. The number of elements used to model the disc rim
was 50 for all the reconstructions in this thesis. This number was
chosen to be high enough to effectively reproduce the orbital hump and
low enough so that the orbital modulation was reasonably smooth.

\subsubsection{The contribution of the secondary star}
The flux from the secondary star can have a detrimental effect on the
quality of the reconstructed map. This is because the relative eclipse
depth is anti-correlated with the eclipse width; any uneclipsed light
breaks this anti-correlation. The additional light is placed by the
maximum entropy reconstruction in the parts of the map which are
constrained the least by the eclipse data, such as the parts of the
accretion disc farthest from the red dwarf (the `back' of the
disc). This is discussed in detail by
\citet{rutten92b,rutten94a,baptista95,baptista96}. These
authors find that the contribution from the secondary can be estimated
by maximum entropy methods. First, the light curve can be offset by
varying amounts and fitted using the usual eclipse mapping
procedure. The offset which allows the largest entropy to be achieved
is then the uneclipsed light. Second, and equivalently, an additional
`virtual pixel' can be introduced into the grid. If the visibility
function for this pixel is unity at all phases, and its weight (refer
to equation~\ref{eq:default}) with respect to all the other pixels is
zero, and unity with respect to itself, then from the definition of
the $\chi^{2}$ statistic (equation~\ref{eq:chisquared}) the
contribution of this pixel to the entropy measure is zero, and the
iterative procedure to maximise the entropy proceeds as usual. The
only effect of this virtual pixel on the intensities of all other
pixels in the reconstruction is an offset, the magnitude of which is
determined by the maximum entropy procedure itself. This method of
determining the contribution from the secondary star fails for highly
asymmetric accretion discs, as noted by \citet{baptista96}. The
spurious structure introduced in the reconstructed map by the
uneclipsed component mixes with the asymmetric (bright spot) emission,
forming a more azimuthally symmetric structure in the disc. Due to the
choice of an azimuthally symmetric default map, this has the effect of
increasing the entropy of the reconstructed map. The map with the
largest entropy value is therefore not that with the correct offset
due to the presence of uneclipsed light. This problem is particularly
severe in the case of the faint, short-period dwarf nov\ae\ such as
XZ~Eri and DV~UMa studied in this thesis, as the disc is effectively
invisible in these objects, meaning that almost all of the emission
originates from the bright spot region: the disc emission is {\em
highly\/} asymmetric.

I have therefore subtracted the contribution to the total light from
the secondary star from the light curve prior to fitting. In the cases
of XZ~Eri and DV~UMa, the secondary contribution was determined from
the {\sc lfit} procedure. For all other objects, the contribution of
the secondary star was first estimated from the mid-eclipse flux
level, and fine-tuned by computing a series of eclipse maps with
different offset values and selecting the map with the least spurious
structure. These offsets are given in table ~\ref{tab:parameters}.

\subsubsection{The iterative procedure}

The {\sc memsys} code iteratively adjusts the intensities of each
element $h_{j}$. The procedure begins with a uniform intensity
distribution, in which each element has an intensity equal to the
maximum value of the light curve divided by the total number of
elements. This obviously gives a poor fit to the data. The code then
iterates until the desired $\chi^{2}_{\rm R}$, known as $\chi^{2}_{\rm
R,aim}$, is achieved, subsequent iterations serving to maximise the entropy
while keeping $\chi^{2}_{\rm R}$ fixed. The exit criterion, which when
satisfied signals that the final solution has been reached, is
$TEST<10^{-3}$, where
\begin{equation}
\label{eq:test}
TEST = \frac{1}{2}\left | \frac{\Delta \chi^{2}_{\rm R}}{|\Delta
  \chi^{2}_{\rm R}|} -
\frac{\Delta S}{|\Delta S|} \right |,
\end{equation}
provided that $|\chi^{2}_{\rm R}-\chi^{2}_{\rm R,aim}|<1$ and that the
entropy during the last five iterations has not decreased by more than
0.5~per~cent. When the entropy has been maximised, $TEST$ should be
zero, since in this case both $\Delta \chi^{2}_{\rm R}$ and $\Delta S$ will
necessarily be small. This definition of $TEST$ is equivalent to that
used internally by {\sc memsys}: $TEST=1-\cos\theta$, where $\theta$
is the angle between the gradients of entropy $S$ and $\chi^{2}_{\rm R}$. The
reconstructed images are those that are closest to the default map
from those that are consistent with the data to $\chi^{2}_{\rm R,aim}$.

The choice of $\chi^{2}_{\rm R,aim}$ affects the properties of the
reconstructed map. If too large a value is chosen, the map will not be
well-constrained by the data (i.e.\ entropy will dominate), and the
features present will be smeared out by the effects of the default map
adopted. On the other hand, if too small a value of $\chi^{2}_{\rm
R,aim}$ is used, then the reconstructed map will be noisy, with a
characteristic `grainy' texture, the result of trying to fit noise in
the data. Note that the optimum value $\chi^{2}_{\rm R,aim}$ is usually
greater than unity due to the presence of flickering in the light
curve resulting in the scatter of the data points being greater than
that implied by the errors on the data points.

To perform a maximum entropy reconstruction of the accretion disc, the
program was first run with $\chi^{2}_{\rm R,aim}=1$. In almost all cases
this $\chi^{2}_{\rm R,aim}$ was never achieved, so the program was
re-started with a value for $\chi^{2}_{\rm R,aim}$ adopted that {\em
was\/} reached in the previous attempt. The reconstructed image was
visually inspected, the value of $\chi^{2}_{\rm R,aim}$ adjusted and the
code re-run until the resulting intensity map was judged to be
`non-grainy' and the spurious arcs were minimised.

\subsubsection{Testing of the eclipse mapping code}
The eclipse mapping code was tested using the data for XZ~Eri and
DV~UMa described in table~\ref{tab:journal} and
chapter~\ref{ch:xzeridvuma}. The positions of the white dwarf and
bright spot and the absence of a significantly luminous accretion disc
formed the testing criteria. The XZ~Eri and DV~UMa data were used
instead of creating fake light curves due to their high quality
(especially for XZ~Eri) and the fact that the system parameters were
determined to a high degree of accuracy from my previous work using
{\sc lfit} (see \S~\ref{sec:xzeridvuma_techniques_model}). The results
of the eclipse mapping experiments are presented and discussed in
subsequent chapters. For ease of reference, the parameters used in
each reconstruction are given in table~\ref{tab:parameters} below.

\begin{sidewaystable}
\begin{center}
\caption[The parameters used in the maximum entropy reconstructions of
  the disc intensities.]{The parameters used in the maximum entropy
  reconstructions of the disc intensities. For each reconstruction,
  the default map is number~4 with $\Delta_{\theta}=0.7$~radians and
  $\Delta_{R}=0.01a$ (see table~\ref{tab:defaults}). The size of the
  grid in each case is $0.6a\times0.6a$. The format of the quoted grid
  dimensions is {\em x\/} tiles $\times$ {\em y\/} tiles $+$ disc rim
  tiles. The tiles in the $x,y$ grid are subdivided as described in
  \S~\ref{sec:em_practice}. Some light curves were rebinned by the
  factor shown (using a weighted mean) in order to reduce flickering.}
\vspace{0.3cm}
\small
\begin{tabular}{ccccccccccc}
Object & Cycle & Filter & $\chi^{2}_{\rm R,aim}$ & Grid & Sub- &
Iterations & Phase & Secondary & Binning & Radius of \\
 & & & & dimensions & divisions & & range &
(mJy) & factor & disc rim ($a$) \\
\hline

XZ Eri & all & {\em u\/}$^{\prime}$ & 1.15 & 
 $59\times59+50$ & 2 & 33 & all & 0.0020 & --  & 0.3 \\ 
 & all & {\em g\/}$^{\prime}$ & 1.4 & 
 $59\times59+50$ & 2 & 32 & all & 0.0029 & --  & 0.3 \\ 
 & all & {\em r\/}$^{\prime}$ & 2.5 & 
 $59\times59+50$ & 2 & 36 & all & 0.0064 & --  & 0.3 \\ 

DV UMa & all & {\em u\/}$^{\prime}$ & 1.4 & 
 $71\times71+50$ & 6 & 60 & $-0.06$ to $0.13$ & 0.0027 & -- & 0.31805 \\
 & all & {\em g\/}$^{\prime}$ & 1.3 & 
 $71\times71+50$ & 6 & 61 & $-0.06$ to $0.13$ & 0.00531 & -- & 0.31805 \\ 
 & all & {\em i\/}$^{\prime}$ & 7.0 & 
 $71\times71+50$ & 6 & 87 & $-0.06$ to $0.13$ & 0.0680 & -- & 0.31805 \\ 

HT Cas & 2002 data & {\em u\/}$^{\prime}$ & 2.8 & 
 $75\times75+50$ & 7 & 28 & $-0.09$ to $0.09$ & 0.15 & 3 & 0.28 \\
 & 2002 data & {\em g\/}$^{\prime}$ & 24 & 
 $75\times75+50$ & 7 & 29 & $-0.09$ to $0.09$ & 0.09 & 3 & 0.28 \\ 
 & 2002 data & {\em i\/}$^{\prime}$ & 7 & 
 $75\times75+50$ & 7 & 31 & $-0.09$ to $0.09$ & 0.32 & 0 & 0.28 \\

HT Cas & 2003 data & {\em u\/}$^{\prime}$ & 1.4 & 
 $75\times75+50$ & 7 & 71 & $-0.09$ to $0.09$ & 0.24 & 3 & 0.26 \\
 & 2003 data & {\em g\/}$^{\prime}$ & 15 & 
 $75\times75+50$ & 7 & 74 & $-0.09$ to $0.09$ & 0.14 & 3 & 0.26 \\ 
 & 2003 data & {\em i\/}$^{\prime}$ & 6 & 
 $75\times75+50$ & 7 & 134 & $-0.09$ to $0.09$ & 0.42 & 3 & 0.26 \\

OU~Vir & 2002 5 18 & {\em u\/}$^{\prime}$ & 1 & 
 $70\times70+50$ & 2 & 200 & $-0.06$ to $0.1$ & 0.06 & -- & 0.2315 \\
 & 2002 5 18 & {\em g\/}$^{\prime}$ & 4.5 & 
 $70\times70+50$ & 2 & 30 & $-0.06$ to $0.1$ & 0.06 & -- & 0.2315 \\ 
 & 2002 5 18 & {\em r\/}$^{\prime}$ & 2.5 & 
 $70\times70+50$ & 2 & 29 & $-0.06$ to $0.1$ & 0.11 & -- & 0.2315 \\

OU~Vir & 2003 5 22 & {\em u\/}$^{\prime}$ & 1.5 & 
 $70\times70+50$ & 2 & 28 & $-0.06$ to $0.1$ & 0.04 & -- & 0.2315 \\
 & 2003 5 22 & {\em g\/}$^{\prime}$ & 9 & 
 $70\times70+50$ & 2 & 54 & $-0.06$ to $0.1$ & 0.024 & -- & 0.2315 \\ 
 & 2003 5 22 & {\em i\/}$^{\prime}$ & 3.0 & 
 $70\times70+50$ & 2 & 26 & $-0.06$ to $0.1$ & 0.09 & -- & 0.2315 \\

IR~Com & all & {\em u\/}$^{\prime}$ & 15 
& $65\times65+50$ & 5 & 71 & $-0.06$ to $0.1$ & 0.27 & 2 & 0.3 \\
 & all & {\em g\/}$^{\prime}$ & 250 
& $65\times65+50$ & 5 & 73 & $-0.06$ to $0.1$ & 0.12 & 2 & 0.3 \\
 & all & {\em i\/}$^{\prime}$ & 120 
& $65\times65+50$ & 5 & 71 & $-0.06$ to $0.1$ & 0.35 & 2 & 0.3 \\

GY~Cnc & all & {\em u\/}$^{\prime}$ & 8 & 
$74\times74+50$ & 9 & 49 & all & 0.35 & -- & 0.3 \\
 & all & {\em g\/}$^{\prime}$ & 60 & 
$74\times74+50$ & 9 & 51 & all & 0.37 & -- & 0.3 \\
 & 2003 5 23 & {\em i\/}$^{\prime}$ & 11 & 
$74\times74+50$ & 9 & 53 & all & 2.1 & -- & 0.3 \\
 & 2003 5 19 & {\em z\/}$^{\prime}$ & 4.0 & 
$74\times74+50$ & 9 & 56 & all & 2.2 & -- & 0.3 \\

\end{tabular}
\normalsize
\label{tab:parameters}
\end{center}
\end{sidewaystable}

%% file: ouvir_results.tex
\chapter{OU~Vir}
\label{ch:ouvir}

The contents of this chapter have been published in the Monthly
Notices of the Royal Astronomical Society, {\bf 347}, 1173 and {\bf
354}, 1279 as {\em ULTRACAM photometry of the eclipsing cataclysmic
variable OU~Vir\/} by \citet*{feline04b,feline04c}. The exceptions to
this are the eclipse mapping results presented in
\S~\ref{sec:ouvir_eclipse_mapping}, and the observations of 2003 May
18, which were overlooked for publication due to an error in the
hand-written observing logs. The data were analysed using the
derivative technique only, and not the {\sc lfit} method, as the
significant night-to-night and short-term (flickering) variability
meant that the modelling approach of the latter technique was found to
be unsuitable in this case. The reduction and analysis of the data are
all my own, as is the text below. Dr.~Vik Dhillon supervised all work
presented here.

OU~Vir is a faint (V$\sim18$~mag; \citealp{mason02}) eclipsing CV with a
period of 1.75~hr which has been seen in outburst and probably
superoutburst \citep{vanmunster00}, marking it as an SU~UMa dwarf
nova. \citet{mason02} presented time-resolved, multi-colour
photometry and spectroscopy of OU~Vir, concluding that the eclipse is
of the bright spot and disc, but not the white dwarf.

\begin{figure}
\centerline{\includegraphics[width=10cm,angle=-90.]{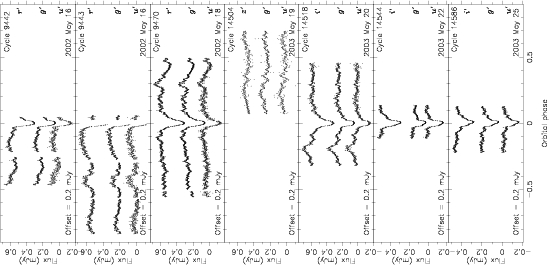}}
\caption[The light curve of OU~Vir.]{The light curve of OU~Vir. The {\em
r}$^{\prime}$, {\em z\/}$^{\prime}$ and {\em i\/}$^{\prime}$ data are
offset vertically upwards and the {\em u\/}$^{\prime}$ data are offset
vertically downwards.}
\label{fig:ouvir_lightcurve}
\end{figure}

The observations of OU~Vir are summarised in table~\ref{tab:journal},
and the data reduction procedure is detailed in
\S~\ref{sec:reduction}. The light curves of OU~Vir are shown in
figure~\ref{fig:ouvir_lightcurve}. The data of 2002 May were obtained
during the first night of commissioning of {\sc ultracam} and hence
were adversely affected by typical commissioning problems, chiefly
excess noise in the {\em u\/}$^{\prime}$ band and limited time
resolution due to the dead-time between exposures. The data taken in
2003 had no such problems.

\section{Light curve morphology}

\citet{mason02} found that for OU~Vir out of eclipse and during
quiescence, $V=18.08$~mag and $B-V=0.14$~mag, which corresponds to {\em
g}$^{\prime}\sim0.2$ mJy \citep{smith02}. \citet{vanmunster00} quote
an outburst amplitude of approximately 4 magnitudes (corresponding to
a peak {\em g\/}$^{\prime}$ flux of $\sim8.4$ mJy). OU~Vir was observed
at quiescence in both 2002 and 2003, as can be readily seen by
inspection of figure~\ref{fig:ouvir_lightcurve}. It is worth noting,
however, that the 2003 observations took place about 18~days after a
superoutburst of OU~Vir was first reported, on 2003 May 2 \citep{kato03}.
The light curves of 2002 May 16 (at phase $\sim-0.45$) and 18 (phase
$\sim0.3$) and 2003 May 20 (phase $\sim-0.2$) shown in
figure~\ref{fig:ouvir_lightcurve} show a feature which strongly
resembles a superhump, and suggests that the system may have recently
undergone a superoutburst in 2002 as well as 2003.

\begin{table}
\begin{center}
\caption[Mid-eclipse timings of OU~Vir.]{Mid-eclipse timings of
  OU~Vir, accurate to $\pm4\times10^{-5}$~days.}
\vspace{0.3cm}
\small
\begin{tabular}{ccccc}
\hline
Date & \multicolumn{4}{c}{HJD}\\
 & {\em u\/}$^{\prime}$ & {\em g\/}$^{\prime}$ & {\em r\/}$^{\prime}$ &
 {\em i\/}$^{\prime}$\\
\hline
2002 05 16 & 2452411.523977 & 2452411.523921 & 2452411.523892 & --\\
2003 05 20 & 2452780.580157 & 2452780.580278 & -- & 2452780.580217\\
2003 05 22 & 2452782.470534 & 2452782.470558 & -- & 2452782.470509\\
2003 05 25 & 2452785.524083 & 2452785.524083 & -- & 2452785.524255\\
\hline
\end{tabular}
\normalsize
\label{tab:ouvir_eclipse_times}
\end{center}
\end{table}

\begin{sidewaystable}
\begin{center}
\caption[White dwarf contact phases and flux of OU~Vir.]{White dwarf
  contact phases, accurate to $\pm0.0006$, and flux, accurate to
  $\pm0.01$~mJy, of OU~Vir.}
\vspace{0.3cm}
\small
\begin{tabular}{ccccccccc}
\hline
Date & Band & $\phi_{\rm{w}1}$ & $\phi_{\rm{w}2}$ & $\phi_{\rm{w}3}$ &
$\phi_{\rm{w}4}$ & $\phi_{\rm{wi}}$ & $\phi_{\rm{we}}$ & Flux (mJy)\\
\hline
2002 05 16 & {\em u\/}$^{\prime}$ & $-0.024414$ & $-0.017578$ &
$0.018555$ & $0.025391$ & $-0.020508$ & $0.022461$ & --\\
      & {\em g\/}$^{\prime}$ & $-0.027344$ & $-0.018555$ &
$0.016602$ & $0.025391$ & $-0.022461$ & $0.021484$ & --\\
      & {\em r\/}$^{\prime}$ & $-0.028320$ & $-0.013672$ & $0.012695$
& $0.027344$ & $-0.020508$ & $0.020508$ & --\\
2003 05 20 & {\em u\/}$^{\prime}$ & $-0.023438$ & $-0.016602$ & $0.018555$
& $0.025491$ & $-0.019531$ & $0.022461$ & --\\
      & {\em g\/}$^{\prime}$ & $-0.025391$ & $-0.017578$ & $0.017578$
& $0.024414$ & $-0.021484$ & $0.021484$ & --\\
      & {\em i\/}$^{\prime}$ & $-0.023438$ & $-0.014648$ & $0.016602$
& $0.025391$ & $-0.018555$ & $0.021484$ & --\\
2003 05 22 & {\em u\/}$^{\prime}$ & $-0.024414$ & $-0.016602$ & $0.017578$
& $0.025391$ & $-0.020508$ & $0.021484$ & --\\
      & {\em g\/}$^{\prime}$ & $-0.026367$ & $-0.018555$ & $0.016602$
& $0.024414$ & $-0.022461$ & $0.020508$ & --\\
      & {\em i\/}$^{\prime}$ & $-0.021484$ & $-0.018555$ & $0.016602$
& $0.020508$ & $-0.019531$ & $0.018555$ & --\\
2003 05 25 & {\em u\/}$^{\prime}$ & $-0.025391$ & $-0.016602$ & $0.014648$
& $0.023438$ & $-0.020508$ & $0.019531$ & $0.0537$\\
      & {\em g\/}$^{\prime}$ & $-0.025391$ & $-0.019531$ & $0.016602$
& $0.022461$ & $-0.022461$ & $0.019531$ & $0.0519$\\
      & {\em i\/}$^{\prime}$ & $-0.022461$ & $-0.018555$ & $0.017578$ &
$0.022461$ & $-0.020508$ & $0.020508$ & $0.0146$\\
\hline
\end{tabular}
\normalsize
\label{tab:ouvir_wd_times}
\end{center}

\vspace{2.0cm}

\begin{center}
\caption[Bright spot contact phases of OU~Vir.]{Bright spot contact
  phases of OU~Vir, accurate to $\pm0.0006$. The weakness of the
  bright spot feature and the
  presence of flickering in the light curves meant that it was not
  possible to accurately determine the bright spot contact phases for
  all the light curves.}
\vspace{0.3cm}
\small
\begin{tabular}{cccccccc}
\hline
Date & Band & $\phi_{\rm{b}1}$ & $\phi_{\rm{b}2}$ & $\phi_{\rm{b}3}$ &
$\phi_{\rm{b}4}$ & $\phi_{\rm{bi}}$ & $\phi_{\rm{be}}$\\
\hline
2003 05 22 & {\em g\/}$^{\prime}$ & $-0.002930$ & $0.002930$ & -- & -- &
$0.000977$ & --\\
2003 05 25 & {\em u\/}$^{\prime}$ & $-0.007813$ & 0$.002930$ &
$0.060547$ & $0.065430$ & $-0.002930$ & $0.062500$\\
      & {\em g\/}$^{\prime}$ & $-0.007813$ & $0.002930$ &
$0.058594$ & $0.068359$ & $-0.002930$ & $0.062500$\\
\hline
\end{tabular}
\normalsize
\label{tab:ouvir_bs_times}
\end{center}

\end{sidewaystable}

\section{Eclipse contact phases}

\begin{figure}
\centerline{\includegraphics[width=14cm,angle=-90.]{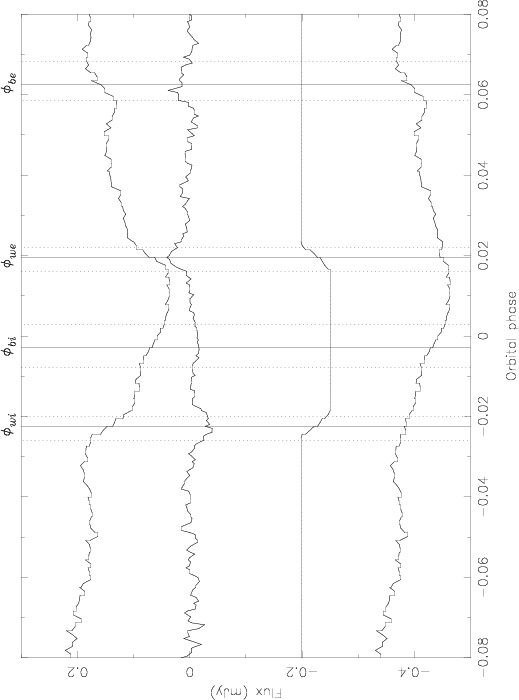}}
\caption[White dwarf deconvolution of the {\em g\/}$^{\prime}$ band
  light curve of OU~Vir of 2003 May 25.]{White dwarf deconvolution of
  the {\em g\/}$^{\prime}$ band light curve of OU~Vir of 2003 May
  25. Top to bottom: The data after smoothing by a median filter; the
  derivative after smoothing by a box car filter and subtraction of
  the spline fit to this, multiplied by a factor of 5 for clarity; the
  reconstructed white dwarf light curve, shifted downwards by 0.25
  mJy; the original light curve minus the white dwarf light curve
  after smoothing by a median filter, shifted downwards by 0.5
  mJy. The vertical lines show the contact phases of the white dwarf
  and bright spot eclipses (see \S~\ref{sec:derivative}), the dotted
  lines corresponding to $\phi_{\rm{w}1}\ldots\phi_{\rm{w}4}$,
  $\phi_{\rm{b}1}\ldots\phi_{\rm{b}4}$ and the solid lines (labelled)
  to $\phi_{\rm{wi}}$, $\phi_{\rm{we}}$ and $\phi_{\rm{bi}}$,
  $\phi_{\rm{be}}$. The bright spot ingress and egress are plainly
  visible, quickly following the white dwarf ingress and egress
  respectively.}
\label{fig:ouvir_deconvolution}
\end{figure}

The white dwarf eclipse contact phases (defined in
\S~\ref{sec:derivative}) given in tables~\ref{tab:ouvir_eclipse_times}
and \ref{tab:ouvir_wd_times} were determined using the techniques
described in \S~\ref{sec:derivative}. Once the white dwarf eclipse
contact phases had been found, the white dwarf light curve was
reconstructed (as discussed in \S~\ref{sec:deconvolution}) and
subtracted from the overall light curve, as illustrated in
figure~\ref{fig:ouvir_deconvolution}. The out-of-eclipse white dwarf
fluxes thus found are given in table~\ref{tab:ouvir_wd_times}. The
white dwarf flux can be used to determine its temperature and
distance; see \S~\ref{sec:fitting}. Once this has been done the bright
spot eclipse contact phases (given in table~\ref{tab:ouvir_bs_times})
can be determined by a similar method \citep{wood89a} and its light
curve removed from that of the disc eclipse. If successful, this
process can be used to determine the bright spot
temperature. Unfortunately I was unsuccessful in my attempts to do
this, probably because flickering hindered accurate determination of
the bright spot flux and contact phases.

\section{Orbital ephemeris}

A linear least-squares fit to the times of mid-eclipse given in
table~\ref{tab:ouvir_eclipse_times} (calculated using the techniques
described in \S~\ref{sec:derivative} and taking the midpoint of the
white dwarf eclipse as the point of mid-eclipse) and those of
\citet[][private communication]{vanmunster00} gives the following
ephemeris:
\begin{displaymath}
\begin{array}{ccrcrl}
\\ HJD & = & 2451725.03283 & + & 0.072706113 & E.\\ & & 7 & \pm & 5 &
\end{array} 
\end{displaymath} Errors of $\pm 4\times 10^{-5}$ days were used for
 the {\sc ultracam} data, and errors of $\pm 7\times 10^{-4}$ days for
 the \citet[][private communication]{vanmunster00} data. This
 ephemeris was used to phase all of the data.

I do not present an $O-C$ (observed times of mid-eclipse minus times
of mid-eclipse caluclated using an ephemeris) diagram as first, the
mid-eclipse times
reported by \citet[][private communication]{vanmunster00} were times
{\em predicted\/} by their ephemeris for the cycle numbers they
observed\footnote{This does not present a major problem with using
them to determine the ephemeris, however, because of the relative
weightings of the times.}, and second, the time resolution of their
data was too poor (between 30--160~seconds). Any $O-C$ diagram would
therefore be meaningless.

\section{System parameters}
\label{sec:ouvir_parameters}

\begin{figure}
\centerline{\includegraphics[width=12.0cm,angle=-90.]{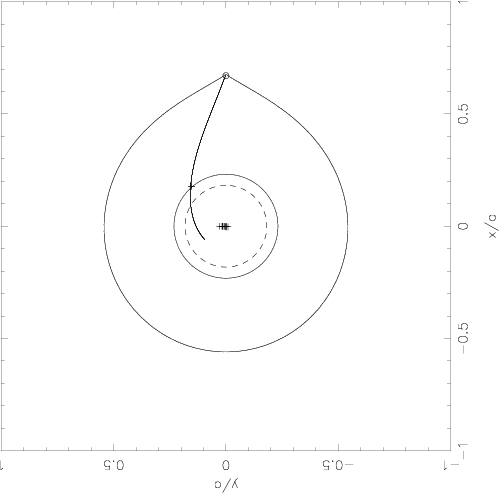}}
\caption[Trajectory of the gas stream from the secondary star for
  OU~Vir.]{Trajectory of the gas stream from the secondary star for
  OU~Vir, with $q=0.175$ and $i=79.2^{\circ}$. Top: The system with
  the primary Roche lobe, L$_{1}$ point and disc of radius
  $R_{\rm{d}}=0.2315a$ plotted. The positions of the white dwarf and
  bright spot light centres corresponding to the observed ingress and
  egress phases for $q$ and $i$ as above are also plotted. The
  circularisation radius \citep[][their equation 13]{verbunt88} of
  $R_{\rm{circ}}=0.1820a$ is shown as a dashed circle. The stream
  passes through the bright spot points (note that the timings of
  2003 May 22 are of the ingress only which prevents it from being
  plotted).}
\label{fig:ouvir_mass}
\end{figure}

\begin{figure}
\centerline{\includegraphics[width=10.0cm,angle=-90.]{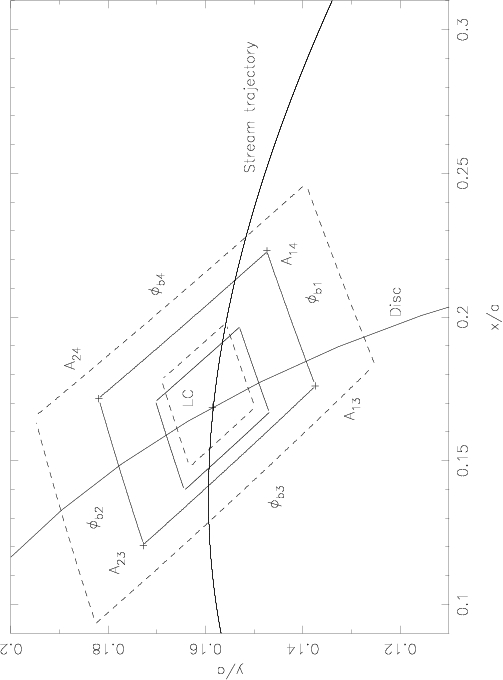}}
\caption[Horizontal structure of the bright spot of
  OU~Vir.]{Horizontal structure of the bright spot of OU~Vir for
  $q=0.175$, showing the region on the orbital plane within which the
  bright spot lies. The light centre {\em LC\/} is marked by a cross,
  surrounded by the inner solid box which corresponds to the rms
  variations in position. The phase arcs which correspond to the
  bright spot contact phases are shown as the outer solid box, with
  the rms variations in position shown as the two dashed boxes. As all
  the timings of $\phi_{\rm{b}2}$ and $\phi_{\rm{be}}$ are identical,
  the rms variations of $\phi_{\rm{b}1}$ and $\phi_{\rm{bi}}$,
  respectively, have been used instead. Intersections of the phase arcs
  $\phi_{{\rm b}j}$ and $\phi_{{\rm b}k}$ are marked $A_{jk}$, with
  crosses. The stream trajectory and disc of radius
  $R_{\rm{d}}=0.2315a$ are also plotted as solid curves.}
\label{fig:ouvir_bs_horizontal}
\end{figure}

\begin{figure}
\centerline{\includegraphics[width=10.0cm,angle=-90.]{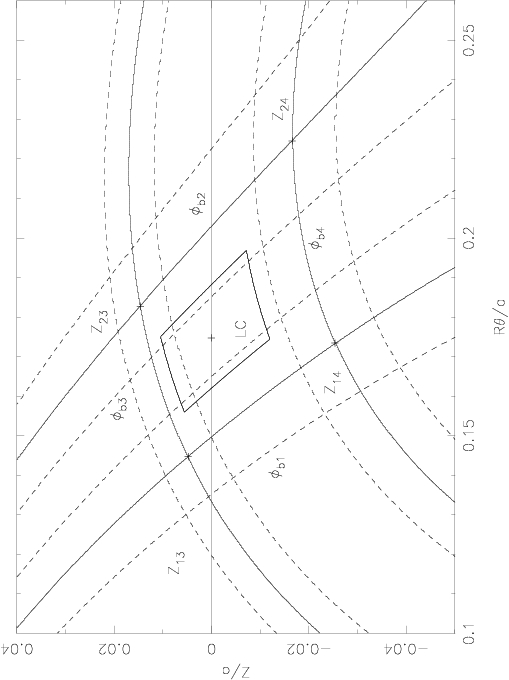}}
\caption[Vertical structure of the bright spot of OU~Vir.]{As
  figure~\ref{fig:ouvir_bs_horizontal}, but showing the vertical
  structure of the bright spot. The phase arcs are projected onto a
  vertical cylinder of radius $0.2315a$ (equal to that of the disc),
  i.e.\ the {\em x\/}-axis is stepping around the edge of the
  disc. $\theta$ is in radians. The intersections of the phase arcs
  $\phi_{{\rm b}j}$ and $\phi_{{\rm b}k}$ are marked $Z_{jk}$, with
  crosses.}
\label{fig:ouvir_bs_vertical}
\end{figure}

The derivation of the system parameters of OU~Vir proceeds as
discussed in \S~\ref{sec:derivative}. Figure~\ref{fig:ouvir_mass}
shows the theoretical gas stream trajectory for
$q=0.175$. As illustrated in figures~\ref{fig:ouvir_bs_horizontal} and
\ref{fig:ouvir_bs_vertical}, which show expanded views of the bright spot
region, I constrain the light centre of the
bright spot to be the point where the gas stream and outer edge of the
disc intersect. The 2003 bright spot timings thus yield a mass ratio of
$q=0.175 \pm 0.025$ and an inclination of $i=79.2^{\circ}\pm0.7^{\circ}$ for
an eclipse phase width $\Delta\phi = 0.0416\pm0.0008$. The errors are
determined by the rms variations in the measured contact
phases.

Figures~\ref{fig:ouvir_bs_horizontal} and \ref{fig:ouvir_bs_vertical}
show the eclipse constraints on the structure of the bright spot. I
use these to determine upper limits on the angular size and the radial
and vertical extent of the bright spot, with $\Delta\theta$, $\Delta
R_{\rm{d}}$, $\Delta Z$ and $\Delta Z_{2}$ defined as in
equations~\ref{eq:theta}--\ref{eq:z2}. The mean position and extent of
the bright spot are given in table~\ref{tab:ouvir_bs}. From
figures~\ref{fig:ouvir_bs_horizontal} and \ref{fig:ouvir_bs_vertical}
I estimate visually that the gas stream passing through the light
centre of the
bright spot could just touch the phase arcs corresponding to
$\phi_{\rm{b}1}$ and $\phi_{\rm{b}4}$ for a stream of circular
cross-section with a radius $\varepsilon /a = 0.0175 \pm 0.0025$. The
bright spot appears to be more extended azimuthally than radially,
which can be understood by the shock front extending both up-disc and
down-disc from the point of impact. The size of the stream is similar
to that expected from theoretical studies \citep{lubow75, lubow76} and
that obtained by studies of similar objects \citep{wood86b,wood89a},
so the assumption that the stream passes through the light centre of
the bright spot is reasonable.

\begin{figure}
\centerline{\includegraphics[width=12.0cm,angle=-90.]{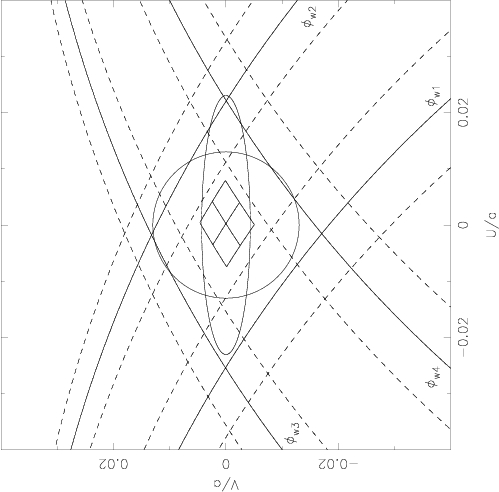}}
\caption[Projection of the white dwarf phase arcs of OU~Vir onto the
  plane perpendicular to the line of sight.]{Projection of the white
  dwarf phase arcs of OU~Vir onto the plane perpendicular to the line
  of sight. $U$ and $V$ are orthogonal co-ordinates perpendicular to
  the line of sight, $U$ being parallel to the binary plane. Solid
  curves correspond to the contact phases of the white dwarf, dotted
  curves to the rms variations of the phase arcs. The light centre is
  also plotted surrounded by the solid box corresponding to the rms
  variations in phase. The projection of the white dwarf and accretion
  belt centred on $U,V=0$ are shown for $R_{1}=0.013a$ and
  $R_{\rm{belt}}=0.023a$.}
\label{fig:ouvir_whitedwarf}
\end{figure}

Figure~\ref{fig:ouvir_whitedwarf} shows the eclipse constraints on the
radius of the white dwarf. Using the mass ratio and orbital
inclination derived earlier, I find that the white dwarf has a radius
of $R_{1}=0.013\pm0.004a$. An alternative possibility is that the
sharp eclipse is caused by a bright inner disc region or boundary
layer of radius $R_{\rm{belt}}=0.023\pm0.010a$ surrounding the white
dwarf like a belt. These errors are calculated using the rms
variations in the measured contact phases. Another possibility is that
the lower hemisphere of the white dwarf is obscured by an optically
thick accretion disc, which would result in the white dwarf radius
being $R_{1}\geq0.013a$. This can be seen from inspection of
figure~\ref{fig:ouvir_whitedwarf} and considering that the phase arcs
$\phi_{\rm w1}$ and $\phi_{\rm w4}$ in this case correspond to the
lowest unobscured sections of the white dwarf. As discussed in
\S~\ref{sec:derivative}, for a symmetrical light distribution centred
on the origin, as we would expect of a white dwarf, then the contact phases
$\phi_{\rm{wi}}$ and $\phi_{\rm{we}}$ lie half-way through the white
dwarf ingress and egress. Figure~\ref{fig:ouvir_whitedwarf}
illustrates that this is indeed the case, and so it is probable that
the lower half of the white dwarf remains unobscured.

The determination of the absolute system parameters assumes that the
eclipse is solely of a white dwarf. If the eclipse is actually of a belt
and the white dwarf itself is not visible, then the white dwarf radius
must be smaller than $R_{\rm{belt}}=0.023a$. If the white dwarf does
contribute significantly to the eclipsed light, then we have the
additional constraint that its radius must be $R_{1}\leq0.013a$, so
that the white dwarf mass given in table~\ref{tab:ouvir_parameters} is
actually a lower limit. The only way to verify the assumption that the
central light source is the white dwarf alone is to measure the
semi-amplitude of the radial velocity curve of the secondary star,
$K_{\rm{2}}$, and compare it to that predicted by the photometric
model in table~\ref{tab:ouvir_parameters}. One could also check if
this assumption is true using a longer baseline of quiescent
observations, as one might expect eclipse timings of an accretion belt
to be much more variable than those of a white dwarf. I note, however,
that the white dwarf mass given in table~\ref{tab:ouvir_parameters} is
consistent with the mean white dwarf mass of $0.69\pm0.13\;{\rm
M}_{\odot}$ for CVs below the period gap \citep{smith98a}. Also, as
discussed in \S~\ref{sec:derivative}, short-period dwarf nov\ae\ like
OU Vir tend to accrete directly onto the white dwarf, whereas
longer-period dwarf nov\ae\ usually have boundary layers. The system
parameters of OU~Vir, assuming that the central eclipsed object is
indeed a white dwarf, are given in
table~\ref{tab:ouvir_parameters}.

The superhump period of OU~Vir is $P_{\rm sh}=0.078\pm0.002$~days
\citep{vanmunster00}, which means that OU~Vir lies $5\sigma$ off the
superhump period excess--mass ratio relation of \citet[][his equation
8]{patterson98}, with the superhump period excess $\epsilon=(P_{\rm
sh}-P_{\rm orb})/P_{\rm orb}\sim0.073$. However, it does not lie on
the superhump period excess-orbital period relation either, perhaps
indicating that the current estimate of the superhump period $P_{\rm
sh}$ is inaccurate.

\begin{table}
\begin{center}
\caption[Mean position and extent of the bright spot of OU~Vir.]{Mean
  position and extent of the bright spot of OU~Vir as defined by
  equations~\ref{eq:theta}--\ref{eq:z}. $\Delta Z_{2}$ is calculated
  according to the definition used by \citet{wood86b}, for ease of
  comparison.} 
\vspace{0.3cm}
\small
\begin{tabular}{cc}
\hline
$\Delta R_{\rm{d}}/a$ & 0.0417\\
$\Delta \theta$ & $15.17^{\circ}$\\
$\Delta Z/a$ & 0.0200\\
$\Delta Z_{2}/a$ & 0.0147\\
$R_{\rm{d}}/a$ & 0.2315\\
$\theta$ & $43.24^{\circ}$\\
\hline
\end{tabular}
\normalsize
\label{tab:ouvir_bs}
\end{center}
\end{table}

\begin{table}
\begin{center}
\caption[System parameters of OU~Vir.]{System parameters of
  OU~Vir. The secondary radius given is the volume radius of the
  secondary's Roche lobe (Eggleton 1983), as defined by
  equation~\ref{eq:RL2}. Parameters left blank in the right-hand
  column are independent of the model used. The radial velocities
  quoted are estimated from the derived system parameters.} 
\vspace{0.3cm}
\small
\begin{tabular}{lcc}
\hline
Parameter & Nauenberg (cold) & Nauenberg ($21\,700$~K)\\
\hline
Inclination $i$ & $79.2^{\circ} \pm 0.7^{\circ}$ & \\
Mass ratio $q=M_{2}/M_{1}$ & $0.175 \pm 0.025$ & \\
White dwarf mass $M_{1}/{\rm M}_{\odot}$ & $0.85 \pm 0.20$ & $0.90 \pm 0.19$\\
Secondary mass $M_{2}/{\rm M}_{\odot}$ & $0.15 \pm 0.04$ & $0.16 \pm 0.04$\\
White dwarf radius $R_{1}/R_{\odot}$ & $0.0095 \pm 0.0030$ & $0.0097 \pm 0.0031$\\
Secondary radius $R_{2}/{\rm R}_{\odot}$ & $0.177 \pm 0.024$ & $0.181 \pm 0.024$\\
Separation $a/{\rm R}_{\odot}$ & $0.73 \pm 0.06$ & $0.75 \pm 0.05$\\
White dwarf radial velocity $K_{1}/\rm{km\,s^{-1}}$ & $75 \pm 12$ & $76 \pm 12$\\
Secondary radial velocity $K_{2}/\rm{km\,s^{-1}}$ & $426 \pm 6$ & $434 \pm 6$\\
Outer disc radius $R_{\rm{d}}/a$ & $0.2315 \pm 0.0150$ & \\
Distance $D/\rm{pc}$ & $650 \pm 210$ & \\
White dwarf temperature $T_{1}/\rm{K}$ & $21\,700 \pm 1200$ & \\
\hline
\end{tabular}
\normalsize
\label{tab:ouvir_parameters}
\end{center}
\end{table}

\section{Eclipse mapping}
\label{sec:ouvir_eclipse_mapping}

\begin{figure}
\begin{tabular}{cc}
\begin{minipage}{2.5in}
\includegraphics[width=6.0cm,angle=0.]{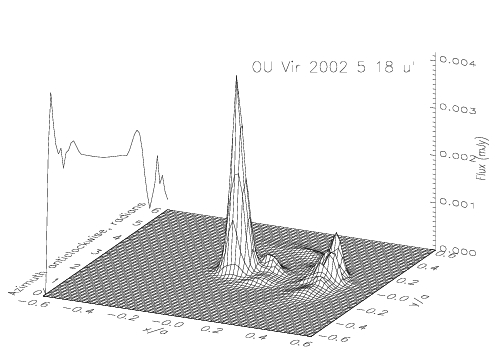} \end{minipage}&
\begin{minipage}{2.5in}
\includegraphics[width=4.0cm,angle=-90.]{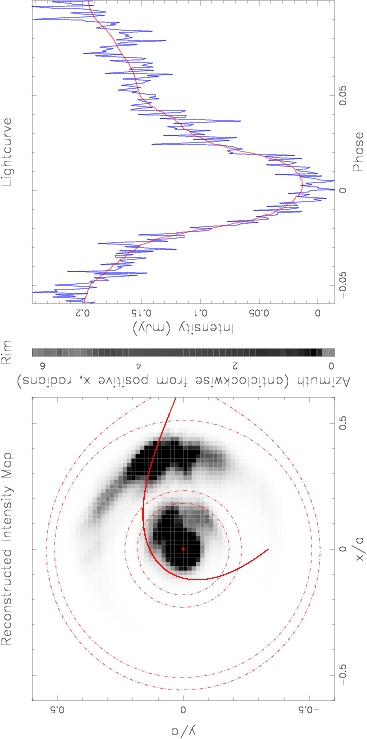} \end{minipage}\\
\begin{minipage}{2.5in}
\includegraphics[width=6.0cm,angle=0.]{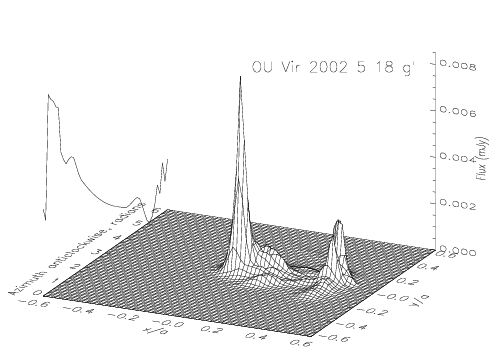} \end{minipage}&
\begin{minipage}{2.5in}
\includegraphics[width=4.0cm,angle=-90.]{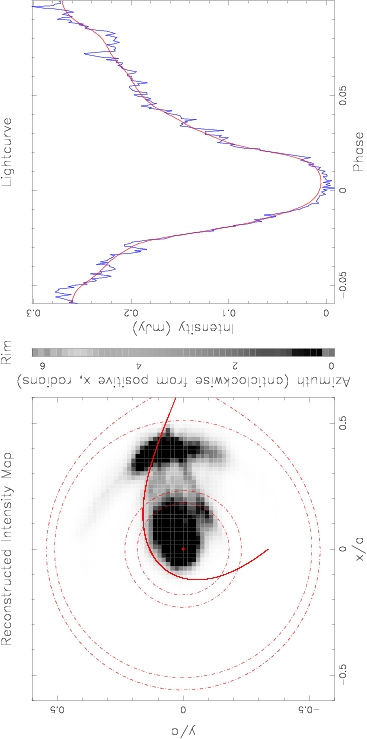} \end{minipage}\\
\begin{minipage}{2.5in}
\includegraphics[width=6.0cm,angle=0.]{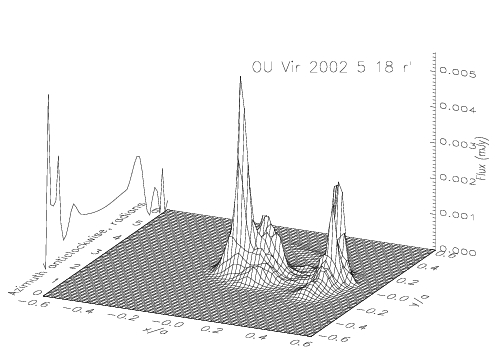} \end{minipage}&
\begin{minipage}{2.5in}
\includegraphics[width=4.0cm,angle=-90.]{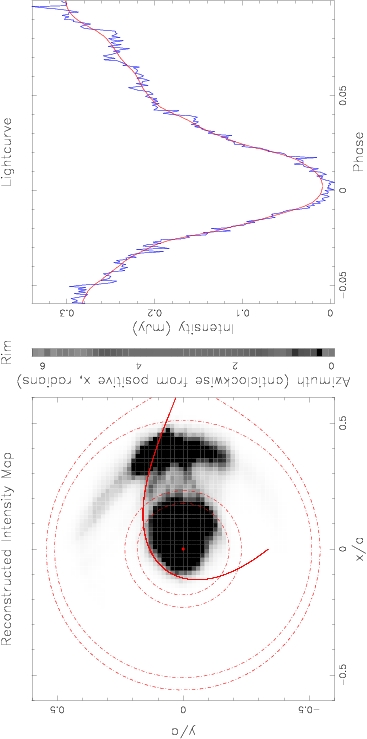} \end{minipage}\\
\end{tabular}
\caption[Eclipse maps for the {\em u}$^{\prime}$, {\em g\/}$^{\prime}$
  and {\em i\/}$^{\prime}$ 2002 May 18 light curves of OU~Vir.]{Top
  row. Left: a three-dimensional representation of the reconstructed
  accretion disc for the  2002 May 18 {\em u\/}$^{\prime}$ data of
  OU~Vir. The white dwarf is at the origin, and the red dwarf at
  $(x,y)=(a,0)$. The rim intensity is shown on the edge of the
  grid. Centre: a two-dimensional view of the reconstructed accretion
  disc as before. The dot-dashed red lines are, from the centre out,
  the circularisation radius \citep[][their equation 13]{verbunt88},
  the disc radius (the same radius as the rim), the tidal radius
  \citep{paczynski77} and the primary star's Roche lobe. The solid red
  line is the trajectory of the gas stream. The rim intensity is
  shown to the right. The scale is linear from zero to five per cent
  of the maximum, with areas with intensities greater than 5 per cent
  of the maximum shown in black. Dark regions are brighter. Right: The
  light curve (blue) with the fit (red)
  corresponding to the intensity distribution shown on the left
  overlaid. The middle and bottom rows are as the top, but for the
  {\em g\/}$^{\prime}$ and {\em i\/}$^{\prime}$ data,
  respectively. The system parameters adopted for the reconstruction
  are $q=0.175$, $i=79.2^{\circ}$ and $R_{\rm d}=0.2315a$ (the
  parameters given in table~\ref{tab:ouvir_parameters}).}
\label{fig:em_ouvir_18}
\end{figure}

\begin{figure}
\centerline{\includegraphics[width=14cm,angle=0]{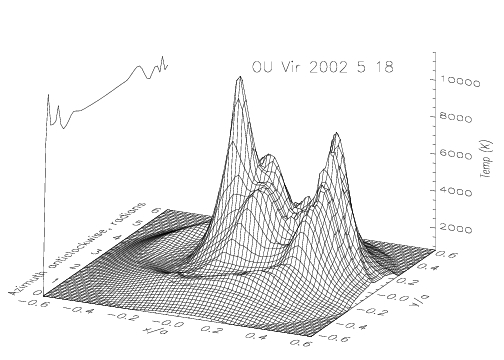}}
\caption[3-dimensional plot of a blackbody fit to the reconstructed
  2002 May 18 disc intensities of OU~Vir.]{3-dimensional plot of a
  blackbody fit to the reconstructed 2002 May 18 disc intensities of
  OU~Vir shown in figure~\ref{fig:em_ouvir_18}.}
\label{fig:em_ouvir_temp_18}
\end{figure}

\begin{figure}
\begin{tabular}{cc}
\begin{minipage}{2.5in}
\includegraphics[width=6.0cm,angle=0.]{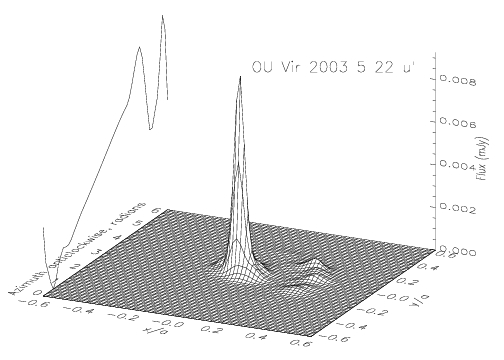} \end{minipage}&
\begin{minipage}{2.5in}
\includegraphics[width=4.0cm,angle=-90.]{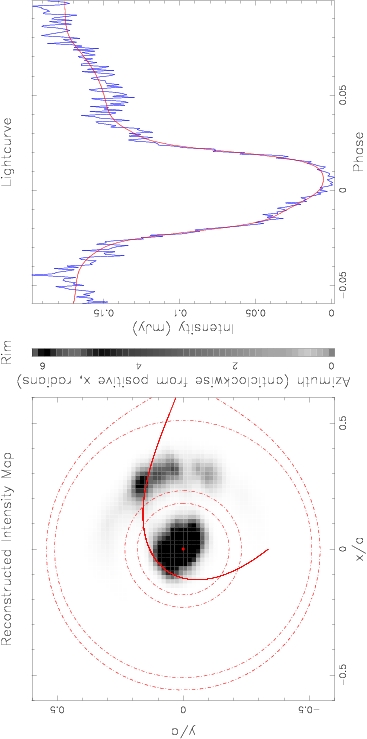} \end{minipage}\\
\begin{minipage}{2.5in}
\includegraphics[width=6.0cm,angle=0.]{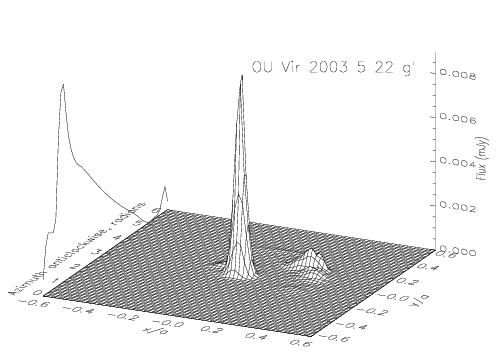} \end{minipage}&
\begin{minipage}{2.5in}
\includegraphics[width=4.0cm,angle=-90.]{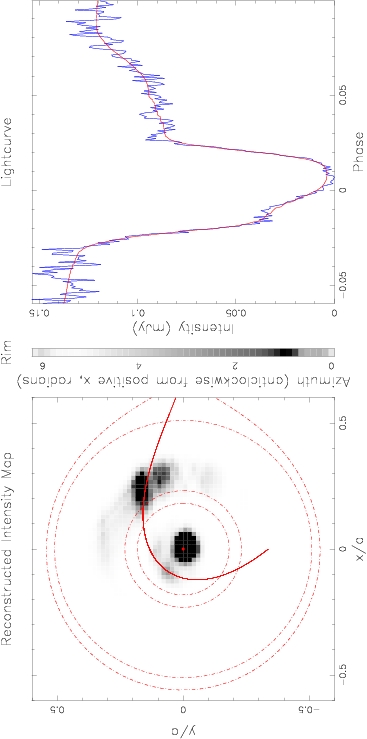} \end{minipage}\\
\begin{minipage}{2.5in}
\includegraphics[width=6.0cm,angle=0.]{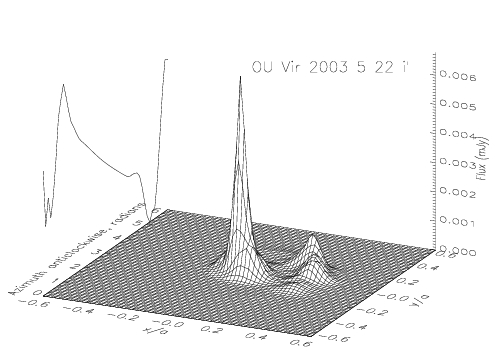} \end{minipage}&
\begin{minipage}{2.5in}
\includegraphics[width=4.0cm,angle=-90.]{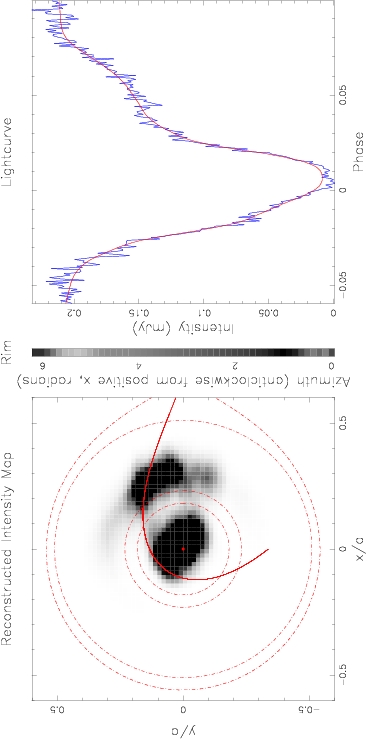} \end{minipage}\\
\end{tabular}
\caption[Eclipse maps for the {\em u}$^{\prime}$, {\em g\/}$^{\prime}$
  and {\em i\/}$^{\prime}$ 2003 May 22 light curves of OU~Vir.]{As
  figure~\ref{fig:em_ouvir_18}, but for, from top, the {\em
  u}$^{\prime}$, {\em g\/}$^{\prime}$ and {\em i\/}$^{\prime}$ 2003 May
  22 data of OU~Vir. The system parameters adopted for the
  reconstruction are $q=0.175$, $i=79.2^{\circ}$ and $R_{\rm
  d}=0.2315a$ (the parameters given in
  table~\ref{tab:ouvir_parameters}).}
\label{fig:em_ouvir_22}
\end{figure}

\begin{figure}
\centerline{\includegraphics[width=14cm,angle=0]{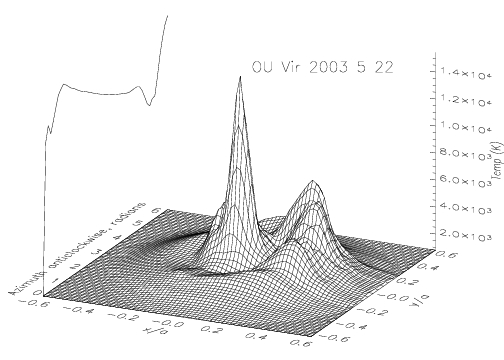}}
\caption[3-dimensional plot of a blackbody fit to the reconstructed
  2003 May 22 disc intensities of OU~Vir.]{3-dimensional plot of a
  blackbody fit to the reconstructed 2003 May 22 disc intensities of
  OU~Vir shown in figure~\ref{fig:em_ouvir_22}.}
\label{fig:em_ouvir_temp_22}
\end{figure}

The eclipse maps of OU~Vir presented in
figures~\ref{fig:em_ouvir_18}--\ref{fig:em_ouvir_temp_22} were
produced using the parameters given in
table~\ref{tab:ouvir_parameters} and derived in
\S~\ref{sec:ouvir_parameters}. The data of 2002~May~18 and 2003~May~22
were chosen as the eclipses are complete and relatively uncontaminated
by flickering. The data of different nights could not have been
phase-folded since there were significant differences between them. As
expected for such short-period systems, none of the reconstructed maps
show evidence for disc emission. The reconstructed maps of 2002~May~18
show a white dwarf at the centre of the disc and a bright spot, which
appears to be at a larger radius than that derived in the previous
section (which used the 2003 data). I speculate that this is due to
an enlarged disc radius because of the system being in decline from
outburst. The feature near the central white dwarf is probably
spurious, due to the intersection of two phase arcs at that point
which intersect the white dwarf and bright spot. The reconstructed
maps of 2003~May~22 also show the central white dwarf and the bright
spot. The bright spot position (assuming that it is located at the
intersection of the disc and the gas stream) for these data
corresponds to a disc radius smaller than that of 2002 May 18, but still
slightly greater than derived in \S~\ref{sec:ouvir_parameters}. This
is possibly an effect of the entropy smearing out the intensity
distribution during the reconstruction process, but is more likely
indicative of the uncertainty in estimating the eclipse contact phases
in \S~\ref{sec:ouvir_parameters}, and possibly a change in the radius
of the accretion disc over the course of the 2003 observations.

\newpage

\pagebreak

\newpage

%% file: xzeridvuma.tex
\chapter{XZ~Eri and DV~UMa}
\label{ch:xzeridvuma}

The contents of this chapter have been published in the Monthly
Notices of the Royal Astronomical Society, {\bf 355}, 1 as {\em
ULTRACAM photometry of the eclipsing cataclysmic variables XZ~Eri and
DV~UMa\/} by \citet*{feline04a}. The exceptions to this are the eclipse
mapping results presented at the end of this chapter, in
\S~\ref{sec:gycncircomhtcas_temp_maps}. The system parameters were
estimated using both the derivative and {\sc lfit} techniques, as the
light curves of both objects were largely free of flickering and were
excellent candidates for model fitting. The reduction
and analysis of the data are all my own, as is the text below. Dr.~Vik
Dhillon supervised all work presented here.

XZ~Eri was first noted to be variable by \citet{shapley34}. Until
recently \citep{howell91,szkody92}, however, XZ~Eri had been rather
poorly studied. The presence of eclipses in the light curve of XZ~Eri
was discovered by \citet{woudt01}. More recently, \citet{uemura04}
observed superhumps in the outburst light curve of XZ~Eri, confirming
its classification as an SU~UMa star.

Previous observations of DV~UMa are summarised by \citet{nogami01},
who also present light curves obtained during the 1995 outburst and
the 1997 superoutburst. \citet{patterson00} present superoutburst and
quiescent photometry from which they derive the system
parameters. \citet{mukai90} estimated the spectral type of the
secondary star to be $\sim$M4.5 from spectroscopic observations.

In this chapter I present simultaneous three-colour, high-speed
photometry of XZ~Eri and DV~UMa. I derive the system parameters via
two separate methods---timings of the eclipse contact phases and
fitting a parameterized model of the eclipse---and discuss the
relative merits of each.

\begin{figure}
\centerline{\includegraphics[height=7.3cm,angle=-90.]{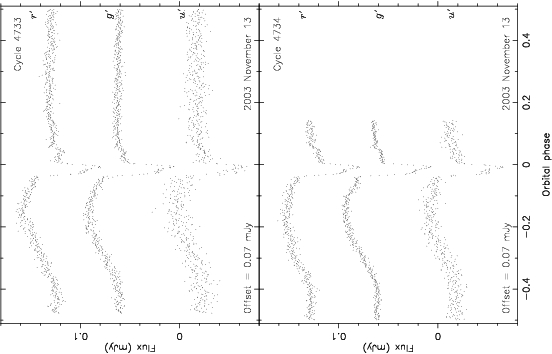}}
\caption[The light curve of XZ~Eri.]{The light curve of XZ~Eri. The
  data are contiguous. The {\em r\/}$^{\prime}$ data are offset
  vertically upwards and the {\em u\/}$^{\prime}$ data are offset
  vertically downwards.}
\label{fig:xzeri_lightcurve}
\end{figure}

\begin{figure}
\centerline{\includegraphics[height=7.3cm,angle=-90.]{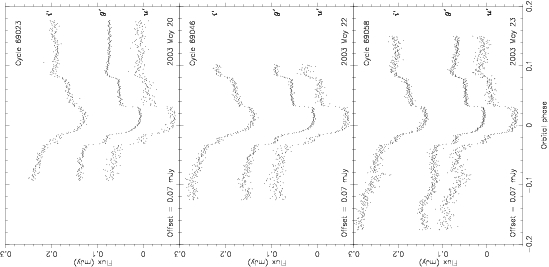}}
\caption[The light curve of DV~UMa.]{The light curve of DV~UMa. The
  {\em  i\/}$^{\prime}$ data are offset vertically upwards and the {\em
  u\/}$^{\prime}$ data are offset vertically downwards.}
\label{fig:dvuma_lightcurve}
\end{figure}

The observations of XZ~Eri and DV~UMa are summarised in
table~\ref{tab:journal}, and the data reduction procedure is detailed
in \S~\ref{sec:reduction}. The light curves of XZ~Eri and DV~UMa are
shown in figures~\ref{fig:xzeri_lightcurve} and
\ref{fig:dvuma_lightcurve}, respectively. The observations of XZ~Eri
began at high airmass (1.8)---this is evident in the improved quality
of the second cycle, which was observed at lower airmass. Note also
that the XZ~Eri data of 2003 November
13 have significantly worse time-resolution than those of DV~UMa,
despite both objects being of similar magnitude. This is due to the
higher brightness of the sky on 2003 November 13.

\section{Light curve morphology}
\label{sec:xzeridvuma_morphology}

The light curve of XZ~Eri shown in figure~\ref{fig:xzeri_lightcurve} is a
classic example of an eclipsing dwarf nova. Between phase $-0.4$ and
the start of eclipse, the orbital hump is clearly visible, with a
brightening in {\em g\/}$^{\prime}$ flux of 0.025 mJy (0.5 mag). The
light curve clearly shows separate eclipses of the white dwarf and
bright spot (see figure~\ref{fig:contactphases}) in all three colour
bands.

During my observations XZ~Eri had {\em g}$^{\prime}\sim19.5$~mag, falling
to {\em g}$^{\prime}\sim21.5$~mag in mid-eclipse. Comparing this to the
previous (quiescent) observations of \citet{woudt01}, who observed the
system at $\rm{V}\sim19.2$~mag, confirms that XZ~Eri was in quiescence at the
time of our observations.

The light curve of DV~UMa is presented in
figure~\ref{fig:dvuma_lightcurve}. Although the phase coverage is less
complete than for XZ~Eri, the eclipse morphology is again typical of
eclipsing short-period dwarf nov\ae. The white dwarf and bright spot
ingress and egress are both clear and distinct. The orbital hump in
DV~UMa is much less pronounced than in XZ~Eri.

\citet{howell88} quoted $\rm{V}\sim19.2$~mag in quiescence for DV~UMa. This
compares to {\em g}$^{\prime}\sim19$~mag at the time of our observations,
which fell to {\em g}$^{\prime}\sim22$~mag during eclipse. DV~UMa was
therefore in quiescence over the course of our observations.

\section{Orbital ephemerides}
\label{sec:xzeridvuma_ephemeris}

The times of mid-eclipse $T_{\rm{mid}}$
given in table~\ref{sec:xzeridvuma_eclipse_times} were determined 
as described in \S~\ref{sec:derivative}, taking the midpoint of the
white dwarf eclipse as the the point of mid-eclipse.

To determine the orbital ephemeris of XZ~Eri I used the one
mid-eclipse time of \citet{woudt01}, the 25 eclipse timings of
\citet{uemura04} and the six times of mid-eclipse given in
table~\ref{sec:xzeridvuma_eclipse_times}. I used errors of
$\pm5\times10^{-4}$~days for the \citet{woudt01} data,
$\pm1\times10^{-4}$~days 
for the \citet{uemura04} timings and $\pm4\times10^{-5}$~days for the
ULTRACAM data. A linear least-squares fit to these times gives:
\begin{displaymath}
\begin{array}{ccrcrl}
\\ HJD & = & 2452668.04099 & + & 0.061159491 & E.  \\
 & & 2 & \pm & 5 &
\end{array} 
\end{displaymath}

\begin{table}
\begin{center}
\caption[Mid-eclipse timings of XZ~Eri and
  DV~UMa.]{Mid-eclipse timings $({\rm HJD}+2\,452\,000)$ of XZ~Eri and
  DV~UMa, accurate to $\pm4\times10^{-5}$~days.}
\vspace{0.3cm}
\small
\begin{tabular}{cccc}
\hline
XZ~Eri cycle & {\em u\/}$^{\prime}$ & {\em g\/}$^{\prime}$ & {\em r\/}$^{\prime}$\\
\hline
4733 & 957.508910 & 957.508789 & 957.508870\\
4734 & 957.570081 & 957.570000 & 957.570000\\
\hline
DV~UMa cycle & {\em u\/}$^{\prime}$ & {\em g\/}$^{\prime}$ & {\em i\/}$^{\prime}$\\
\hline
69023 & 780.469225 & 780.469225 & 780.469225\\
69046 & 782.443801 & 782.443829 & 782.443801\\
69058 & 783.474062 & 783.474040 & 783.474040\\
\hline
\end{tabular}
\normalsize
\label{sec:xzeridvuma_eclipse_times}
\end{center}
\end{table}

The orbital ephemeris of DV~UMa was determined in a similar way using
the 18 mid-eclipse timings of \citet{nogami01}, the 12 timings of
\citet{howell88}, the 12 timings of \citet{patterson00} and the nine
times of mid-eclipse given in table
\ref{sec:xzeridvuma_eclipse_times}, with errors of $\pm5\times10^{-4}$~days
assigned to the data of \citet{nogami01} and \citet{howell88},
$\pm1\times10^{-4}$~days to the data of \citet{patterson00} and
$\pm4\times10^{-5}$~days to the ULTRACAM data. A linear least-squares fit to
these times gives:
\begin{displaymath}
\begin{array}{ccrcrl}
\\ HJD & = & 2446854.66157 & + & 0.0858526521 & E.
\\ & & 9 & \pm & 14 &
\end{array} 
\end{displaymath}

These ephemerides were used to phase all of the data.

\begin{figure}
\centerline{\includegraphics[height=12cm,angle=-90.]{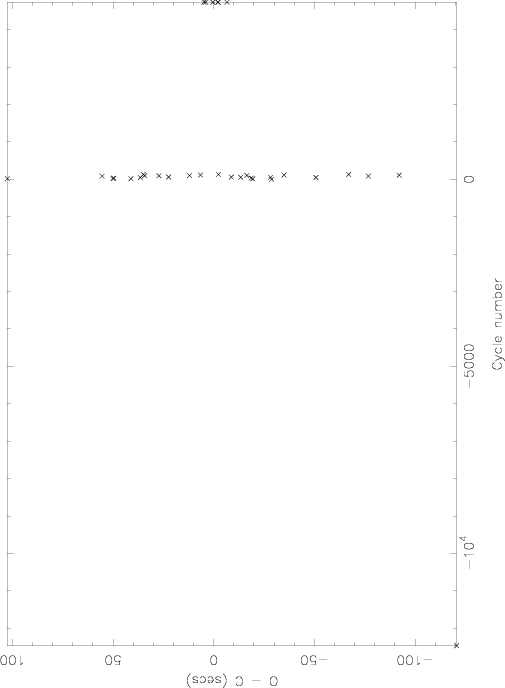}}
\caption[The $O-C$ diagram of XZ~Eri produced using a linear
  ephemeris.]{The $O-C$ diagram of XZ~Eri produced using a linear
  ephemeris as described in the text.}
\label{fig:xzeri_oc}
\end{figure}

\begin{figure}
\centerline{\includegraphics[height=12cm,angle=-90.]{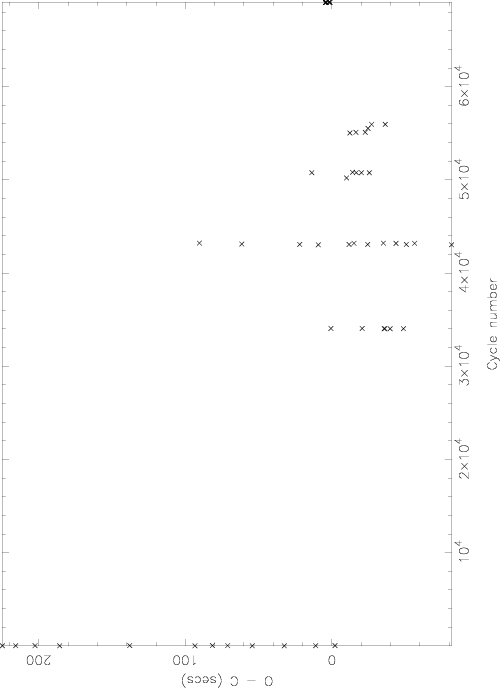}}
\caption[The $O-C$ diagram of DV~UMa produced using a linear
  ephemeris.]{The $O-C$ diagram of DV~UMa produced using a linear
  ephemeris as described in the text.}
\label{fig:dvuma_oc}
\end{figure}

\begin{figure}
\centerline{\includegraphics[height=12cm,angle=-90.]{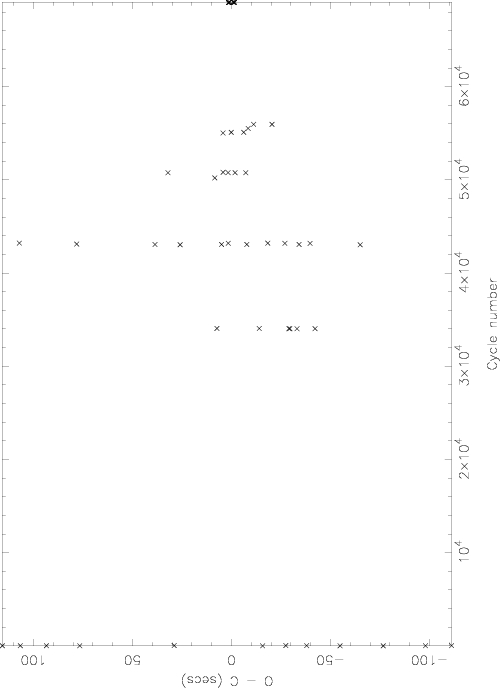}}
\caption[The $O-C$ diagram of DV~UMa produced using a quadratic
  ephemeris.]{The $O-C$ diagram of DV~UMa produced using a quadratic
  ephemeris as described in the text.}
\label{fig:dvuma_quad_oc}
\end{figure}

The $O-C$ diagrams for XZ~Eri and DV~UMa produced using the above
ephemerides and times of mid-eclipse are shown in
figures~\ref{fig:xzeri_oc} and \ref{fig:dvuma_oc},
respectively. Given the distribution of the data points for XZ~Eri, it
is not possible to determine if the system is undergoing period change
or not. DV~UMa, however, does show some evidence for period change,
with a quadratic least-squares fit to the above times giving
\begin{displaymath}
\begin{array}{ccrcrlcll}
\\ HJD & = & 2446854.66283 & + & 0.085852592 & E & + &
6.2\times 10^{-13} & E^{2}. \\
 & & 14 & \pm & 5 & & \pm & 0.5 &
\end{array} 
\end{displaymath}
The $O-C$ diagram produced using the quadratic ephemeris for DV~UMa is
shown in figure~\ref{fig:dvuma_quad_oc}. If accurate, the quadratic
ephemeris implies a time scale for the period change of DV~UMa of
\begin{displaymath}
P/\dot{P} = 3.25\times10^{7}\:{\rm yr}.
\end{displaymath}

I believe that the period change of DV~UMa is unlikely to be real in
the sense that it represents the long-term evolutionary mean,
and have therefore opted to use the linear ephemeris for DV~UMa
throughout this thesis. The reasons for this are two-fold. First, the
eclipse timings of CVs often exhibit considerable `wander'
\citep[e.g.\ IP~Peg,][]{wood89b}, the cause of which is
unclear\footnote{This could be due to the presence of a third body in
the system, a circumbinary disc or solar-type activity cycles of the
secondary star.}. Second, the time resolution of the observations of
\citet{howell88}, which are the critical evidence for period change in
the system, is comparable to the $O-C$ residuals for their
observations shown in figure~\ref{fig:dvuma_oc}.

\begin{figure}
\begin{tabular}{cc}
\includegraphics[height=6.0cm,angle=-90.]{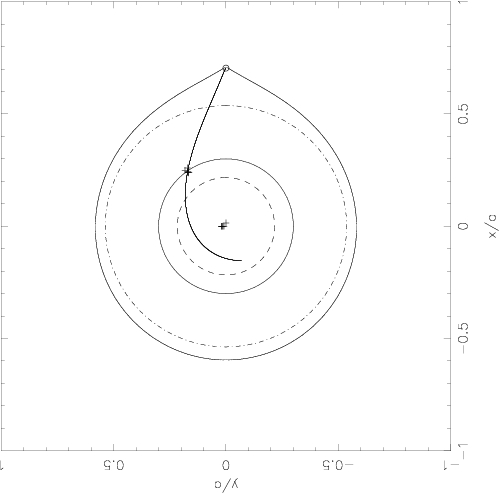} &
\includegraphics[height=6.0cm,angle=-90.]{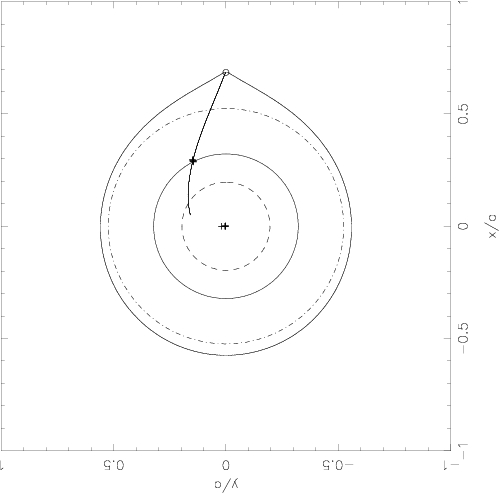}
\\
\includegraphics[height=7.2cm,angle=-90.]{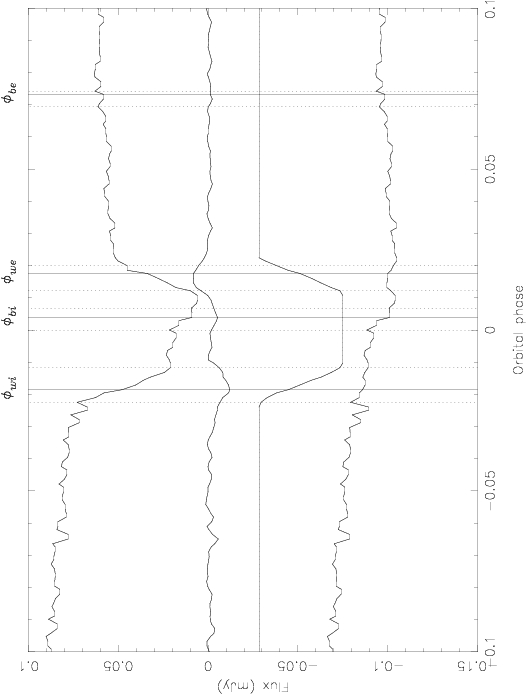} &
\includegraphics[height=7.2cm,angle=-90.]{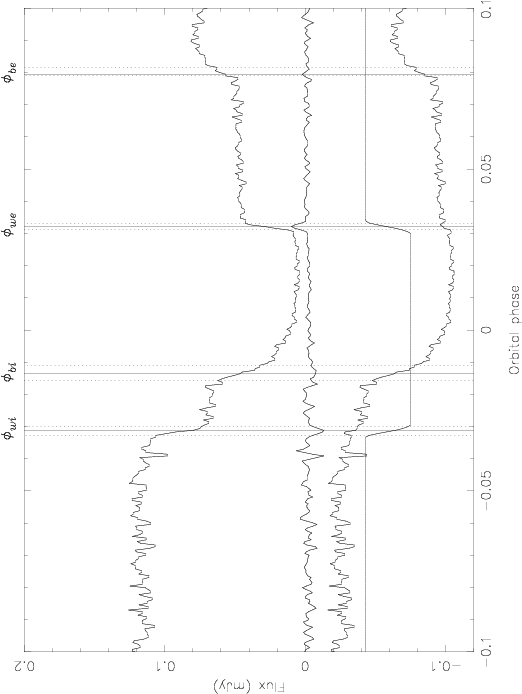}
\end{tabular}
\caption[Trajectory of the gas stream from the secondary star and the
  white dwarf light curve deconvolution for XZ~Eri and DV~UMa,.]{Top
  row: Trajectory of the gas stream from the secondary star for (left)
  XZ~Eri ($q=0.117$, $i=80.3^{\circ}$, $R_{\rm{d}}/a=0.300$ and
  $R_{\rm{circ}}/a=0.217$) and (right) for DV~UMa ($q=0.148$,
  $i=84.4^{\circ}$, $R_{\rm{d}}/a=0.322$ and
  $R_{\rm{circ}}/a=0.196$). The Roche lobe of the primary, the
  position of the inner Lagrangian point $\rm{L}_{1}$ and the disc of
  radius $R_{\rm{d}}$ are all plotted. The positions of the white
  dwarf and bright spot light centres corresponding to the observed
  ingress and egress phases are also plotted. The circularisation
  radius $R_{\rm{circ}}$ \citep[][their equation 13]{verbunt88} is
  shown as a dashed circle, and the tidal radius \citep{paczynski77}
  as a dot-dashed circle. Bottom row: White dwarf light curve
  deconvolution of (left) the {\em g\/}$^{\prime}$ band light curve of
  XZ~Eri on 2003 November 13 (cycle 4733) and (right) the {\em
  g\/}$^{\prime}$ band light curve of DV~UMa on 2003 May 23. Top to
  bottom: The data after smoothing by a median filter; the derivative
  after smoothing by a box-car filter and subtraction of a spline fit
  to this, multiplied by a factor of 1.5 for clarity; the
  reconstructed white dwarf light curve, shifted downwards by
  0.075~mJy; the original light curve minus the white dwarf light
  curve after smoothing by a median filter, shifted downwards by
  0.11~mJy. The vertical lines show the contact phases of the white
  dwarf and bright spot eclipses, the dotted lines corresponding to
  $\phi_{\rm{w}1},\ldots,\phi_{\rm{w}4}$,
  $\phi_{\rm{b}1},\ldots,\phi_{\rm{b}4}$ and the solid lines
  (labelled) to $\phi_{\rm{wi}}$, $\phi_{\rm{we}}$ and
  $\phi_{\rm{bi}}$, $\phi_{\rm{be}}$. The bright spot ingress and
  egress are plainly visible in the light curves of both objects,
  following the white dwarf ingress and egress, respectively.}
\label{fig:xzeridvuma_timings}
\end{figure}

\begin{figure}
\centerline{\includegraphics[height=12.0cm,angle=-90.]{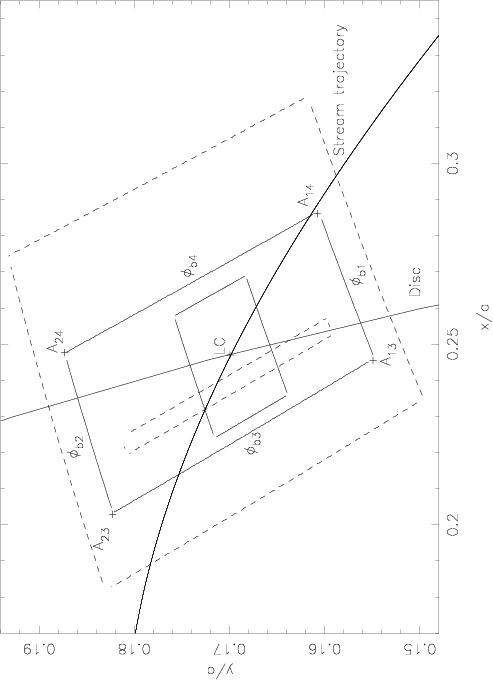}}
\caption[Horizontal structure of the bright spot of
  XZ~Eri.]{Horizontal structure of the bright spot of XZ~Eri for
  $q=0.117$, showing the region on the orbital plane within which the
  bright spot lies. As figure~\ref{fig:ouvir_bs_horizontal}, except
  that the disc radius is $R_{\rm{d}}=0.3a$.}
\label{xzeri_bs_h}
\end{figure}

\begin{figure}
\centerline{\includegraphics[height=12.0cm,angle=-90.]{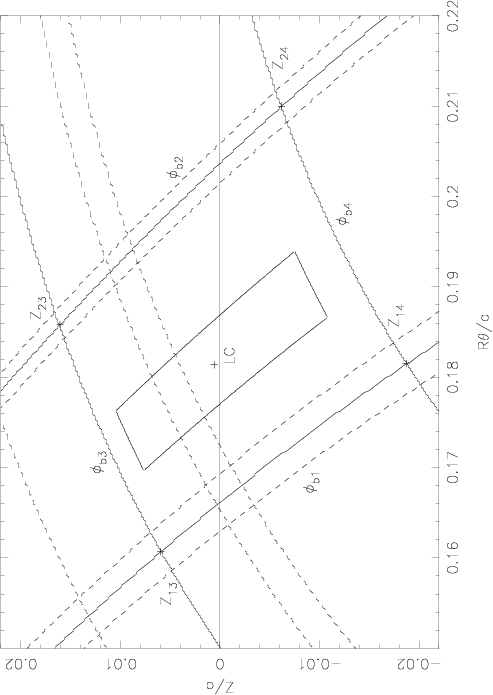}}
\caption[Vertical structure of the bright spot of XZ~Eri.]{Vertical
  structure of the bright spot of XZ~Eri for $q=0.117$. The phase arcs
  are projected onto a vertical cylinder of radius $0.3a$ (equal to
  that of the disc). Otherwise as figure~\ref{fig:ouvir_bs_vertical}.}
\label{xzeri_bs_v}
\end{figure}

\begin{figure}
\centerline{\includegraphics[height=12.0cm,angle=-90.]{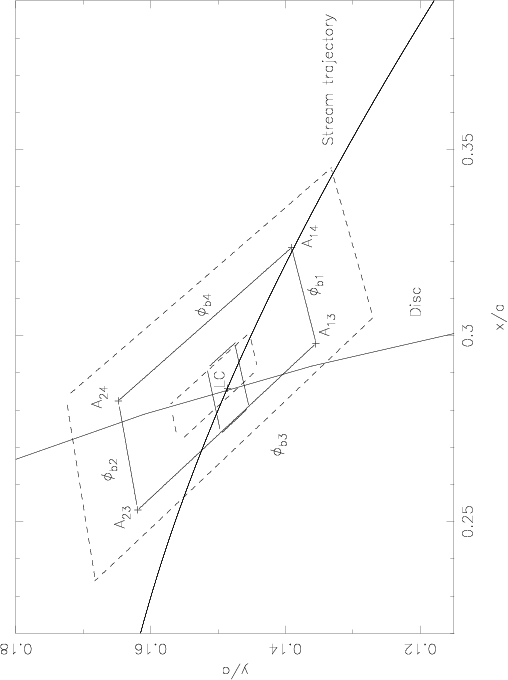}}
\caption[Horizontal structure of the bright spot of
  DV~UMa.]{Horizontal structure of the bright spot of DV~UMa for
  $q=0.148$, showing the region on the orbital plane within which the
  bright spot lies. As figure~\ref{fig:ouvir_bs_horizontal}, except
  that the disc radius is $R_{\rm{d}}=0.322a$.}
\label{dvuma_bs_h}
\end{figure}

\begin{figure}
\centerline{\includegraphics[height=12.0cm,angle=-90.]{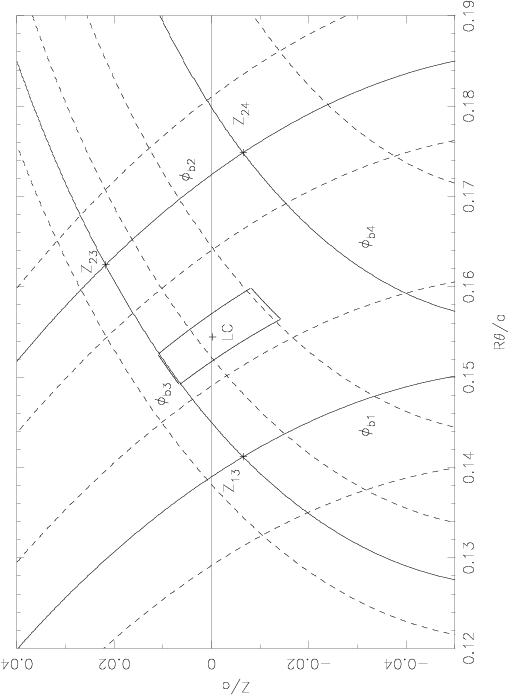}}
\caption[Vertical structure of the bright spot of DV~UMa.]{Vertical
  structure of the bright spot of DV~UMa for $q=0.148$. The phase arcs
  are projected onto a vertical cylinder of radius $0.322a$ (equal to
  that of the disc). Otherwise as figure~\ref{fig:ouvir_bs_vertical}.}
\label{dvuma_bs_v}
\end{figure}

\section{Light curve decomposition}

\subsection{The derivative method}
The eclipse contact phases given in
tables~\ref{tab:xzeridvuma_wd_times} and \ref{tab:xzeridvuma_bs_times}
were determined using the derivative of the light curve, as described
in \S~\ref{sec:derivative}. In the following analysis I have combined
the timings of all three colour bands for each target in order to
increase the accuracy of the results.

Using the techniques described in \S~\ref{sec:derivative}, for the
mean eclipse phase width of $\Delta\phi = 0.0359\pm0.0008$, the eclipse
timings of XZ~Eri (tables~\ref{tab:xzeridvuma_wd_times} and
\ref{tab:xzeridvuma_bs_times}) yield the mass ratio, inclination and
relative disc radius given in
table~\ref{tab:xzeridvuma_parameters}. The results for DV~UMa for the
mean eclipse phase width of $\Delta\phi = 0.0636\pm0.0007$ are also given in
table~\ref{tab:xzeridvuma_parameters}.  The errors on these parameters
are determined by the rms variations in the measured contact phases. I
use the bright spot eclipse timings to determine upper limits on the
angular size and the radial and vertical extent of the bright spots,
defining $\Delta \theta$, $\Delta R_{\rm{d}}$, $\Delta Z$ and $\Delta
Z_{2}$ as in equations~\ref{eq:theta}--\ref{eq:z2}. The mean position
and extent of the bright spots thus derived are given in
table~\ref{tab:xzeridvuma_bs}. The eclipse constraints on the
horizontal and vertical extent of the bright spots of XZ~Eri and
DV~UMa are given in figures~\ref{xzeri_bs_h}--\ref{dvuma_bs_v}.

Using the mass ratio and orbital inclination given in
table~\ref{tab:xzeridvuma_parameters} and the eclipse constraints on
the radius of the white dwarf (table~\ref{tab:xzeridvuma_wd_times}), I
find that the white dwarf in XZ~Eri has a radius of
$R_{1}/a=0.012 \pm 0.002$. For DV~UMa I obtain
$R_{1}/a=0.0075 \pm 0.0020$. I will continue under the assumption
that the eclipsed central object is a bare white dwarf. This
assumption and its consequences are discussed in more detail in
\S~\ref{sec:derivative} and \S~\ref{sec:conc_xzeridvuma}.

The fluxes given in table~\ref{tab:xzeridvuma_wd_times} were fitted to
the hydrogen-rich, $\log g=8$ white dwarf model atmospheres of
\citet{bergeron95}, as discussed in \S~\ref{sec:fitting}. The colour
indices quoted therein were converted to the SDSS system using the
observed transformations of \citet{smith02}. The white dwarf
temperatures $T_{\rm{1}}$ thus calculated are given in
table~\ref{tab:xzeridvuma_parameters}.

To determine the remaining system parameters of XZ~Eri and DV~UMa I
used the Nauenberg mass--radius relation
(equation~\ref{eq:nauenberg}) and approximated the secondary radius by
its volume radius, as described in \S~\ref{sec:massdet}. Because the
Nauenberg mass-radius relation assumes a cold white dwarf, I have
attempted to correct this to a temperature of $T_{\rm{1}}\sim 15\,000$~K
for XZ~Eri and to $T_{\rm{1}}\sim 20\,000$~K for DV~UMa, the approximate
temperatures given by the model atmosphere fit. The radius of the
white dwarf at $10\,000$~K is about 5 per~cent larger than for a cold
($0$~K) white dwarf \citep{koester86}. To correct from $10\,000$~K to
the appropriate temperature, the white dwarf cooling curves of
\citet{wood95} for $M_{1}/{\rm M}_{\odot}=1.0$, the approximate masses
given by the Nauenberg relation, were used. This gave total radial
corrections of $6.0$ and $7.0$ per~cent for XZ~Eri and DV~UMa,
respectively.

\begin{sidewaystable}
\begin{center}
\caption[White dwarf contact phases and out-of-eclipse white dwarf
  fluxes of XZ~Eri and DV~UMa.]{White dwarf contact phases, accurate
  to $\pm0.0006$ (XZ~Eri) and $\pm0.0005$ (DV~UMa), and out-of-eclipse
  white dwarf fluxes of XZ~Eri and DV~UMa. The errors on the fluxes
  are $\pm0.001$~mJy.}
\vspace{0.3cm}
\small
\begin{tabular}{ccccccccc}
\hline
Cycle & Band & $\phi_{\rm{w}1}$ & $\phi_{\rm{w}2}$ &
$\phi_{\rm{w}3}$ & $\phi_{\rm{w}4}$ &
$\phi_{\rm{wi}}$ & $\phi_{\rm{we}}$ & Flux (mJy)\\
\hline
XZ~Eri\\
4733 & {\em u\/}$^{\prime}$ & --0.020996 & --0.015625 & 0.016113
 & 0.022461 & --0.018555 & 0.018555 & 0.0434 \\
 & {\em g\/}$^{\prime}$ & --0.022461 & --0.011719 & 0.012207
 & 0.020020 & --0.018555 & 0.017578 & 0.0466 \\
 & {\em r\/}$^{\prime}$ & --0.022461 & --0.013184 & 0.013184
 & 0.020020 & --0.019531 & 0.016113 & 0.0441 \\
4734 & {\em u\/}$^{\prime}$ & --0.018066 & --0.014160 & 0.010742
 & 0.020020 & --0.017090 & 0.017578 & 0.0341 \\
 & {\em g\/}$^{\prime}$ & --0.022461 & --0.013184 & 0.013672
 & 0.021484 & --0.019531 & 0.017578 & 0.0531 \\
 & {\em r\/}$^{\prime}$ & --0.022461 & --0.013184 & 0.015137
 & 0.020020 & --0.017090 & 0.017578 & 0.0375 \\
\hline
DV~UMa\\
69023 & {\em u\/}$^{\prime}$ & --0.033850 & --0.030644 & 0.030276
 & 0.033482 & --0.031445 & 0.032680 & 0.0465 \\
 & {\em g\/}$^{\prime}$ & --0.033850 & --0.030644 & 0.030276
 & 0.033482 & --0.031445 & 0.032680 & 0.0373 \\
 & {\em i\/}$^{\prime}$ & --0.033048 & --0.029842 & 0.030276 
 & 0.033482 & --0.030644 & 0.031879 & 0.0245 \\
69046  & {\em u\/}$^{\prime}$ & --0.032798 & --0.030132 & 0.031210
 & 0.033210 & --0.031465 & 0.031877 & 0.0451 \\
 & {\em g\/}$^{\prime}$ & --0.033467 & --0.030132 & 0.031210
 & 0.033879 & --0.032132 & 0.032543 & 0.0435 \\
 & {\em i\/}$^{\prime}$ & --0.032798 & --0.030132 & 0.030544 
 & 0.033210 & --0.030799 & 0.031877 & 0.0239 \\
69058 & {\em u\/}$^{\prime}$ & --0.032691 & --0.030564 & 0.030612
 & 0.032206 & --0.030564 & 0.032206 & 0.0356 \\
 & {\em g\/}$^{\prime}$ & --0.033224 & --0.030564 & 0.030612
 & 0.032739 & --0.031097 & 0.032206 & 0.0318 \\
 & {\em i\/}$^{\prime}$ & --0.033755 & --0.030564 & 0.031142
 & 0.033270 & --0.032161 & 0.032739 & 0.0312 \\
\hline
\end{tabular}
\normalsize
\label{tab:xzeridvuma_wd_times}
\end{center}
\end{sidewaystable}

\begin{table}
\begin{center}
\caption[Bright spot contact phases of XZ~Eri and DV~UMa.]{Bright spot
  contact phases of XZ~Eri and DV~UMa, accurate to $\pm0.0006$ (XZ~Eri)
  and $\pm0.0005$ (DV~UMa).}
\vspace{0.3cm}
\small
\begin{tabular}{cccccccc}
\hline
Cycle & Band & $\phi_{\rm{b}1}$ & $\phi_{\rm{b}2}$ &
$\phi_{\rm{b}3}$ & $\phi_{\rm{b}4}$ &
$\phi_{\rm{bi}}$ & $\phi_{\rm{be}}$\\
\hline
XZ~Eri\\
4733 & {\em u\/}$^{\prime}$ & --0.000977 & 0.006836 & 0.064941
& 0.067871 & 0.002930 & 0.066406 \\
  & {\em g\/}$^{\prime}$ & 0.000000 & 0.006836 & 0.069336
& 0.074219 & 0.003906 & 0.073242 \\
  & {\em r\/}$^{\prime}$ & --0.000977 & 0.006836 & 0.069336
& 0.073242 & 0.001465 & 0.070313 \\
4734 & {\em u\/}$^{\prime}$ & --0.000977 & 0.006836 & 0.063965
& 0.081055 & 0.002930 & 0.070801 \\
  & {\em g\/}$^{\prime}$ & --0.000977 & 0.006836 & 0.065430
& 0.070801 & 0.001465 & 0.067871 \\
  & {\em r\/}$^{\prime}$ & 0.000488 & 0.005859 & 0.065430
& 0.069336 & 0.002930 & 0.067871 \\
\hline
DV~UMa\\
69023 & {\em u\/}$^{\prime}$ & --0.018620 & --0.009803 & 0.079171
 & 0.083179 & --0.013811 & 0.082378 \\ 
 & {\em g\/}$^{\prime}$ & --0.016215 & --0.011406 & 0.079171
 & 0.083179 & --0.013811 & 0.080775 \\
 & {\em i\/}$^{\prime}$ & --0.017017 & --0.009001 & 0.078370
 & 0.085584 & --0.013009 & 0.079973 \\ 
69046 & {\em u\/}$^{\prime}$ & --0.016797 & --0.012129 & 0.079884
 & 0.085885 & --0.014131 & 0.080553 \\ 
 & {\em g\/}$^{\prime}$ & --0.018130 & --0.010129 & 0.079884
 & 0.084552 & --0.014131 & 0.081886 \\ 
 & {\em i\/}$^{\prime}$ & --0.022131 & --0.006127 & 0.081220
 & 0.083219 & --0.014797 & 0.081886 \\ 
69058 & {\em u\/}$^{\prime}$ & --0.015671 & --0.009819 & 0.079019
 & 0.080613 & --0.014074 & 0.079550 \\ 
 & {\em g\/}$^{\prime}$ & --0.015671 & --0.010883 & 0.079019
 & 0.081680 & --0.013541 & 0.079550 \\ 
 & {\em i\/}$^{\prime}$ & --0.014604 & --0.010883 & 0.077955 
 & 0.083274 & --0.013010 & 0.078486 \\ 
\hline

\end{tabular}
\normalsize
\label{tab:xzeridvuma_bs_times}
\end{center}
\end{table}

\subsection{A parameterized model of the eclipse}
\label{sec:xzeridvuma_techniques_model}
The system parameters of XZ~Eri and DV~UMa were also determined by
fitting the phase-folded light curves using a parameterized model of the
eclipse. The {\sc lfit} code, developed by \citet{horne94} and
described in detail in \S~\ref{sec:lfit}, was used.

The data were not good enough to determine the limb-darkening
coefficient $U_{\rm{1}}$ accurately, so this was held at a typical
value of 0.5 for each fit. The disc parameter for DV~UMa was held
fixed at $b=1.0$ as it was too faint to be well constrained.

The procedure discussed in \S~\ref{sec:lfit} failed to find the likely
error for the disc exponent $b$ of the {\em u\/}$^{\prime}$ band of
XZ~Eri, as the disc flux is small in this case and the light curve
noisy, so perturbation of the parameter made virtually no difference
to the $\chi^{2}$ of the fit.

\begin{figure}
\begin{tabular}{cc}
\includegraphics[height=7.3cm,angle=-90.]{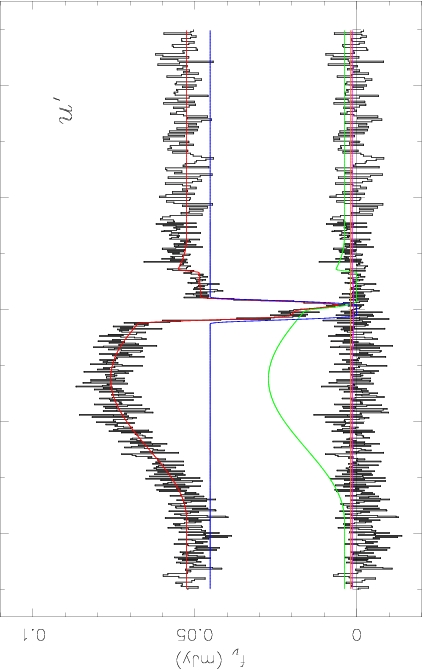} &
\includegraphics[height=7.2cm,angle=-90.]{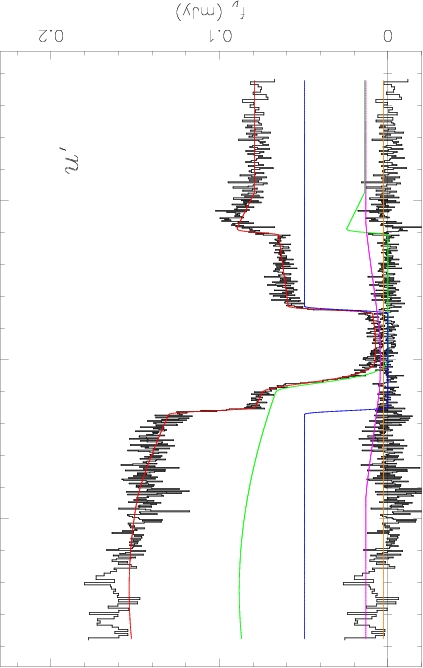} \\
\includegraphics[height=7.3cm,angle=-90.]{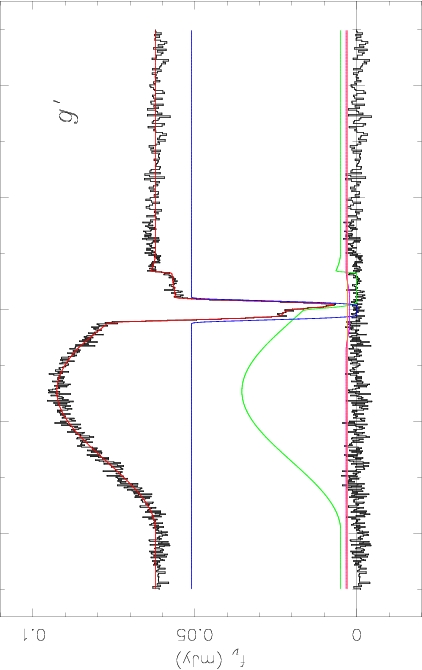} &
\includegraphics[height=7.2cm,angle=-90.]{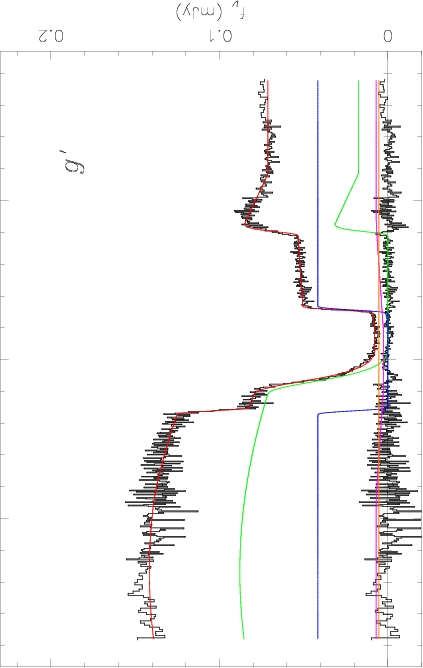} \\
\includegraphics[height=7.3cm,angle=-90.]{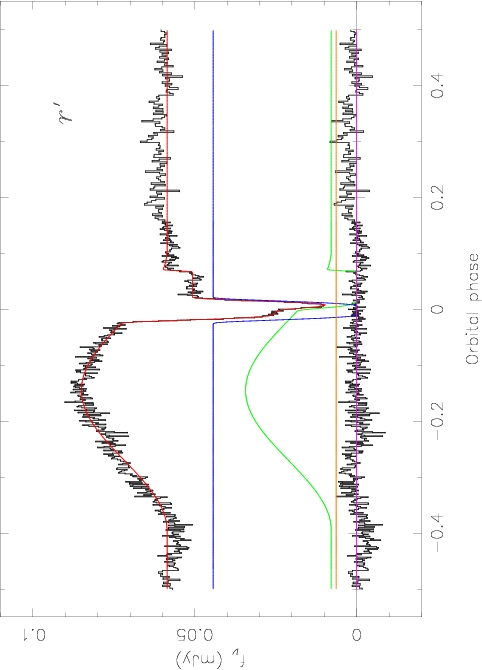} &
\includegraphics[height=7.2cm,angle=-90.]{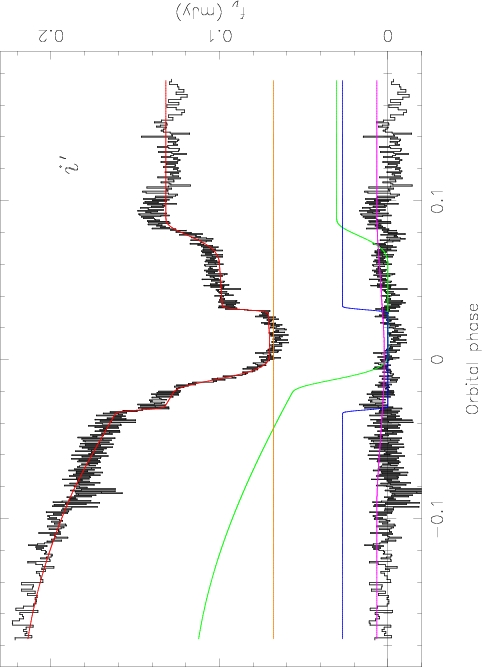}
\end{tabular}
\caption[The phase-folded {\em u\/}$^{\prime}$, {\em g\/}$^{\prime}$ and
  {\em r\/}$^{\prime}$ light curves of XZ~Eri and DV~UMa as fitted by
  {\sc lfit}.]{Left: the phase-folded {\em u\/}$^{\prime}$, {\em
  g\/}$^{\prime}$ and {\em r\/}$^{\prime}$ light curves of XZ~Eri, fitted
  separately using the model described in
  Section~\ref{sec:xzeridvuma_techniques_model}. Right: the
  phase-folded {\em u\/}$^{\prime}$, {\em g\/}$^{\prime}$ and {\em
  i\/}$^{\prime}$ light curves of DV~UMa. The data (black) are shown
  with the fit (red) overlaid and the residuals plotted below
  (black). Below are the separate light curves of the white dwarf
  (blue), bright  spot (green), accretion disc (purple) and the
  secondary star (orange). Note that the disc in both objects is very
  faint, as is the secondary (except for the {\em i\/}$^{\prime}$ band
  of DV~UMa). The differing phase coverage of each night's 
  observations of DV~UMa accounts for the uneven appearance of the
  data.}
\label{fig:xzeridvuma_lfit}
\end{figure}

The results of the model fitting are given in
tables~\ref{tab:xzeri_parameters_lfit} and
\ref{tab:dvuma_parameters_lfit}, and are shown in
figure~\ref{fig:xzeridvuma_lfit}. Each passband was fitted
independently, as there were found to be significant differences
between many of the optimum parameters for each band. This is to be
expected for parameters such as the bright spot scale $SB$, where one
would anticipate that the cooler regions are more extended than the
hotter ones (as seen for DV~UMa). We would, of course, expect the mass
ratio to remain constant in all three passbands for each object, which
it indeed does.

The results of a white dwarf model atmosphere fit \citep{bergeron95}
to the fluxes fit by the model in each passband are given in
table~\ref{tab:xzeridvuma_parameters}.  I have used the white dwarf
cooling curves of \citet{wood95} for $M_{1}/{\rm M}_{\odot}=0.75$
(interpolating between 0.7 and 0.8) and $M_{1}/{\rm M}_{\odot}=1.0$,
the approximate masses found using the Nauenberg relation for XZ~Eri
and DV~UMa, to give radial corrections of 7.6 and 7.0 per~cent,
respectively. These were used to determine the absolute system
parameters given in table~\ref{tab:xzeridvuma_parameters}.

I note that the higher signal-to-noise light curves of the {\em
i\/}$^{\prime}$, {\em r\/}$^{\prime}$ and {\em g\/}$^{\prime}$ bands
have $\chi^{2}/\nu\gg1$ (see tables~\ref{tab:xzeri_parameters_lfit}
and \ref{tab:dvuma_parameters_lfit}). This is because these data are
dominated by flickering, not photon noise, unlike the {\em
u\/}$^{\prime}$ data. If we had enough cycles to completely remove the
effects of flickering we would expect, for an accurate model, to
achieve $\chi^{2}/\nu=1$.

The quality of the light curve fits (figure~\ref{fig:xzeridvuma_lfit}
and tables~\ref{tab:xzeri_parameters_lfit} and
\ref{tab:dvuma_parameters_lfit}) are generally excellent.  Apart from
longer-term variations not allowed for by the model  (for example, an
overall brightening of the accretion disc), the only feature not
satisfactorily reproduced is the bright spot egress of DV~UMa. I
suspect that this is due to the presence of ellipsoidal variations of
the secondary star, which has the greatest contribution in this
filter. Since the amplitude of the ellipsoidal modulation is at a
minimum at phase zero, this will tend to enhance the amplitude of the
orbital hump and shift the peak of the orbital hump towards the peak of
the ellipsoidal modulation, at phase $-0.5$. In the case of the {\em
i}$^{\prime}$ data of DV~UMa shown in
figure~\ref{fig:xzeridvuma_lfit} the result of the
algorithm trying to fit the increased flux at phases $\lesssim-0.1$ is
to worsen the fit to the part of the orbital hump immediately
following the egress of the bright spot. The contribution of the (M4.5;
\citealp{mukai90}) secondary star in the {\em i}$^{\prime}$ band of
DV~UMa is comparable to that of the (M5.4; \citealp{marsh90c})
secondary star of HT~Cas (see also chapter~\ref{ch:gycncircomhtcas}).
Due to the reason outlined above, and since not all of the orbital
cycle of DV~UMa was observed, the 
parameters that are constrained by the orbital hump are rather more
uncertain for DV~UMa than they are for XZ~Eri. This may introduce some
systematic errors into the estimation of the bright spot orientation
$\theta_{\rm{B}}$, the isotropic flux fraction of the bright spot
$f_{\rm iso}$ and the bright spot flux.

\begin{table}
\begin{center}
\caption[Mean position and extent of the bright spots of XZ~Eri and
  DV~UMa.]{Mean position and extent of the bright spots of XZ~Eri and
  DV~UMa, as defined in \S~\ref{sec:derivative}.}
\vspace{0.3cm}
\small
\begin{tabular}{ccc}
\hline
 & XZ~Eri & DV~UMa\\
\hline
$\Delta R_{\rm{d}}/a$ & 0.0378 & 0.0258 \\
$\Delta \theta$ & $8.73^{\circ}$ & $7.57^{\circ}$ \\
$\Delta Z/a$ & 0.0174 & 0.0399 \\
$\Delta Z_{2}/a$ & 0.0161 & 0.0217 \\
$R_{\rm{d}}/a$ & 0.300 & 0.322 \\
$\theta$ & $34.53^{\circ}$ & $27.47^{\circ}$\\
\hline
\end{tabular}
\normalsize
\label{tab:xzeridvuma_bs}
\end{center}
\end{table}

\begin{table}
\begin{center}
\caption[Parameters of XZ~Eri fitted using {\sc
    lfit}.]{Parameters of XZ~Eri fitted using a modified
    version of the {\sc lfit} model of \citet{horne94} described in
    \S~\ref{sec:lfit}. The fluxes of each component are also
    shown. XZ~Eri has been fitted by phase-folding the two eclipses
    and binning by two data points. Note that the orbital inclination
    $i$ is not a fit parameter but is calculated using $q$ and
    $\Delta\phi$.}
\vspace{0.3cm}
\small
\begin{tabular}{lccc}
\hline
Parameter & \multicolumn{3}{c}{XZ~Eri} \\
 & {\em u\/}$^{\prime}$ & {\em g\/}$^{\prime}$ & {\em r\/}$^{\prime}$ \\
\hline
Inclination $i$ & $80.4^{\circ}\pm0.8^{\circ}$ &
$80.1^{\circ}\pm0.1^{\circ}$ & $80.4^{\circ}\pm0.2^{\circ}$ \\
Mass ratio $q$ & $0.11\pm0.02$ & $0.116\pm0.003$ & $0.107\pm0.002$\\
Eclipse phase & $0.0342$ & $0.03362$ & $0.0333$ \\
\hspace{0.1cm} width $\Delta\phi$ & $\pm0.0007$ & $\pm0.00021$
& $\pm0.0003$ \\
Outer disc \\
\hspace{0.1cm} radius $R_{\rm{d}}/a$ & $0.307\pm0.011$ & $0.295\pm0.003$
& $0.316\pm0.005$ \\
White dwarf \\
\hspace{0.1cm} limb & 0.5 & 0.5 & 0.5 \\
\hspace{0.1cm} darkening $U_{\rm{1}}$ \\
White dwarf \\
\hspace{0.1cm} radius $R_{1}/a$ & $0.019\pm0.002$ &
$0.0175\pm0.0006$ & $0.0195\pm0.0010$ \\
Bright spot \\
\hspace{0.1cm} scale $SB/a$ & $0.014\pm0.010$ & $0.013\pm0.002$ &
$0.0147\pm0.0008$ \\
Bright spot \\
\hspace{0.1cm} orientation $\theta_{\rm{B}}$ &
$134.1^{\circ}\pm1.0^{\circ}$ & $141.9^{\circ}\pm0.3^{\circ}$ &
$141.4^{\circ}\pm0.3^{\circ}$ \\
Isotropic flux \\
\hspace{0.1cm} fraction $f_{\rm iso}$ & $0.14\pm0.03$ & $0.140\pm0.008$ &
$0.2294\pm0.0015$ \\
Disc exponent $b$ & 0.74965 & $0.4\pm2.1$ &
$0.3\pm0.3$ \\
Phase offset $\phi_{0}$ &
$16\times10^{-4}$ & $16.3\times10^{-4}$ & $17.0\times10^{-4}$ \\
 &
$\pm3\times10^{-4}$ & $\pm0.8\times10^{-4}$ & $\pm1.2\times10^{-4}$ \\
$\chi^{2}$ of fit & 656 & 897 & 1554 \\
Number of \\
\hspace{0.1cm} datapoints $\nu$ & 611 & 611 & 611 \\
\hline
Flux (mJy) \\
\hspace{0.1cm} White dwarf & $0.0453\pm0.0011$ & $0.0510\pm0.0004$ &
$0.0443\pm0.0004$ \\
\hspace{0.1cm} Accretion disc & $0.001\pm0.003$ & $0.0033\pm0.0009$ &
$0.0000\pm0.0010$ \\
\hspace{0.1cm} Secondary & $0.0020\pm0.0019$ & $0.0029\pm0.0006$ &
$0.0064\pm0.0007$ \\
\hspace{0.1cm} Bright spot & $0.0273\pm0.0005$ & $0.03545\pm0.00018$ &
$0.0343\pm0.0002$ \\
\hline
\end{tabular}
\normalsize
\label{tab:xzeri_parameters_lfit}
\end{center}
\end{table}

\begin{table}
\begin{center}
\caption[Parameters of DV~UMa fitted using {\sc lfit}.]{Parameters of
    DV~UMa fitted using {\sc lfit}. DV~UMa has been fitted by
    phase-folding all three eclipses and binning by two data
    points. Otherwise as table~\ref{tab:xzeri_parameters_lfit}.}
\vspace{0.3cm}
\small
\begin{tabular}{lccc}
\hline
Parameter & \multicolumn{3}{c}{DV~UMa} \\
 & {\em u\/}$^{\prime}$ & {\em g\/}$^{\prime}$ & {\em i\/}$^{\prime}$ \\
\hline
Inclination $i$ & $83.8^{\circ}\pm0.2^{\circ}$ &
$84.3^{\circ}\pm0.1^{\circ}$ & $84.3^{\circ}\pm0.1^{\circ}$ \\ 
Mass ratio $q$ & $0.159\pm0.003$ & $0.1488\pm0.0011$ & $0.153\pm0.002$ \\ 
Eclipse phase & $0.06346$ & $0.06352$ & $0.06307$\\
\hspace{0.1cm} width $\Delta\phi$ & $\pm 0.00017$ & $\pm0.00007$ &
$\pm0.00015$ \\
Outer disc \\
\hspace{0.1cm} radius $R_{\rm{d}}/a$ & $0.317\pm0.004$ &
$0.32278\pm0.00016$ & $0.31272\pm0.00017$\\
White dwarf \\
\hspace{0.1cm} limb & 0.5 & 0.5 & 0.5 \\
\hspace{0.1cm} darkening $U_{\rm{1}}$ \\
White dwarf \\
\hspace{0.1cm} radius $R_{1}/a$ & $0.0091\pm0.0016$ &
$0.0092\pm0.0004$ & $0.0082\pm0.0014$ \\
Bright spot \\
\hspace{0.1cm} scale $SB/a$ & $0.0150\pm0.0010$ & $0.0211\pm0.0002$ &
$0.049\pm0.003$ \\
Bright spot \\
\hspace{0.1cm} orientation $\theta_{\rm{B}}$ &
$142.0^{\circ}\pm0.8^{\circ}$ &
$137.75^{\circ}\pm0.09^{\circ}$ & $169.4^{\circ}\pm0.6^{\circ}$\\ 
Isotropic flux \\
\hspace{0.1cm} fraction $f_{\rm iso}$ & $0.157\pm0.009$ &
$0.1989\pm0.0019$ & $0.262\pm0.004$\\
Disc exponent $b$ & 1. & 1. & 1.\\
Phase offset $\phi_{0}$ & $2.5\times10^{-4}$ & $5.48\times10^{-4}$ &
$1.7\times10^{-4}$\\
 & $\pm0.9\times10^{-4}$ & $\pm0.10\times10^{-4}$ &
$\pm0.7\times10^{-4}$\\ 
$\chi^{2}$ of fit & 1059 & 6873 & 4332\\ 
Number of \\
\hspace{0.1cm} datapoints $\nu$ & 636 & 636 & 636\\
\hline
Flux (mJy) \\
\hspace{0.1cm} White dwarf & $0.0496\pm0.0008$ & $0.0415\pm0.0002$ &
$0.0269\pm0.0004$\\
\hspace{0.1cm} Accretion disc & $0.0131\pm0.0015$ & $0.0069\pm0.0004$
& $0.0065\pm0.0007$\\
\hspace{0.1cm} Secondary & $0.0027\pm0.0007$ & $0.00531\pm0.00018$ &
$0.0680\pm0.0003$\\
\hspace{0.1cm} Bright spot & $0.0882\pm0.0005$ & $0.0879\pm0.00014$ &
$0.1157\pm0.0004$\\ 
\hline
\end{tabular}
\normalsize
\label{tab:dvuma_parameters_lfit}
\end{center}
\end{table}

\subsection{Comparison of methods}
\label{sec:xzeridvumacomp}
I have determined the system parameters of the eclipsing dwarf nov\ae
XZ~Eri and DV~UMa through two methods: the derivative method of
\citet{wood86b} and the parameterized model technique of
\citet{horne94}. I now proceed to compare these two techniques, first
noting that the system parameters determined by each (given in
table~\ref{tab:xzeridvuma_parameters}) are generally in good agreement.

Given data with an excellent signal-to-noise ratio (S/N) and covering
many phase-folded cycles, the measurement of the contact phases from
the light curve derivative is capable of producing accurate and
reliable results \citep[e.g.][]{wood89a}. It is less dependable with
only a few cycles, however, even if they are individually of high
S/N. This is due to flickering having the effect of partially masking
the exact location of the contact phases
$\phi_{1},\ldots,\phi_{4}$. This problem will affect the values for
the deconvolved fluxes of each component and the constraints on the
size of the white dwarf and bright spot, which are used to determine
the individual component masses. The mid-points of ingress and egress,
especially those of the white dwarf, are generally still well
determined, since the signal (a peak in the derivative of the light
curve) is large due to the rapid ingress and egress of the eclipsed
body. This makes the determination of the mass ratio and the orbital
inclination relatively simple and reliable. It also means that this
technique is well suited to determining the times of mid-eclipse in
order to calculate the ephemeris.

I believe that the differences between the component masses and radii
of XZ~Eri determined by each technique
(table~\ref{tab:xzeridvuma_parameters}) are due to the above effect of
flickering. The mass ratios quoted are consistent with each other, but
the relative white dwarf radius estimated from the derivative method
is somewhat smaller than that determined from the parameterized model
($R_{1}/a=0.012\pm0.002$ and $R_{1}/a=0.0181\pm0.0005$,
respectively). This also affects the estimates of the absolute radii
and masses.

For the purpose of determining the system parameters I prefer the
parameterized model technique over the derivative method. This is
because the former constrains the parameters using all the points in
the light curve to minimise $\chi^{2}$. This procedure has several
advantages:
\begin{enumerate}
\item The value of $\chi^{2}$ provides a reliable estimate of the
  goodness of fit which is used to optimise the parameter
  estimates. The measurement of the contact phases and subsequent
  deconvolution of the light curves in the derivative method is not
  unique (it is affected by the choice of box-car and median filters,
  for instance), and this technique lacks a comparable merit function.
\item Rapid flickering and photon noise during the ingress and/or
  egress phases are less problematic for the parameterized model as the
  light curves are evaluated using all the data points, not just the
  few during ingress and egress.
\item The above points indicate that the parameterized model technique
  requires fewer cycles to obtain accurate results. This is indeed
  what I found in practice, meaning that this method could be applied
  to each passband separately to investigate the temperature
  dependence of each parameter, if any.
\item The bright spot egress in particular is often faint (due to
  foreshortening) and difficult to reconstruct using the derivative
  method. The parameterized model method is also likely to be easier to
  apply to cases where the ingress of the white dwarf and bright spot
  are merged, as seen in IP~Peg \citep{wood86b} and EX~Dra
  \citep{baptista00}.
\end{enumerate}

For these reasons, I believe that the results given by the
parameterized model of the eclipse are better determined than those of
the derivative technique. However, the former method does have some
disadvantages. Ideally, it requires observations of most of the
orbital cycle, as the orbital hump is needed to fit some parameters
reliably, particularly the bright spot orientation $\theta_{\rm B}$,
the isotropic flux fraction $f_{\rm iso}$ of the bright spot and the
bright spot flux. Longer time-scale flickering can also cause some
problems if
only a few cycles are available. As with any such technique, the key
weakness of the parameterized model method is the need for an accurate
model. However, as figure~\ref{fig:xzeridvuma_lfit} shows, apart from
the {\em i\/}$^{\prime}$ band of DV~UMa, the residual from the fit
shows no large peaks in areas such as the ingress and egress of the
white dwarf or bright spot. Such peaks would be expected if the model
were not adequately fitting the data.

I note that that the system parameters derived for DV~UMa
(table~\ref{tab:xzeridvuma_parameters}) are consistent with the
superhump period-mass ratio relation of \citet[][his equation
8]{patterson98}. XZ~Eri, however, lies $5\sigma$ off this
relation. Here I have used the superhump periods $P_{\rm
sh}=0.062808\pm0.000017$~days for XZ~Eri \citep{uemura04} and $P_{\rm
sh}=0.08870\pm0.00008$~days for DV~UMa \citep{patterson98}.

\begin{sidewaystable}
\begin{center}
\caption[System parameters of XZ~Eri and DV~UMa derived using the
  Nauenberg mass--radius relation.]{System parameters of XZ~Eri and
  DV~UMa derived using the Nauenberg mass--radius relation corrected
  to the appropriate $T_{\rm{1}}$. $R_{2}$ is the volume radius
  of the secondary's Roche lobe \citep{eggleton83}, and $R_{\rm{min}}$
  is as defined by \citet[][their equation 13]{verbunt88}. The weighted
  means of the appropriate values from
  tables~\ref{tab:xzeri_parameters_lfit} and
  \ref{tab:dvuma_parameters_lfit} are used for the model
  parameters. One column of parameters is calculated using the
  derivative method, the other derived using the parameterized model
  technique.}
\vspace{0.3cm}
\small
\begin{tabular}{lcccc}
\hline
 & \multicolumn{2}{c}{XZ~Eri} & \multicolumn{2}{c}{DV~UMa} \\
Parameter & Derivative & Model & Derivative & Model \\
\hline
Inclination $i$ & $80.3^{\circ}\pm0.6^{\circ}$ &
 $80.16^{\circ}\pm0.09^{\circ}$ & $84.4^{\circ}\pm0.8^{\circ}$&
 $84.24^{\circ}\pm0.07^{\circ}$ \\
Mass ratio $q=M_{2}/M_{1}$ & $0.117 \pm 0.015$ & $0.1098
 \pm 0.0017$ & $0.148 \pm 0.013$ & $0.1506\pm0.0009$ \\
White dwarf mass $M_{1}/{\rm M}_{\odot}$ & $1.01 \pm 0.09$ & $0.767 \pm
 0.018$ & $1.14 \pm 0.12$ & $1.041 \pm 0.024$ \\
Secondary mass $M_{2}/{\rm M}_{\odot}$ & $0.119 \pm 0.019$ & $0.0842 \pm
 0.0024$ & $0.169\pm0.023$ & $0.157\pm0.004$ \\
White dwarf radius $R_{1}/{\rm R}_{\odot}$ & $0.0082 \pm 0.0014$ &
 $0.0112 \pm 0.0003$ & $0.0067\pm0.0018$ & $0.0079\pm0.0004$ \\
Secondary radius $R_{2}/{\rm R}_{\odot}$ & $0.147 \pm 0.015$ & $0.1315
 \pm 0.0019$ & $0.207\pm0.016$ & $0.2022\pm0.0018$ \\
Separation $a/{\rm R}_{\odot}$ & $0.680 \pm 0.021$ & $0.619 \pm 0.005$ &
 $0.90\pm0.03$ & $0.869\pm0.007$ \\
White dwarf radial velocity $K_{\rm{1}}/\rm{km\,s^{-1}}$ & $58 \pm
 8$ & $49.9 \pm 0.9$ & $68\pm6$ & $66.7\pm0.7$ \\
Secondary radial velocity $K_{\rm{2}}/\rm{km\,s^{-1}}$ & $496.9 \pm
 2.0$ & $454.7 \pm 0.4$ & $457.5\pm2.6$ & $443.2\pm0.5$ \\
Outer disc radius $R_{\rm{d}}/a$ & $0.300 \pm 0.017$ & $0.3009 \pm
 0.0025$ & $0.322\pm0.011$ & $0.31805\pm0.00012$ \\
Minimum circularisation radius $R_{\rm{min}}/a$ & $0.217\pm0.013$ &
 $0.2229\pm0.0014$ & $0.196\pm0.008$ & $0.1948\pm0.0005$ \\
White dwarf temperature $T_{\rm{1}}/\rm{K}$ & $15\,000 \pm 500$ &
 $17\,000 \pm 500$ & $20\,000\pm1500$ & $20\,000\pm1500$ \\
\hline
\end{tabular}
\normalsize
\label{tab:xzeridvuma_parameters}
\end{center}
\end{sidewaystable}

\section{Eclipse mapping}
\label{sec:gycncircomhtcas_temp_maps}

\begin{figure}
\begin{tabular}{cc}
\begin{minipage}{2.5in}
\includegraphics[width=6.0cm,angle=0.]{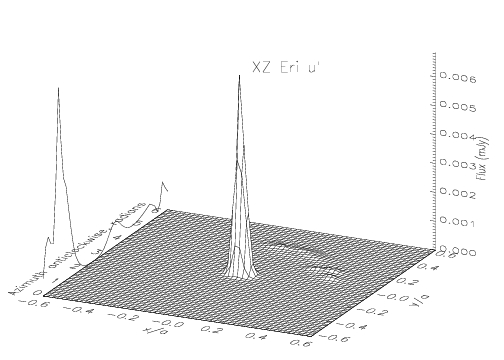} \end{minipage}&
\begin{minipage}{2.5in}
\includegraphics[height=8.0cm,angle=-90.]{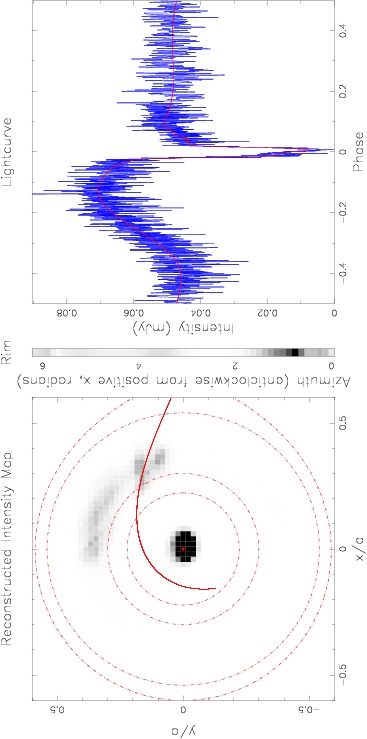} \end{minipage}\\
\begin{minipage}{2.5in}
\includegraphics[width=6.0cm,angle=0.]{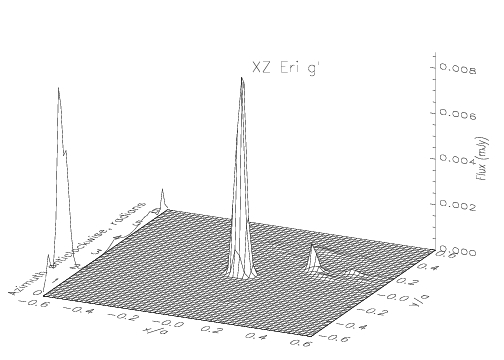} \end{minipage}&
\begin{minipage}{2.5in}
\includegraphics[height=8.0cm,angle=-90.]{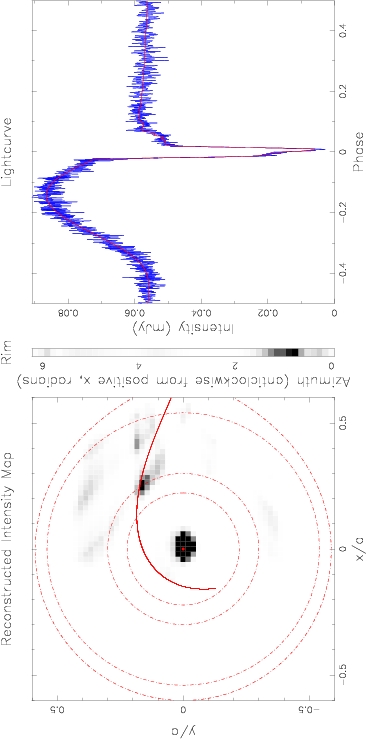} \end{minipage}\\
\begin{minipage}{2.5in}
\includegraphics[width=6.0cm,angle=0.]{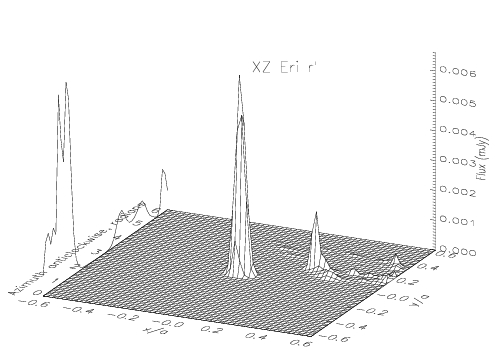} \end{minipage}&
\begin{minipage}{2.5in}
\includegraphics[height=8.0cm,angle=-90.]{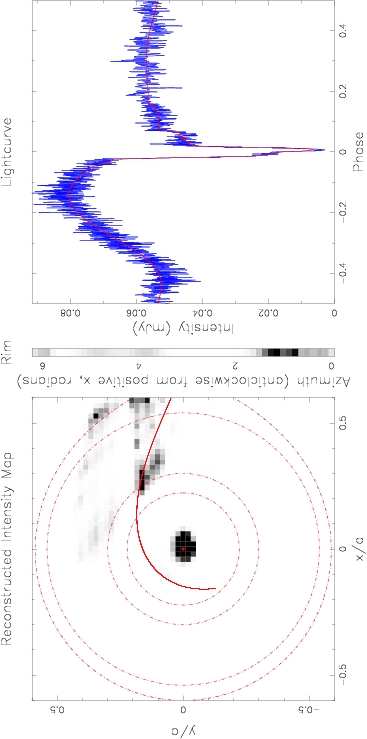} \end{minipage}\\
\end{tabular}
\caption[Eclipse maps for the {\em u\/}$^{\prime}$, {\em g\/}$^{\prime}$
  and {\em r\/}$^{\prime}$ light curves of XZ~Eri.]{As
  figure~\ref{fig:em_ouvir_18}, but for, from top, the {\em
  u\/}$^{\prime}$, {\em g\/}$^{\prime}$ and {\em r\/}$^{\prime}$ data of
  XZ~Eri. The system parameters adopted for the reconstruction
  are $q=0.1098$, $i=80.16^{\circ}$ and $R_{\rm d}=0.3a$}
\label{fig:em_xzeri}
\end{figure}

The eclipse mapping results for XZ~Eri are shown in
figure~\ref{fig:em_xzeri}. The main features of the reconstructed disc
are as follows. The white dwarf and bright spot are evident, with the
white dwarf peak flux much greater than that of the bright spot in all
three passbands. The disc itself is effectively invisible (the
numerous small-scale features are most likely spurious remnants of the
arcs discussed in \S~\ref{sec:em}). The intensity distribution along
the disc rim (of radius $0.3a$; see tables~\ref{tab:xzeridvuma_bs},
\ref{tab:xzeri_parameters_lfit} and \ref{tab:dvuma_parameters_lfit})
shows a peak at the location of the bright spot, with smaller features
probably due to flickering in the light curve. The bright spot is
clearly visible in the {\em g\/}$^{\prime}$ and {\em r\/}$^{\prime}$
band reconstructions, but is severely distorted in the {\em
u\/}$^{\prime}$ map. This is due to the relative faintness of the
bright spot coupled with the increased noise in the latter filter.

\begin{figure}
\centerline{\includegraphics[width=14cm,angle=0]{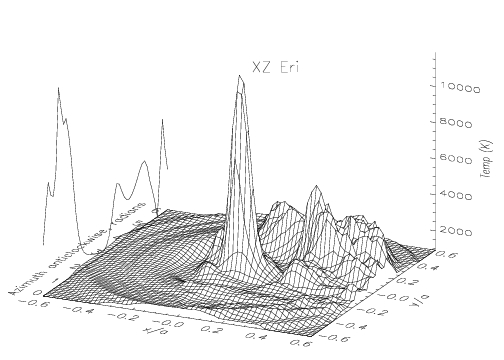}}
\caption[3-dimensional plot of a blackbody fit to the reconstructed
  disc intensities of XZ~Eri.]{3-dimensional plot of a blackbody fit
  to the reconstructed disc intensities of XZ~Eri shown in
  figure~\ref{fig:em_xzeri}.}
\label{fig:em_xzeri_temp_3d}
\end{figure}

The temperature map of the disc of XZ~Eri shown in
figure~\ref{fig:em_xzeri_temp_3d} was produced by fitting a blackbody
function convolved through the filter response functions to each point
in the relevant reconstructed maps. The orbital separation and
distance to the system adopted were those derived in
\S~\ref{sec:xzeridvuma_techniques_model}.

\begin{figure}
\begin{tabular}{cc}
\begin{minipage}{2.5in}
\includegraphics[width=6.0cm,angle=0.]{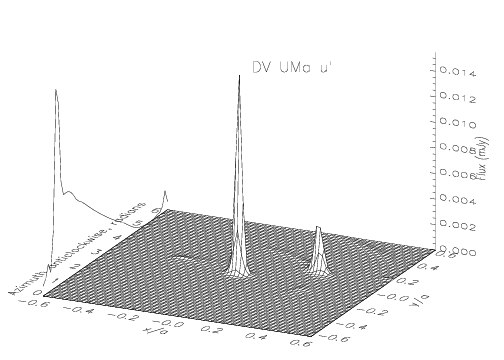} \end{minipage}&
\begin{minipage}{2.5in}
\includegraphics[height=8.0cm,angle=-90.]{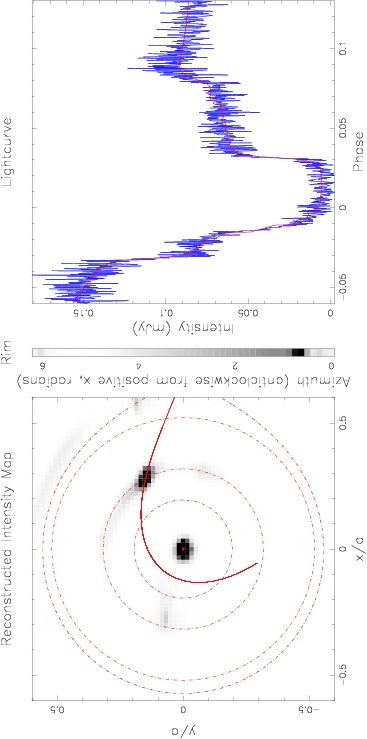} \end{minipage}\\
\begin{minipage}{2.5in}
\includegraphics[width=6.0cm,angle=0.]{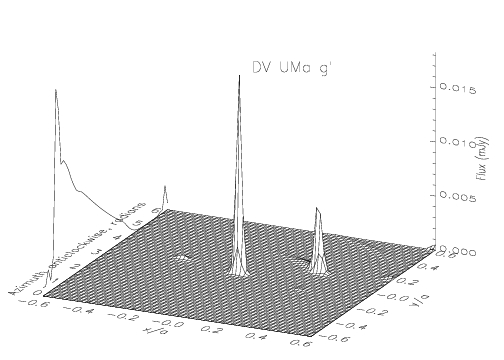} \end{minipage}&
\begin{minipage}{2.5in}
\includegraphics[height=8.0cm,angle=-90.]{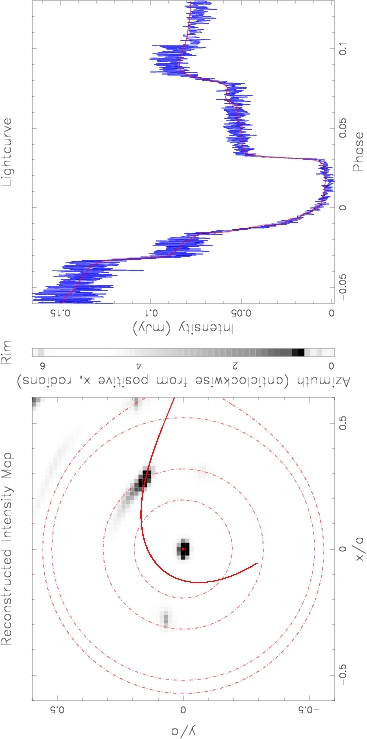} \end{minipage}\\
\begin{minipage}{2.5in}
\includegraphics[width=6.0cm,angle=0.]{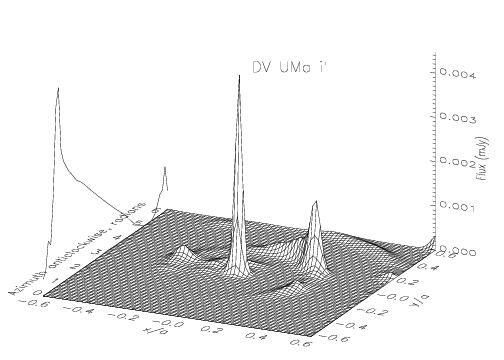} \end{minipage}&
\begin{minipage}{2.5in}
\includegraphics[height=8.0cm,angle=-90.]{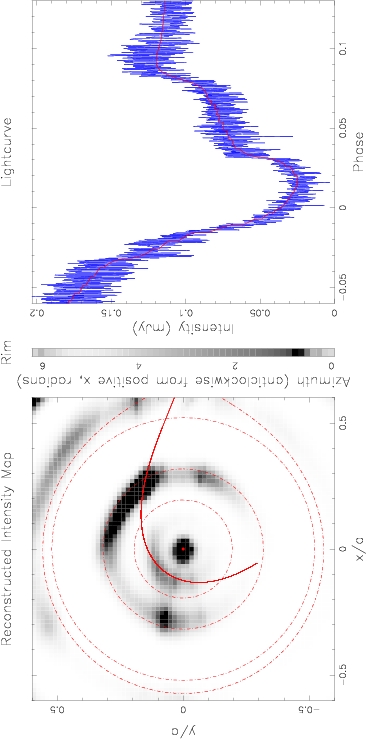} \end{minipage}\\
\end{tabular}
\caption[Eclipse maps for the {\em u\/}$^{\prime}$, {\em g\/}$^{\prime}$
  and {\em i\/}$^{\prime}$ light curves of DV~UMa.]{As
  figure~\ref{fig:em_xzeri}, but for, from top, the {\em
  u\/}$^{\prime}$, {\em g\/}$^{\prime}$ and {\em i\/}$^{\prime}$ data of
  DV~UMa. The system parameters adopted for the reconstruction
  are $q=0.1506$, $i=84.24^{\circ}$ and $R_{\rm d}=0.31805a$}
\label{fig:em_dvuma}
\end{figure}

The eclipse mapping results for DV~UMa are shown in
figures~\ref{fig:em_dvuma}. The main features of the reconstructed
disc are as follows. The white dwarf and bright spot are evident, with
the white dwarf being significantly the brighter of the two. The disc
itself is effectively invisible (the numerous small-scale features are
most likely spurious remnants of the arcs discussed in
\S~\ref{sec:em}). The intensity distribution along the disc rim (of
radius $0.31805a$; see tables~\ref{tab:xzeridvuma_bs},
\ref{tab:xzeri_parameters_lfit} and \ref{tab:dvuma_parameters_lfit})
shows a peak at the location of the bright spot, with smaller features
probably due to flickering in the light curve.

\begin{figure}
\centerline{\includegraphics[width=14cm,angle=0]{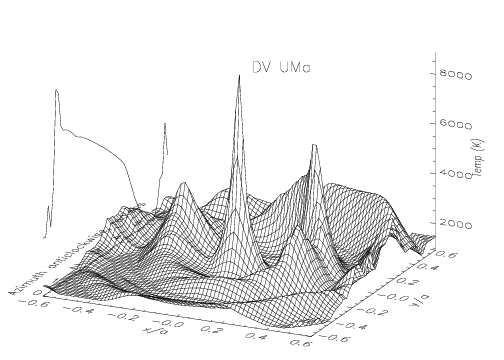}}
\caption[3-dimensional plot of a blackbody fit to the reconstructed
  disc intensities of DV~UMa.]{3-dimensional
  plot of a blackbody fit to the reconstructed disc intensities of
  DV~UMa shown in figure~\ref{fig:em_dvuma}.}
\label{fig:em_dvuma_temp_3d}
\end{figure}

The temperature map of the disc of DV~UMa shown in
figure~\ref{fig:em_dvuma_temp_3d} was produced by fitting a blackbody
function convolved through the filter response functions to each point
in the relevant reconstructed maps. The orbital separation and
distance to the system adopted were those derived in
\S~\ref{sec:xzeridvuma_techniques_model}.

%% file: gycncircomhtcas.tex
\chapter{GY~Cnc, IR~Com and HT~Cas}
\label{ch:gycncircomhtcas}

The contents of this chapter have been accepted for publication in
the Monthly Notices of the Royal Astronomical Society as {\em ULTRACAM
photometry of the eclipsing cataclysmic variables GY~Cnc, IR~Com and
HT~Cas\/} by \citet*{feline05}. The exceptions to this are the eclipse
mapping results for GY~Cnc and IR~Com presented at the end of this
chapter in \S~\ref{sec:htcasem}. The reduction and
analysis of the data are all my own, as is the text below. Dr.~Vik
Dhillon supervised all work presented here.

GY~Cnc (= RX~J0909.8+1849 = HS~0907+1902) is a $V\sim16$~mag eclipsing
dwarf nova with an orbital period  $P_{{\rm orb}}=4.2$~hr. GY~Cnc was
detected in both the Hamburg Schmidt objective prism survey
\citep{hagen95} and the ROSAT Bright Source catalogue \citep{voges99},
and identified as a possible CV by \citet{bade98}. Spectroscopic and
photometric follow-up observations by \citet{gansicke00} confirmed the
status of GY~Cnc as an eclipsing dwarf nova by detecting it in both
outburst and quiescence. \citet{shafter00} used multi-colour
photometric observations of GY~Cnc to determine the temperatures of
the white dwarf, bright spot and accretion disc and the disc power-law
temperature exponent, which they found to be largely independent of
the mass ratio assumed. Spectroscopic and photometric observations
obtained by \citet{thorstensen00} constrain the mass ratio
$q=0.41\pm0.04$ and the orbital inclination
$i=77.0^{\circ}\pm0.9^{\circ}$ (after applying corrections to the
radial velocity of the secondary star $K_{2}$). The spectral type of
the secondary star has been estimated  as M$3\pm1.5$
\citep{gansicke00,thorstensen00}. GY~Cnc was observed during decline
from outburst in 2001 November by \citet{kato02b}, who suggest that
GY~Cnc is an ``above-the-gap counterpart'' to the dwarf nova HT~Cas.

IR~Com (= S~10932~Com) was discovered as the optical counterpart to
the ROSAT X-ray source RX~J1239.5 \citep{richter95}. IR~Com exhibits
high (photographic magnitude ${\rm m}_{\rm ph}=16.5$~mag) and low
(${\rm m}_{\rm ph}=18.5$~mag) brightness states
\citep{richter95,richter97}, with outburst amplitudes of ${\rm m}_{\rm
ph}\sim4.5$~mag
(\citealp{richter97}; \citealp{kato02a}). \citet{wenzel95} detected
eclipses in the light curve of IR~Com and determined an orbital period
of 2.1~hr, just below the period gap. \citet{richter97} present
photometric and spectroscopic observations of IR~Com, which illustrate
the highly variable nature of the target. \citet{kato02a} reported
observations of IR~Com in, and during the decline from, outburst. None
of the published light curves of IR~Com show much evidence for the
presence of an orbital hump before eclipse, or for asymmetry of the
eclipse itself (although the limited time-resolution of the
observations may mask such asymmetries to an extent). \citet{kato02a}
again suggest that IR~Com is a twin of HT~Cas.

HT~Cas is a well-known and well-studied eclipsing dwarf nova. It
has a quiescent magnitude of $V\sim16.4$~mag and an orbital period of
1.77~hr. The literature on HT~Cas is extensive; here I only discuss a
selection of relevant work. The system parameters of HT~Cas have been
well-determined by \citet{horne91b} using simultaneous {\em U, B, V\/}
and {\em R\/} observations in conjunction with those of
\citet{patterson81}: $q=0.15\pm0.03$ and
$i=81.0^{\circ}\pm1.0^{\circ}$. In a companion paper, \citet{wood92}
determined the temperature of the white dwarf ($T=14\,000\pm1000$~K)
and estimated the distance to the system ($D=125\pm8$~pc). They also
eclipse-mapped the accretion disc, illustrating the flat radial
temperature profile typical of quiescent dwarf
nov\ae. \citet{vrielmann02} have recently reconstructed the
temperatures and surface densities of the quiescent accretion disc of
HT~Cas using physical parameter eclipse mapping. This method also
yields an estimate of the distance, $D=207\pm10$~pc. \citet{marsh90c}
detected the secondary star in HT~Cas using low-resolution spectra,
estimating the spectral type as M$5.4\pm0.3$. \citet{marsh90c} found
the secondary star to be consistent with main-sequence values for the
mass, radius and luminosity. \citet{robertson96} discuss the long-term
quiescent light curve of HT~Cas, with particular regard to the
(unusual) presence of high- and low-states (at 16.4 and 17.7~mag,
respectively). \citet{wood95a} detected an X-ray eclipse of HT~Cas
using ROSAT observations during one of the system's low-luminosity
states. The X-rays are believed to originate in a boundary layer
between the white dwarf and inner accretion disc.

In this chapter, I present light curves of GY~Cnc, IR~Com and HT~Cas,
obtained with {\sc ultracam}. The data for GY~Cnc and IR~Com are of the
highest time-resolution yet obtained, and are the first simultaneous,
three-colour light curves for these objects. I present eclipse maps of
HT~Cas in quiescence in 2002 and 2003, which show distinct changes in
the structure of the accretion disc which are related to the overall
brightness of the system.

\begin{figure}
\begin{tabular}{ccc}
\includegraphics[width=9.0cm,angle=-90.]{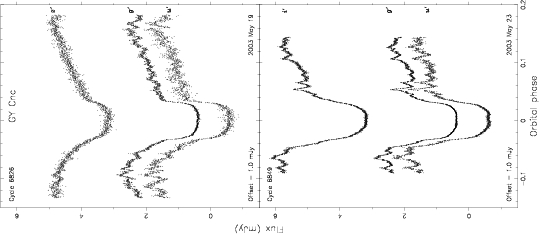} &
\includegraphics[width=9.0cm,angle=-90.]{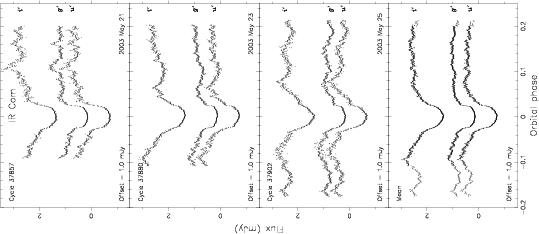} &
\includegraphics[width=9.0cm,angle=-90.]{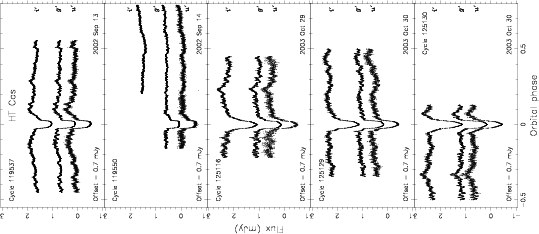} 
\end{tabular}
\caption[The light curves of GY~Cnc, IR~Com and HT~Cas.]{Left. The
  light curves of GY~Cnc. Centre. The light curves of
  IR~Com. Right. The light curves of HT~Cas. The {\em i\/}$^{\prime}$
  and {\em z\/}$^{\prime}$ data are offset vertically upwards and the
  {\em u\/}$^{\prime}$ data are offset vertically downwards by the
  amount specified in the relevant plot. Note that the {\em
  i\/}$^{\prime}$ HT~Cas data of 2002 September 14 were lost due to a
  technical problem with this CCD. The mean light curve of IR~Com is
  also shown.}
\label{fig:gycncircomhtcas_lightcurves}
\end{figure}

The observations of GY~Cnc, IR~Com and HT~Cas are summarised in
table~\ref{tab:journal}, and the data reduction procedure is detailed
in \S~\ref{sec:reduction}. The light curves of GY~Cnc, IR~Com and
HT~Cas are shown in figure~\ref{fig:gycncircomhtcas_lightcurves}. A
technical problem with the {\em i}$^{\prime}$ band CCD destroyed the
data in this band on 2002 September 14 (HT~Cas). Note also that the
timing data of 2003 October 29 (HT~Cas) were corrupted. The relative
timings remained precise, however, enabling the accurate phasing of
the data (see \S~\ref{sec:ephemerides}).

\section{Orbital ephemerides}
\label{sec:ephemerides}

The times of mid-eclipse $T_{{\rm mid}}$
given in table~\ref{tab:gycncircomhtcas_eclipse_times} were determined
as decribed in \S~\ref{sec:derivative}, taking the midpoint of the
white dwarf eclipse as the point of mid-eclipse. If the sharp
eclipse is caused by the obscuration of the bright spot rather than
the white dwarf, then phase zero, as defined by the ephemerides below,
may not necessarily correspond to the conjunction of the white and red
dwarf components. As discussed in \S~\ref{sec:gycnc}--\ref{sec:htcas},
however, it is probable that the sharp eclipse in all three objects
(and certainly HT~Cas) is of the white dwarf.

The orbital ephemeris of GY~Cnc was determined using the seven eclipse
timings of \citet{gansicke00}, the eight timings of \citet{shafter00},
the seven timings of \citet{kato00}, the two timings of
\citet{vanmunster4210}, the four timings of \citet{kato02b} and the
six {\sc ultracam} timings determined in this chapter and given in
table~\ref{tab:gycncircomhtcas_eclipse_times}. Errors adopted were
$\pm1\times10^{-4}$~days for the data of \citet{gansicke00} and
\citet{shafter00}, $\pm5\times10^{-5}$~days for the data of
\citet{kato00}, \citet{vanmunster4210} and \citet{kato02b} and
$\pm1\times10^{-5}$~days for the {\sc ultracam} data. A linear least
squares fit to these times gives the following orbital ephemeris for
GY~Cnc:
\begin{displaymath}
\begin{array}{ccrcrl}
\\ HJD & = & 2451581.826653 & + & 0.1754424988 & E.  \\
 & & 14 & \pm & 21 &
\end{array} 
\end{displaymath}

\begin{table}
\begin{center}
\caption[Mid-eclipse timings of GY~Cnc, IR~Com and
  HT~Cas.]{Mid-eclipse timings of GY~Cnc, IR~Com and HT~Cas. The cycle
  numbers were determined from the ephemerides described in
  \S~\ref{sec:ephemerides}. Note that a technical problem with the
  {\em i\/}$^{\prime}$ CCD corrupted the data during eclipse in this
  band on 2002 September 14. The timings are accurate to
  $\pm1\times10^{-5}$~days (GY~Cnc), $\pm2\times10^{-5}$~days (IR~Com)
  and $\pm5\times10^{-6}$~days (HT~Cas).}
\vspace{0.3cm}
\small
\begin{tabular}{lccccccc}
\hline
Target & UT date at & Cycle & \multicolumn{3}{c}{${\rm HJD}+2\,452\,530$}\\
 & start of night & & {\em u}$^{\prime}$ & {\em g}$^{\prime}$ & {\em
  i}$^{\prime}$ & {\em z}$^{\prime}$\\
\hline
GY~Cnc & 2003 5 19 & 6826 & 249.397235 & 249.397223 & -- &
249.397248 \\
GY~Cnc & 2003 5 23 & 6849 & 253.432215 & 253.432254 &
253.432254 & -- \\

IR~Com & 2003 5 21 & 37857 & 251.503269 & 251.503250 &
251.503194 & -- \\ 
IR~Com & 2003 5 23 & 37880 & 253.505096 & 253.505153 &
253.505153 & -- \\
IR~Com & 2003 5 25 & 37902 & 255.419890 & 255.419966 &
255.419909 & -- \\ 

HT~Cas & 2002 9 13 & 119537 & 1.503015 & 1.503035 & 1.502995 & -- \\
HT~Cas & 2002 9 14 & 119550 & 2.460443 & 2.460477 & -- & -- \\
HT~Cas & 2003 10 30 & 125129 & 413.338089 & 413.338159 & 413.338199 &
-- \\ 
HT~Cas & 2003 10 30 & 125130 & 413.411832 & 413.411792 & 413.411769 &
-- \\ 
\hline
\end{tabular}
\normalsize
\label{tab:gycncircomhtcas_eclipse_times}
\end{center}
\end{table}

To determine the orbital ephemeris of IR~Com, I used the 24 timings of
\citet[as listed in \citealp{kato02a}]{richter97}, the 14 eclipse
timings of \citet{kato02a} and those nine, given in
table~\ref{tab:gycncircomhtcas_eclipse_times}, determined from the
{\sc ultracam} data. The errors adopted for the data of \citet{richter97}
and \citet{kato02a} were $\pm1\times10^{-3}$~days for cycles $-134516$,
$-51035$, $-42189$, $-29597$ and $-21531$ and
$\pm5\times10^{-5}$~days for subsequent cycles, except where stated
otherwise by \citet{kato02a}.  Those adopted for the {\sc ultracam} timings
were $\pm2\times10^{-5}$~days. The orbital ephemeris of IR~Com was
determined by a linear least squares fit to the above timings, and is
\begin{displaymath}
\begin{array}{ccrcrl}
\\ HJD & = & 2449486.4818691 & + & 0.08703862787 & E.  \\
 & & 26 & \pm & 20 &
\end{array} 
\end{displaymath}

To determine the orbital ephemeris of HT~Cas I used the 11 mid-eclipse
times of \citet{patterson81}, the 23 times of \citet{zhang86}, the 15
times of \citet{horne91b} and the 11 {\sc ultracam} times given in
table~\ref{tab:gycncircomhtcas_eclipse_times}. The times of
\citet{patterson81}, \citet{zhang86} and \citet{horne91b} were
assigned errors of $\pm5\times10^{-5}$~days and the times in
table~\ref{tab:gycncircomhtcas_eclipse_times} assigned errors of
$\pm5\times10^{-6}$~days. A linear least squares fit to these times gives
the following orbital ephemeris for HT~Cas:
\begin{displaymath}
\begin{array}{ccrcrl}
\\ HJD & = & 2443727.937290 & + & 0.07364720309 & E.\\ & & 8 & \pm
 & 7 &
\end{array} 
\end{displaymath}

The loss of accurate timings for the 2003 October 29 HT~Cas data meant
that these data were phased according to the orbital period derived
above, but with a different zero-point. The zero-point used instead
was the mid-point of the observed eclipse. Note that the relative
timings for this data remained accurate; the times were merely out by
a constant, unknown, offset. The cycle number was accurately
determined from the times in the hand-written observing log. This may
result in a slight fixed time offset for these data due to the
uncertainty in determining the point of mid-eclipse.

These ephemerides were used to phase all of the data.

\begin{figure}
\centerline{\includegraphics[height=12cm,angle=-90.]{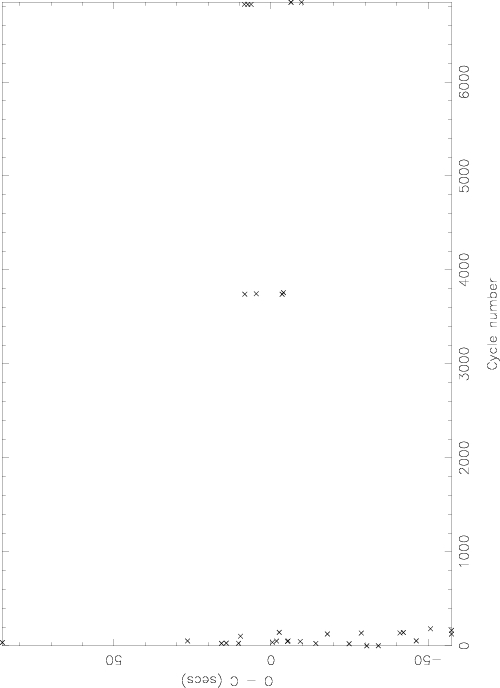}}
\caption[The $O-C$ diagram of GY~Cnc produced using a linear
  ephemeris.]{The $O-C$ diagram of GY~Cnc produced using a linear
  ephemeris as described in the text.}
\label{fig:gycnc_oc}
\end{figure}

\begin{figure}
\centerline{\includegraphics[height=12cm,angle=-90.]{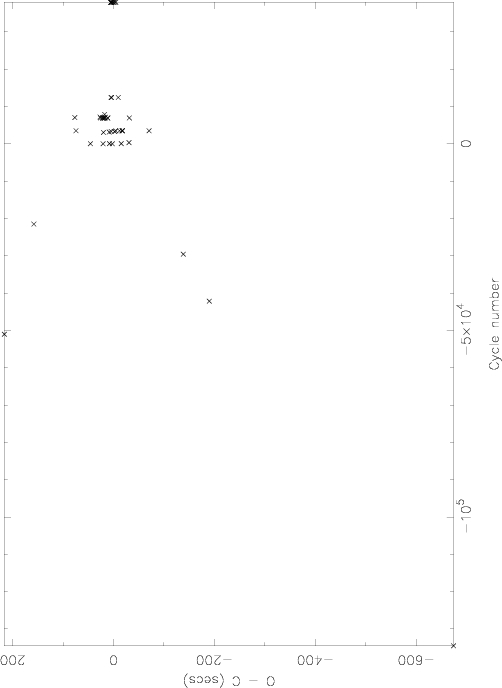}}
\caption[The $O-C$ diagram of IR~Com produced using a linear
  ephemeris.]{The $O-C$ diagram of IR~Com produced using a linear
  ephemeris as described in the text.}
\label{fig:ircom_oc}
\end{figure}

\begin{figure}
\centerline{\includegraphics[height=12cm,angle=-90.]{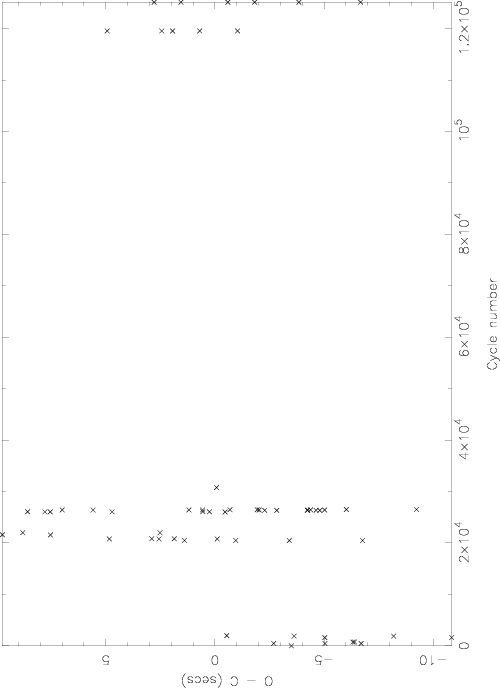}}
\caption[The $O-C$ diagram of HT~Cas produced using a linear
  ephemeris.]{The $O-C$ diagram of HT~Cas produced using a linear
  ephemeris as described in the text.}
\label{fig:htcas_oc}
\end{figure}

The $O-C$ diagrams for GY~Cnc, IR~Com and HT~Cas produced using the above
ephemerides and times of mid-eclipse are shown in
figures~\ref{fig:gycnc_oc}, \ref{fig:ircom_oc} and \ref{fig:htcas_oc},
respectively. None shows any convincing evidence for period
change.\footnote{Note that the data point at the lower left of
figure~\ref{fig:ircom_oc} was observed with very poor time-resolution
of 1800~sec, greater than the $O-C$ residual, and therefore does not
(alone; the other points are consistent with the derived period)
constitute significant evidence for period change.}

%________________________________________________________________

\section{GY~Cnc}
\label{sec:gycnc}

In keeping with previous observations (summarised in
the introduction to this chapter), the light curve of GY~Cnc shown in
figure~\ref{fig:gycncircomhtcas_lightcurves} shows a deep primary
eclipse, with the {\em g\/}$^{\prime}$ flux dropping from a peak value
of approximately 3~mJy (15.2~mag) to about 0.6~mJy (17.0~mag) at
mid-eclipse. This places the system slightly above its quiescent
brightness of $V=16$~mag, shortly after an outburst which reached
twelfth-magnitude on 2003 May 13 (Waagen, private communication;
observed by the amateur organisation the American Association  of
Variable Star Observers, AAVSO). The system was therefore likely to
still be in decline from outburst. The eclipse morphology
appears to be that of a gradual disc eclipse with a sharp eclipse of
the white dwarf or bright spot superimposed thereon. The sharp eclipse
is most probably that of a white dwarf, not the bright spot, as in
both cycles the ingress and egress are of the same order in terms of
both duration and depth. The eclipse is flat-bottomed, suggesting that
the disc and white dwarf are completely obscured at these phases. The
eclipse of the disc appears to be asymmetric, with the ingress being
rather sharper than the more gradual egress. This is indicative of
asymmetry in the disc structure, possibly due to an extended bright
spot at the disc rim. The changing foreshortening of the bright spot,
the cause of the orbital hump often observed in other dwarf nov\ae,
would also account for the rather greater flux before eclipse than
after. Indeed, the ingress observed on 2003 May 23 appears to show two
steps, which I attribute to first the bright spot and then the white
dwarf entering eclipse. The likely presence of an extended bright spot
is another reason why I suspect that the sharp, discrete eclipse
visible in both nights' data is that of the white dwarf. The light
curve of figure~\ref{fig:gycncircomhtcas_lightcurves} is
morphologically similar to quiescent light curves in the literature
\citep{gansicke00,shafter00,thorstensen00}. These data were not
suitable to use for determining the system parameters via either the
derivative or {\sc lfit} techniques. In the first case, the bright
spot egress is contaminated by flickering and is too gradual to
produce a clear peak in the derivative of the light curve. In the
second case, the presence of large-scale flickering prevented the
program from locating the ingress and egress points reliably.

\begin{table}
\begin{center}
\caption[White dwarf contact phases of IR~Com.]{White dwarf contact
phases of IR~Com, as defined in the text, accurate to $\pm0.00023$.}
\vspace{0.3cm}
\small
\begin{tabular}{lccccccc}
\hline
Date & Band & $\phi_{{\rm w}1}$ & $\phi_{{\rm w}2}$ & $\phi_{{\rm
    w}3}$ & $\phi_{{\rm w}4}$ & $\phi_{{\rm wi}}$ & $\phi_{{\rm we}}$
\\
\hline
2003 5 21 & {\em u\/}$^{\prime}$ & --0.031748 & --0.020917 &
0.022841 & 0.028474 & --0.025249 & 0.024575 \\
          & {\em g\/}$^{\prime}$ & --0.026983 & --0.021351 & 
0.021108 & 0.027174 & --0.023517 & 0.025440 \\
          & {\em i\/}$^{\prime}$ & --0.030015 & --0.018318 & 
0.019808 & 0.028907 & --0.025249 & 0.024575 \\
2003 5 23 & {\em u\/}$^{\prime}$ & --0.027904 & --0.021404 & 
0.021487 & 0.026686 & --0.026170 & 0.025388 \\
          & {\em g\/}$^{\prime}$ & --0.026603 & --0.021837 & 
0.021487 & 0.026253 & --0.024003 & 0.025388 \\
          & {\em i\/}$^{\prime}$ & --0.028770 & --0.020971 & 
0.023221 & 0.025820 & --0.021837 & 0.025388 \\
2003 5 25 & {\em u\/}$^{\prime}$ & --0.027680 & --0.020744 & 
0.020848 & 0.027346 & --0.026376 & 0.024311 \\
          & {\em g\/}$^{\prime}$ & --0.026376 & --0.021610 & 
0.022579 & 0.026913 & --0.024645 & 0.025176 \\
          & {\em i\/}$^{\prime}$ & --0.029844 & --0.016411 & 
0.020415 & 0.026047 & --0.022914 & 0.022579 \\
Mean      & {\em u\/}$^{\prime}$ & --0.027811 & --0.022901 & 
0.022448 & 0.027358 & --0.025788 & 0.025046 \\
\hspace{0.1cm} light & {\em g\/}$^{\prime}$ & --0.026077 & --0.021744 & 
0.021292 & 0.027069 & --0.023478 & 0.024758 \\
\hspace{0.1cm} curve & {\em i\/}$^{\prime}$ & --0.028965 & --0.019722 & 
0.021003 & 0.027358 & --0.027231 & 0.025625 \\
\hline
\end{tabular}
\normalsize
\label{tab:ircom_timings}
\end{center}
\end{table}

\section{IR~Com}
\label{sec:ircom}

The light curve of IR~Com, shown in
figure~\ref{fig:gycncircomhtcas_lightcurves}, also exhibits a deep
primary eclipse. The light curve is highly variable outside of
eclipse, with a maximum {\em g\/}$^{\prime}$ flux of about 1.6~mJy
(15.9~mag) and a minimum during eclipse of approximately 0.16~mJy
(18.4~mag). The average out-of-eclipse {\em g\/}$^{\prime}$ flux level
of IR~Com during my observations was 1.0~mJy (16.4~mag), consistent
with the system being in quiescence. From the light curve of IR~Com
shown in figure~\ref{fig:gycncircomhtcas_lightcurves} it is clear that
the light curve of this object is highly variable outside of
eclipse. There is a
clear eclipse of a compact structure, either the white dwarf or bright
spot, as evidenced by the sharpness of the ingress and egress. The
highly variable nature of the light curve of IR~Com makes it difficult
to determine whether the sharp eclipse is of the white dwarf or bright
spot. The mean light curve of IR~Com shown in
figure~\ref{fig:gycncircomhtcas_lightcurves} shows the main features
of the light curve much more clearly, as flickering is much
reduced. The sharp eclipse is revealed to be nearly symmetric, with
evidence for an eclipse of the disc in the V-shaped eclipse bottom and
the slopes before and after the sharp eclipse. No sign of the eclipse
of another compact object is seen, so the sharp eclipse must be of the
white dwarf (in which case the bright spot is extremely faint) or an
eclipse of the bright spot (in which case the white dwarf remains
visible at all phases).

Contact phases of the sharp eclipse of IR~Com were determined using
the derivative of the light curve, as described in
\S~\ref{sec:derivative}. These timings, given in
table~\ref{tab:ircom_timings}, do not show any evidence for asymmetry
in the duration of ingress and egress (as is frequently the case with
the eclipse of a bright spot, where the ingress is of a longer
duration than the egress). Additionally, using the \citet{nauenberg72}
mass--radius relation for a cold, rotating white dwarf with Kepler's
third law (as described in \S~\ref{sec:massdet}) for reasonable values
of $q$ shows that the eclipse contact phases are entirely consistent
with the eclipsed object being of the correct size for a white
dwarf. As \citet{kato02a} point out, their mid-eclipse timings show no
significant differences between outburst and quiescence, implying that
in both quiescence and outburst the brightness distribution is centred
on the white dwarf. These points lead me to believe that the primary
eclipse is of the white dwarf, rather than of the bright spot.

No unambiguous bright spot feature is visible in either the individual
or mean light curves of IR~Com shown in
figure~\ref{fig:gycncircomhtcas_lightcurves}. The absence of a
bright spot eclipse prevented the determination of the system
parameters via either the derivative or {\sc lfit} techniques. From
the absence of flickering during primary eclipse, it appears that the
flickering is confined to the inner regions of the accretion disc or
the white dwarf itself. The origin of the flickering in IR~Com is most
likely the boundary layer between the white dwarf and accretion disc
\citep[eclipse mapping of the flickering sources in CVs is discussed
in, for example,][]{bruch00b,baptista04,baptista04b}. The absence of a
bright spot eclipse prevented the determination of the system
parameters via either the derivative or {\sc lfit} techniques.

\begin{figure}
\centerline{\includegraphics[height=12cm,angle=-90.]{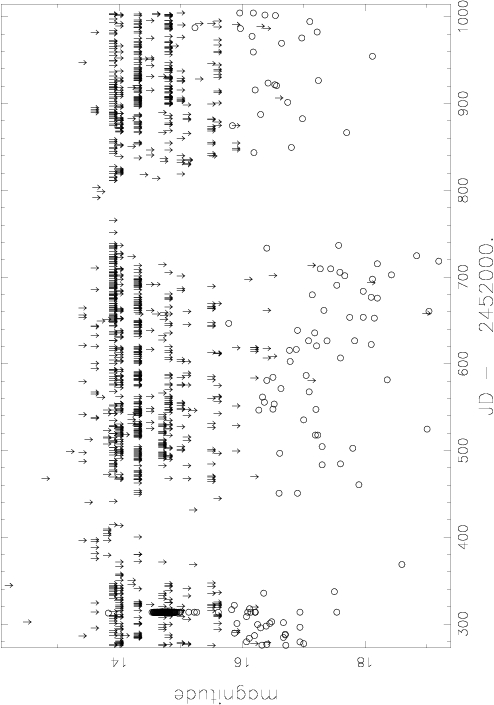}}
\caption[The long-term light curve of HT~Cas.]{The long-term light
  curve of HT~Cas, courtesy of the AAVSO (Waagen, private
  communication). Open circles are V band observations; arrows mark
  upper limits on the magnitude of the system. The Julian date scale
  corresponds to calendar dates from 2002 January 1 to 2004 January
  1. Note the outburst in 2002 February, which peaked on 2002 February
  6 ($={\rm JD}\;2\,452\,312$).}
\label{fig:htcas_aavso}
\end{figure}

\begin{figure}
\begin{tabular}{cc}
\begin{minipage}{2.5in}
\includegraphics[width=6.0cm,angle=0.,clip]{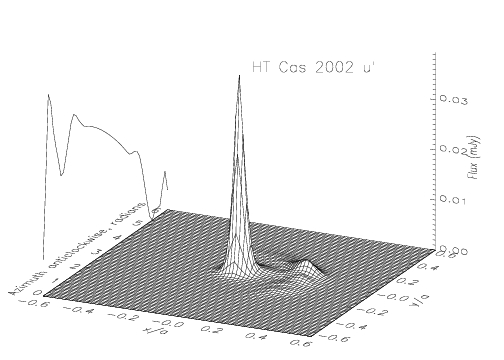} 
\end{minipage} &
\begin{minipage}{2.5in}
\includegraphics[width=4.0cm,angle=-90.,clip]{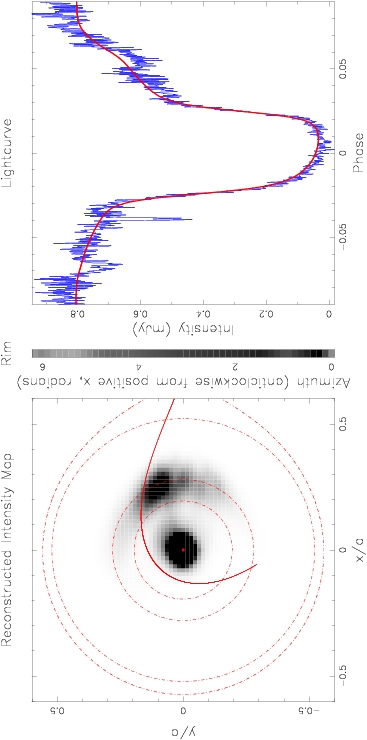} 
\end{minipage} \\
\begin{minipage}{2.5in}
\includegraphics[width=6.0cm,angle=0.,clip]{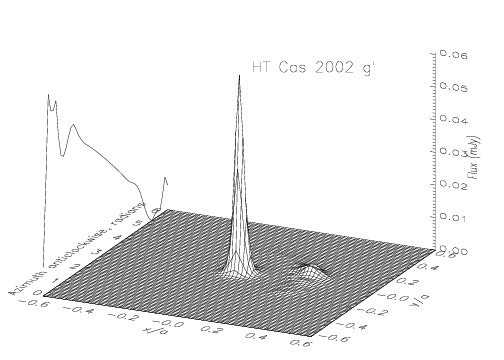} 
\end{minipage} &
\begin{minipage}{2.5in}
\includegraphics[width=4.0cm,angle=-90.,clip]{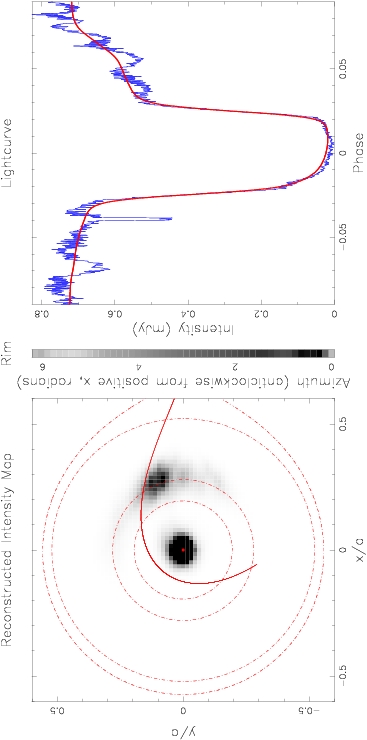} 
\end{minipage} \\
\begin{minipage}{2.5in}
\includegraphics[width=6.0cm,angle=0.,clip]{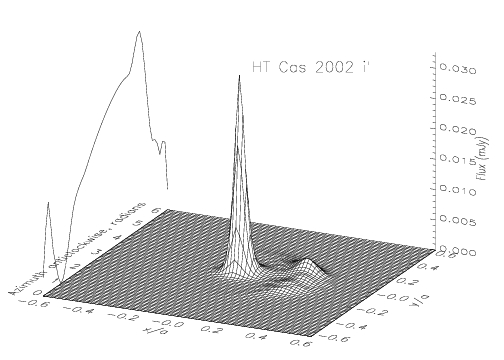} 
\end{minipage} &
\begin{minipage}{2.5in}
\includegraphics[width=4.0cm,angle=-90.,clip]{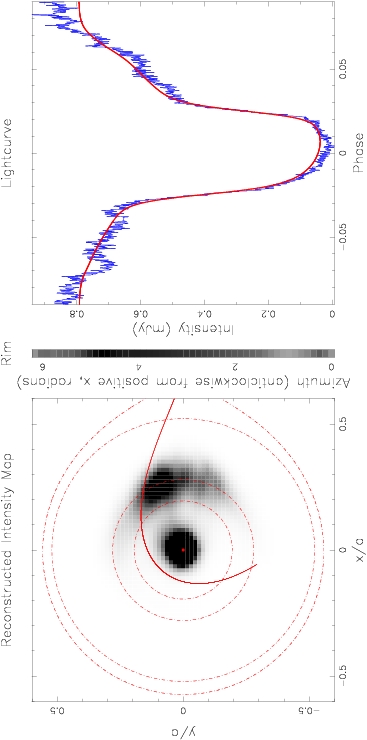} 
\end{minipage} \\
\end{tabular}
\caption[Eclipse maps for the {\em u\/}$^{\prime}$, {\em
  g\/}$^{\prime}$ and {\em i\/}$^{\prime}$ 2002 September 13--14 light
  curves of HT~Cas.]{As figure~\ref{fig:em_xzeri}, but for, from top,
  the {\em u\/}$^{\prime}$, {\em g\/}$^{\prime}$ and {\em
  i\/}$^{\prime}$ 2002 September 13--14 light curves of HT~Cas. The
  parameters adopted for these reconstructions were those derived by
  \citet{horne91b}, $q=0.15$ and $i=81.0^{\circ}$. The disc radius was
  estimated from the position of the reconstructed bright spot, and is
  $R_{\rm d}=0.28a$. Prior to fitting, a (constant) offset was
  subtracted from the light curves. This offset was 0.15, 0.09 and
  0.32~mJy for the {\em u}$^{\prime}$, {\em g}$^{\prime}$ and {\em
  i}$^{\prime}$ data, respectively. This offset was 0.15, 0.09 and
  0.32~mJy for the {\em u\/}$^{\prime}$, {\em g\/}$^{\prime}$ and {\em
  i\/}$^{\prime}$ data, respectively. The sharp dip visible at about
  phase $-0.04$ in the {\em u\/}$^{\prime}$ and {\em g\/}$^{\prime}$
  light curves is due to a short gap between observing runs on 2003
  September 13 coinciding with a dip in the 2003 September 14 light
  curve.}
\label{fig:htcas_2002}
\end{figure}

\begin{figure}
\begin{tabular}{cc}
\begin{minipage}{2.5in}
\includegraphics[width=6.0cm,angle=0.]{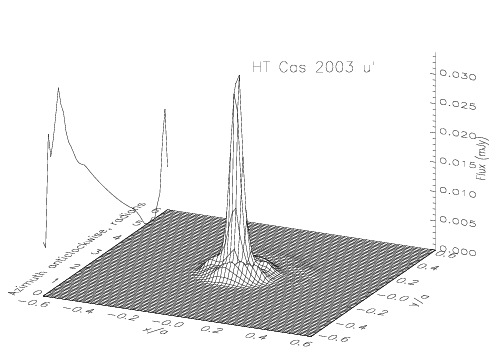} 
\end{minipage}&
\begin{minipage}{2.5in}
\includegraphics[width=4.0cm,angle=-90.]{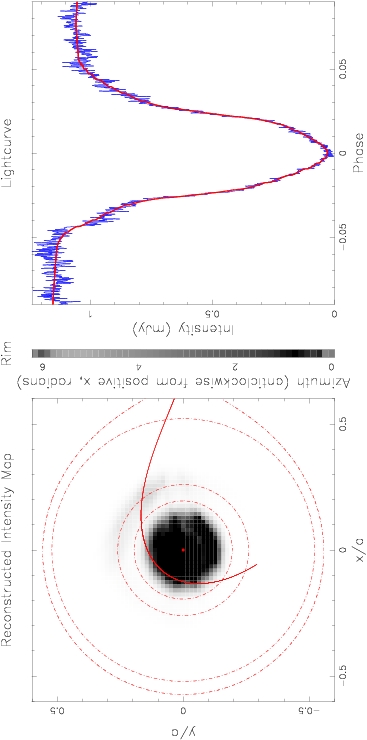} 
\end{minipage}\\
\begin{minipage}{2.5in}
\includegraphics[width=6.0cm,angle=0.]{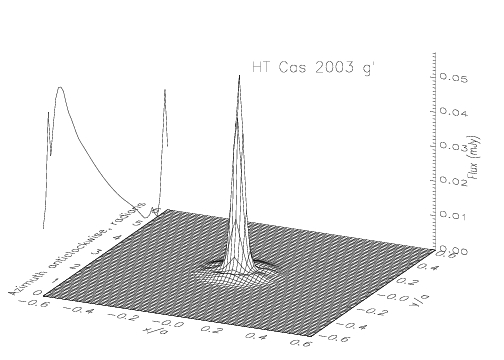} 
\end{minipage}&
\begin{minipage}{2.5in}
\includegraphics[width=4.0cm,angle=-90.]{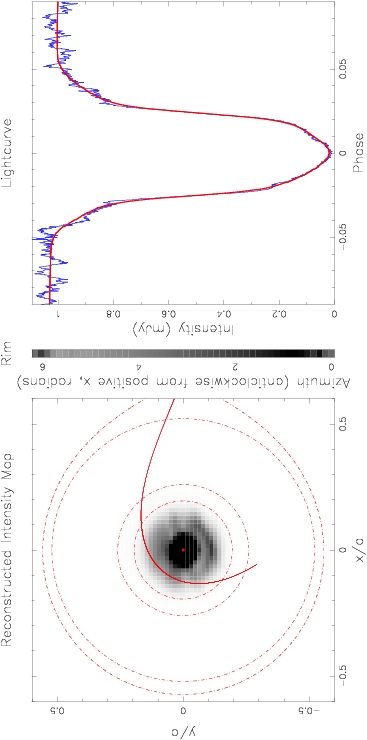} 
\end{minipage}\\
\begin{minipage}{2.5in}
\includegraphics[width=6.0cm,angle=0.]{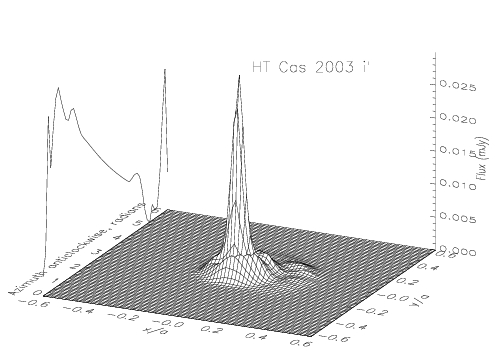} 
\end{minipage} &
\begin{minipage}{2.5in}
\includegraphics[width=4.0cm,angle=-90.]{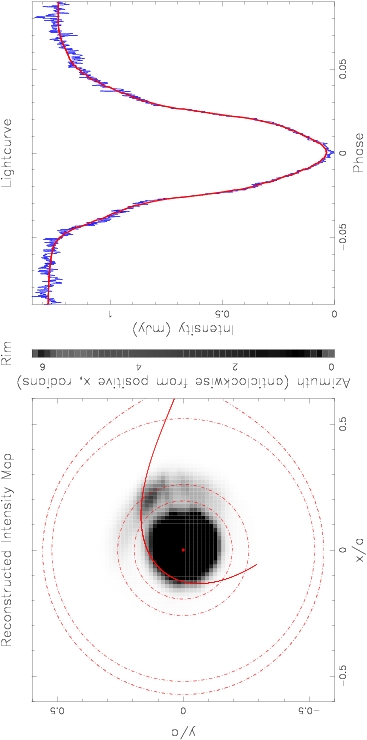} 
\end{minipage}\\
\end{tabular}
\caption[Eclipse maps for the {\em u\/}$^{\prime}$, {\em
  g\/}$^{\prime}$ and {\em i\/}$^{\prime}$ 2003 October 29-30 light
  curves of HT~Cas.]{As figure~\ref{fig:em_xzeri}, but for the {\em
  u\/}$^{\prime}$ (top), {\em g\/}$^{\prime}$ (middle) and {\em
  i\/}$^{\prime}$ (bottom) 2003 October 29--30 light curves of
  HT~Cas. The radius of the disc (the same as that of the disc rim)
  was estimated from the position of the bright spot in the eclipse
  maps, and is $0.26a$. The {\em u}$^{\prime}$, {\em g}$^{\prime}$ and
  {\em i}$^{\prime}$ light curves were offset vertically by 0.24, 0.14
  and 0.42~mJy, respectively.}
\label{fig:htcas_2003}
\end{figure}

\section{HT~Cas}
\label{sec:htcas}
\subsection{Light curve morphology}

The light curves of HT~Cas shown in
figure~\ref{fig:gycncircomhtcas_lightcurves} are typical of those
found in the literature \citep[e.g.][]{patterson81, horne91b}. The
typical out-of-eclipse {\em g\/}$^{\prime}$ flux for the 2002 data is
0.7~mJy (16.8~mag), and the typical mid-eclipse {\em g\/}$^{\prime}$
flux is 0.1~mJy (18.9~mag; see also figure~\ref{fig:htcas_2002}). The
peak {\em g\/}$^{\prime}$ flux in the 2002 dataset is approximately
1.0~mJy (16.4~mag). HT~Cas is slightly brighter in the 2003 data: the
typical out-of-eclipse {\em g\/}$^{\prime}$ flux is 1.2~mJy
(16.2~mag), and the mid-eclipse {\em g\/}$^{\prime}$ flux is again
about 0.1~mJy (18.9~mag; see also figure~\ref{fig:htcas_2003}). The
2003 data set is entirely consistent with HT~Cas being in its high
(brighter) quiescent state (HT~Cas exhibits much variability in
its quiescent magnitude; see \citealp{robertson96}), but the 2002
data appears to be
somewhere between the high and low states described therein. The data
from 2002 show other clear differences from the data of 2003. First,
the eclipse bottoms in 2002 were much flatter than those of 2003,
implying that the brightness distribution in 2002 was more centrally
concentrated than in 2003. Second, the eclipse depth in 2003 was
greater than in 2002, which, as we shall see, is due to increased
disc emission and not an increase in the brightness of the white
dwarf. There is also visible in the 2002 data a clear shoulder during
egress. This appears to be the egress of the bright spot, similar in
appearance to the feature seen in the light curves of
\citet{patterson81}. Unfortunately for the aim of determining the
system parameters from the eclipse contact phases, no bright spot
ingress is visible. This variability of the light curve is typical of
HT~Cas \citep[e.g.][]{patterson81,horne91b,robertson96}.  The fact
that the flux increase between 2002 and 2003 is slight illustrates
that these brightness and morphological variations are not due to an
overwhelming increase in the disc flux `drowning out' the bright spot,
as occurs during outburst. Observations of HT~Cas by the AAVSO
(Waagen, private communication; figure~\ref{fig:htcas_aavso}) show no
evidence for  an outburst of HT~Cas before the 2003 October
observations---the change in the light curve is not due to the system
being on the way up or down from an outburst. I term the 2003 and 2002
data the `high' and `low' quiescent states, respectively (although I
caution that these probably differ in both underlying cause and
detailed observational properties from the various high and low
quiescent states discussed in the literature).

%________________________________________________________________

\subsection{Eclipse mapping}
\label{sec:htcasem}
The accretion disc of HT~Cas was mapped using the techniques described
in detail in \S~\ref{sec:em}. The data of 2002 and 2003 were mapped
independently, since subsequent analysis showed there to be
significant differences, discussed below, between the two data epochs.

I have estimated the radius of the disc rim from the position of the
reconstructed bright spot, since the disc radius, of $0.28a$ in 2002
and $0.26a$ in 2003, is larger than that derived by \citet{horne91b}
of $0.23a$. I have chosen not to deconvolve and remove the white dwarf
from the light curves as the presence of flickering and the lack of a
clear distinction between the eclipses of the white dwarf, bright spot
and accretion disc make this difficult to do so reliably (see,
for example, chapter~\ref{ch:xzeridvuma}). Besides, clear evidence for
the features reproduced in the eclipse maps and discussed below can be
seen directly in the light curves themselves.

As only a few light curves were used, the noise in the light curves is
dominated by flickering rather than by photon noise. Iterating to a
reduced $\chi^{2}=1$ is therefore inappropriate in this case and leads
to the noise in the light curves (flickering) being transposed to the
eclipse maps. Consequently, the eclipse maps were computed by
progressively relaxing the $\chi^{2}$ constraint until the noise in
the eclipse maps was satisfactorily ameliorated (as judged by visual
inspection).

\begin{figure}
\begin{tabular}{c}
\centerline{\includegraphics[width=8.0cm,angle=-90.]{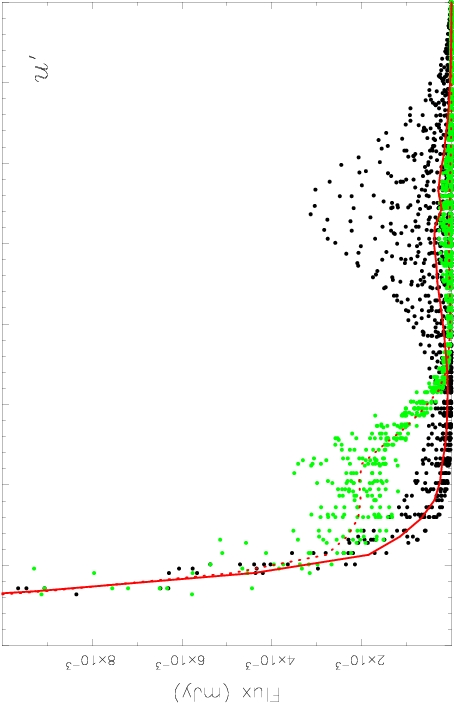}} \\
\centerline{\includegraphics[width=8.0cm,angle=-90.]{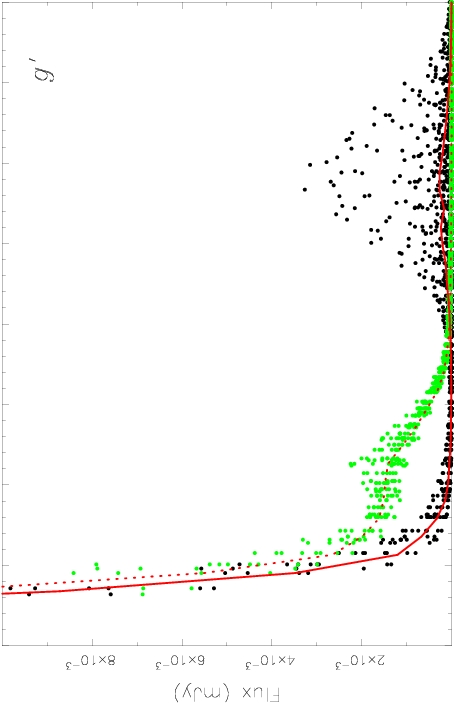}} \\
\centerline{\includegraphics[width=8.1cm,angle=-90.]{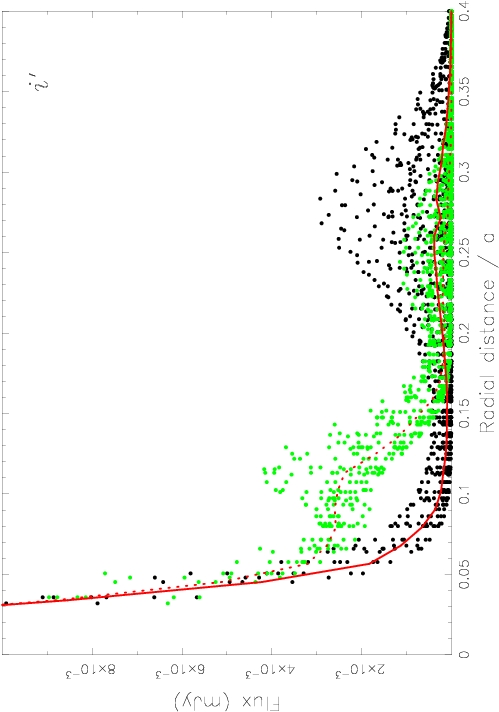}} \\
\end{tabular}
\caption[The radial flux distribution of the reconstructed accretion
  disc of HT~Cas for the 2002 and 2003 data.]{The radial flux
  distribution of the reconstructed accretion disc of HT~Cas for the
  2002 data (black dots and solid red line) and the 2003 data (green
  dots and dotted red line). The dots represent the flux and radius of
  the individual grid elements; the red lines represent the mean flux
  in concentric annuli. The {\em u\/}$^{\prime}$ and {\em g\/}$^{\prime}$
  flux distributions were determined using the data of 2002 September
  13--14 and 2003 October 29--30, whereas the {\em i\/}$^{\prime}$
  distributions were determined using the data of 2002 September 13
  and 2003 October 29--30 (due to a loss of sensitivity in the {\em
  i\/}$^{\prime}$ chip on 2002 September 14.)}
\label{fig:htcas_radial}
\end{figure}

The reconstructed eclipse maps of HT~Cas shown in
figures~\ref{fig:htcas_2002} and \ref{fig:htcas_2003} show distinct
morphological changes from 2002 to 2003. In both 2002 and 2003 the
flux comes primarily from the white dwarf, but in 2002 there is clear
evidence for a faint bright spot in the outer regions of the accretion
disc. The 2003 eclipse maps show a very weak bright spot in the {\em
u}$^{\prime}$ and {\em i}$^{\prime}$ passbands only. The absence of a
bright spot in the 2003 {\em g}$^{\prime}$ band may, however, be a
contrast effect, as the white dwarf is approximately twice as bright
in {\em g}$^{\prime}$ as in the other bands. Not only is the bright
spot much fainter/absent in the 2003 reconstructed maps, but there is
evidence for emission from the inner portions of the accretion
disc. This is best illustrated by the radial flux profiles shown in
figure~\ref{fig:htcas_radial}. Comparing the two radial profiles to
the light curves shown in figure~\ref{fig:gycncircomhtcas_lightcurves}
demonstrates that the increased emission from the inner disc
corresponds to a higher overall brightness state. The greater flux in
2003 is not, upon inspection of the eclipse maps, due to increased
emission from the white dwarf, which is actually fainter in 2003 than
in 2002, but due to a brighter inner disc.

By summing the flux from each element of each eclipse map whose centre
lies within $0.03a$ of the centre of the white dwarf and fitting the
resulting colours to the hydrogen-rich, $\log g=8$ white dwarf model
atmospheres of \citet{bergeron95}, converted to the SDSS system using
the observed transformations of \citet{smith02}, the temperature of
the white dwarf was determined to be $T_{\rm 1}=15\,000\pm1000$~K in
2002 and $T_{\rm 1}=14\,000\pm1000$~K in 2003. Varying the distance
from the white dwarf over which the summation took place between
$0.01a$--$0.07a$ did not significantly affect the colours and
therefore did not significantly affect these temperature
estimates. These temperatures are consistent with those found by
\citet[$T_{\rm 1}=14\,000\pm1000$~K]{wood92} and by \citet[$T_{\rm
1}=15\,500$~K]{vrielmann02} from the same set of quiescent photometric
observations. The effect of the variable nature of the accretion disc of HT~Cas is
evident in the white dwarf temperatures of HT~Cas derived by \citet{wood95a},
of $T_{\rm 1}=13\,200\pm1200$~K during a low state (which is
consistent with these results) and $T_{\rm 1}=18\,700\pm1800$~K during
a normal state (which differs from these results by $\sim2\sigma$).

\begin{figure}
\begin{tabular}{c}
\centerline{\includegraphics[width=10.0cm,angle=-90.]{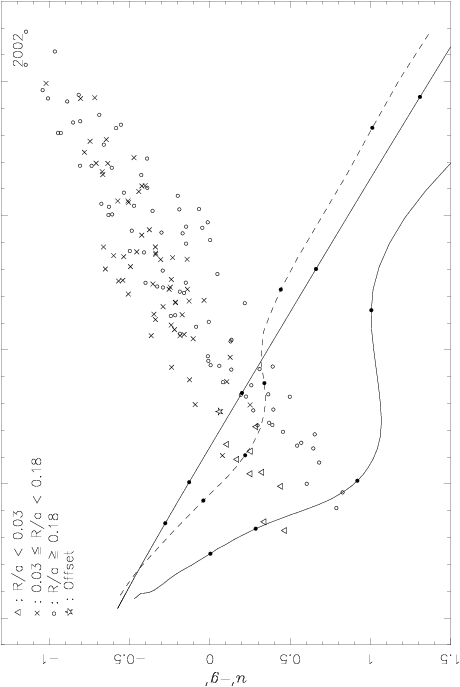}} \\
\centerline{\includegraphics[width=10.0cm,angle=-90.]{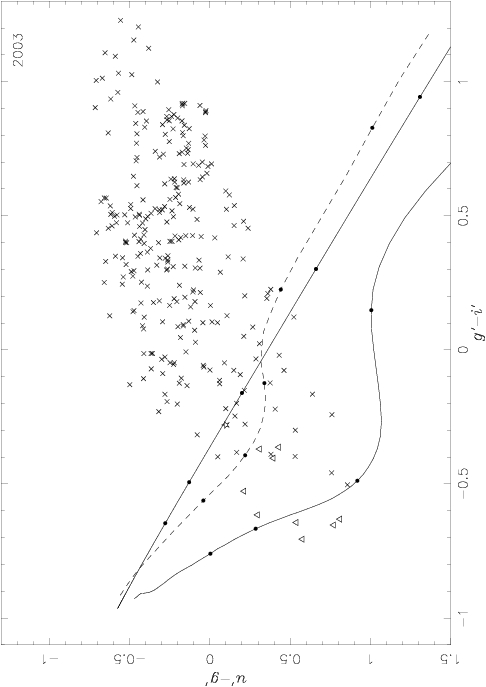}}
\end{tabular}
\caption[Colour-colour diagrams of the accretion disc of HT~Cas in
  2002 and 2003.]{Colour-colour diagrams of the accretion disc of
  HT~Cas in (top) 2002 and (bottom) 2003. The solid straight line is a
  blackbody relationship, the solid curve is the main-sequence
  relationship of \citet{girardi04} and the dashed curve is the white
  dwarf model atmosphere relation of \citet{bergeron95} described in
  section~\ref{sec:fitting}. The filled circles superimposed upon each
  of these lines indicate temperatures of $20\,000$, $15\,000$,
  $10\,000$, $7000$ and $5000$~K, with the hotter temperatures located
  at the upper left of the plots (the $5000$~K point for the
  main-sequence curve lies off the plot). Each of the other points
  represents one element
  of the eclipse  map. Elements at different radial distances $R$ from
  the centre of the white dwarf are plotted using different markers,
  as indicated in the figure. The position of the mid-eclipse (offset)
  flux is also plotted. In the interests of clarity, only points where
  the flux in all three passbands was greater than
  $5\times10^{-4}$~mJy were plotted.}
\label{fig:htcas_colours}
\end{figure}

In figure~\ref{fig:htcas_colours} I present the colour-colour diagrams
for the 2002 and 2003 eclipse maps of HT~Cas. In both 2002 and 2003
the scatter of the data points from the central regions of the disc
($R/a<0.03$), comprising the white dwarf and boundary layer, is
consistent with an increase in the {\em g}$^{\prime}$ flux over that
expected from a lone white dwarf shifting the position of the data
point down and to the left on figure~\ref{fig:htcas_colours}. This
suggests a contribution to this
flux from the boundary layer surrounding the white dwarf. This excess
{\em g}$^{\prime}$ flux is (marginally) more pronounced in 2003 than
in 2002, as might be expected given the differences between the
distribution of the disc flux for these dates. The emission from both
the inner ($0.03\leq R/a<0.18$) and outer ($R/a\geq0.18$) regions of
the disc are concentrated to the right of the blackbody relation in
figure~\ref{fig:htcas_colours}, possibly due to Balmer emission in the
{\em u}$^{\prime}$ band, suggesting that the disc is optically
thin \citep[e.g.][]{horne85a,wood92,baptista96}. There seems to be no
significant difference between the colours of the inner and outer
discs of 2002 (in 2003 the outer disc was too faint to be plotted on
figure~\ref{fig:htcas_colours}). Interestingly, in both 2002 and 2003,
the offset colours (the colours of the flux subtracted from the light
curves prior to fitting) lie on the blackbody relation rather than
being on the main-sequence curve. A blackbody fit to the 2002 and 2003
offset colours gave $T\sim11\,000$~K and $T\sim11\,200$~K,
respectively. 

GY~Cnc was eclipse mapped using the system parameters of $q=0.421$
and $i=77.4^{\circ}$. These were calculated from the
secondary mass-orbital period relation of \citet{smith98a}, given in
equation~\ref{eq:massradius2}, assuming a primary mass of
$M_{1}/{\rm M}_{\odot}=1.0$. The orbital inclination was then determined
from the eclipse phase width of $\Delta\phi=0.06364\pm0.00008$ determined
from the eclipse timings. For the purposes of modelling the orbital
hump using the disc rim, the radius of the accretion disc was assumed
to be $R_{{\rm d}}=0.3a$. This value was, arbitrarily, chosen on the
basis that it is similar to those of the other dwarf nov\ae\ discussed
in this thesis, OU~Vir, XZ~Eri, DV~UMa and HT~Cas.

As the light curves of IR~Com are highly variable, IR~Com was eclipse
mapped using the mean light curve. The system parameters adopted were
$q=0.153$ and $i=81.1^{\circ}$. These were again calculated from the
secondary mass-orbital period relation of \citet{smith98a}, assuming a
primary mass of $M_{1}/{\rm M}_{\odot}=1.0$. The orbital inclination was
then determined from the eclipse phase width of
$\Delta\phi=0.0506\pm0.0003$ determined from the eclipse timings. The
radius of the accretion disc was assumed to be $R_{{\rm d}}=0.3a$.

\begin{figure}
\begin{tabular}{cc}
\begin{minipage}{2.5in}
\includegraphics[width=6.0cm,angle=0.]{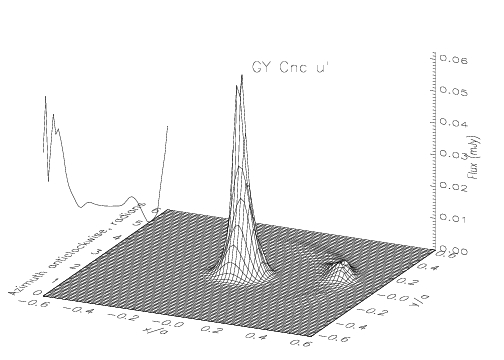} 
\end{minipage}&
\begin{minipage}{2.5in}
\includegraphics[width=4.0cm,angle=-90.]{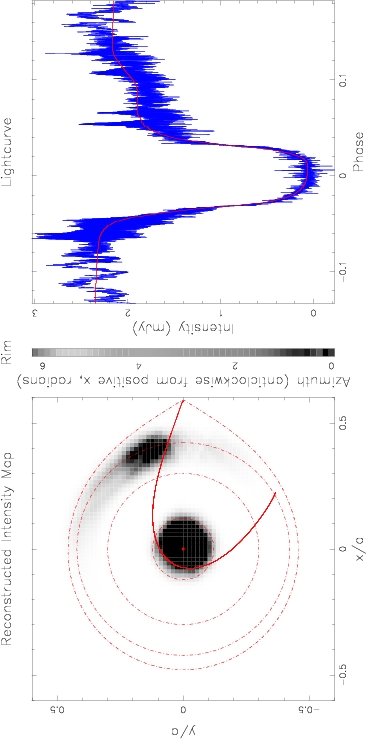} 
\end{minipage}\\
\begin{minipage}{2.5in}
\includegraphics[width=6.0cm,angle=0.]{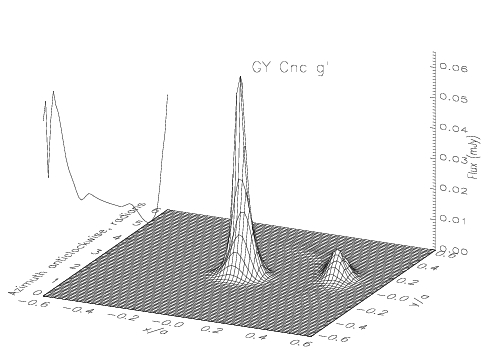} \end{minipage}&
\begin{minipage}{2.5in}
\includegraphics[width=4.0cm,angle=-90.]{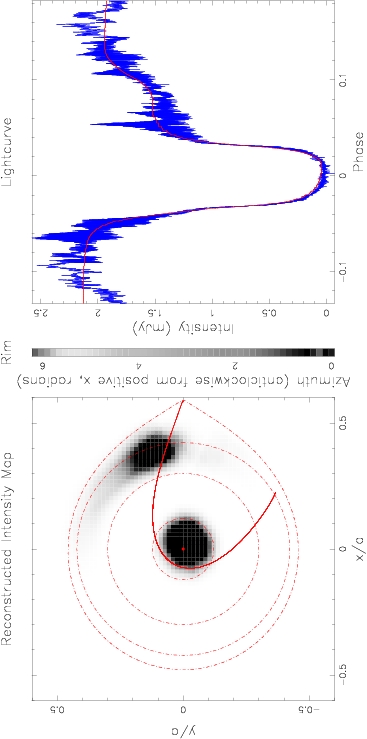} \end{minipage}\\
\begin{minipage}{2.5in}
\includegraphics[width=6.0cm,angle=0.]{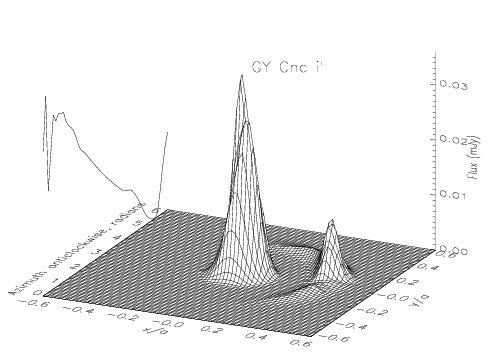} \end{minipage}&
\begin{minipage}{2.5in}
\includegraphics[width=4.0cm,angle=-90.]{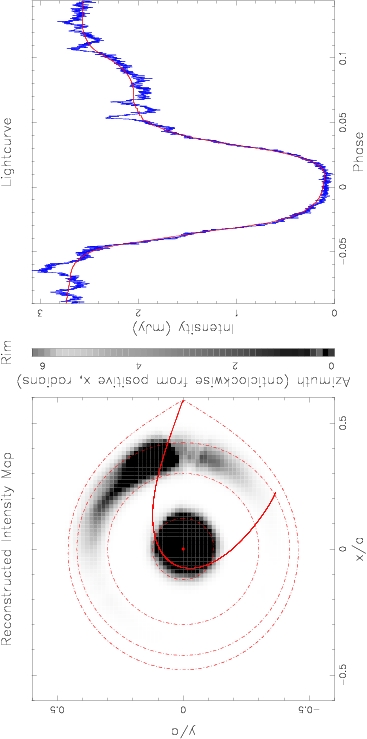} \end{minipage}\\
\begin{minipage}{2.5in}
\includegraphics[width=6.0cm,angle=0.]{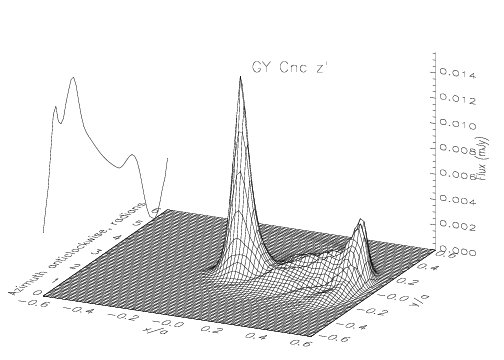} \end{minipage}&
\begin{minipage}{2.5in}
\includegraphics[width=4.0cm,angle=-90.]{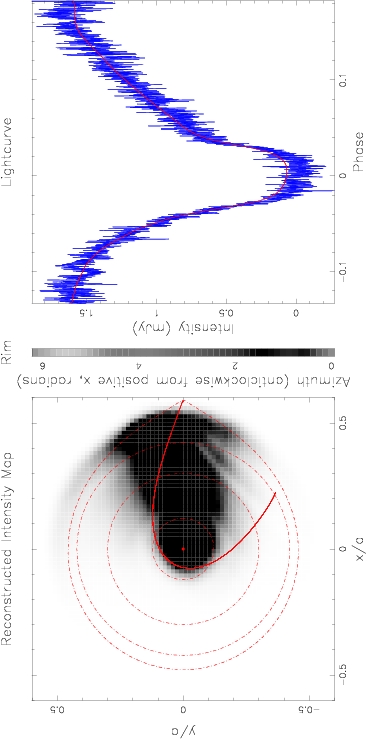} \end{minipage} \\
\end{tabular}
\caption[Eclipse maps for the {\em u\/}$^{\prime}$, {\em
  g\/}$^{\prime}$, {\em i\/}$^{\prime}$ and {\em z\/}$^{\prime}$ light
  curves of GY~Cnc.]{As figure~\ref{fig:em_xzeri}, but for, from top,
  the {\em u\/}$^{\prime}$, {\em g\/}$^{\prime}$, {\em i\/}$^{\prime}$
  and {\em z\/}$^{\prime}$ light curves of GY~Cnc. The {\em
  u\/}$^{\prime}$ and {\em g\/}$^{\prime}$ light curves are of 2003
  May 19 and 23 combined, whereas the {\em i\/}$^{\prime}$ and {\em
  z\/}$^{\prime}$ light curves are of 2003 May 23 and 19,
  respectively. The system parameters adopted for the reconstruction
  are $q=0.421$, $i=77.4^{\circ}$ and $R_{\rm d}=0.3a$ (see text for
  details).}
\label{fig:gycnc}
\end{figure}

\begin{figure}
\begin{tabular}{cc}
\begin{minipage}{2.5in}
\includegraphics[width=6.0cm,angle=0.]{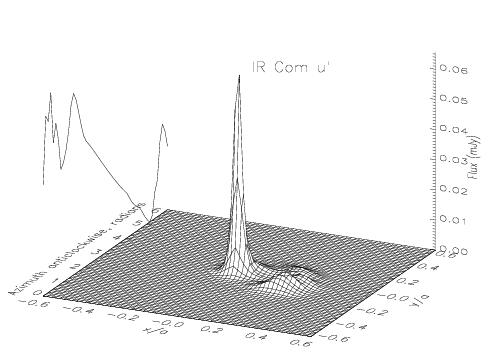} \end{minipage}&
\begin{minipage}{2.5in}
\includegraphics[width=4.0cm,angle=-90.]{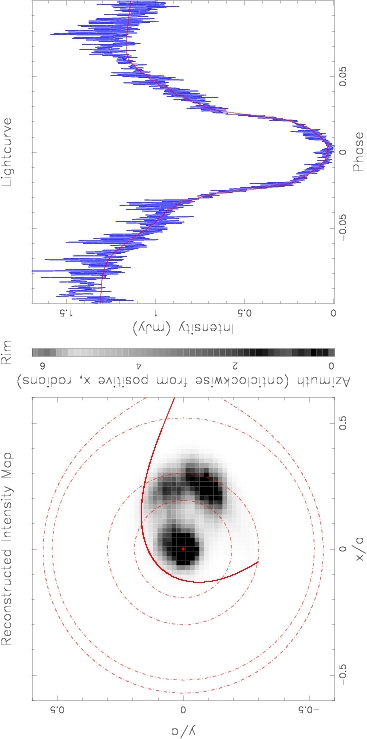} \end{minipage}\\
\begin{minipage}{2.5in}
\includegraphics[width=6.0cm,angle=0.]{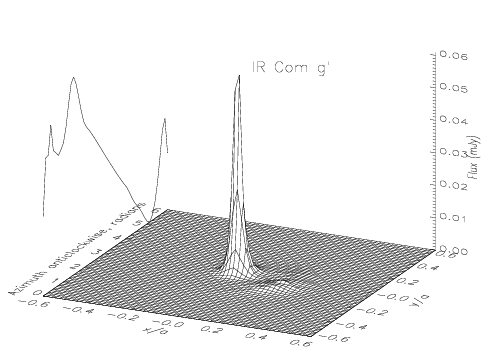} \end{minipage}&
\begin{minipage}{2.5in}
\includegraphics[width=4.0cm,angle=-90.]{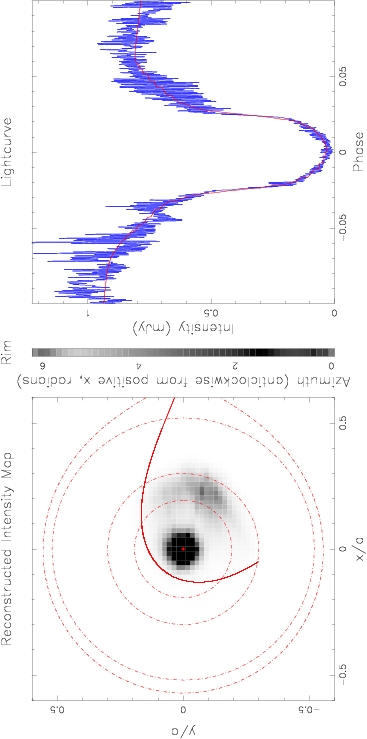} \end{minipage}\\
\begin{minipage}{2.5in}
\includegraphics[width=6.0cm,angle=0.]{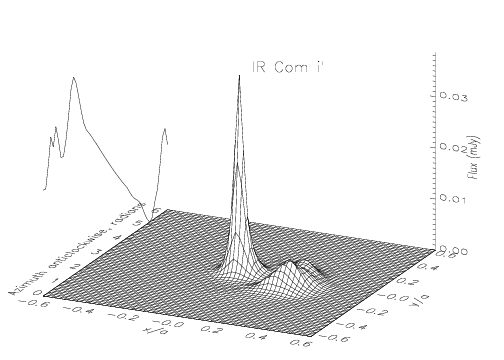} \end{minipage}&
\begin{minipage}{2.5in}
\includegraphics[width=4.0cm,angle=-90.]{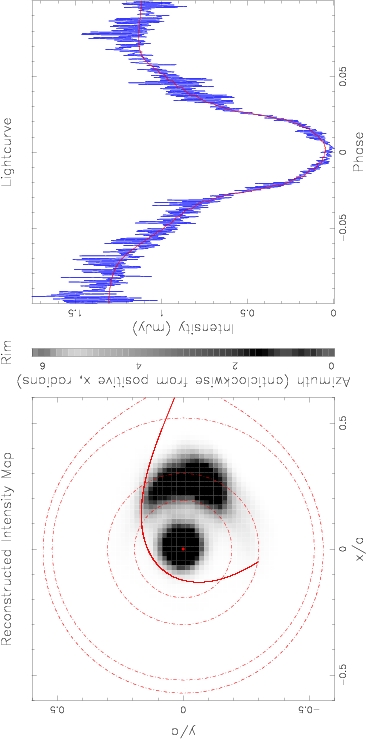} \end{minipage}\\
\end{tabular}
\caption[Eclipse maps for the {\em u\/}$^{\prime}$, {\em
  g\/}$^{\prime}$ and {\em i\/}$^{\prime}$ mean light curves of
  IR~Com.]{As figure~\ref{fig:em_xzeri}, but for, from top, the {\em
  u\/}$^{\prime}$, {\em g\/}$^{\prime}$ and {\em i\/}$^{\prime}$ mean
  light curves of IR~Com. The system parameters adopted for the
  reconstruction are $q=0.153$, $i=81.1^{\circ}$ and $R_{\rm d}=0.3a$
  (see text for details).}
\label{fig:ircom}
\end{figure}

The resulting eclipse maps of GY~Cnc and IR~Com are shown in
figures~\ref{fig:gycnc} and \ref{fig:ircom}, respectively. They show
that in both systems the white dwarf dominates the emission from the
system, with the accretion disc virtually invisible.

%________________________________________________________________

\subsection{Temperature maps}

\begin{figure}
\centerline{\includegraphics[width=14cm,angle=0]{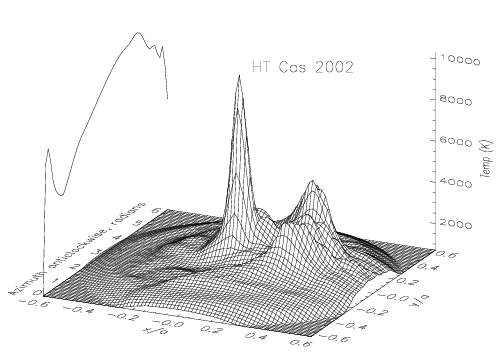}}
\caption[3-dimensional plot of a blackbody fit to the reconstructed
  disc intensities of HT~Cas of the 2002 data.]{3-dimensional plot of
  a blackbody fit to the reconstructed 2002 disc intensities of HT~Cas
  shown in figure~\ref{fig:htcas_2002}.}
\label{fig:htcas_temp_3d_2002}
\end{figure}

\begin{figure}
\centerline{\includegraphics[width=14cm,angle=0]{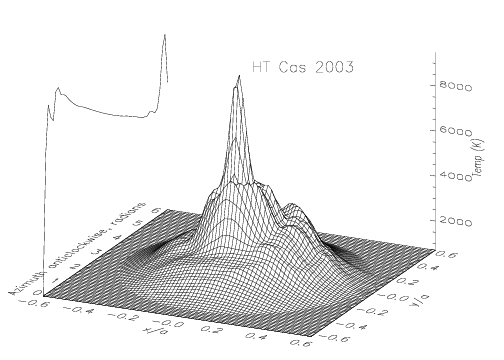}}
\caption[3-dimensional plot of a blackbody fit to the reconstructed
  disc intensities of HT~Cas of the 2003 data.]{3-dimensional plot of
  a blackbody fit to the reconstructed 2003 disc intensities of HT~Cas
  shown in figure~\ref{fig:htcas_2003}.}
\label{fig:htcas_temp_3d_2003}
\end{figure}

The temperature maps of HT~Cas shown in
figures~\ref{fig:htcas_temp_3d_2002} and
\ref{fig:htcas_temp_3d_2003} were produced by fitting a blackbody
function convolved through the filter response functions to each point
in the relevant reconstructed maps. The orbital separation adopted was
that derived by \citet{horne91b}, $a/{\rm R_{\odot}} = 0.670\pm0.019$,
and the distance to the system was that derived by \citet{wood92}, $D
= 125\pm 8 {\rm pc}$. Note that, as figure~\ref{fig:htcas_colours}
illustrates, the disc of HT~Cas is not a good approximation to a
blackbody in many regions, so the temperature scale in
figures~\ref{fig:htcas_temp_3d_2002} and \ref{fig:htcas_temp_3d_2003}
should be regarded cautiously.

It was not possible to produce meaningful temperature maps of the
accretion discs of GY~Cnc and IR~Com because both the orbital
separations and the distances to the systems are currently unknown.

%% file: conclusions.tex
\chapter{Discussion, conclusions and future work}
\label{ch:conclusions}

\section{Discussion and conclusions}
\subsection{OU~Vir}

I have presented an analysis of 5 eclipses of OU Vir. These eclipses
have been used to make the first determination of the system
parameters, given in table~\ref{tab:ouvir_parameters}. My main
conclusions are as follows:
\begin{enumerate}
\item Eclipses of both the white dwarf and bright spot were observed
  during quiescence. The identification of the bright spot ingress and
  egress appears unambiguous.
\item By requiring the gas stream to pass directly through the light
  centre of the bright spot the mass ratio and orbital inclination
  were found to be $q=0.175 \pm 0.025$ and $i=79.2^{\circ}\pm0.7^{\circ}$.
\item Assuming that the central eclipsed object is circular, that its
  size accurately reflects that of the white dwarf and that it obeys
  the \citet{nauenberg72} approximation to the \citet{hamada61}
  mass-radius relationship, adjusted to $T=21\,700$~K, I find that
  the white dwarf radius is $R_{1}=0.0097\pm0.0031\;{\rm R}_{\odot}$
  and its mass is $M_{1}=0.90\pm0.19\;{\rm M}_{\odot}$. This is in good
  agreement with the mean mass of CV white dwarfs found by
  \citet{smith98a} of $\bar{M_{1}}=0.69\pm0.13\;{\rm M}_{\odot}$.
\item With the same assumptions, I find that the volume radius of the
  secondary star is $R_{2}=0.181\pm0.024\;{\rm R}_{\odot}$ and that its mass
  is $M_{2}=0.16\pm0.04\;{\rm M}_{\odot}$. The secondary star is therefore
  consistent with the empirical mass-radius relation for the
  main-sequence secondary stars in CVs of \citet{smith98a}.
\item A blackbody fit to the white dwarf flux gives a temperature
  $T_{1}=21\,700\pm1200$~K and a distance $D=650\pm210$~pc
  with the same assumptions as above. These are purely formal errors
  from the least-squares fit using estimated errors of $\pm0.01$ mJy
  for each flux measurement. Given that I use data from only one
  eclipse, with a single measurement of the flux from each passband,
  the actual uncertainties are likely to be significantly larger.
\item The accretion disc radius of $R_{{\rm d}}/a=0.2315\pm0.0150$ is
  similar in size to that of HT Cas, for which \citet{horne91b}
  derived $R_{{\rm d}}/a=0.23\pm0.03$. This is an unusually small disc
  radius compared to many other dwarf nov\ae\ (e.g.\ Z Cha, which has
  $R_{{\rm d}}/a=0.334$; \citealp{wood89a}), but larger than the
  circularisation radius \citep[][their equation 13]{verbunt88} of
  $R_{{\rm circ}}=0.1820a$. This small disc radius is especially
  surprising as it was determined from observations obtained only
  20 days after the superoutburst was first reported \citep{kato03}.
\item The superhump period of OU Vir is $P_{{\rm
  sh}}=0.078\pm0.002$~days \citep{vanmunster00}, which means OU Vir
  lies $5\sigma$ off the superhump period excess-mass ratio relation
  of \citet[][his equation 8]{patterson98}, with the superhump period
  excess $\epsilon=(P_{{\rm sh}}-P_{{\rm orb}})/P_{{\rm
  orb}}\sim0.073$. However, it does not lie on the superhump period
  excess-orbital period relation either, perhaps indicating that the
  current estimate of the superhump period $P_{{\rm sh}}$ is
  inaccurate.
\item The eclipse maps of OU~Vir shown in
  figures~\ref{fig:em_ouvir_18}--\ref{fig:em_ouvir_temp_22} show the
  white dwarf and bright spot to be the main sources of luminosity in
  the system, and no evidence for the presence of accretion disc
  emission. They also illustrate the difficulties present in eclipse
  mapping discrete sources of emission using the maximum entropy
  technique, especially when combined with the presence of flickering
  in a single cycle.
\end{enumerate}

\subsection{XZ~Eri and DV~UMa}
\label{sec:conc_xzeridvuma}

I have presented an analysis of two quiescent eclipses of XZ~Eri and
three quiescent eclipses of DV~UMa. These eclipses have been used to
determine the system parameters, given in
table~\ref{tab:xzeridvuma_parameters}, via two independent
methods. The first of these is through analysis of the light curve
derivative \citep{wood85,wood86b} and the second by fitting a
parameterized model of the eclipse \citep{horne94}. This is the first
determination of the system parameters of XZ~Eri. My main conclusions
follow:
\begin{enumerate}
\item For both objects, separate eclipses of the white dwarf and
  bright spot were observed. The identification of the bright spot
  ingress and egress is unambiguous in each case. The eclipse maps of
  XZ~Eri and DV~UMa shown in
  figures~\ref{fig:em_xzeri}--\ref{fig:em_dvuma_temp_3d} show the
  white dwarf and bright spot to be the main sources of luminosity in
  these systems, and there is no evidence for the presence of
  accretion disc emission. Compared to the eclipse maps of OU~Vir
  (figures~\ref{fig:em_ouvir_18}--\ref{fig:em_ouvir_temp_22}), they
  also demonstrate the advantages of reducing flickering by
  phase-folding multiple (similar) cycles.
\item By requiring the gas stream to pass directly through the light
  centre of the bright spot the mass ratio and orbital inclination
  were found to be $q=0.117 \pm 0.015$ and
  $i=80.3^{\circ}\pm0.6^{\circ}$ for XZ~Eri and $q=0.148 \pm 0.013$
  and $i=84.4^{\circ}\pm0.8^{\circ}$ for DV~UMa. The parameterized
  model of the eclipse yielded $q=0.1098 \pm 0.0017$ and
  $i=80.16^{\circ}\pm0.09^{\circ}$ for XZ~Eri and $q=0.1506 \pm
  0.0009$ and $i=84.24^{\circ}\pm0.07^{\circ}$ for DV~UMa. The two
  techniques therefore produce results that are in good agreement with
  each other. The system parameters of DV~UMa have also been estimated
  by \citet{patterson00} using eclipse deconvolution. My analysis is
  consistent with their findings. The mass ratio I derive for XZ~Eri,
  $q=0.1098\pm0.0017$, is consistent with XZ~Eri being an SU~UMa star
  \citep{whitehurst88,whitehurst91}, as indicated by its
  (super)outburst history \citep{woudt01,uemura04}. DV~UMa is already
  confirmed as an SU~UMa star after the observation of superhumps in
  its outburst light curve by \citet{nogami01}.
\item The empirical mass-radius and mass-period relations for the
  secondary stars of CVs of \citet{smith98a} are in good agreement
  with the values determined here. The results from the parameterized
  model of XZ~Eri give a very low secondary star mass of
  $M_{2}/{\rm M}_{\odot}=0.0842\pm0.0024$. This is close to the upper limit
  on the mass of a brown dwarf, which is $0.072\;{\rm M}_{\odot}$ for objects
  with solar composition, but can be up to $0.086\;{\rm M}_{\odot}$ for
  objects with zero metallicity \citep{basri00}. The only dwarf nova
  with an accurately determined secondary star mass that is less than
  this is the well-studied system OY~Car, which has $M_{2}/{\rm
  M}_{\odot}=0.070\pm0.002$ \citep{wood89a}. I note also that the
  orbital period and mass ratio of XZ~Eri are similar to those of
  OY~Car \citep{wood89a}. As \citet{patterson00} note, the spectral
  type of the secondary star in DV~UMa (M4.5; \citealp{mukai90})
  implies $M_{2}/{\rm M}_{\odot}=0.12-0.18$ for a main-sequence star
  of solar metallicity \citep{chabrier97,henry99}, consistent with my
  results of $0.169\pm0.023\;{\rm M}_{\odot}$ and $0.157\pm0.009\;{\rm
  M}_{\odot}$ for the derivative and model techniques, respectively.
\item \citet{mukai90} derive the primary temperature and radius of
  DV~UMa from spectroscopic observations by assuming that the white
  dwarf emits a blackbody spectrum. The temperature they derive,
  $T_{1}=22\,000\pm1500$~K, is consistent with my result of
  $20\,000\pm1500$~K. The primary radius ($R_{1}=26\,000-7700$~km)
  \citet{mukai90} calculate is only marginally consistent with my
  results for the derivative technique
  ($R_{1}=0.0067\pm0.0018\;{\rm R}_{\odot}$) and not consistent with the
  results of the parameterized model
  ($R_{1}=0.0079\pm0.0004\;{\rm R}_{\odot}$). This is probably due to the
  limitation of assuming a blackbody spectrum \citep{mukai90}. The
  white dwarf in DV~UMa is unusually massive. My assumption that we
  are observing a bare white dwarf and not a boundary layer around the
  primary cannot explain this, as the white dwarf mass derived would
  in this case be a lower limit \citep[e.g.][]{feline04b}. The mass of
  the white dwarf in XZ~Eri is, however, consistent with the mean mass
  of white dwarfs in dwarf nov\ae\ below the period gap derived by
  \citet{smith98a}.
\item The bright spot scale $SB$ of XZ~Eri is constant over all three
  colour bands. In DV~UMa, however, it increases in size as the colour
  becomes redder. The latter result can be interpreted as follows: the
  material cools as it moves farther from the impact region between
  the accretion disc and the gas stream. These results imply that
  either the time-scale for cooling of the bright spot material is
  greater for DV~UMa than for XZ~Eri or that the post-impact material
  spreads more quickly into the surrounding disc of DV~UMa than of
  XZ~Eri. The time-scale for cooling and/or migration of shock-heated
  material in the bright spot is likely to be affected by factors such
  as the density and composition of the disc material, the mass ratio
  of the system and the disc radius.
\item Finally, I note that the system parameters I derive for DV~UMa
  are consistent with the superhump period-mass ratio relation of
  \citet[][his equation 8]{patterson98}. XZ~Eri, however, lies $5\sigma$
  off this relation. I use here the superhump periods
  $P_{\rm{sh}}=0.062808\pm0.000017$~days for XZ~Eri \citep{uemura04}
  and $P_{\rm{sh}}=0.08870\pm0.00008$~days for DV~UMa
  \citep{patterson00}.
\end{enumerate}

\subsection{GY~Cnc, IR~Com and HT~Cas}
I have found that the dwarf nov\ae\ GY~Cnc and IR~Com both exhibit
eclipses of the white dwarf, and have a bright spot which is faint
(GY~Cnc) or undetected (IR~Com). I have determined updated ephemerides
for both of these objects. IR~Com, with its short orbital period,
significant flickering, high/low quiescent states
\citep{richter95,richter97} and lack of orbital hump or bright spot
strongly resembles HT~Cas in terms of its photometric behaviour
\citep[see also][]{kato02a}.

The colours of the offset flux of HT~Cas shown in
figure~\ref{fig:htcas_colours}, which is estimated from the flux at
mid-eclipse, suggest that it does not originate solely from the
secondary star. \citet{marsh90c} detected the secondary star in
HT~Cas, and found it to be indistinguishable from a main-sequence star
of spectral type M$5.4\pm0.25$, which lies off to the bottom right of
the plot of figure~\ref{fig:htcas_colours} on the main-sequence
relation. It is unlikely that the mid-eclipse flux is from outer
regions of the accretion disc at the back of the disc which remain
uneclipsed at phase zero, since examination of the eclipse maps shown
in figures~\ref{fig:htcas_2002} and \ref{fig:htcas_2003} reveals that the
emission from the rest of the disc is restricted to either the bright
spot (in 2002) or the inner disc (in 2003). Given this, my preferred
explanation for the mid-eclipse colours of HT~Cas is that they are a
combination of flux from the secondary star (which dominates in the
{\em i}$^{\prime}$ band) and flux from a vertically extended,
optically thin disc wind, whose Balmer emission causes it to dominate
in the {\em u}$^{\prime}$ band. I note that the offset flux level
seems to be correlated with the flux from the inner regions of the
accretion disc, which supports the hypothesis of a disc wind
originating from the inner disc region or boundary layer of HT~Cas. I
caution, however, that systematic errors may affect the offset fluxes
due to the technique used to determine them, and that this conclusion
is therefore tentative. (As the offset flux is a small fraction of the
total light, except in the {\em i}$^{\prime}$ band, any systematic
errors present in the offset fluxes will not significantly affect the
rest of these results.)

The eclipse maps of HT~Cas shown in figures~\ref{fig:htcas_2002} and
\ref{fig:htcas_2003} and the radial flux profiles shown in
figure~\ref{fig:htcas_radial} clearly demonstrate that the accretion
disc of HT~Cas was in two distinct states during the 2002 and 2003
observations. In the 2002 data the disc provided a negligible
contribution to the total light, except for the presence of a bright
spot in its outer regions. In 2003 the bright spot was much fainter,
but the inner disc was luminous, causing the overall system brightness
to be $\sim0.6$~mJy brighter than in 2002. The uneclipsed component
was also slightly brighter in 2003 than 2002 (probably due to
variability of the secondary star; see captions to
figures~\ref{fig:htcas_2002} and \ref{fig:htcas_2003}), but was not
the major cause of the differences in the flux. I proceed to review
previous observations and to discuss various possible explanations for
this behaviour.

The most likely reasons for the observed changes in the intensity
distribution of the quiescent accretion disc of HT~Cas lie in
variability of the secondary star (which supplies the disc with
material) or some property of the accretion disc itself. We can exclude
the white dwarf as the cause of the variability since the only
plausible way that this could affect the majority of the disc is
via a magnetic field. HT~Cas is a confirmed dwarf nova, whose white
dwarfs do not have magnetic fields strong enough to significantly
influence the motion of gas in the disc \citep[e.g.][]{warner95}.

The most obvious mechanism for the accretion disc to produce the
observed behaviour of HT~Cas is via some relationship to the outburst
cycle. \citet{baptista01} reported changes in the structure of the
accretion disc of EX~Dra (a dwarf nova above the period gap) through
its outburst cycle, specifically the presence of a low-brightness
state immediately after outburst during which the disc and bright spot
were exceptionally faint. In EX~Dra, the low-brightness state is due
to reduced emission from all parts of the disc and white dwarf; the
eclipse maps of HT~Cas presented in
figures~\ref{fig:htcas_2002}--\ref{fig:htcas_radial}, however,
demonstrate that the quiescent luminosity of HT~Cas is affected by
which areas of the disc are luminous. \citet{robertson96} find that
both the transition between the quiescent high and low states and the
duration of the low state in HT~Cas occur on time-scales of days to
months compared to the outburst cycle length of $\sim400$~days
\citep{wenzel87}. \citet*{truss04} proposed a (slowly cooling) hot
inner region of the disc in order to explain the constant quiescent
brightness observed in (most) dwarf nov\ae, which is contrary to the
increase of 1--3 magnitudes predicted by most disc instability models
(see \citealp{lasota01} for a review). This fails to account for the
observed changes in the outer accretion disc of HT~Cas, but does
provide a plausible explanation for the variability of the inner
regions of the disc. This model, however, necessitates an outburst
between the two sets of observations reported here, which amateur
observations (figure~\ref{fig:htcas_aavso}) can neither confirm nor
refute. I conclude that this latter scenario is the only plausible way
in which the changes in the accretion disc of HT~Cas could be related
to its position in the outburst cycle.

The secondary star can also induce changes in the accretion disc. For
example, variability of the rate of mass transfer from the secondary
star is often (plausibly) cited as a mechanism to explain the
quiescent variability of dwarf nov\ae\ (and other
CVs). \citet{baptista04} observed the short-period dwarf nova
V2051~Oph in high and low quiescent states. Eclipse maps showed that
the increased emission in the high state was due to greater emission
from the bright spot and gas stream region, implying a higher mass
transfer rate from the secondary star. Interestingly, this is the
opposite to what I find for HT~Cas.

Variability of the secondary star is in fact the mechanism usually
proposed to explain the well-documented presence of high/low
quiescence states in HT~Cas (e.g.\ \citealp{berriman87},
\citealp{wood95,robertson96}). The most frequently cited explanation
is that suggested by \citet{livio94}, of star spots passing over the
inner Lagrangian point temporarily lowering the mass transfer rate
from the secondary star. Another possible causal process is magnetic
variability of the secondary star. \citet{ak01} found cyclical
variations in the quiescent magnitudes and outburst intervals of 22
CVs, which they attributed to solar-type magnetic activity cycles of
the secondary stars. This can result in an increased mass transfer
rate from the secondary star as well as the removal of angular
momentum from the outer regions of the disc, causing material to
accumulate in the inner regions of the disc rather than in the outer
regions. The magnetic activity cycle of the secondary stars derived by
\citet{ak01} is, however, on the wrong time-scale (years) to explain
the frequency of the high/low state transitions and durations
(days/months) observed by \citet{robertson96}.

In summary, variations in the rate of mass transfer from the secondary
star can explain the variability of the bright spot, but fail to
account for the changes in the inner disc. These can be explained by a
larger mass transfer rate through the accretion disc (possibly due to
a rise in the disc viscosity and/or the scenario proposed by
\citealp{truss04}) increasing the emission from the inner disc via
viscous dissipation.

I conclude that the variability of the quiescent accretion disc of
HT~Cas is caused by variations both in the rate of mass transfer from
the secondary star and through the accretion disc. An increase in the
viscosity of the accretion disc leading to an increase in the rate of
mass transfer through the disc cannot explain these results
alone. Such a viscosity increment would spread material both inwards
and outwards, meaning that, although this scenario could explain the
rise in the 2003 inner disc flux, one would still expect to observe a
bright spot at the intersection of the outer disc and gas stream. In
the 2002 observations then, the rate of mass transfer through the disc
was lower and the rate of mass transfer from the secondary star
greater than in 2003. It is clearly desirable to undertake long-term
monitoring of HT~Cas (or a similar object, e.g.\ IR~Com) with the aim
of eclipse mapping the changes that occur in the disc during
quiescence and especially during a transition between the high and low
states in order to determine the triggers and physical mechanisms
underlying this behaviour.

\section{Overview}

\begin{figure}
\centerline{\includegraphics[width=12cm,angle=0]{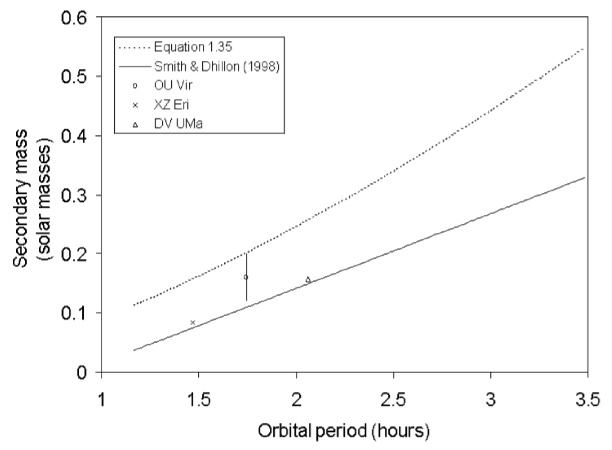}}
\caption[The secondary star masses and orbital periods for OU~Vir,
  XZ~Eri and DV~UMa, compared to two mass-period relations.]{The
  secondary star masses and orbital periods for OU~Vir, XZ~Eri and
  DV~UMa determined in this thesis, compared to the empirical
  mass-period relations given in equation~\ref{eq:massradius} and
  obtained by \citet[][equation~\ref{eq:massradius2}]{smith98a}. The
  error bars on the orbital periods of OU~Vir, XZ~Eri and DV~UMa and
  on the secondary star masses of XZ~Eri and DV~UMa are of the order
  of the size of the symbols used to plot the points, and so are not
  plotted.}
\label{fig:massradius}
\end{figure}

The work contained in this thesis has increased the number of
accurately known CV masses by three. As of 1998, there were only 14
reliable mass determinations of CV secondary stars
\citep{smith98a}. This work forms part of a long term project at
Sheffield that aims to increase this number
\citep[e.g.][]{smith98a,thoroughgood01,thoroughgood04,thoroughgood05,
thoroughgood05a,feline04a,feline04b,feline04c}.  Since the review of
\citet{smith98a}, there have been at least eleven accurate mass
determinations of CVs, to whit: OU~Vir\footnote[1]{Estimated using
eclipse timings.} \citep{feline04b,feline04c}; XZ~Eri\footnotemark[1]
and DV~UMa\footnotemark[1] \citep{feline04a};
U~Sco\footnote[2]{Estimated using the radial velocities of the primary
and secondary stars.} \citep{thoroughgood01};
AC~Cnc\footnote[3]{Estimated using the radial and rotational
($v_{2}\sin i$) velocities of the secondary star with the eclipse
phase width measured from photometry.} and V363~Aur\footnotemark[3]
\citep{thoroughgood04}; V347~Pup\footnotemark[3]
\citep{thoroughgood05}; WZ~Sge\footnotemark[1] \citep{skidmore02};
IY~UMa\footnotemark[1] \citep{steeghs03}; HS~0907+1902\footnotemark[2]
\citep{thorstensen00}; EX~Dra\footnotemark[1] \citep{baptista00}; and
U~Gem\footnotemark[2]
\citep{long99,naylor05}. Figure~\ref{fig:massradius} shows the results
from the work contained in this thesis for the secondary star masses
and orbital periods of OU~Vir, XZ~Eri and DV~UMa compared to the
empirical mass-period relations given in equation~\ref{eq:massradius}
and obtained by \citet[][equation~\ref{eq:massradius2}]{smith98a}.

As discussed in \S~\ref{sec:massdetermination}, accurate measurements
of the masses of CVs are important for many reasons.

The common envelope theory of stellar evolution (see
\S~\ref{sec:commonenvelope} and \citealp{iben93}), for example,
requires reliable endpoint masses---those of CVs---in order to test
computational models. \citet{kolb93} finds that 70~per~cent of all CVs
should be systems that have evolved to contain brown dwarfs, whereas
the models of \citet{politano04} predict that $\sim18$~per~cent of the
zero-age CV population will contain a brown dwarf secondary star. The
models may differ in their predictions of the exact fraction of
systems with brown dwarf secondary stars, but they agree that the
number is significant. This figure is highly dependent on
the efficiency of the common envelope in removing the orbital energy
of the pre-CV, the initial mass ratio distribution of the CV
population and the details of the disrupted magnetic braking
model. The detection of a brown dwarf secondary star would be a
crucial step in understanding the evolutionary processes that have
formed the current CV population. Unfortunately, searches for CVs
with brown dwarf secondaries \citep[e.g][]{mennickent04} have yet to
find direct evidence for such systems, although several candidates
exist for which there is ``significant indirect evidence'' for a brown
dwarf secondary star \citep{littlefair03}. The secondary stars in CVs
are often faint and undetectable spectroscopically
\citep[see][]{littlefair03}, rendering unlikely the measurement of the
radial velocity of the secondary star.\footnote[4]{The detection of
lithium would be very strong evidence of a brown dwarf secondary star,
since it is only expected to be present in the spectra of young stars
or brown dwarfs, but this is again unlikely due to the faintness of
the secondary.} One would expect systems that have evolved to near the
minimum period to have a low mass-transfer rate and hence a faint
accretion disc, allowing the bright spot and white dwarf eclipses to
be clearly observed. The results contained in this thesis (especially
those for XZ~Eri) demonstrate that the photometric technique works well
for systems near the period minimum where we expect to find these
objects. The photometric method of mass determination described in
this thesis is the only unambiguous method of detecting a brown dwarf
secondary star in a CV.

\begin{figure}
\centerline{\includegraphics[width=12cm,angle=0]{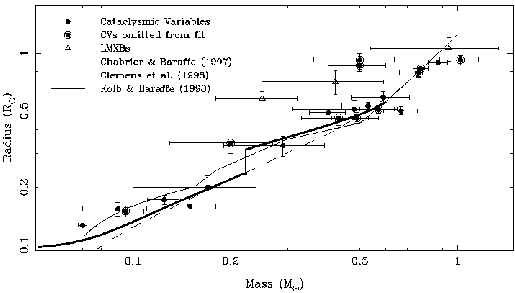}}
\caption[The masses and radii of the secondary stars in CVs and
  LMXBs.]{The masses and radii of the secondary stars in CVs and
  LMXBs. The filled circles are data for CVs; those which are ringed
  were excluded from the fit.  The open triangles are data for low
  mass X-ray binaries (LMXBs). The theoretical models of
  \citet[][the dashed line]{chabrier97} and the empirical relation
  derived by \citet[][the thin solid line]{clemens98} are plotted. The
  thick solid line shows the secular evolution of the mass and radius
  of the secondary star computed by \citet[][described in the figure
  as Kolb \& Baraffe 1998]{kolb99}. From \citet{smith98a}. The
  secondary star masses and radii determined in this thesis are, for
  comparison: OU~Vir, $M_{2}/{\rm M}_{\odot}=0.16\pm0.04$ and
  $R_{2}/{\rm R}_{\odot}=0.177\pm0.024$; XZ~Eri, $M_{2}/{\rm
  M}_{\odot}=0.0842\pm0.0024$ and $R_{2}/{\rm
  R}_{\odot}=0.1315\pm0.0019$; DV~UMa, $M_{2}/{\rm
  M}_{\odot}=0.157\pm0.004$ and $R_{2}/{\rm
  R}_{\odot}=0.2022\pm0.0018$. All three lie within the scatter of the
  points.}
\label{fig:smithdhillon98}
\end{figure}

The disrupted magnetic braking model (discussed in
\S~\ref{sec:disrupted}), the currently favoured mechanism for the
origin of the period gap, also needs accurate secondary star masses
for observational confirmation. In this model, the secondary stars in
CVs change from being out of thermal equilibrium and hence
under-massive for their radii above the period gap to thermally relaxed
(i.e.\ in thermal equilibrium) below the period gap. This should
produce a `kink' in the mass--radius plot for the secondary stars of
CVs (see figure~\ref{fig:smithdhillon98} and \citealp{smith98a}), which
cannot currently be seen due to two reasons: first, the scarcity of
mass measurements, and second, the inaccuracy of many of these
measurements. The three mass determinations contained within this
thesis are consistent with the secondary star mass-radius relation of
\citet[see also figure~\ref{fig:smithdhillon98}]{smith98a}.

\section{Future work}

More candidate systems for photometric mass determinations are
constantly being discovered. The Sloan Digital Sky Survey (SDSS) has
already uncovered many excellent candidates
\citep{szkody02,szkody03,szkody04}. The number of previously unknown
CVs discovered by the SDSS is expected to number about 400 systems,
most of which will be faint, low mass transfer rate systems, ideal for
the above purposes. Candidate systems are also being uncovered by the
Hamburg Quasar Survey \citep{hagen95,araujobetancor05}. This thesis
has demonstrated that the photometric method is an essential tool for
accurate mass determinations of the many faint, short period,
eclipsing dwarf nov\ae\ that will be uncovered by surveys such as
these.

An attractive future project with {\sc ultracam} is the long-term
monitoring of dwarf nov\ae. Future plans for {\sc ultracam} include
mounting it on the 2-m class Aristarchos and Isaac Newton (INT)
telescopes in Greece and La Palma, respectively. The combination of
{\sc ultracam} with either of these telescopes would be ideal for such
long-term monitoring. Observing time for such a long-term project
would be difficult to obtain on larger telescopes, and the above
combinations of {\sc ultracam} and either Aristarchos or the INT would
in any case be able to achieve sufficient signal-to-noise and time
resolution to map the accretion discs of bright CVs in detail. Of
particular interest are changes in the accretion disc through the
outburst cycle \citep[e.g.][]{baptista01} and the physical mechanisms
underlying the existence of high and low quiescent states in several
short-period dwarf nov\ae\ (see chapter~\ref{ch:gycncircomhtcas} and
\citealp{feline05}). It might be hoped that detailed eclipse mapping
through the outburst cycle of a dwarf nova would reveal
heating/cooling fronts running from the outer to the inner regions of
the accretion disc or {\em vice versa} \citep[see][and references
therein for a detailed discussion]{warner95}. Doppler tomography, a
technique which produces Doppler maps of the system, also holds
promise in this regard. For example, \citet{steeghs04} observed WZ~Sge
through its 2001 outburst, producing movies of the evolving accretion
flow. The detection of such heating and cooling  fronts would provide
an invaluable observational framework in which to hang improved theory
of the underlying physics of accretion discs.

%% file: thesis_2side.bbl
\begin{thebibliography}{210}
\expandafter\ifx\csname natexlab\endcsname\relax\def\natexlab#1{#1}\fi

\bibitem[{Ak {et~al.}(2001)Ak, Ozkan, \& Mattei}]{ak01}
Ak, T., Ozkan, M.~T., \& Mattei, J.~A. 2001, A\&A, 369, 882

\bibitem[{Allan {et~al.}(1996)Allan, Hellier, \& Ramseyer}]{allan96}
Allan, A., Hellier, C., \& Ramseyer, T.~F. 1996, MNRAS, 282, 699

\bibitem[{Andronov \& Pinsonneault(2004)}]{andronov04}
Andronov, N. \& Pinsonneault, M. 2004, ApJ, 614, 326

\bibitem[{Andronov {et~al.}(2003)Andronov, Pinsonneault, \& Sills}]{andronov03}
Andronov, N., Pinsonneault, M., \& Sills, A. 2003, ApJ, 582, 358

\bibitem[{Araujo-Betancor {et~al.}(2005)Araujo-Betancor, G{\"{a}}nsicke, Hagen,
  Marsh, Harlaftis, Thorstensen, Fried, Schmeer, \& Engels}]{araujobetancor05}
Araujo-Betancor, S., G{\"{a}}nsicke, B.~T., Hagen, H.-J., Marsh, T.~R.,
  Harlaftis, E.~T., Thorstensen, J., Fried, R.~E., Schmeer, P., \& Engels, D.
  2005, A\&A, 430, 629

\bibitem[{Baba {et~al.}(2002)Baba, Sadakane, Norimoto, Ayani, Ioroi, Matsumoto,
  Nogami, Makita, \& Taichi}]{baba02}
Baba, H., Sadakane, K., Norimoto, Y., Ayani, K., Ioroi, M., Matsumoto, K.,
  Nogami, D., Makita, M., \& Taichi, K. 2002, PASJ, 54, L7

\bibitem[{Bade {et~al.}(1998)Bade, Engels, Voges, Beckmann, Boller, Cordis,
  Dahlem, Englhauser, Molthagen, Nass, \& Reimers}]{bade98}
Bade, N., Engels, D., Voges, W., Beckmann, V., Boller, T., Cordis, L., Dahlem,
  M., Englhauser, J., Molthagen, K., Nass, P.and~Studt, J., \& Reimers, D.
  1998, A\&AS, 127, 145

\bibitem[{Bailey(1979)}]{bailey79}
Bailey, J.~A. 1979, MNRAS, 187, 645

\bibitem[{Balbus \& Hawley(1991)}]{balbus91}
Balbus, S.~A. \& Hawley, J.~F. 1991, AJ, 376, 214

\bibitem[{Baptista(2001)}]{baptista01a}
Baptista, R. 2001, in Astrotomography: Indirect Imaging Methods in
  Observational Astronomy, ed. H.~M.~J. Boffin, D.~Steeghs, \& J.~Cuypers
  (Berlin: Springer-Verlag Lecture Notes in Physics), 307

\bibitem[{Baptista(2004)}]{baptista04b}
Baptista, R. 2004, 325, 181

\bibitem[{Baptista \& Bortoletto(2004)}]{baptista04}
Baptista, R. \& Bortoletto, A. 2004, AJ, 128, 411

\bibitem[{Baptista \& Catal\'an(2001)}]{baptista01}
Baptista, R. \& Catal\'an, M.~S. 2001, MNRAS, 324, 599

\bibitem[{Baptista {et~al.}(2000)Baptista, Catal\'an, \& Costa}]{baptista00}
Baptista, R., Catal\'an, M.~S., \& Costa, L. 2000, MNRAS, 316, 529

\bibitem[{Baptista {et~al.}(1998)Baptista, Catal\'an, Horne, \&
  Zilli}]{baptista98}
Baptista, R., Catal\'an, M.~S., Horne, K., \& Zilli, D. 1998, MNRAS, 300, 233

\bibitem[{Baptista {et~al.}(1995)Baptista, Horne, Hilditch, Mason, \&
  Drew}]{baptista95}
Baptista, R., Horne, K., Hilditch, R.~W., Mason, K.~O., \& Drew, J.~E. 1995,
  ApJ, 448, 395

\bibitem[{Baptista \& Steiner(1993)}]{baptista93}
Baptista, R. \& Steiner, J.~E. 1993, A\&A, 277, 331

\bibitem[{Baptista {et~al.}(1996)Baptista, Steiner, \& Horne}]{baptista96}
Baptista, R., Steiner, J.~E., \& Horne, K. 1996, MNRAS, 282, 99

\bibitem[{Basri(1987)}]{basri87}
Basri, G. 1987, ApJ, 316, 377

\bibitem[{Basri(2000)}]{basri00}
---. 2000, ARA\&A, 38, 485

\bibitem[{Bath \& Pringle(1981)}]{bath81}
Bath, G.~T. \& Pringle, J.~E. 1981, MNRAS, 194, 967

\bibitem[{Beard {et~al.}(2002)Beard, Vick, Atkinson, Dhillon, Marsh, McLay,
  Stevenson, \& Tierney}]{beard02}
Beard, S., Vick, A., Atkinson, D., Dhillon, V., Marsh, T., McLay, S.,
  Stevenson, M., \& Tierney, C. 2002, in Proc. SPIE 4848, Advanced Telescope
  and Instrumentation Control Software II

\bibitem[{Beekman {et~al.}(2000)Beekman, Somers, Naylor, \&
  Hellier}]{beekman00}
Beekman, G., Somers, M., Naylor, T., \& Hellier, C. 2000, MNRAS, 318, 9

\bibitem[{Bergeron {et~al.}(1995)Bergeron, Wesemael, \& Beauchamp}]{bergeron95}
Bergeron, P., Wesemael, F., \& Beauchamp, A. 1995, PASP, 107, 1047

\bibitem[{Berriman {et~al.}(1983)Berriman, Beattie, Gatley, Lee, Mochnacki, \&
  Szkody}]{berriman83}
Berriman, G., Beattie, D.~H., Gatley, I., Lee, T.~J., Mochnacki, S.~W., \&
  Szkody, P. 1983, MNRAS, 204, 1105

\bibitem[{Berriman {et~al.}(1987)Berriman, Kenyon, \& Boyle}]{berriman87}
Berriman, G., Kenyon, S., \& Boyle, C. 1987, AJ, 94, 1291

\bibitem[{Bobinger {et~al.}(1997)Bobinger, Horne, Mantel, \& Wolf}]{bobinger97}
Bobinger, A., Horne, K., Mantel, K.-H., \& Wolf, S. 1997, A\&A, 327, 1023

\bibitem[{Bruch(2000)}]{bruch00b}
Bruch, A. 2000, A\&A, 359, 998

\bibitem[{Caillault \& Patterson(1990)}]{caillaut90}
Caillault, J.-P. \& Patterson, J. 1990, AJ, 100, 825

\bibitem[{Cannizzo {et~al.}(1988)Cannizzo, Shafter, \& Wheeler}]{cannizzo88}
Cannizzo, J.~K., Shafter, A.~W., \& Wheeler, J.~C. 1988, ApJ, 333, 227

\bibitem[{Chabrier \& Baraffe(1997)}]{chabrier97}
Chabrier, G. \& Baraffe, Y. 1997, A\&A, 327, 1039

\bibitem[{Claret \& Hauschildt(2003)}]{claret03}
Claret, A. \& Hauschildt, P.~H. 2003, A\&A, 412, 241

\bibitem[{Clemens {et~al.}(1998)Clemens, Reid, Gizis, \& O'Brien}]{clemens98}
Clemens, J.~C., Reid, I.~N., Gizis, J.~E., \& O'Brien, M. 1998, ApJ, 496, 352

\bibitem[{de~Kool(1992)}]{dekool92}
de~Kool, M. 1992, A\&A, 261, 188

\bibitem[{Dhillon(1990)}]{dhillon90}
Dhillon, V.~S. 1990, PhD thesis, University of Sussex

\bibitem[{Dhillon \& Marsh(2001)}]{dhillon01b}
Dhillon, V.~S. \& Marsh, T.~R. 2001, New~Ast.~Rev., 45, Issue 1-2, 91

\bibitem[{Dhillon {et~al.}(2005)Dhillon, Marsh, \& et~al.}]{dhillon05}
Dhillon, V.~S., Marsh, T.~R., \& et~al. 2005, MNRAS, in preparation

\bibitem[{Dhillon {et~al.}(1991)Dhillon, Marsh, \& Jones}]{dhillon91}
Dhillon, V.~S., Marsh, T.~R., \& Jones, D. H.~P. 1991, MNRAS, 252, 342

\bibitem[{Duerbeck(1992)}]{duerbeck92}
Duerbeck, H.~W. 1992, MNRAS, 258, 629

\bibitem[{Eggleton(1983)}]{eggleton83}
Eggleton, P.~P. 1983, ApJ, 268, 368

\bibitem[{Feline {et~al.}(2004a)Feline, Dhillon, Marsh, \&
  Brinkworth}]{feline04a}
Feline, W.~J., Dhillon, V.~S., Marsh, T.~R., \& Brinkworth, C.~S. 2004a, MNRAS,
  355, 1

\bibitem[{Feline {et~al.}(2004b)Feline, Dhillon, Marsh, Stevenson, Watson, \&
  Brinkworth}]{feline04b}
Feline, W.~J., Dhillon, V.~S., Marsh, T.~R., Stevenson, M.~J., Watson, C.~A.,
  \& Brinkworth, C.~S. 2004b, MNRAS, 347, 1173

\bibitem[{Feline {et~al.}(2004c)Feline, Dhillon, Marsh, Stevenson, Watson, \&
  Brinkworth}]{feline04c}
---. 2004c, MNRAS, 354, 1279

\bibitem[{Feline {et~al.}(2005)Feline, Dhillon, Marsh, Watson, \&
  Littlefair}]{feline05}
Feline, W.~J., Dhillon, V.~S., Marsh, T.~R., Watson, C.~A., \& Littlefair,
  S.~P. 2005, MNRAS, in press

\bibitem[{Fiedler {et~al.}(1997)Fiedler, Barwig, \& Mantel}]{fiedler97}
Fiedler, H., Barwig, H., \& Mantel, K.~H. 1997, A\&A, 327, 173

\bibitem[{Flannery(1975)}]{flannery75}
Flannery, B.~P. 1975, MNRAS, 170, 325

\bibitem[{Frank {et~al.}(1985)Frank, King, \& Raine}]{frank85}
Frank, J., King, A.~R., \& Raine, D.~J. 1985, Accretion Power in Astrophysics
  (Cambridge: Cambridge University Press)

\bibitem[{Friend {et~al.}(1990)Friend, Martin, Smith, \& Jones}]{friend90a}
Friend, M.~T., Martin, J.~S., Smith, R.~C., \& Jones, D. H.~P. 1990, MNRAS,
  246, 637

\bibitem[{Fukugita {et~al.}(1996)Fukugita, Ichikawa, Gunn, Doi, Shimasaku, \&
  Schneider}]{fukugita96}
Fukugita, M., Ichikawa, T., Gunn, J.~E., Doi, M., Shimasaku, K., \& Schneider,
  D.~P. 1996, AJ, 111, 1748

\bibitem[{G{\"{a}}nsicke {et~al.}(2000)G{\"{a}}nsicke, Fried, Hagen, Beuermann,
  Engels, Hessman, Nogami, \& Reinsch}]{gansicke00}
G{\"{a}}nsicke, B.~T., Fried, R.~E., Hagen, H.-J., Beuermann, K., Engels, D.,
  Hessman, F.~V., Nogami, D., \& Reinsch, K. 2000, A\&A, 356, L79

\bibitem[{Girardi {et~al.}(2004)Girardi, Grebel, Odenkirchen, \&
  Chiosi}]{girardi04}
Girardi, L., Grebel, E.~K., Odenkirchen, M., \& Chiosi, C. 2004, A\&A, 422, 205

\bibitem[{Groot(2001)}]{groot01}
Groot, P.~J. 2001, ApJ, 551, L89

\bibitem[{Gull \& Skilling(1989)}]{gull89}
Gull, S.~F. \& Skilling, J. 1989, Quantified Maximum Entropy MEMSYS 3 Users'
  Manual, User manual, Maximum Entropy Data Consultants Ltd

\bibitem[{Gull \& Skilling(1991)}]{gull91}
---. 1991, Quantified Maximum Entropy MEMSYS 5 Users' Manual, User manual,
  Maximum Entropy Data Consultants Ltd

\bibitem[{Hagen {et~al.}(1995)Hagen, Groote, Engels, \& Reimers}]{hagen95}
Hagen, H.-J., Groote, D., Engels, D., \& Reimers, D. 1995, A\&AS, 111, 195

\bibitem[{Hamada \& Salpeter(1961)}]{hamada61}
Hamada, T. \& Salpeter, E.~E. 1961, ApJ, 134, 683

\bibitem[{Hawley \& Balbus(1991)}]{hawley91}
Hawley, J.~F. \& Balbus, S.~A. 1991, AJ, 376, 223

\bibitem[{Hellier(2001)}]{hellier01}
Hellier, C. 2001, Cataclysmic Variable Stars: How and Why They Vary (UK:
  Springer-Verlag)

\bibitem[{Henry {et~al.}(1999)Henry, Franz, Wasserman, Benedict, Shelus, Ianna,
  Kirkpatrick, \& McCarthy}]{henry99}
Henry, T.~J., Franz, O.~G., Wasserman, L.~H., Benedict, G.~F., Shelus, P.~J.,
  Ianna, P.~A., Kirkpatrick, J.~D., \& McCarthy, D.~W. 1999, ApJ, 512, 864

\bibitem[{Hind(1856)}]{hind1856}
Hind, J.~R. 1856, MNRAS, 16, 56

\bibitem[{Horne(1985)}]{horne85}
Horne, K. 1985, MNRAS, 213, 129

\bibitem[{Horne \& Cook(1985)}]{horne85a}
Horne, K. \& Cook, M.~C. 1985, MNRAS, 214, 307

\bibitem[{Horne {et~al.}(1994)Horne, Marsh, Cheng, Hubeny, \& Lanz}]{horne94}
Horne, K., Marsh, T.~R., Cheng, F.~H., Hubeny, I., \& Lanz, T. 1994, ApJ, 426,
  294

\bibitem[{Horne \& Stiening(1985)}]{horne85b}
Horne, K. \& Stiening, R.~F. 1985, MNRAS, 216, 933

\bibitem[{Horne {et~al.}(1986)Horne, Wade, \& Szkody}]{horne86b}
Horne, K., Wade, R.~A., \& Szkody, P. 1986, MNRAS, 219, 791

\bibitem[{Horne {et~al.}(1993)Horne, Welsh, \& Wade}]{horne93}
Horne, K., Welsh, W.~F., \& Wade, R.~A. 1993, ApJ, 410, 357

\bibitem[{Horne {et~al.}(1991)Horne, Wood, \& Steining}]{horne91b}
Horne, K., Wood, J.~H., \& Steining, R.~F. 1991, ApJ, 378, 271

\bibitem[{Howell {et~al.}(1988)Howell, Mason, Reichart, Warnock, \&
  Kreidl}]{howell88}
Howell, S.~B., Mason, K.~O., Reichart, G.~A., Warnock, A., \& Kreidl, T.~J.
  1988, MNRAS, 233, 79

\bibitem[{Howell {et~al.}(1991)Howell, Szkody, Kreidl, \& Dobrzycka}]{howell91}
Howell, S.~B., Szkody, P., Kreidl, T.~J., \& Dobrzycka, D. 1991, PASP, 103, 300

\bibitem[{Iben \& Livio(1993)}]{iben93}
Iben, I. \& Livio, M. 1993, PASP, 105, 1373

\bibitem[{Joergens {et~al.}(2000)Joergens, Spruit, \& Rutten}]{joergens00}
Joergens, V., Spruit, H.~C., \& Rutten, R. G.~M. 2000, A\&A, 356, L33

\bibitem[{Kato(2003)}]{kato03}
Kato, T. 2003, vsnet-alert, No. 7733

\bibitem[{Kato {et~al.}(2002a)Kato, Baba, \& Nogami}]{kato02a}
Kato, T., Baba, H., \& Nogami, D. 2002a, PASJ, 54, 79

\bibitem[{Kato {et~al.}(2002b)Kato, Ishioka, \& Uemura}]{kato02b}
Kato, T., Ishioka, R., \& Uemura, M. 2002b, PASJ, 54, 1023

\bibitem[{Kato {et~al.}(2000)Kato, Uemura, Schmeer, Garradd, Martin, Maehara,
  Kinnunen, \& Watanabe}]{kato00}
Kato, T., Uemura, M., Schmeer, P., Garradd, G., Martin, B., Maehara, H.,
  Kinnunen, T., \& Watanabe, T. 2000, Inf.\ Bull.\ var.\ Stars, 4873

\bibitem[{King(1988)}]{king88}
King, A.~R. 1988, QJRAS, 29, 1

\bibitem[{King(1985)}]{king85}
King, D.~L. 1985, Atmospheric Extinction at the Roque de los Muchachos
  Observatory, La Palma, RGO/La Palma Technical Note~31

\bibitem[{Knigge {et~al.}(2000)Knigge, Long, Hoard, Szkody, \&
  Dhillon}]{knigge00}
Knigge, C., Long, K.~S., Hoard, D.~W., Szkody, P., \& Dhillon, V.~S. 2000, ApJ,
  539, L49

\bibitem[{{Koester} \& {Sch\"{o}nberner}(1986)}]{koester86}
{Koester}, D. \& {Sch\"{o}nberner}, D. 1986, A\&A, 154, 125

\bibitem[{Kolb(1993)}]{kolb93}
Kolb, U. 1993, A\&A, 271, 149

\bibitem[{Kolb \& Baraffe(1999)}]{kolb99}
Kolb, U. \& Baraffe, I. 1999, MNRAS, 309, 1034

\bibitem[{Kraft(1959)}]{kraft59}
Kraft, R.~P. 1959, ApJ, 130, 110

\bibitem[{Kraft(1962)}]{kraft62}
---. 1962, ApJ, 135, 408

\bibitem[{Kristensen(1998)}]{kristensen98}
Kristensen, L.~K. 1998, Astron.~Nachr., 3, 193

\bibitem[{Kruszewski(1966)}]{kruszewski66}
Kruszewski, A. 1966, Adv. Astr. Astrophys., 4, 233

\bibitem[{Lampton {et~al.}(1976)Lampton, Margon, \& Bowyer}]{lampton76}
Lampton, M., Margon, B., \& Bowyer, S. 1976, ApJ, 208, 177

\bibitem[{Landau \& Lifschitz(1958)}]{landau58}
Landau, L. \& Lifschitz, E. 1958, The Classical Theory of Fields (Oxford:
  Pergamon)

\bibitem[{Lasota(2001)}]{lasota01}
Lasota, J.~P. 2001, New~Ast.~Rev., 45, 449

\bibitem[{Littlefair(2001)}]{littlefair02}
Littlefair, S.~P. 2001, PhD thesis, University of Sheffield

\bibitem[{Littlefair {et~al.}(2003)Littlefair, Dhillon, \&
  Mart{\'{\i}}n}]{littlefair03}
Littlefair, S.~P., Dhillon, V.~S., \& Mart{\'{\i}}n, E.~L. 2003, MNRAS, 340,
  264

\bibitem[{Livio \& Pringle(1994)}]{livio94}
Livio, M. \& Pringle, J.~E. 1994, ApJ, 427, 956

\bibitem[{Long \& Gilliland(1999)}]{long99}
Long, K.~S. \& Gilliland, R.~L. 1999, ApJ, 511, 916

\bibitem[{Lubow \& Shu(1975)}]{lubow75}
Lubow, S.~H. \& Shu, F.~H. 1975, ApJ, 198, 383

\bibitem[{Lubow \& Shu(1976)}]{lubow76}
---. 1976, ApJ, 207, L53

\bibitem[{Lucy(1967)}]{lucy67}
Lucy, L.~B. 1967, Z. Astrophysik, 65, 89

\bibitem[{Lynden-Bell \& Pringle(1974)}]{lyndenbell74}
Lynden-Bell, D. \& Pringle, J.~E. 1974, MNRAS, 168, 603

\bibitem[{Marsh(1990)}]{marsh90c}
Marsh, T.~R. 1990, ApJ, 357, 621

\bibitem[{Marsh {et~al.}(1987)Marsh, Horne, \& Shipman}]{marsh87}
Marsh, T.~R., Horne, K., \& Shipman, H.~L. 1987, MNRAS, 225, 551

\bibitem[{Mason {et~al.}(2002)Mason, Howell, Szkody, Harrison, Holtzman, \&
  Hoard}]{mason02}
Mason, E., Howell, S.~B., Szkody, P., Harrison, T.~E., Holtzman, J.~A., \&
  Hoard, D.~W. 2002, A\&A, 396, 633

\bibitem[{McClintock {et~al.}(1983)McClintock, Petro, Remillard, \&
  Ricker}]{mcclintock83}
McClintock, J.~E., Petro, L.~D., Remillard, R.~A., \& Ricker, G.~R. 1983, ApJ,
  266, L27

\bibitem[{Mennickent {et~al.}(2004)Mennickent, Diaz, \& Tappert}]{mennickent04}
Mennickent, R.~E., Diaz, M.~P., \& Tappert, C. 2004, MNRAS, 347, 1180

\bibitem[{Mineshige \& Wood(1989)}]{mineshige89}
Mineshige, S. \& Wood, J.~H. 1989, MNRAS, 241, 259

\bibitem[{Moffat(1969)}]{moffat69}
Moffat, A. F.~J. 1969, A\&A, 3, 455

\bibitem[{Morales-Rueda(2004)}]{moralesrueda04}
Morales-Rueda, L. 2004, Astron.~Nachr., 325, 193

\bibitem[{Mukai {et~al.}(1990)Mukai, Mason, Howell, Allington-Smith, Callanan,
  Charles, Hassall, Machin, Naylor, Smale, \& van Paradijs}]{mukai90}
Mukai, K., Mason, K.~O., Howell, S.~B., Allington-Smith, J., Callanan, P.~J.,
  Charles, P.~A., Hassall, B. J.~M., Machin, G., Naylor, T., Smale, A.~P., \&
  van Paradijs, J. 1990, MNRAS, 245, 385

\bibitem[{Nauenberg(1972)}]{nauenberg72}
Nauenberg, M. 1972, ApJ, 175, 417

\bibitem[{Naylor(1998)}]{naylor98}
Naylor, T. 1998, MNRAS, 296, 339

\bibitem[{Naylor {et~al.}(2005)Naylor, Allan, \& Long}]{naylor05}
Naylor, T., Allan, A., \& Long, K.~S. 2005, MNRAS, in press

\bibitem[{Neustroev {et~al.}(2002)Neustroev, Borisov, Barwig, Bobinger, Mantel,
  {\u{S}}imi{\'{c}}, \& Wolf}]{neustroev02a}
Neustroev, V.~V., Borisov, N.~V., Barwig, H., Bobinger, A., Mantel, K.~H.,
  {\u{S}}imi{\'{c}}, D., \& Wolf, S. 2002, A\&A, 393, 239

\bibitem[{Nogami {et~al.}(2001)Nogami, Kato, Baba, Nov\'{a}k, Lockley, \&
  Somers}]{nogami01}
Nogami, D., Kato, T., Baba, H., Nov\'{a}k, R., Lockley, J., \& Somers, M. 2001,
  MNRAS, 322, 79

\bibitem[{O'Donoghue(1990)}]{odonoghue90}
O'Donoghue, D. 1990, MNRAS, 246, 29

\bibitem[{Oke \& Gunn(1983)}]{oke83}
Oke, J.~B. \& Gunn, J.~E. 1983, ApJ, 266, 713

\bibitem[{Osaki(1974)}]{osaki74}
Osaki, Y. 1974, PASJ, 26, 429

\bibitem[{Paczy\'{n}ski(1965)}]{paczynski65}
Paczy\'{n}ski, B. 1965, Acta Astron., 15, 89

\bibitem[{Paczy\'{n}ski(1971)}]{paczynski71}
---. 1971, ARA\&A, 9, 183

\bibitem[{Paczy\'{n}ski(1977)}]{paczynski77}
---. 1977, ApJ, 216, 822

\bibitem[{Pantazis \& Niarchos(1998)}]{pantazis98}
Pantazis, G. \& Niarchos, P.~G. 1998, A\&A, 335, 199

\bibitem[{Parkhurst(1897)}]{parkhurst1897}
Parkhurst, J.~A. 1897, Pop.~Astr., 5, 164

\bibitem[{Patterson(1981)}]{patterson81}
Patterson, J. 1981, ApJS, 45, 517

\bibitem[{Patterson(1998)}]{patterson98}
---. 1998, PASP, 110, 1132

\bibitem[{Patterson {et~al.}(2000)Patterson, Vanmunster, Skillman, Jensen,
  Stull, Martin, Cook, Kemp, \& Knigge}]{patterson00}
Patterson, J., Vanmunster, T., Skillman, D.~R., Jensen, L., Stull, J., Martin,
  B., Cook, L.~M., Kemp, J., \& Knigge, C. 2000, PASJ, 112, 1584

\bibitem[{Payne-Gaposchkin \& Gaposchkin(1938)}]{payne38}
Payne-Gaposchkin, C. \& Gaposchkin, S. 1938, in Variable Stars (Cambridge,
  Mass.: Harv.~Obs.~Mono.~No.~5)

\bibitem[{Penny \& Dickens(1986)}]{penny86}
Penny, A.~J. \& Dickens, R.~J. 1986, MNRAS, 220, 845

\bibitem[{Politano(2004)}]{politano04}
Politano, M. 2004, ApJ, 604, 817

\bibitem[{Press {et~al.}(1986)Press, Flannery, Teukolsky, \&
  Vetterling}]{press86}
Press, W.~H., Flannery, B.~P., Teukolsky, S.~A., \& Vetterling, W.~T. 1986,
  Numerical Recipes in Fortran (Cambridge: Cambridge University Press)

\bibitem[{Pringle(1985)}]{pringle85}
Pringle, J.~E. 1985, in Interacting Binary Stars, ed. J.~E. Pringle \& R.~A.
  Wade (Cambridge: Cambridge University Press), 1

\bibitem[{Pylyser \& Savonije(1988{\natexlab{a}})}]{pylyser88a}
Pylyser, R.~E. \& Savonije, G.~J. 1988{\natexlab{a}}, A\&A, 191, 57

\bibitem[{Pylyser \& Savonije(1988{\natexlab{b}})}]{pylyser88b}
---. 1988{\natexlab{b}}, A\&A, 208, 52

\bibitem[{Rappaport {et~al.}(1983)Rappaport, Verbunt, \& Joss}]{rappaport83}
Rappaport, S., Verbunt, F., \& Joss, P.~C. 1983, ApJ, 275, 713

\bibitem[{Richter \& Greiner(1995)}]{richter95}
Richter, G.~A. \& Greiner, J. 1995, in Proc.\ Abano-Padova Conf.\ on
  Cataclysmic Variables, ed. A.~Bianchini, M.~Della~Valle, \& M.~Orio, 205, 177

\bibitem[{Richter {et~al.}(1997)Richter, Kroll, Greiner, Wenzel, Luthardt, \&
  Schwarz}]{richter97}
Richter, G.~A., Kroll, P., Greiner, J., Wenzel, W., Luthardt, R., \& Schwarz,
  R. 1997, A\&A, 325, 994

\bibitem[{Ritter \& Kolb(1998)}]{ritter98}
Ritter, H. \& Kolb, U. 1998, A\&AS, 129, 83

\bibitem[{Robertson \& Honeycutt(1996)}]{robertson96}
Robertson, J.~W. \& Honeycutt, R.~K. 1996, AJ, 112, 2248

\bibitem[{Robinson(1973)}]{robinson73}
Robinson, E.~L. 1973, ApJ, 180, 121

\bibitem[{Robinson(1976)}]{robinson76}
---. 1976, ApJ, 203, 485

\bibitem[{Robinson {et~al.}(1981)Robinson, Barker, Cochran, Cochran, \&
  Nather}]{robinson81}
Robinson, E.~L., Barker, E.~S., Cochran, A.~L., Cochran, W.~D., \& Nather,
  R.~E. 1981, ApJ, 251, 611

\bibitem[{Russell(1945)}]{russell45}
Russell, H.~N. 1945, ApJ, 102, 1

\bibitem[{Rutten(1998)}]{rutten98}
Rutten, R. G.~M. 1998, A\&AS, 127, 581

\bibitem[{Rutten {et~al.}(1994)Rutten, Dhillon, Horne, \& Kuulkers}]{rutten94a}
Rutten, R. G.~M., Dhillon, V.~S., Horne, K., \& Kuulkers, E. 1994, A\&A, 283,
  441

\bibitem[{Rutten {et~al.}(1993)Rutten, Dhillon, Horne, Kuulkers, \& van
  Paradijs}]{rutten93}
Rutten, R. G.~M., Dhillon, V.~S., Horne, K., Kuulkers, E., \& van Paradijs, J.
  1993, Nat, 362, 518

\bibitem[{Rutten {et~al.}(1992)Rutten, van Paradijs, \& Tinbergen}]{rutten92b}
Rutten, R. G.~M., van Paradijs, J., \& Tinbergen, J. 1992, A\&A, 260, 213

\bibitem[{Sawada {et~al.}(1986{\natexlab{a}})Sawada, Matsuda, \&
  Hachisu}]{sawada86a}
Sawada, K., Matsuda, T., \& Hachisu, I. 1986{\natexlab{a}}, MNRAS, 219, 75

\bibitem[{Sawada {et~al.}(1986{\natexlab{b}})Sawada, Matsuda, \&
  Hachisu}]{sawada86b}
---. 1986{\natexlab{b}}, MNRAS, 221, 679

\bibitem[{Schneider \& Young(1980)}]{schneider80}
Schneider, D.~P. \& Young, P.~J. 1980, ApJ, 238, 946

\bibitem[{Schoembs \& Hartmann(1983)}]{schoembs83}
Schoembs, R. \& Hartmann, K. 1983, A\&A, 128, 37

\bibitem[{Schreiber {et~al.}(2003)Schreiber, Hameury, \& Lasota}]{schreiber03}
Schreiber, M.~R., Hameury, J.-M., \& Lasota, J.-P. 2003, MNRAS, 410, 239

\bibitem[{Shafter {et~al.}(2000)Shafter, Clark, Holland, \&
  Williams}]{shafter00}
Shafter, A.~W., Clark, L.~L., Holland, J., \& Williams, S.~J. 2000, PASP, 112,
  1467

\bibitem[{Shahbaz {et~al.}(1994)Shahbaz, Naylor, \& Charles}]{shahbaz94}
Shahbaz, T., Naylor, T., \& Charles, P.~A. 1994, MNRAS, 268, 756

\bibitem[{Shakura \& Sunyaev(1973)}]{shakura73}
Shakura, N.~I. \& Sunyaev, R.~A. 1973, A\&A, 24, 337

\bibitem[{Shapley \& Hughes(1934)}]{shapley34}
Shapley, H. \& Hughes, E.~M. 1934, Ann.~Harvard~Coll.~Obser., 90, 163

\bibitem[{Sherrington {et~al.}(1982)Sherrington, Jameson, Bailey, \&
  Giles}]{sherrington82}
Sherrington, M.~R., Jameson, R.~F., Bailey, J., \& Giles, A.~B. 1982, MNRAS,
  200, 861

\bibitem[{Sion {et~al.}(1998)Sion, Cheng, Szkody, Sparks, G{\"{a}}nsicke,
  Huang, \& Mattei}]{sion98}
Sion, E.~M., Cheng, F.~H., Szkody, P., Sparks, W., G{\"{a}}nsicke, B., Huang,
  M., \& Mattei, J. 1998, ApJ, 496, 449

\bibitem[{Sisan {et~al.}(2004)Sisan, Mujica, Tillotson, Huang, Dorland, Hassam,
  Antonsen, \& Lathrop}]{sisan04}
Sisan, D.~R., Mujica, N., Tillotson, W.~A., Huang, Y.-M., Dorland, W., Hassam,
  A.~B., Antonsen, T.~M., \& Lathrop, D.~P. 2004, Phys. Rev. Lett., 93, 114502

\bibitem[{Skidmore {et~al.}(2002)Skidmore, Wynn, Leach, \&
  Jameson}]{skidmore02}
Skidmore, W., Wynn, G.~A., Leach, R., \& Jameson, R.~F. 2002, MNRAS, 336, 1223

\bibitem[{Skilling \& Bryan(1984)}]{skilling84}
Skilling, J. \& Bryan, R.~K. 1984, MNRAS, 211, 111

\bibitem[{Smak(1971)}]{smak71}
Smak, J. 1971, Acta Astronomica, 21, 15

\bibitem[{Smith \& Dhillon(1998)}]{smith98a}
Smith, D.~A. \& Dhillon, V.~S. 1998, MNRAS, 301, 767

\bibitem[{Smith {et~al.}(2002)Smith, Tucker, \& et~al.}]{smith02}
Smith, J.~A., Tucker, D.~L., \& et~al. 2002, AJ, 123, 2121

\bibitem[{Steeghs(2004)}]{steeghs04}
Steeghs, D. 2004, Astron.~Nachr., 325, 185

\bibitem[{Steeghs {et~al.}(1997)Steeghs, Harlaftis, \& Horne}]{steeghs97}
Steeghs, D., Harlaftis, E.~T., \& Horne, K. 1997, MNRAS, 290, L28

\bibitem[{Steeghs {et~al.}(1998)Steeghs, Harlaftis, \& Horne}]{steeghs98}
---. 1998, MNRAS, 296, 463

\bibitem[{Steeghs {et~al.}(2003)Steeghs, Perryman, Reynolds, de~Bruijne, Marsh,
  Dhillon, \& Peacock}]{steeghs03}
Steeghs, D., Perryman, M. A.~C., Reynolds, A., de~Bruijne, J. H.~J., Marsh, T.,
  Dhillon, V.~S., \& Peacock, A. 2003, MNRAS, 339, 810

\bibitem[{Stetson(1987)}]{stetson87}
Stetson, P.~B. 1987, PASP, 99, 191

\bibitem[{Stevenson(2005)}]{stevenson05}
Stevenson, M.~J. 2005, PhD thesis, University of Sheffield

\bibitem[{Still {et~al.}(1998)Still, Buckley, \& Garlick}]{still98}
Still, M.~D., Buckley, D. A.~H., \& Garlick, M.~A. 1998, MNRAS, 299, 545

\bibitem[{Stover(1981)}]{stover81}
Stover, R.~J. 1981, ApJ, 249, 673

\bibitem[{Stover {et~al.}(1980)Stover, Robinson, Nather, \&
  Montemayor}]{stover80}
Stover, R.~J., Robinson, E.~L., Nather, R.~E., \& Montemayor, T.~J. 1980, ApJ,
  240, 597

\bibitem[{Szkody {et~al.}(2002)Szkody, Anderson, \& et~al.}]{szkody02}
Szkody, P., Anderson, S.~F., \& et~al. 2002, AJ, 123, 430

\bibitem[{Szkody {et~al.}(2003)Szkody, Fraser, \& et~al.}]{szkody03}
Szkody, P., Fraser, O., \& et~al. 2003, AJ, 126, 1499

\bibitem[{Szkody {et~al.}(2004)Szkody, Henden, \& et~al.}]{szkody04}
Szkody, P., Henden, A., \& et~al. 2004, AJ, 128, 1882

\bibitem[{Szkody \& Howell(1992)}]{szkody92}
Szkody, P. \& Howell, S.~B. 1992, ApJS, 78, 537

\bibitem[{Taylor \& Weisberg(1982)}]{taylor82}
Taylor, J.~H. \& Weisberg, J.~M. 1982, ApJ, 253, 908

\bibitem[{Thoroughgood(2005)}]{thoroughgood05a}
Thoroughgood, T.~D. 2005, PhD thesis, University of Sheffield

\bibitem[{Thoroughgood {et~al.}(2001)Thoroughgood, Dhillon, Littlefair, Marsh,
  \& Smith}]{thoroughgood01}
Thoroughgood, T.~D., Dhillon, V.~S., Littlefair, S.~P., Marsh, T.~R., \& Smith,
  D.~A. 2001, MNRAS, 327, 1323

\bibitem[{Thoroughgood {et~al.}(2005)Thoroughgood, Dhillon, Steeghs, Watson,
  Buckley, Littlefair, Smith, Still, van~der Heyden, \&
  Warner}]{thoroughgood05}
Thoroughgood, T.~D., Dhillon, V.~S., Steeghs, D., Watson, C.~A., Buckley, D.
  A.~H., Littlefair, S.~P., Smith, D.~A., Still, M., van~der Heyden, K.~J., \&
  Warner, B. 2005, MNRAS, 357, 881

\bibitem[{Thoroughgood {et~al.}(2004)Thoroughgood, Dhillon, Watson, Buckley,
  Steeghs, \& Stevenson}]{thoroughgood04}
Thoroughgood, T.~D., Dhillon, V.~S., Watson, C.~A., Buckley, D. A.~H., Steeghs,
  D., \& Stevenson, M.~J. 2004, MNRAS, 353, 1135

\bibitem[{Thorstensen(2000)}]{thorstensen00}
Thorstensen, J.~R. 2000, PASP, 112, 1269

\bibitem[{Truss {et~al.}(2004)Truss, Wynn, \& Wheatley}]{truss04}
Truss, M.~R., Wynn, G.~A., \& Wheatley, P.~J. 2004, MNRAS, 347, 569

\bibitem[{Uemura {et~al.}(2004)Uemura, Kato, Ishioka, Bolt, Cook, Monard,
  Stubbings, Torii, Kiyota, Nogami, Tanabe, Starkey, \& Miyashita}]{uemura04}
Uemura, M., Kato, T., Ishioka, R., Bolt, G., Cook, L., Monard, B., Stubbings,
  R., Torii, K., Kiyota, S., Nogami, D., Tanabe, K., Starkey, D., \& Miyashita,
  A. 2004, PASJ, 56, S141

\bibitem[{Unda-Sanzana(2005)}]{undasanzana05}
Unda-Sanzana, E. 2005, PhD thesis, University of Southampton

\bibitem[{Vande~Putte {et~al.}(2003)Vande~Putte, Smith, Hawkins, \&
  Martin}]{vandeputte03}
Vande~Putte, D., Smith, R.~C., Hawkins, N.~A., \& Martin, J.~S. 2003, MNRAS,
  342, 151

\bibitem[{Vanmunster(2000)}]{vanmunster4210}
Vanmunster, T. 2000, vsnet-alert, No. 4210

\bibitem[{Vanmunster {et~al.}(2000)Vanmunster, Velthuis, \&
  McCormick}]{vanmunster00}
Vanmunster, T., Velthuis, F., \& McCormick, J. 2000, Inf.\ Bull.\ var.\ Stars,
  4955

\bibitem[{Verbunt(1984)}]{verbunt84}
Verbunt, F. 1984, MNRAS, 209, 227

\bibitem[{Verbunt \& Rappaport(1988)}]{verbunt88}
Verbunt, F. \& Rappaport, S. 1988, ApJ, 332, 193

\bibitem[{Voges {et~al.}(1999)Voges, Aschenbach, Boller, Br{\"{a}}uninger,
  Briel, Burkert, Dennerl, Englhauser, Gruber, Haberl, Hartner, Hasinger,
  K{\"{u}}rster, Pfeffermann, Pietsch, Predehl, Rosso, Schmitt, Tr{\"{u}}mper,
  \& U.}]{voges99}
Voges, W., Aschenbach, B., Boller, T., Br{\"{a}}uninger, H., Briel, U.,
  Burkert, W., Dennerl, K., Englhauser, J., Gruber, R., Haberl, F., Hartner,
  G., Hasinger, G., K{\"{u}}rster, M., Pfeffermann, E., Pietsch, W., Predehl,
  P., Rosso, C., Schmitt, J.~H.~M.~M., Tr{\"{u}}mper, J., \& U., Z.~H. 1999,
  A\&A, 349, 389

\bibitem[{Vogt(1982)}]{vogt82}
Vogt, N. 1982, ApJ, 252, 653

\bibitem[{Vrielmann {et~al.}(2002)Vrielmann, Hessman, \& Horne}]{vrielmann02}
Vrielmann, S., Hessman, F.~V., \& Horne, K. 2002, MNRAS, 332, 176

\bibitem[{Warner(1995)}]{warner95}
Warner, B. 1995, Cataclysmic Variable Stars (Cambridge: Cambridge University
  Press)

\bibitem[{Warner \& Nather(1971)}]{warner71}
Warner, B. \& Nather, R.~E. 1971, MNRAS, 152, 219

\bibitem[{Warner \& Peters(1972)}]{warner72}
Warner, B. \& Peters, W.~L. 1972, MNRAS, 160, 15

\bibitem[{Watson(2002)}]{watson02a}
Watson, C.~A. 2002, PhD thesis, University of Sheffield

\bibitem[{Watson {et~al.}(2003)Watson, Dhillon, Rutten, \& Schwope}]{watson03}
Watson, C.~A., Dhillon, V.~S., Rutten, R. G.~M., \& Schwope, A.~D. 2003, MNRAS,
  341, 129

\bibitem[{Watts {et~al.}(1986)Watts, Bailey, Hill, Greenhill, McCowage, \&
  Carty}]{watts86}
Watts, D.~J., Bailey, J., Hill, P.~W., Greenhill, J.~G., McCowage, C., \&
  Carty, T. 1986, A\&A, 154, 197

\bibitem[{Welsh \& Wood(1995)}]{welsh95}
Welsh, W.~F. \& Wood, J.~H. 1995, in Flares and Flashes, Proc. IAU Coll. 151.,
  ed. J.~Greiner, H.~W. Deuerback, \& G.~R. E. (Berlin: Springer-Verlag Lecture
  Notes in Physics), 300

\bibitem[{Wenzel(1987)}]{wenzel87}
Wenzel, W. 1987, 308, 75

\bibitem[{Wenzel {et~al.}(1995)Wenzel, Richter, \& Luthardt}]{wenzel95}
Wenzel, W., Richter, G.~A., \& Luthardt, R. 1995, Inf.\ Bull.\ var.\ Stars,
  4182

\bibitem[{Whitehurst(1988)}]{whitehurst88}
Whitehurst, R. 1988, MNRAS, 232, 35

\bibitem[{Whitehurst \& King(1991)}]{whitehurst91}
Whitehurst, R. \& King, A. 1991, MNRAS, 249, 25

\bibitem[{Wood \& Crawford(1986)}]{wood86a}
Wood, J.~H. \& Crawford, C.~S. 1986, MNRAS, 222, 645

\bibitem[{Wood \& Horne(1990)}]{wood90}
Wood, J.~H. \& Horne, K. 1990, MNRAS, 242, 606

\bibitem[{Wood {et~al.}(1989{\natexlab{a}})Wood, Horne, Berriman, \&
  Wade}]{wood89a}
Wood, J.~H., Horne, K., Berriman, G., \& Wade. 1989{\natexlab{a}}, ApJ, 314,
  974

\bibitem[{Wood {et~al.}(1986)Wood, Horne, Berriman, Wade, O'Donoghue, \&
  Warner}]{wood86b}
Wood, J.~H., Horne, K., Berriman, G., Wade, R., O'Donoghue, D., \& Warner, B.
  1986, MNRAS, 219, 629

\bibitem[{Wood {et~al.}(1992)Wood, Horne, \& Vennes}]{wood92}
Wood, J.~H., Horne, K., \& Vennes, S. 1992, ApJ, 385, 294

\bibitem[{Wood {et~al.}(1985)Wood, Irwin, \& Pringle}]{wood85}
Wood, J.~H., Irwin, M.~J., \& Pringle, J.~E. 1985, MNRAS, 214, 475

\bibitem[{Wood {et~al.}(1989{\natexlab{b}})Wood, Marsh, Robinson, Stiening,
  Horne, Stover, Schoembs, Allen, Bond, Jones, Grauer, \& Ciardullo}]{wood89b}
Wood, J.~H., Marsh, T.~R., Robinson, E.~L., Stiening, R.~F., Horne, K., Stover,
  R.~J., Schoembs, R., Allen, S.~L., Bond, H.~E., Jones, D. H.~P., Grauer,
  A.~D., \& Ciardullo, R. 1989{\natexlab{b}}, MNRAS, 239, 809

\bibitem[{Wood {et~al.}(1995)Wood, Naylor, Hassall, \& Ramseyer}]{wood95a}
Wood, J.~H., Naylor, T., Hassall, B.~J.~M., \& Ramseyer, T.~F. 1995, MNRAS,
  273, 772

\bibitem[{Wood(1995)}]{wood95}
Wood, M.~A. 1995, in European Workshop on White Dwarfs,, ed. D.~Koester \&
  K.~Werner (Volume 443: Lecture Notes in Physics), 41

\bibitem[{Woudt \& Warner(2001)}]{woudt01}
Woudt, P.~A. \& Warner, B. 2001, MNRAS, 328, 159

\bibitem[{Zhang {et~al.}(1986)Zhang, Robinson, \& Nather}]{zhang86}
Zhang, E.-H., Robinson, E.~L., \& Nather, R.~E. 1986, ApJ, 305, 740

\end{thebibliography}
